\documentclass[a4paper, 11pt]{article}

\usepackage[english]{babel}

\usepackage[margin=1in]{geometry}

\usepackage{amsmath}
\usepackage{amsfonts}
\usepackage{bm}
\usepackage{graphicx}
\usepackage[hidelinks]{hyperref}
\usepackage[authoryear,round]{natbib}
\usepackage{float}
\usepackage{booktabs}
\usepackage{longtable}
\usepackage{array}
\usepackage{enumitem}
\usepackage{xcolor}
\usepackage{rotating}
\usepackage[nodisplayskipstretch]{setspace}
\usepackage[bottom,hang,flushmargin]{footmisc}

\usepackage[ruled,vlined,linesnumbered]{algorithm2e}
\makeatletter
\newcommand{\nosemic}{\renewcommand{\@endalgocfline}{\relax}}
\newcommand{\dosemic}{\renewcommand{\@endalgocfline}{\algocf@endline}}
\newcommand{\pushline}{\Indp}
\newcommand{\popline}{\Indm\dosemic}
\let\oldnl\nl
\newcommand{\nonl}{\renewcommand{\nl}{\let\nl\oldnl}}
\makeatother

\begin{document}
\title{Multivariate GARCH and portfolio variance prediction: a forecast reconciliation perspective}
\bigskip
\bigskip

\author{Massimiliano Caporin\thanks{Corresponding author: Department of Statistical Sciences, University of Padova, Via C. Battisti 241, 35121 Padova, Italy - email: massimiliano.caporin@unipd.it - phone: +39-049-827-4199.\\
We thank Eduardo Rossi, Juan-Angel Jimenez-Martin, Alfonso Novales, Esther Ruiz, Helena Veiga, Paolo Santucci de Magistris, and the participants to the Italian Conference of Economic Statistics (SMEA) 2025, the Computational and Financial Econometrics conference 2025, and the seminars at Univrsidad Carlos III Madrid and Universidad Complutense de Madrid for the suggestions, comments and stimulating discussions. The authors acknowledge financial support from project PRIN2022 PRICE: A New Paradigm for High-Frequency Finance, 2022C799SX. This study was funded by the European Union-NextGeneration EU, Mission 4, Component 2, in the framework of the GRINS-Growing Resilient, INclusive and Sustainable project [grant numbers GRINS PE00000018-CUP C93C22005270001].}\\
Department of Statistical Sciences\\
University of Padova, Italy\\
massimiliano.caporin@unipd.it\\
and\\
Daniele Girolimetto\\
Department of Statistical Sciences\\
University of Padova, Italy\\
daniele.girolimetto@unipd.it\\
and\\
Emanuele Lopetuso\\
Department of Statistical Sciences\\
University of Padova, Italy\\
emanuele.lopetuso@unipd.it}
\maketitle

\begin{abstract} 
\singlespacing
We assess the advantage of combining univariate and multivariate portfolio risk forecasts with the aid of forecast reconciliation techniques. In our analyzes, we assume knowledge of portfolio weights, a standard for portfolio risk management applications. With an extensive simulation experiment, we show that, if the true covariance is known, forecast reconciliation improves over a standard multivariate approach, in particular when the adopted multivariate model is misspecified. However, if noisy proxies are used, correctly specified models and the misspecified ones (for instance, neglecting spillovers) turn out to be, in several cases, indistinguishable, with forecast reconciliation still providing improvements, even though smaller. The noise in the covariance proxy plays a crucial role both in driving the improvement of forecast reconciliation and in the identification of the optimal model. An empirical analysis shows how forecast reconciliation can be adopted with real data to improve traditional GARCH-based portfolio variance forecasts.\\


\end{abstract}
\bigskip
\doublespacing

\section{Introduction}
	
Multivariate conditional heteroskedastic models belonging to the Generalized Auto Regressive Conditional Heteroskedasticity (MGARCH) class are standard tools in financial econometrics. They represent classic instruments for a range of applications, from risk management to hedging and asset allocation; see surveys by \cite{bauwensetal2005}, \cite{SilvennoinenTerasvirta2009}, and \cite{FrancqZakoian2019}. Among the several specifications proposed by the literature (see the surveys previously cited for details), only a few are frequently adopted, and these include the Dynamic Conditional Correlation (DCC) model of \cite{Engle2002}, the Orthogonal GARCH (OGARCH) model of \cite{alexander2002}, and the Scalar BEKK of \cite{DingEngle2001}, which are feasible even in large dimensional cases. A relevant challenge for MGARCH models is, in fact, related to the so-called \textit{curse of dimensionality}, that is, the large number of parameters of the most flexible models, which is sensibly limiting their use. For instance, the BEKK model of \cite{englekroner1995} is commonly considered only under strong parametric restrictions, such as in the just mentioned Scalar-BEKK case, while the VECH specification (see again \cite{englekroner1995}) has received limited attention in empirical analyzes. However, restrictions imply a reduction in model flexibility, which generally corresponds to a limit in the interdependence across assets shocks and/or covariances (or correlations). This is in contrast with the empirical evidence suggesting that interdependence is present and relevant, as demonstrated by the various works re-introducing limited flexibility in MGARCH specifications; see, among many others, \cite{CaporinParuolo2015} and \cite{billioetal2023}. From a different perspective, the existence of interdependence is also at the base of the growing literature on systemic risk and financial connectedness, originating from the seminal contribution of \cite{DieboldYilmaz2009}; see also \citep{DieboldYilmaz2023} and references therein cited. 

An aspect not yet fully explored by the literature is the possibility of indirectly taking into account the interdependence in a MGARCH framework and of exploiting its potential in forecasting, thus paving the way for possible applications either in risk management and portfolio allocation. Our intuition takes the point of view of a risk manager that has to evaluate the risk (i.e., to predict the variance) of an existing portfolio. By construction, the weights of the assets are known, and if sufficient historical data are available for the returns or all the assets included in the portfolio, backward simulated, synthetic, portfolio returns can be derived. The portfolio returns are then accounting for the interdependence across the assets risk (either in terms of shocks or variance spillovers), and this will be captured implicitly by any univariate GARCH-type model we might fit on the portfolio returns. Of course, a univariate model might represent a too restrictive specification, not taking properly into account the heterogeneity in the dynamic of assets variances, covariances and/or correlations, something that could be provided only by a MGARCH model. Fortunately, the forecasting literature includes tools for combining a fit of a univariate model in a synthetic portfolio returns series, with a MGARCH model adapted to the collection of portfolio assets returns; we refer to forecast reconciliation approaches \citep{Hyndman2011, Wickramasuriya2019, Panagiotelis2021, DifonzoGirolimetto2022, Girolimetto2024, Girolimettoetal2023, Athanasopoulos2024}.

In this work, we show that the prediction of the variance of a univariate GARCH model on portfolio returns can be efficiently combined with the prediction of the covariance of an MGARCH specification to improve the prediction of a given portfolio risk. We thus contribute both to the literature on forecast reconciliation, extending the application of its tools in finance (see \citealp{Caporin2024} and \cite{Matteraetal2024}), and to the MGARCH literature, showing how assets interdependence can be accounted for. 
For what concerns the forecast reconciliation contribution, we design a reconciliation approach tailored to the portfolio-variance setting, adapting shrinkage-based methods to the aggregation constraints implied by portfolio weights. Our procedure ensures coherence and preserves the structure of the covariance matrix, making the reconciled forecasts more accurate and economically interpretable.
 Using simulations, we show how the combination of univariate and multivariate forecasts improves prediction accuracy, in line with the findings of \cite{Wickramasuriya2019} and \cite{Panagiotelis2021}. Specifically, when focusing on the financial econometrics side, in addition to the introduction of interdependence by means of a univariate fit, we provide further contributions. First, we show how the predictive performance of MGARCH models might deteriorate under misspecification. Second, and more relevant, we demonstrate that the use of noisy proxies masks the possible presence of interdependence, making MGARCH models neglecting interdependence equivalent to the correctly specified ones.

The paper proceeds as follows. Section 2 is devoted to the methods that show how forecast reconciliation can be used for portfolio variances, describing both the MGARCH models and the forecast comparison tools that we will consider. Section 3 reports the simulation design and the results, while Section 4 includes an empirical example with real data. Section 5 concludes the paper. An extensive Online Appendix includes detailed results for the simulations and the empirical study.

\section{Methods}

\subsection{Forecast reconciliation for portfolio variance} \label{sec:recop}

As we noted in the introduction, the portfolio variance can be forecast either through a univariate approach applied directly to portfolio returns or via a bottom-up strategy leveraging a multivariate model of asset returns. This dual structure provides an ideal setting for the application of forecast reconciliation techniques.

We first define the framework we refer to. Assume that the interest is in the prediction of the portfolio variance for the next period, a common need in financial risk management. If this is the case, the portfolio composition is assumed to be known. Therefore, in a setting where $N$ assets are available, we denote by $\bm{\omega}$ the (known) $N-$dimensional vector of portfolio weights. We also assume that the returns\footnote{For simplicity, we assume that the returns are de-meaned.} of the $N$ assets are available over a common sample, that is, the $N$-dimensional vectors $\boldsymbol r_t \in \mathbb{R}^N$ are observed for $t=1,2,\ldots, T$.

The goal is to produce a forecast of the portfolio variance for $T+1$. Two approaches are commonly pursued. In the first, one computes the portfolio returns $r_{p,t} = \bm{\omega}^{\prime}\boldsymbol r_t$ and fits a univariate GARCH-type model to obtain the one-step-ahead forecast $\sigma_{p,T+1}^2$ of the portfolio variance.\footnote{For simplicity, we do not denote predicted quantities with hats or conditioning on the available information set, unless it is needed for clarity of the discussion.} In the second approach, a multivariate GARCH (MGARCH) model is fitted to the asset return vector $\boldsymbol r_t$, providing a forecast $\bm{\Sigma}_{T+1}$ of the conditional covariance matrix. Then, we recover the predicted portfolio variance by aggregating as $\gamma_{p,T+1}^2=\bm{\omega}^{\prime} \bm{\Sigma}_{T+1}\bm{\omega}$.

The presence of a reference forecast recovered from a univariate approach and of a second forecast obtained by aggregation of a collection of elements (the predictions of asset variances and covariances) represents a standard situation in forecast reconciliation. Similar settings appear in non-financial applications such as hierarchical forecasting in demography \citep{Shang2017, Li2019}, macroeconomic \citep{Petropoulos2014, Mircetic2022} energy \citep{Silva2018, Wang2021, Abolghasemi2025}. In the financial literature, notable examples are  \cite{Caporin2024} and \cite{Matteraetal2024}. To simplify notation, all forecast quantities are implicitly assumed to refer to the next period, and the time subscript $T+1$ is omitted throughout the discussion of this subsection: $\sigma_{p,T+1}^2 \equiv \widehat{\sigma}_{p}^2$ and $\gamma_{p,T+1}^2 \equiv \gamma_{p}^2$, $\bm{\Sigma}_{T+1} \equiv\widehat{\bm{\Sigma}}$.

Moving back to our setting, the prediction of portfolio variance using a univariate approach and of the covariance matrix, $\widehat{\sigma}_{p}^2$ and $\widehat{\bm{\Sigma}}$, represent the so-called \textit{base} forecasts. The construction of the portfolio variance prediction by aggregation gives the \textit{bottom-up} alternative $\gamma_{p}^2$, exploiting the informative content of the risk predictions coming from the single components of the portfolio and accounting for their interdependence (measured by both the correlations and the links between variances and covariances). In general, the base forecast and the bottom-up forecast do not coincide, that is, $\widehat{\sigma}_{p}^2 \neq \gamma_{p}^2$.

To connect our framework with the results of the forecast reconciliation literature, we express the bottom-up forecast as a linear combination of the univariate underlying elements (variances and covariances) of $\widehat{\bm{\Sigma}}$:
\begin{equation}
	\label{eq:gamma}
	\gamma_{p}^2 = \bm{\omega}^{\prime} \widehat{\bm{\Sigma}} \bm{\omega} = \left( \bm{\omega}^{\prime} \otimes \bm{\omega}^{\prime} \right) \bm{D}_N \mathit{vech}(\widehat{\bm{\Sigma}}) = \bm{a}' \widehat{\bm{\sigma}},
\end{equation}
where $\mathit{vech}(\cdot)$ denotes the half-vectorization operator, 
$\bm{D}_N$ is the duplication matrix,\footnote{The duplication matrix satisfies $\mathit{vec}\left(\widehat{\bm{\Sigma}}\right)=\bm{D}_N \mathit{vech}\left(\widehat{\bm{\Sigma}}\right)$ with $\mathit{vec}$ being the matrix vectorization operator and has size $\left(N^2 \times \frac{N(N+1)}{2}\right)$.} and $\widehat{\bm{\sigma}}$ is the vector of dimension $m = \frac{N(N+1)}{2}$ collecting the different elements of $\widehat{\bm{\Sigma}}$. The $(m \times 1)$ vector $\bm{a} = \left[(\bm{\omega}^{\prime} \otimes \bm{\omega}^{\prime}) \bm{D}_N\right]^{\prime} = \bm{D}^{\prime}_N (\bm{\omega} \otimes \bm{\omega}) $ is also called the \textit{aggregation} \citep[or linear combination,][]{Girolimetto2024} vector that is equivalent to the aggregation matrix in the reconciliation literature \citep{Hyndman2011}.
Then, the portfolio variance forecasts obtained from the univariate GARCH-type approach and from the MGARCH model can be reconciled by solving the following generalized least squares problem:
\begin{equation}
	\label{eq:ott}
	\operatorname{argmin}_{\bm{y}} \left(\bm{y} - \widehat{\bm{y}} \right)^{\prime} \bm{\Omega}^{-1} \left( \bm{y} - \widehat{\bm{y}} \right) \quad \text{subject to} \quad \bm{c}^{\prime} \bm{y} = \bm{0},
\end{equation}
with the solution given by \citep{Stone1942, Byron1978, Byron1979}:
$$
\widetilde{\bm{y}} = \left[\bm{I}_{m+1} - \bm{\Omega} \bm{c} (\bm{c}^{\prime} \bm{\Omega} \bm{c})^{-1} \bm{c}^{\prime} \right] \widehat{\bm{y}},
$$
where
$$
\bm{y} = \begin{bmatrix} \sigma_{p}^2 \\ \bm{\sigma} \end{bmatrix}, \qquad
\widehat{\bm{y}} = \begin{bmatrix} \widehat{\sigma}_{p}^2 \\ \widehat{\bm{\sigma}} \end{bmatrix}, \qquad
\widetilde{\bm{y}} = \begin{bmatrix} \widetilde{\sigma}_{p}^2 \\ \widetilde{\bm{\sigma}} \end{bmatrix}, \qquad \text{and} \qquad
\bm{c} = \begin{bmatrix} 1 \\ -\bm{a} \end{bmatrix},
$$
are, respectively, the $[(m+1) \times 1]$ stacked variance vector, the base forecast vector collecting the incoherent variance forecasts from both the univariate and multivariate models, the reconciled forecast vector, and the constraints vector. In addition, $\bm{\Omega}$ is a $[(m+1) \times (m+1)]$ positive definite matrix representing the covariance matrix of the (time series of) forecast errors of both the bottom-level components (i.e., the elements of vector $\widehat{\bm{\sigma}}$) and the portfolio series (i.e., $\widehat{\sigma}_{p}^2$).

Several approaches have been proposed in the literature to estimate $\bm{\Omega}$, see  \cite{Athanasopoulos2024}. However, in this work, we consider the state-of-the-art shrinkage estimators proposed by \cite{Wickramasuriya2019}, based on the in-sample residuals of the individual forecasting models \citep[see, e.g.,][]{Hyndman2016, Panagiotelis2021, Caporin2024}. 

\subsubsection{Reconciliation procedures and correlation matrix coherence}\label{sec:optionAlg} 

After reconstructing the reconciled covariance matrix as $\widetilde{\bm{\Sigma}} = \mathit{vech}^{-1}(\widetilde{\bm{\sigma}})$, the corresponding correlation matrix $\widetilde{\bm{R}}$ may not satisfy the required properties of a proper correlation matrix. In particular, it may contain off-diagonal elements with absolute values that exceed one, that is, $|\widetilde{\rho}_{i,j}| > 1$, thus violating the mathematical definition of a correlation coefficient.
Therefore, we propose two novel reconciliation strategies designed to specifically address this issue:

\begin{enumerate}[label=\textbf{(\Alph*)}]
	\item \textbf{Non-linear constrained optimization.}\\ Non-linear inequality constraints are introduced in the optimization problem (\ref{eq:ott}) such that
	\begin{equation}\label{eq:nlott}
		\begin{aligned}
			\operatorname{argmin}_{\bm{y}} &\left(\bm{y} - \widehat{\bm{y}} \right)^{\prime} \bm{\Omega}^{-1} \left( \bm{y} - \widehat{\bm{y}} \right)\\
			&\mathit{s.t.} \quad  \bm{c}'\bm{y} = \bm{0} \quad \text{and} \quad \left\{\begin{aligned}
				g(\bm{y})_{i\neq j} &\leq 1 \\[-0.25\baselineskip] g(\bm{y})_{i=i} &= 1,
			\end{aligned}\right.\quad \text{with} \quad 	i,j = 1,...,N
		\end{aligned}
	\end{equation}
	where $g(\bm{y})$ is a function that extracts the correlation matrix from the covariance matrix $\bm{\Sigma} = \operatorname{vech}^{-1}(\bm{\sigma})$ implied by $\bm{y}$ and returns the absolute values of its elements in vectorized form:
	$$
	g(\bm{y}) = \left| vech\left( cor \left(vech^{-1}(\bm{\sigma}) \right) \right) \right|.
	$$
	This non-linear problem might be solved, for instance, using the  \textsf{R} package \texttt{Rsolnp} \citep{Rsolnp}. 
	
	\item \textbf{Reconciliation via correlation decomposition.} \\
	Starting from equation~\eqref{eq:gamma}, we decompose the \textit{bottom-up} portfolio variance exploiting the correlation matrix:
	\begin{equation}\label{eq:Ai}
		\gamma_{p}^2 =  \bm{\omega}^{\prime} \widehat{\bm{\Sigma}} \bm{\omega} 
		=  \bm{\omega}^{\prime} \widehat{\bm{S}} \widehat{\bm{R}} \widehat{\bm{S}} \bm{\omega} 
		= \left(\bm{\omega}^{\prime} \otimes \bm{\omega}^{\prime}\right) \left( \widehat{\bm{S}} \otimes \widehat{\bm{S}} \right)\bm{D}_N \, \mathit{vech}(\widehat{\bm{R}}) 
		= \widehat{\bm{a}}_{\bm{\sigma}}^{\prime} \widehat{\boldsymbol\rho},
	\end{equation}
	where $\widehat{\bm{\Sigma}} = \widehat{\bm{S}} \widehat{\bm{R}} \widehat{\bm{S}}$ is the forecast conditional covariance matrix, $\widehat{\bm{S}} = \mathrm{diag}(\widehat{\sigma}_1,\widehat{\sigma}_2,\ldots,\widehat{\sigma}_N)$ is the diagonal matrix of forecast conditional standard deviations of the $N$ assets, 
	and $\widehat{\boldsymbol\rho} = \mathit{vech}(\widehat{\bm{R}})$ is the vector of forecast correlations. 
	Note that, unlike the aggregation vector $\bm{a}$ in equation~\eqref{eq:gamma}, the vector $\widehat{\bm{a}}_{\bm{\sigma}} = \bm{D}_N^{\prime} \left( \widehat{\bm{S}} \otimes \widehat{\bm{S}} \right) (\bm{\omega} \otimes \bm{\omega})$ is not fixed, as it depends on the forecast standard deviations through $\widehat{\bm{S}}$.
	
	Thus, once the reconciled vector $\widetilde{\bm{\sigma}}$ is obtained from equation~\eqref{eq:ott}, we can reconstruct the reconciled covariance matrix $\widetilde{\bm{\Sigma}} = \mathit{vech}^{-1}\left(\widetilde{\bm{\sigma}}\right)$ and compute:
	\begin{equation*}
		\widetilde{\bm{S}} = \mathrm{diag}\!\left(\sqrt{\mathrm{diag}(\widetilde{\bm{\Sigma}})}\right), \quad
		\widetilde{\bm{a}}_{\bm{\sigma}} = \bm{D}_N^{\prime} \left(\widetilde{\bm{S}} \otimes \widetilde{\bm{S}} \right) (\bm{\omega} \otimes \bm{\omega}), \quad \text{and} \quad
		\widetilde{\bm{c}}_{\bm{\sigma}} = \begin{bmatrix} 1 \\ -\widetilde{\bm{a}}_{\bm{\sigma}} \end{bmatrix}.
	\end{equation*}
	
	Reconciliation is then performed directly on correlations:
	\begin{equation}\label{eq:ottcor}
		\begin{aligned}
			\operatorname{argmin}_{\bm{x}} &\left(\bm{x} - \widehat{\bm{x}}\right)^{\prime} \bm{W}^{-1} \left(\bm{x} - \widehat{\bm{x}}\right) \\
			&\mathit{s.t.} \quad  \widetilde{\bm{c}}_{\bm{\sigma}}^{\;\prime} \bm{x} = \bm{0} \quad \text{and}\quad
			\left\{\begin{aligned}
				-1 \leq  {\boldsymbol{\rho}}_{i\neq j} &\leq 1 \\[-0.25\baselineskip]  {\boldsymbol{\rho}}_{i= j} &= 1
			\end{aligned}\right. \quad\text{with} \quad i,j = 1,...,N
		\end{aligned}
	\end{equation}
	where
	$$
	\widehat{\bm{x}} = \begin{bmatrix} \widehat{\sigma}_{p}^2 \\ \widehat{\boldsymbol\rho} \end{bmatrix}, \qquad
	\widetilde{\bm{x}} = \begin{bmatrix} \widetilde{\sigma}_{p}^2 \\ \widetilde{\boldsymbol\rho} \end{bmatrix},
	$$
	are, respectively, the base forecast vector collecting the portfolio variance forecast and the disaggregated correlation forecasts, and the reconciled forecast vector. $\bm{W}$ is a positive definite weighting matrix representing the forecast error variance of the portfolio and the $N$ assets.
	The quadratic programming problem with linear equality and inequality constraints~\eqref{eq:ottcor} can be solved using standard numerical optimization techniques, for instance those available in the \textsf{R} package \texttt{FoReco} \citep{FoReco}, which provides a flexible implementation of reconciliation procedures within this framework.
\end{enumerate}

Taking into account the previous elements, we propose a method for reconciling portfolio variance forecasts to enhance the accuracy of risk estimation by combining univariate and multivariate approaches. The steps of this reconciliation procedure are described in Algorithm~\ref{alg:FR}.

We stress that the main objective of our research is to determine if forecast reconciliation tools lead to potential improvements in the prediction of portfolio variance. Moreover, we are also interested in determining if, by exploiting the information contained in the direct portfolio variance prediction (based on univariate methods), we will improve the prediction based on the commonly adopted MGARCH models, which might be miss-specified due to the lack of variance interdependence. In both cases, the evaluation will compare the baseline portfolio variance forecast with the \textit{bottom-up} one and forecast reconciliation, for a given pair of fitted univariate GARCH and MGARCH specifications.

\begin{algorithm}[!tbh]
	\label{alg:FR}
	\setstretch{1}
	\caption{Forecast reconciliation of portfolio variance}
	\SetAlgoLined
	\KwIn{Portfolio weights $\bm{\omega} \in \mathbb{R}^N$, univariate (base) forecast for the portfolio variance $\widehat{\sigma}_p^2$, multivariate (base) forecast for the $N$ assets covariance matrix $\widehat{\bm{\Sigma}}$}
	\KwOut{Reconciled portfolio variance $\widetilde{\sigma}_p^2$, reconciled covariance matrix $\widetilde{\bm{\Sigma}}$}
	
	Compute $\widehat{\bm{\sigma}} \leftarrow \mathit{vech}(\widehat{\bm{\Sigma}})$\;
	Compute aggregation vector: $\bm{a} \leftarrow \bm{D}_N^{\prime}(\bm{\omega} \otimes \bm{\omega})$\;
	
	Define the base forecast vector: $\widehat{\bm{y}} \leftarrow \begin{bmatrix} \widehat{\sigma}_p^2 \\ \widehat{\bm{\sigma}} \end{bmatrix}$\;
	Define constraint vector: $\bm{c} \leftarrow \begin{bmatrix} 1 \\ -\bm{a} \end{bmatrix}$\;
	
	Solve the generalized least squares problem (\ref{eq:ott}): $\widetilde{\bm{y}} \leftarrow \begin{bmatrix} \widetilde{\sigma}_p^2 \\ \widetilde{\bm{\sigma}} \end{bmatrix}$\;
	Reconstruct reconciled covariance matrix: $\widetilde{\bm{\Sigma}} \leftarrow \mathit{vech}^{-1}(\widetilde{\bm{\sigma}})$\;
	Compute correlation matrix: $\widetilde{\bm{R}} \leftarrow \operatorname{cor}(\widetilde{\bm{\Sigma}})$\;
	
	\If{any off-diagonal element of $\widetilde{\bm{R}}$ violates $|\widetilde{\rho}_{i,j}| > 1$}{
		\tcc{Option A: Non-linear constrained optimization}
		\pushline Solve the non-linear constrained optimization problem (\ref{eq:nlott}): $\widetilde{\bm{y}} \leftarrow \begin{bmatrix} \widetilde{\sigma}_p^2 \\ \widetilde{\bm{\sigma}} \end{bmatrix}$\;
		Reconstruct reconciled covariance matrix: $\widetilde{\bm{\Sigma}} \leftarrow \mathit{vech}^{-1}(\widetilde{\bm{\sigma}})$\;
		\nonl \nosemic \;
		\popline \dosemic \tcc{Option B: Reconciliation via correlation decomposition}
		\pushline Compute $\widehat{\boldsymbol\rho} \leftarrow \mathit{vech}(\widehat{\bm{R}})$ with $\widehat{\bm{R}} \leftarrow \operatorname{cor}(\widehat{\bm{\Sigma}})$\;
		Compute $\widetilde{\bm{S}} \leftarrow \mathrm{diag}\!\left(\sqrt{\mathrm{diag}(\widetilde{\bm{\Sigma}})}\right)$\;
		Compute modified aggregation vector: $\widetilde{\bm{a}}_{\bm{\sigma}} \leftarrow \bm{D}_N^{\prime} \left(\widetilde{\bm{S}} \otimes \widetilde{\bm{S}} \right) (\bm{\omega} \otimes \bm{\omega})$\;
		Define the base forecast vector: $\widehat{\bm{x}} \leftarrow \begin{bmatrix} \widehat{\sigma}_p^2 \\ \widehat{\boldsymbol\rho} \end{bmatrix}$\;
		Define constraint vector: $\widetilde{\bm{c}}_{\bm{\sigma}} \leftarrow \begin{bmatrix} 1 \\ -\widetilde{\bm{a}}_{\bm{\sigma}} \end{bmatrix}$\;
		Solve the constrained optimization problem (\ref{eq:ottcor}): $\widetilde{\bm{x}} \leftarrow \begin{bmatrix} \widetilde{\sigma}_p^2 \\ \widetilde{\boldsymbol\rho} \end{bmatrix}$\;
		\nosemic Reconstruct reconciled correlation and covariance matrix: \;
		\nonl \hspace*{3cm}$\widetilde{\bm{R}} \leftarrow \mathit{vech}^{-1}(\widetilde{\boldsymbol\rho})$ and $\widetilde{\bm{\Sigma}} \leftarrow \widetilde{\bm{S}} \widetilde{\bm{R}} \widetilde{\bm{S}}$\;
	}
	\Return $\widetilde{\sigma}_p^2$ and $\widetilde{\bm{\Sigma}}$\;
\end{algorithm}
	
\subsection{Univariate and Multivariate GARCH models}

As mentioned in the previous section, the construction of reconciled forecasts for the portfolio variance, starting from either portfolio or asset returns, requires the specification of a univariate model for the former and of a multivariate model for the latter.

For simplicity, in the case of the portfolio returns, we specify a simple GARCH(1,1) model:

\begin{equation}
\sigma_{p,t}^2=\omega + \alpha r_{p,t-1}^2 + \beta \sigma_{p,t-1}^2.
\end{equation}

We are aware that a more standard approach is now given by a model with asymmetry in the variances, but we prefer to maintain now the coherence in the features captured by the model providing the base forecast and the multivariate model behind the bottom-up forecast. Further generalizations are left for future work.

Moving to the Multivariate GARCH models used to produce forecasts, we restrict our attention to two specific cases: the BEKK model of \cite{englekroner1995} and the Dynamic Conditional Correlation (DCC) model of \cite{Engle2002}. For the BEKK model, we consider the simplest specification:

\begin{equation}
\boldsymbol \Sigma_t =  \mathcal{C} \mathcal{C}^{\prime} + \mathcal{A} \boldsymbol r_{t-1} \boldsymbol r_{t-1}^{\prime} \mathcal{A} + \mathcal{B} \boldsymbol \Sigma_{t-1} \mathcal{B}^{\prime},
\label{eq:FBEKK}
\end{equation}

where $\mathcal{C}$ is lower triangular, while $\mathcal{A}$ and $\mathcal{B}$ should be full matrices. However, in empirical studies, to deal with the so-called \textit{curse of dimensionality}, both $\mathcal{A}$ and $\mathcal{B}$ are restricted to be diagonal or even driven by a single parameter, see, among many others, \cite{englekroner1995,DingEngle2001,CaporinMcAleer2008}.

The second model builds on the DCC dynamic for correlations, accompanied by a peculiar dynamic over marginals to allow, potentially, for variance interdependence. The model we consider might be seen as a special case of the Vector ARMA-GARCH of \cite{LingMcAleer2003} with DCC dynamic, and has been proposed, in the case of constant conditional correlation, by \cite{HeTerasvirta2004}, and by \cite{Caporale2014} for the DCC model. Our target is to maintain model feasibility (that is, allowing for parameter estimation in a two-step procedure, first univariate GARCH on the marginals and then the correlation dynamic) and to allow for interdependence, limiting attention to positive spillover effects; the generalization allowing for negative parameters as in \cite{ConradKaranasos2010} is left to future research. In our simulations, the univariate models for the single assets are thus specified as follows:
\begin{equation}
\sigma_{i,t}^2=\omega_i + \beta_i \sigma_{p,t-1}^2  + \sum_{j=1}^n \alpha_j r_{j,t-1}^2, \quad i=1,2,\ldots n.
\label{eq:DCC1}
\end{equation}

We define the standardized innovations $\eta_{i,t}=\sigma_{i,t}^{-1}r_{i,t-1}$ for $i=1,2,\ldots n$ and then specify a DCC model
\begin{align}
\boldsymbol Q_t &= \left(1-\theta_1 - \theta_2\right) \boldsymbol\Gamma + \theta_1 \boldsymbol \eta_{t-1}\boldsymbol \eta_{t-1}^{\prime} + \theta_2 \boldsymbol Q_{t-1} \label{eq:DCC2}\\
\nonumber \boldsymbol \Gamma_t &= \left(\boldsymbol \tilde{Q}_t\right)^{-1} \boldsymbol Q_t \left(\boldsymbol \tilde{Q}_t\right)^{-1}\\
\nonumber \boldsymbol \tilde{Q}_t &=\mathrm{diag}\left(\sigma_{1,t},\sigma_{2,t},\ldots,\sigma_{n,t}\right).
\end{align}
In the following, we refer to this model as Extended DCC (DCC with variance interaction, EDCC).

In the simulations, we will also consider the restricted specifications of both the BEKK and the DCC models. In detail, we will consider the BEKK model with scalar parameters and the baseline DCC model without interactions (i.e., the marginals are simple GARCH(1,1)).

Following the standard practice, the models are estimated by Quasi Maximum Likelihood, and in the case of the DCC specification using a multi-step procedure, starting from the marginals, then the $\boldsymbol\Gamma$ matrix (with a sample correlation estimator on $\boldsymbol\eta_t$), and finally the parameters governing the correlation dynamic.

\subsection{Forecast comparison methods}

In this section, we describe the methodologies used to compare the forecast performance of different portfolio variance forecast approaches. These methods include the accuracy of the point forecast using both absolute and relative indicators, and employ hypothesis testing based on two different well-known statistical procedures: the \cite{Diebold1995} test and the Model Confidence Set \citep{Hansen2011}. 

Five different approaches are compared based on their accuracy in forecasting the portfolio variance. These approaches include both univariate and multivariate models, as well as different reconciliation techniques, aligning with the steps described in Algorithm 1 for forecast reconciliation. The approaches considered are as follows:
\begin{itemize}[leftmargin=1cm, nosep]
	\item[\textit{base}] The univariate model forecast for the portfolio variance, denoted as $\widehat{\sigma}_{p}^2$ in Section \ref{sec:recop}. This approach assumes that the variance of the portfolio can be estimated from the historical data of portfolio returns.

	\item[\textit{bu}] The bottom-up approach using a multivariate model for the portfolio variance, expressed as $\gamma_{p}^2 = \bm{\omega}^{\prime} \widehat{\bm{\Sigma}} \bm{\omega}$ in Section \ref{sec:recop}. It first forecasts the variances and covariances of the individual assets and then combines these to estimate the overall portfolio variance. 

	\item[\textit{shr}] The optimal linear reconciliation (in a least squares sense), computed using the expression (\ref{eq:ott}), denoted as $\widetilde{\sigma}_{p}^2$. This method involves reconciling forecasts from the univariate portfolio variance forecasts and the variance and covariance matrix of the $N$ assets.
	
	\item[\textit{shr}$_{A}$] This is an extension of the reconciliation approach $shr$, which addresses potential issues arising when the correlation matrix $\widetilde{\bm{R}}$ contains off-diagonal elements violating the condition $|\rho_{i,j}| > 1$; see Section~\ref{sec:optionAlg} and Option A in Algorithm~\ref{alg:FR}. It applies a non-linear constrained optimization method to adjust the forecast correlation matrix, ensuring that all correlations remain within the feasible range of $[-1, 1]$.
	
	\item[\textit{shr}$_{B}$] This is an extension of the reconciliation approach $shr$, where reconciliation is performed via a correlation decomposition approach; see Section~\ref{sec:optionAlg}  and Option B in Algorithm~\ref{alg:FR}. Here, we apply reconciliation using the correlation decomposition (\ref{eq:Ai}) and solving the linear optimization problem (\ref{eq:ottcor}). This method provides a more computational efficient alternative to $shr_A$.
\end{itemize}

To evaluate the point forecast accuracy of the different models, we employ three widely used metrics: the Mean Squared Error \citep[MSE,][]{Davydenko2013}, the Mean Absolute Error \citep[MAE,][]{Davydenko2013}, and the Quasi-Likelihood score \citep[QLIKE,][]{Patton2011a, Caporin2024}. Each of these metrics provides a distinct perspective on the accuracy of the predicted variance. The expressions for these measures are given by the following:
\begin{align}
	\text{MSE}_{j,q} &= \frac{1}{M}\sum_{i = 1}^{M} \left(\sigma_{p,i}^2 - h_{i,j,q}^2\right)^2, \\
	\text{MAE}_{j,q} &= \frac{1}{M}\sum_{i = 1}^{M} \left|\sigma_{p,i}^2 - h_{i,j,q}^2\right|, \\
	\text{QLIKE}_{j,q} &= \frac{1}{M}\sum_{i = 1}^{M} \left[\frac{\sigma_{p,i}^2}{h_{i,j,q}^2} - \log\left(\frac{\sigma_{p,i}^2}{h_{i,j,q}^2} \right)-1\right],
\end{align}
where $\sigma_{p,i}^2$ denotes the true portfolio variance at time $i$, $h_{i,j,q}^2$ denotes the corresponding portfolio variance forecast from approach $j$ at replication $q$, $M$ denotes the size of the test set,\footnote{This is set to 250 for simulations in Section~\ref{sec:simu}, and 4037 for applications in Section~\ref{sec:app}.} and $Q$ denotes the number of replications.\footnote{We use 500 for simulations in Section~\ref{sec:simu} and 1 for the application in Section~\ref{sec:app}.} The index $j$ identifies the forecasting approach, with $j \in \{base, bu, shr, shr_A, shr_B\}$. We note that while MSE and MAE are symmetric loss functions, QLIKE is asymmetric, with a larger loss associated to under-prediction.
To obtain the overall accuracy measures, we compute the average of each metric across all replications:
$$
\text{IND}_j =  \frac{1}{Q}\sum_{q = 1}^{Q} \text{IND}_{j,q},
$$
and for relative accuracy \citep{Davydenko2013}, we calculate:
$$
\text{AvgRelIND}_{j,x} = \left(\prod_{q = 1}^{Q} \frac{\text{IND}_{j,q}}{ \text{IND}_{x,q}}\right)^{\frac{1}{Q}},
$$
where $x \in \{base, bu\}$ and $\text{IND}$ represents one of the evaluation metrics (MSE, MAE, or QLIKE). Specifically, $\text{AvgRelIND}_{bu}$ and $\text{AvgRelIND}_{base}$ are two relative indicators that show an improvement in forecast compared to the \textit{bu} and \textit{base} models, respectively.

In addition to these overall and relative measures, we conduct hypothesis testing to compare forecast accuracy across different models. 
We employ the Diebold and Mariano \cite{Diebold1995} (DM) test to evaluate the null hypothesis of equal predictive accuracy (EPA) between competing models, with an overall significance level set at $\alpha=0.05$. To account for multiple pairwise comparisons, the DM test is implemented using the Bonferroni correction \citep{Dunn1961-xm}. In addition, we apply the Model Confidence Set (MCS) procedure proposed by \cite{Hansen2011} to identify the subset of models that cannot be statistically distinguished in terms of forecast accuracy, reporting results at confidence levels of 70\%, 75\%, 80\%, 85\%, 90\% and 95\%. These tests are crucial in determining whether the observed differences in forecast performance are statistically significant or are merely due to randomness in the data.

In the following sections, we will consider the forecast comparison for both simulated and observed data. In the first case, the true portfolio variance $\sigma^2$ is known and will be used to evaluate the competing approaches. However, with real data, the true variance is not observed and a proxy must be used. Two solutions are commonly considered, the first being squared observed de-meaned returns. As shown by \cite{PattonSheppard2009} for the univariate case and \cite{Laurentetal2013} for multivariate models, this noisy proxy might lead to non-robust model selection for some loss functions, in particular for the MAE (while MSE and QLIKE are loss functions robust the presence of noise in the proxy). An alternative proxy might be recovered using high frequency data, and the daily variance (covariance) is estimated using the Realized Variance (Realized Covariance). Given the complexity of simulating high frequency data under a Multivariate GARCH model, we chose to first consider the noisy proxy in the simulation experiments, and then to mimic the existence of a better proxy by contaminating with additive noise the true covariance. In contrast, realized measures will be used, together with the squared returns, as proxies for the unknown variance when considering real data.

\section{Simulations} \label{sec:simu}

\subsection{Simulation design}

We consider several scenarios with varying sample sizes and number of variables. In particular, simulated return series are generated for portfolios comprising $9$ and $24$ assets using four distinct data generating processes: a fully parametrized BEKK, a scalar BEKK, the standard DCC-GARCH, and the EDCC-GARCH (incorporating interactions). We begin by illustrating the DGP used in the simulations for the case of $9$ assets.

The covariance matrix in the full BEKK specification is generated according to equation \eqref{eq:FBEKK} and setting $\boldsymbol r_t = \boldsymbol \Sigma_t^{\frac{1}{2}}\boldsymbol z_t$ with $\boldsymbol z_t$ sampled from a Multivariate Normal density with zero mean and covariance equal to the identity matrix. In terms of parameters, this formulation represents the most general approach, capturing a broad range of interdependencies. In contrast, the scalar BEKK restricts the dynamics by imposing that the matrices $\mathcal{A}$ and $\mathcal{B}$ assume the forms $\sqrt{\alpha} I$ and $\sqrt{\beta} I$, respectively, where $\alpha$ and $\beta$ are scalars and $I$ denotes identity. 

In the scalar BEKK model, covariance stationarity is guaranteed when $\alpha + \beta < 1$. Accordingly, in each simulation we draw $\alpha \sim \mathcal{U}(0.05, 0.20)$ and $\beta \sim \mathcal{U}(0.70, 0.95)$. We then verify the stationarity condition and, if $\alpha + \beta \ge 1$, we redraw the parameters until $\alpha + \beta < 1$ holds.

 In the full BEKK, the covariance stationarity condition depends on the eigenvalues:

$$
P_k\,(A \otimes A)D_k + P_k\,(B \otimes B)D_k
$$

\noindent which are required to be strictly below unity in modulus; in the previous equation $D_k$ represents the duplication matrix of size $k$ and $P_k$ its generalized inverse.\footnote{The duplication matrix $D_k$ satisfies $\mathit{vec}\left(M\right)=D_k \mathit{vech}\left(M\right)$ for a $k-$dimensional square symmetric matrix $M$.}

We considered specific designs for the $\mathcal{B}$ matrix. In particular, we partition $\mathcal{B}$ into $3\times 3$ blocks and, for each diagonal block, we set the entries as follows: the main-diagonal elements are drawn from a uniform distribution on $[0.70,0.95]$, whereas the off-diagonal elements are drawn from a uniform distribution on $[0,0.10]$. In contrast, $\mathcal{A}$ is generated without imposing any block structure, and all of its entries are drawn from a uniform distribution on $[0,0.10]$. After generating $\mathcal{A}$ and $\mathcal{B}$, we check the stationarity conditions; if they are not satisfied, the parameter draw is discarded and the matrices are resampled until the conditions hold. This procedure is repeated independently for each simulation.

For both scalar and full BEKK models, the matrix $\mathcal{C}$ is generated as a lower triangular matrix with random entries, constructed to ensure that the product $\mathcal{C}\mathcal{C}'$ maintains full rank and has a positive diagonal.

The remaining generators are based on the DCC-GARCH framework, as given in \eqref{eq:DCC1} and \eqref{eq:DCC2}. Similarly to the BEKK case, we randomly draw parameters. In each simulation, the parameters governing the correlation dynamics are drawn from uniform distributions, with $\theta_1 \sim \mathcal{U}(0.05,0.30)$ and $\theta_2 \sim \mathcal{U}(0.70,0.85)$. To enforce stationarity, we discard the draw if the condition $\theta_1 + \theta_2 < 1$ is not satisfied and resample the parameters until it holds. Moreover, the elements of $\boldsymbol\Gamma$ are generated as follows: first, we sample a matrix $A$ of random Normal numbers with mean $-0.15$ and standard deviation $0.6$;  then we compute $Q=A^{\prime}A$, and normalize it so that it has unit elements on the main diagonal; finally, we set $\boldsymbol\Gamma=0.5\left(Q+Q^{\prime}\right)$ and retain the simulated matrix only if the smallest eigenvalue is larger than $1e-10$. The initial generation of random numbers ensures that the average correlation level is slightly higher than $0.4$.

The difference between the standard DCC-GARCH and the variant with interactions lies in the specification of individual dynamics. In the simpler DCC model, the evolution of variances follows \eqref{eq:DCC1}, with the DCC-GARCH coefficients $\alpha_i$ and $\beta_i$ sampled from the distributions $\alpha_i \sim U(0.05, 0.15)$ and $\beta_i \sim U(0.7, 0.85)$, with $\beta_j=0$ for $j\neq i$. In the model incorporating interactions, we represent the collection of variances in a matrix form as
$$
\bm{\sigma}^2_t = \boldsymbol{\nu} + A\,r^2_{t-1} + B\,\bm{\sigma}^2_t,
$$
where $\boldsymbol{\nu}$ is the vector of intercepts, $\bm{\sigma}^2_t$ is the vector of variances, $B$ is diagonal and $A$ is unrestricted, permitting nonzero off-diagonal elements. The diagonal entries for $A$ and $B$ are selected from uniform distributions in $[0, 0.2]$ and $[0.7, 0.85]$, respectively, while the off-diagonal elements of $A$ are drawn from $U(0, 0.02)$. The coefficients are generated repeatedly until the eigenvalues of $A+B$ are all within the unit circle, thus guaranteeing covariance stationarity as discussed in \cite{LingMcAleer2003}.

For the simulations with $24$ assets, we adopt fixed coefficient designs. For models such as the full BEKK, drawing fully random parameter matrices that (i) satisfy covariance stationarity and (ii) yield reasonable parameter values (e.g., larger and positive diagonal entries) would typically require repeated draws for each simulation, substantially increasing the computational burden. We therefore fix the parameters of the data-generating processes.

Specifically, in the scalar BEKK we set $\alpha=0.15$ and $\beta=0.80$. In the full BEKK, the coefficient matrices $\mathcal{A}$ and $\mathcal{B}$ are set to
\[
\mathcal{B}= I_{8}\otimes
\begin{bmatrix}
0.80 & 0.05 & 0.05 \\
0.05 & 0.80 & 0.05 \\
0.05 & 0.05 & 0.80
\end{bmatrix},\qquad
\mathcal{A}=
\begin{bmatrix}
0.025\,\boldsymbol{1}_{8\times 8} & 0.0125\,\boldsymbol{1}_{8\times 8} & 0\cdot\boldsymbol{1}_{8\times 8} \\[1mm]
0.0125\,\boldsymbol{1}_{8\times 8} & 0.0187\,\boldsymbol{1}_{8\times 8} & 0.025\,\boldsymbol{1}_{8\times 8} \\[1mm]
0.0187\,\boldsymbol{1}_{8\times 8} & 0.0125\,\boldsymbol{1}_{8\times 8} & 0.0312\,\boldsymbol{1}_{8\times 8}
\end{bmatrix}.
\]

For the DCC--GARCH model, we set $\theta_1=0.15$ and $\theta_2=0.80$, and we fix the univariate GARCH parameters to $\alpha_i=0.15$ and $\beta_i=0.80$ for all $i$. In the model incorporating interactions, we choose $\theta_1$ and $\theta_2$ analogously, and we set $B$ accordingly. Conversely, the matrix $A$ is specified with $0.08$ on the main diagonal and $0.05$ elsewhere.

Given the parameters, we simulate returns sequences from the various data generating processes, also storing the true conditional covariance and correlation matrices. 
Subsequently, on the returns simulated from each DGP, we estimate three multivariate models, the Scalar BEKK, a standard DCC-GARCH, and the extended DCC (EDCC); the last model is not estimated in all cases. In addition, we aggregate the multivariate returns into a portfolio, either using equal weights, $1/n$, where $n$ denotes the total number of assets ($8$ or $24$), or with randomly generated weights (ensuring that their sum equals $1$). On the portfolio returns, we estimate a univariate GARCH model.

For each scenario, $100+T+250$ observations are simulated, the first $100$ are then discarded to avoid dependence on starting values, $T$ are employed for model estimation and the final $250$ reserved for out-of-sample evaluation. All forecasts are one-step ahead, meaning that $\Sigma_{t+1}$ is estimated using the observed return $r_t$. For the portfolios with 24 assets we set $T=1000$, while for the 9-asset scenario we also consider the alternative sample sizes of $T=500$ and $T=2000$. For all simulation designs, we performed $500$ experiments.

\subsection{Results}

In this subsection, we analyze the simulation results. We first focus on univariate versus multivariate models, then highlight the benefits of forecast reconciliation, and later discuss the impact of a noisy proxy. We stress that we are not contrasting fitted DCC to fitted BEKK or EDCC but rather contrasting the portfolio variance forecast from a univariate GARCH to those of a MGARCH model and the forecast reconciliation based on the given MGARCH.

\subsubsection{Base vs. bottom-up: DGPs without interdependence}

We start by contrasting the portfolio variance forecasts made following a bottom-up approach (\textit{bu}), that is, estimating a MGARCH model, with those obtained by directly working on the portfolio returns, that is, with a univariate GARCH model (\textit{base}). At first we consider a correctly specified MGARCH model and look at the relative accuracy indexes setting the univariate model as the reference (with a relative index thus equal to $1$): see Table \ref{tab:res1}, columns labeled \textit{base} and \textit{bu}; full results including the value of the accuracy indexes are reported in the Online Appendix.

\begin{sidewaystable}
\begin{center}
\begin{tabular}{c|ccccc|ccccc|ccccc}
\toprule
 & \multicolumn{5}{c}{$T=500$} & \multicolumn{5}{c}{$T=1000$} & \multicolumn{5}{c}{$T=2000$}\\
\cmidrule{2-16}
Index & \textit{base} & \textit{bu} & \textit{shr} & \textit{shr$_A$} & \textit{shr$_B$} & \textit{base} & \textit{bu} & \textit{shr} & \textit{shr$_A$} & \textit{shr$_B$} & \textit{base} & \textit{bu} & \textit{shr} & \textit{shr$_A$} & \textit{shr$_B$} \\
\midrule
\multicolumn{16}{c}{\textbf{DGP and Fitted model: Scalar BEKK - Portfolio weights $1/N$}}\\
MSE & 1.000 & 0.408 & \textbf{0.156} & \em{0.156} & 0.158 & 1.000 & 0.427 & \em{0.207} & \textbf{0.207} & 0.212 & 1.000 & 0.504 & \em{0.276} & \textbf{0.276} & 0.280\\
MAE & 1.000 & 0.665 & \textbf{0.385} & \em{0.385} & 0.385 & 1.000 & 0.686 & \em{0.458} & \textbf{0.458} & 0.459 & 1.000 & 0.739 & \em{0.541} & \textbf{0.541} & 0.542\\
QLIKE & 1.000 & 0.418 & \textbf{0.158} & \em{0.158} & 0.161 & 1.000 & 0.433 & \em{0.210} & \textbf{0.210} & 0.213 & 1.000 & 0.504 & \em{0.278} & \textbf{0.278} & 0.281\\
\addlinespace
\multicolumn{16}{c}{\textbf{DGP and Fitted model: Scalar BEKK - Portfolio weights: random}}\\
MSE & 1.000 & 0.397 & \textbf{0.153} & \em{0.153} & 0.155 & 1.000 & 0.411 & \textbf{0.189} & \em{0.189} & 0.190 & 1.000 & 0.495 & \textbf{0.276} & \em{0.276} & 0.280\\
MAE & 1.000 & 0.658 & \textbf{0.382} & \em{0.382} & 0.383 & 1.000 & 0.667 & \textbf{0.434} & \em{0.434} & 0.434 & 1.000 & 0.734 & \textbf{0.543} & \em{0.543} & 0.544\\
QLIKE & 1.000 & 0.409 & \textbf{0.157} & \em{0.157} & 0.159 & 1.000 & 0.416 & \textbf{0.192} & \em{0.192} & 0.192 & 1.000 & 0.496 & \textbf{0.279} & \em{0.279} & 0.282\\
\addlinespace
\multicolumn{16}{c}{\textbf{DGP and Fitted model: DCC-GARCH - Portfolio weights: $1/N$}}\\
MSE & 1.000 & 0.285 & \em{0.233} & \textbf{0.233} & 0.234 & 1.000 & 0.200 & \em{0.173} & \textbf{0.173} & 0.173 & 1.000 & 0.126 & \textbf{0.111} & \em{0.111} & 0.111\\
MAE & 1.000 & 0.534 & \em{0.495} & \textbf{0.494} & 0.495 & 1.000 & 0.459 & \em{0.432} & \textbf{0.432} & 0.433 & 1.000 & 0.371 & \textbf{0.352} & \em{0.352} & 0.352\\
QLIKE & 1.000 & 0.278 & \em{0.251} & \textbf{0.251} & 0.251 & 1.000 & 0.209 & \em{0.193} & \textbf{0.193} & 0.193 & 1.000 & 0.142 & \em{0.132} & \textbf{0.132} & 0.133\\
\addlinespace
\multicolumn{16}{c}{\textbf{DGP and Fitted model: DCC-GARCH - Portfolio weights: random}}\\
MSE & 1.000 & 0.291 & \em{0.238} & \textbf{0.238} & 0.238 & 1.000 & 0.212 & \textbf{0.180} & \em{0.180} & 0.180 & 1.000 & 0.142 & \textbf{0.119} & \em{0.119} & 0.119\\
MAE & 1.000 & 0.543 & \em{0.503} & \textbf{0.503} & 0.503 & 1.000 & 0.469 & \em{0.439} & \textbf{0.439} & 0.439 & 1.000 & 0.391 & \textbf{0.365} & \em{0.365} & 0.365\\
QLIKE & 1.000 & 0.292 & \em{0.265} & \textbf{0.265} & 0.265 & 1.000 & 0.216 & \em{0.198} & \textbf{0.198} & 0.198 & 1.000 & 0.154 & \em{0.140} & \textbf{0.140} & 0.140\\
\hline
\end{tabular}
\end{center}
\caption{Average relative accuracy indexes where the reference forecast is the univariate GARCH fitted on simulated portfolio returns, and the DGP is either the Scalar BEKK (first and second panels) or the DCC-GARCH (third and fourth panels). The top rows report the sample size ($T$) and the variance forecast approach, including the univariate GARCH on portfolio returns, \textit{base}, the bottom-up approach using MGARCH models, \textit{bu}, and the three forecast reconciliation cases discussed in the previous section. The first and third panels consider equally weighted portfolios (the $1/N$ case), while the second and fourth consider random portfolio weights. All values are averages across 500 replications. In each row, the best and second best forecast approaches are in bold and italic, respectively.}
\label{tab:res1}
\end{sidewaystable}

Notably, if we estimate a correctly specified MGARCH model, the prediction of portfolio risk is closer to the true value than the one obtained from a univariate GARCH model (miss-specified) fit on portfolio returns, that is, we observe a value lower than $1$ for the \textit{bu} case. Note that, as reported in the Appendix, the accuracy measures decrease with the sample size in all cases, as expected. The pattern observed on relative accuracy, lower \textit{bu} values, does not depend on the portfolio weights (equal or randomly generated), on the sample size, and on the accuracy measure. We observe that, for the Scalar-BEKK DGP, the preference for the bottom-up approach is attenuated as the sample size increases (the \textit{bu} values increase) and is, overall, less pronounced than in the DCC-GARCH case. We interpret this as a consequence of aggregating returns generated under a Scalar-BEKK model with Gaussian innovations. In fact, if returns follow $\boldsymbol r_t \vert \mathcal{I}_{t-1}\sim \mathcal{N}\left(\boldsymbol 0, \bm{\Sigma}_t\right)$, with $\mathcal{I}_{t-1}$ being the time $t-1$ information set, and we set the portfolio weights to be time-invariant, $\bm{\omega}$ (and summing up to one), we have
\begin{align*}
\bm{\omega}^{\prime}\bm{\Sigma}_t\bm{\omega} &= \sigma^2_{p,t}=\bm{\omega}^{\prime}\mathcal{C}\mathcal{C}^{\prime}\bm{\omega} + \alpha\bm{\omega}^{\prime} \boldsymbol r_{t-1}\boldsymbol r_{t-1}^{\prime}\bm{\omega} + \beta\bm{\omega}^{\prime}\bm{\Sigma}_{t-1}\bm{\omega}\\
&= \tilde{\nu} + \alpha \overline{\boldsymbol r}_{t-1}^2 + \beta \sigma^2_{p,t-1}
\end{align*}
with $\overline{\boldsymbol r}_{t-1}$ being the returns of the portfolio (a weighted average). The preference of the MGARCH specification for shorter samples might be linked to the larger amount of information used for parameters estimation ($N$ series against $1$) an effect that tends to disappear asymptotically, as suggested by the closer performances for $T=2000$ with random weights. A similar aggregation result does not hold for the DCC-GARCH model due to the heterogeneity in the volatility dynamic and the standardization required to obtain the dynamic correlations. In line with this pattern, for increasing sample size, the performance of the correctly specified MGARCH improves, with \textit{bu} values decreasing with $T$. Overall, these first outcomes are somewhat expected.

Results start to be more interesting if we consider a misspecified model, still focusing only on DGPs without any form of interdependence; see Table \ref{tab:res2}, again columns labeled \textit{base} and \textit{bu}. In terms of levels of accuracy indexes, they decrease with increasing sample size (as expected); see the Appendix. Moving to relative indicators, if we estimate a Scalar BEKK on series generated from a DCC-GARCH, and focus on small sample sizes, the wrongly specified model is providing superior forecasts compared to a univariate GARCH fitted on portfolio returns. However, as the sample size increases, the \textit{base} and \textit{bu} approaches tend to converge, and then \textit{bu} starts deteriorating with respect to \textit{base}. This behaviour does not depend on portfolio weights. The outcome is different if the DGP is a Scalar BEKK and the fitted model is a DCC-GARCH: the univariate model on portfolio returns is inferior to the misspecified MGARCH irrespective of the sample size (and again this does not depend on the portfolio weights). We link such evidence, again, to the aggregation of Scalar BEKK into a univariate GARCH on portfolio returns. In this case, the DCC-GARCH is clearly miss-specified but its heterogeneity helps in capturing the differences in unconditional variances (as driven by the differences across elements in the intercept).

Extending the cross-sectional dimension from $N=9$ to $N=24$ produces some notable shifts in the simulations. Under correctly specified estimation, when the DGP is DCC-GARCH, the univariate baseline now outperforms the correctly specified multivariate model—reversing the $N=9$ ranking. By contrast, when the DGP is Scalar BEKK, the correctly specified multivariate model continues to dominate the univariate benchmark, in line with the $N=9$ evidence.

Under misspecification, dimensionality also reshuffles the cross-cases. If the DGP is Scalar BEKK but the estimation is performed with DCC-GARCH, the multivariate specification performs better at $N=24$ as in the $N=9$ case. Conversely, if the DGP is DCC-GARCH and the multivariate estimator is a Scalar BEKK, the univariate baseline is now preferred. Overall, moving from nine to twenty-four assets strengthens the bias–variance trade-off and aggregation effects, eroding the advantage of correctly specified high-dimensional DCC in one case while preserving the edge of Scalar BEKK in the other. Results referring to the $N=24$ case are available in the Online Appendix.

\begin{sidewaystable}
\begin{center}
\begin{tabular}{c|ccccc|ccccc|ccccc}
\toprule
 & \multicolumn{5}{c}{$T=500$} & \multicolumn{5}{c}{$T=1000$} & \multicolumn{5}{c}{$T=2000$}\\
\cmidrule{2-16}
Index & \textit{base} & \textit{bu} & \textit{shr} & \textit{shr$_A$} & \textit{shr$_B$} & \textit{base} & \textit{bu} & \textit{shr} & \textit{shr$_A$} & \textit{shr$_B$} & \textit{base} & \textit{bu} & \textit{shr} & \textit{shr$_A$} & \textit{shr$_B$} \\
\midrule
\multicolumn{16}{c}{\textbf{DGP: DCC-GARCH - Fitted model: Scalar BEKK - Portfolio weights $1/N$}}\\
MSE & 1.000 & 0.686 & 0.595 & \em{0.595} & \textbf{0.594} & 1.000 & 0.861 & \em{0.760} & \textbf{0.760} & 0.760 & 1.000 & 1.002 & \textbf{0.851} & \em{0.852} & 0.852\\
MAE & 1.000 & 0.848 & 0.788 & \em{0.788} & \textbf{0.788} & 1.000 & 0.954 & \em{0.888} & \textbf{0.888} & 0.888 & 1.000 & 1.024 & \textbf{0.935} & \em{0.935} & 0.936\\
QLIKE & 1.000 & 0.763 & \em{0.681} & \textbf{0.681} & 0.681 & 1.000 & 0.953 & \em{0.846} & \textbf{0.846} & 0.846 & 1.000 & 1.074 & \em{0.927} & \textbf{0.927} & 0.928\\
\addlinespace
\multicolumn{16}{c}{\textbf{DGP: DCC-GARCH - Fitted model: Scalar BEKK - Portfolio weights: random}}\\
MSE & 1.000 & 0.676 & 0.559 & \em{0.558} & \textbf{0.558} & 1.000 & 0.885 & \textbf{0.738} & \em{0.738} & 0.738 & 1.000 & 1.104 & \textbf{0.857} & \em{0.857} & 0.857\\
MAE & 1.000 & 0.843 & 0.766 & \em{0.766} & \textbf{0.766} & 1.000 & 0.963 & \textbf{0.875} & \em{0.875} & 0.875 & 1.000 & 1.065 & \textbf{0.941} & \em{0.941} & 0.941\\
QLIKE & 1.000 & 0.756 & \em{0.645} & \textbf{0.645} & 0.645 & 1.000 & 0.961 & \em{0.819} & \textbf{0.819} & 0.820 & 1.000 & 1.153 & \em{0.939} & \textbf{0.939} & 0.940\\
\addlinespace
\multicolumn{16}{c}{\textbf{DGP: Scalar BEKK - Fitted model: DCC-GARCH - Portfolio weights: $1/N$}}\\
MSE & 1.000 & 0.773 & \textbf{0.428} & \em{0.428} & 0.444 & 1.000 & 0.784 & \textbf{0.477} & \em{0.477} & 0.492 & 1.000 & 0.780 & \textbf{0.489} & \em{0.489} & 0.495\\
MAE & 1.000 & 0.878 & \textbf{0.646} & \em{0.646} & 0.649 & 1.000 & 0.899 & \textbf{0.696} & \em{0.696} & 0.699 & 1.000 & 0.893 & \em{0.716} & \textbf{0.716} & 0.717\\
QLIKE & 1.000 & 0.786 & \textbf{0.435} & \em{0.435} & 0.452 & 1.000 & 0.799 & \textbf{0.487} & \em{0.487} & 0.496 & 1.000 & 0.786 & \textbf{0.494} & \em{0.494} & 0.498\\
\addlinespace
\multicolumn{16}{c}{\textbf{DGP: Scalar BEKK - Fitted model: DCC-GARCH - Portfolio weights: random}}\\
MSE & 1.000 & 0.839 & \textbf{0.455} & \em{0.455} & 0.459 & 1.000 & 0.788 & \em{0.485} & \textbf{0.485} & 0.492 & 1.000 & 0.815 & \textbf{0.526} & \em{0.526} & 0.527\\
MAE & 1.000 & 0.915 & \textbf{0.666} & \em{0.666} & 0.667 & 1.000 & 0.895 & \em{0.702} & \textbf{0.702} & 0.703 & 1.000 & 0.917 & \textbf{0.742} & \em{0.742} & 0.742\\
QLIKE & 1.000 & 0.857 & \textbf{0.468} & \em{0.468} & 0.473 & 1.000 & 0.800 & \em{0.495} & \textbf{0.495} & 0.503 & 1.000 & 0.824 & \textbf{0.533} & \em{0.533} & 0.534\\
\bottomrule
\end{tabular}
\end{center}
\caption{Average relative accuracy indexes where the reference forecast is the univariate GARCH fitted on simulated portfolio returns, and the DGP is either the DCC-GARCH (first and second panels) or the Scalar BEKK (third and fourth panels). The fitted MGARCH models are misspecified (and still without interdependence): we estimate the Scalar BEKK under the DCC-GARCH DGP, and the DCC-GARCH with data generated from a Scalar BEKK. the The top rows report the sample size ($T$) and the variance forecast approach, including the univariate GARCH on portfolio returns, \textit{base}, the bottom-up approach using MGARCH models, \textit{bu}, and the three forecast reconciliation cases discussed in the previous section. The first and third panels consider equally weighted portfolios (the $1/N$ case), while the second and fourth consider random portfolio weights. All values are averages across 500 replications. In each row, the best and second best forecast approaches are in bold and italic, respectively.}
\label{tab:res2}
\end{sidewaystable}

\subsubsection{Base vs. bottom-up: DGPs with interdependence}

Consider now the two DGPs that include interdependence, namely, the EDCC-GARCH and the BEKK models. Even in this case, we estimate the two specifications without interdependence, the DCC-GARCH and the Scalar BEKK, and we contrast the portfolio variance forecast obtained from a univariate GARCH with that of the misspecified MGARCH models; see Tables \ref{tab:res3} and  \ref{tab:res4}, still focusing only on columns \textit{base} and \textit{bu}. For completeness, we also estimate the correctly specified model under the EDCC-GARCH DGP (the fully parameterized BEKK estimation remains computationally unfeasible). For the BEKK DGP the univariate model is providing superior forecast abilities compared to either the Scalar BEKK and the DCC-GARCH. Moreover, the performance of MGARCH models worsens, in relative terms, with increasing sample size. This is valid irrespective of the loss function considered. On the other hand, for the EDCC-GARCH DGP, the use of a misspecified DCC-GARCH provides better forecasts than a univariate model on portfolio returns. This is in striking contrast with the result of the fitted Scalar BEKK (on the EDCC-GARCH DGP) whose forecasts are clearly inferior to those of the univariate model for sample size above 1000. We link such a finding to the relatively limited size of the coefficients of other asset shocks in the EDCC-GARCH DGP. The correctly specified model clearly favors the bottom-up approach compared to the univariate baseline. All results are equivalent for equally weighted portfolios and randomly generated portfolios. See the Online Appendix for detailed results for both the $N=9$ and the $N=24$ cases.

\begin{sidewaystable}
\begin{center}
\begin{tabular}{c|ccccc|ccccc|ccccc}
\toprule
 & \multicolumn{5}{c}{$T=500$} & \multicolumn{5}{c}{$T=1000$} & \multicolumn{5}{c}{$T=2000$}\\
\cmidrule{2-16}
Index & \textit{base} & \textit{bu} & \textit{shr} & \textit{shr$_A$} & \textit{shr$_B$} & \textit{base} & \textit{bu} & \textit{shr} & \textit{shr$_A$} & \textit{shr$_B$} & \textit{base} & \textit{bu} & \textit{shr} & \textit{shr$_A$} & \textit{shr$_B$} \\
\midrule
\multicolumn{16}{c}{\textbf{DGP: Full BEKK - Fitted model: Scalar BEKK - Portfolio weights $1/N$}}\\
MSE & 1.000 & 19.151 & \em{0.663} & \textbf{0.663} & 0.663 & 1.000 & 23.737 & \em{0.742} & \textbf{0.742} & 0.742 & 1.000 & 27.699 & \textbf{0.786} & \em{0.786} & 0.786\\
MAE & 1.000 & 4.745 & \em{0.897} & \textbf{0.897} & 0.897 & 1.000 & 5.449 & \em{0.941} & \textbf{0.941} & 0.941 & 1.000 & 5.994 & \textbf{0.963} & \em{0.963} & 0.963\\
QLIKE & \textbf{1.000} & 21.092 & \em{1.148} & 1.169 & 1.148 & \textbf{1.000} & 27.143 & 1.229 & \em{1.228} & 1.228 & \textbf{1.000} & 32.571 & \em{1.209} & 1.214 & 1.209\\
\addlinespace
\multicolumn{16}{c}{\textbf{DGP: Full BEKK - Fitted model: Scalar BEKK - Portfolio weights: random}}\\
MSE & 1.000 & 11.402 & \em{0.761} & \textbf{0.761} & 0.761 & 1.000 & 13.696 & \em{0.810} & \textbf{0.810} & 0.810 & 1.000 & 15.417 & \em{0.832} & \textbf{0.832} & 0.832\\
MAE & 1.000 & 3.643 & 0.930 & \em{0.930} & \textbf{0.930} & 1.000 & 4.055 & 0.956 & \textbf{0.956} & \em{0.956} & 1.000 & 4.371 & 0.970 & \textbf{0.970} & \em{0.970}\\
QLIKE & \textbf{1.000} & 11.836 & 1.163 & 1.162 & \em{1.161} & \textbf{1.000} & 14.163 & 1.222 & 1.221 & \em{1.221} & \textbf{1.000} & 16.291 & 1.202 & 1.202 & \em{1.202}\\
\addlinespace
\multicolumn{16}{c}{\textbf{DGP: Full BEKK - Fitted model: DCC-GARCH - Portfolio weights: $1/N$}}\\
MSE & 1.000 & 7.054 & \em{0.557} & \textbf{0.557} & 0.558 & 1.000 & 8.168 & \em{0.637} & \textbf{0.637} & 0.638 & 1.000 & 8.968 & \em{0.689} & \textbf{0.689} & 0.690\\
MAE & 1.000 & 2.700 & \em{0.813} & \textbf{0.813} & 0.813 & 1.000 & 2.930 & \em{0.858} & \textbf{0.858} & 0.858 & 1.000 & 3.103 & \em{0.885} & \textbf{0.885} & 0.886\\
QLIKE & 1.000 & 7.432 & 0.910 & \textbf{0.910} & \em{0.910} & 1.000 & 8.745 & 0.937 & \textbf{0.937} & \em{0.937} & 1.000 & 9.681 & \em{0.928} & \textbf{0.928} & 0.928\\
\addlinespace
\multicolumn{16}{c}{\textbf{DGP: Full BEKK - Fitted model: DCC-GARCH - Portfolio weights: random}}\\
MSE & 1.000 & 4.233 & \em{0.691} & \textbf{0.691} & 0.691 & 1.000 & 4.702 & \em{0.742} & \textbf{0.742} & 0.743 & 1.000 & 5.000 & \em{0.772} & \textbf{0.772} & 0.772\\
MAE & 1.000 & 2.083 & \em{0.880} & \textbf{0.880} & 0.880 & 1.000 & 2.186 & \em{0.908} & \textbf{0.908} & 0.908 & 1.000 & 2.269 & \em{0.923} & \textbf{0.923} & 0.923\\
QLIKE & 1.000 & 4.184 & 0.968 & \em{0.967} & \textbf{0.967} & 1.000 & 4.574 & 0.989 & \em{0.989} & \textbf{0.989} & 1.000 & 4.861 & 0.984 & \textbf{0.984} & \em{0.984}\\
\addlinespace
\multicolumn{16}{c}{\textbf{DGP: Full BEKK - Fitted model: EDCC-GARCH - Portfolio weights: $1/N$}}\\
MSE & 1.000 & 4.714 & \em{0.691} & \textbf{0.691} & 0.692 & 1.000 & 5.485 & \em{0.732} & \textbf{0.732} & 0.733 & 1.000 & 5.916 & \em{0.774} & \textbf{0.773} & 0.774\\
MAE & 1.000 & 2.176 & \em{0.883} & \textbf{0.883} & 0.883 & 1.000 & 2.375 & \em{0.905} & \textbf{0.905} & 0.905 & 1.000 & 2.497 & \em{0.927} & \textbf{0.927} & 0.927\\
QLIKE & \textbf{1.000} & 4.826 & 1.038 & \em{1.034} & 1.043 & \textbf{1.000} & 5.793 & 1.008 & \em{1.008} & 1.009 & 1.000 & 6.360 & \em{0.992} & \textbf{0.992} & 0.993\\
\addlinespace
\multicolumn{16}{c}{\textbf{DGP: Full BEKK - Fitted model: EDCC-GARCH - Portfolio weights: random}}\\
MSE & 1.000 & 2.726 & 0.841 & \em{0.841} & \textbf{0.841} & 1.000 & 3.066 & 0.859 & \em{0.859} & \textbf{0.859} & 1.000 & 3.184 & \em{0.876} & \textbf{0.876} & 0.876\\
MAE & 1.000 & 1.648 & 0.948 & \em{0.948} & \textbf{0.948} & 1.000 & 1.743 & 0.958 & \em{0.958} & \textbf{0.958} & 1.000 & 1.791 & 0.968 & \textbf{0.968} & \em{0.968}\\
QLIKE & \textbf{1.000} & 2.621 & 1.043 & 1.042 & \em{1.041} & \textbf{1.000} & 2.928 & 1.042 & \em{1.041} & 1.042 & \textbf{1.000} & 3.059 & 1.034 & \em{1.034} & 1.034\\
\bottomrule
\end{tabular}
\end{center}
\caption{Average relative accuracy indexes where the reference forecast is the univariate GARCH fitted on simulated portfolio returns, and the DGP is a Full BEKK. The fitted MGARCH models are misspecified: we estimate the Scalar BEKK (first and second panels), and the DCC-GARCH (third and fourth panels). The top rows report the sample size ($T$) and the variance forecast approach, including the univariate GARCH on portfolio returns, \textit{base}, the bottom-up approach using MGARCH models, \textit{bu}, and the three forecast reconciliation cases discussed in the previous section. The first and third panels consider equally weighted portfolios (the $1/N$ case), while the second and fourth consider random portfolio weights. All values are averages across 500 replications. In each row, the best and second best forecast approaches are in bold and italic, respectively.}
\label{tab:res3}
\end{sidewaystable}

\begin{sidewaystable}
\begin{center}
\begin{tabular}{c|ccccc|ccccc|ccccc}
\toprule
 & \multicolumn{5}{c}{$T=500$} & \multicolumn{5}{c}{$T=1000$} & \multicolumn{5}{c}{$T=2000$}\\
\cmidrule{2-16}
Index & \textit{base} & \textit{bu} & \textit{shr} & \textit{shr$_A$} & \textit{shr$_B$} & \textit{base} & \textit{bu} & \textit{shr} & \textit{shr$_A$} & \textit{shr$_B$} & \textit{base} & \textit{bu} & \textit{shr} & \textit{shr$_A$} & \textit{shr$_B$} \\
\midrule
\multicolumn{16}{c}{\textbf{DGP: EDCC-GARCH - Fitted model: Scalar BEKK - Portfolio weights $1/N$}}\\
MSE & 1.000 & 0.995 & \textbf{0.801} & \em{0.802} & 0.802 & 1.000 & 1.107 & \em{0.885} & \textbf{0.885} & 0.885 & 1.000 & 1.186 & \em{0.938} & 0.939 & \textbf{0.938}\\
MAE & 1.000 & 1.037 & \textbf{0.923} & \em{0.923} & 0.923 & 1.000 & 1.102 & \em{0.966} & \textbf{0.966} & 0.966 & 1.000 & 1.141 & \em{0.991} & 0.991 & \textbf{0.991}\\
QLIKE & 1.000 & 1.130 & \textbf{0.917} & \em{0.917} & 0.917 & 1.000 & 1.278 & \em{0.991} & \textbf{0.991} & 0.991 & \textbf{1.000} & 1.356 & 1.029 & 1.029 & \em{1.029}\\
\addlinespace
\multicolumn{16}{c}{\textbf{DGP: EDCC-GARCH - Fitted model: Scalar BEKK - Portfolio weights: random}}\\
MSE & 1.000 & 1.048 & \textbf{0.821} & \em{0.821} & 0.821 & 1.000 & 1.156 & \textbf{0.895} & \em{0.895} & 0.895 & 1.000 & 1.228 & \em{0.943} & 0.943 & \textbf{0.943}\\
MAE & 1.000 & 1.057 & \textbf{0.928} & \em{0.928} & 0.928 & 1.000 & 1.119 & \textbf{0.970} & 0.970 & \em{0.970} & 1.000 & 1.154 & \textbf{0.990} & \em{0.990} & 0.990\\
QLIKE & 1.000 & 1.160 & \em{0.928} & \textbf{0.928} & 0.928 & 1.000 & 1.307 & \em{0.995} & \textbf{0.995} & 0.995 & \textbf{1.000} & 1.361 & 1.025 & 1.025 & \em{1.025}\\
\addlinespace
\multicolumn{16}{c}{\textbf{DGP: EDCC-GARCH - Fitted model: DCC-GARCH - Portfolio weights: $1/N$}}\\
MSE & 1.000 & 0.662 & \textbf{0.547} & \em{0.547} & 0.547 & 1.000 & 0.635 & \em{0.548} & \textbf{0.548} & 0.549 & 1.000 & 0.647 & \em{0.560} & \textbf{0.560} & 0.560\\
MAE & 1.000 & 0.838 & \textbf{0.765} & \em{0.765} & 0.766 & 1.000 & 0.824 & \em{0.765} & \textbf{0.765} & 0.766 & 1.000 & 0.824 & \em{0.769} & \textbf{0.769} & 0.769\\
QLIKE & 1.000 & 0.661 & \textbf{0.592} & \em{0.592} & 0.592 & 1.000 & 0.641 & \em{0.581} & \textbf{0.580} & 0.581 & 1.000 & 0.634 & 0.579 & \textbf{0.579} & \em{0.579}\\
\addlinespace
\multicolumn{16}{c}{\textbf{DGP: EDCC-GARCH - Fitted model: DCC-GARCH - Portfolio weights: random}}\\
MSE & 1.000 & 0.696 & \em{0.588} & \textbf{0.588} & 0.589 & 1.000 & 0.679 & \em{0.606} & \textbf{0.606} & 0.606 & 1.000 & 0.662 & \em{0.590} & \textbf{0.590} & 0.590\\
MAE & 1.000 & 0.857 & \em{0.792} & \textbf{0.792} & 0.792 & 1.000 & 0.844 & \em{0.797} & \textbf{0.797} & 0.797 & 1.000 & 0.831 & \em{0.783} & \textbf{0.783} & 0.783\\
QLIKE & 1.000 & 0.692 & \em{0.632} & \textbf{0.632} & 0.632 & 1.000 & 0.672 & \em{0.628} & \textbf{0.628} & 0.628 & 1.000 & 0.647 & \em{0.595} & \textbf{0.594} & 0.595\\
\addlinespace
\multicolumn{16}{c}{\textbf{DGP: EDCC-GARCH - Fitted model: EDCC-GARCH - Portfolio weights: $1/N$}}\\
MSE & 1.000 & 0.431 & \em{0.309} & \textbf{0.309} & 0.310 & 1.000 & 0.246 & \textbf{0.206} & \em{0.206} & 0.206 & 1.000 & 0.145 & 0.135 & \em{0.135} & \textbf{0.135}\\
MAE & 1.000 & 0.647 & \em{0.565} & \textbf{0.565} & 0.566 & 1.000 & 0.493 & \em{0.459} & \textbf{0.459} & 0.459 & 1.000 & 0.380 & 0.371 & \textbf{0.371} & \em{0.371}\\
QLIKE & 1.000 & 0.380 & \em{0.311} & \textbf{0.311} & 0.311 & 1.000 & 0.221 & \em{0.198} & \textbf{0.198} & 0.198 & 1.000 & 0.134 & \em{0.130} & \textbf{0.130} & 0.130\\
\addlinespace
\multicolumn{16}{c}{\textbf{DGP: EDCC-GARCH - Fitted model: EDCC-GARCH - Portfolio weights: random}}\\
MSE & 1.000 & 0.483 & \em{0.336} & \textbf{0.336} & 0.338 & 1.000 & 0.268 & 0.223 & \em{0.222} & \textbf{0.222} & 1.000 & 0.153 & \textbf{0.141} & \em{0.141} & 0.141\\
MAE & 1.000 & 0.680 & \em{0.586} & \textbf{0.586} & 0.587 & 1.000 & 0.512 & \em{0.475} & \textbf{0.475} & 0.475 & 1.000 & 0.388 & \em{0.376} & \textbf{0.376} & 0.377\\
QLIKE & 1.000 & 0.425 & \em{0.335} & \textbf{0.335} & 0.335 & 1.000 & 0.241 & \em{0.213} & \textbf{0.213} & 0.213 & 1.000 & 0.139 & \em{0.134} & \textbf{0.134} & 0.134\\
\bottomrule
\end{tabular}
\end{center}
\caption{Average relative accuracy indexes where the reference forecast is the univariate GARCH fitted on simulated portfolio returns, and the DGP is a EDCC-GARCH. The fitted MGARCH models are: the Scalar BEKK  (first and second panels), the DCC-GARCH (third and fourth panels), the EDCC-GARCH (fifth and sixth panels). The top rows report the sample size ($T$) and the variance forecast approach, including the univariate GARCH on portfolio returns, \textit{base}, the bottom-up approach using MGARCH models, \textit{bu}, and the three forecast reconciliation cases discussed in the previous section. The first and third panels consider equally weighted portfolios (the $1/N$ case), while the second and fourth consider random portfolio weights. All values are averages across 500 replications. In each row, the best and second best forecast approaches are in bold and italic, respectively.}
\label{tab:res4}
\end{sidewaystable}

Re-running the simulations with $N=24$ assets yields patterns broadly consistent with the $N=9$ case when the DGP is a BEKK: the baseline univariate forecast (base) remains preferable to the bottom-up aggregation (bu), with the gap especially pronounced when the estimated multivariate model is a Scalar BEKK. By contrast, under the EDCC–GARCH DGP, moving to $N=24$ makes the univariate estimate outperform the bottom-up approach both when the multivariate estimator is Scalar BEKK and when it is a standard DCC–GARCH. Overall, higher dimensionality amplifies estimation risk and misspecification costs in parsimonious multivariate structures, keeping the univariate baseline a particularly competitive benchmark at $N=24$.

\subsubsection{Forecast reconciliation}

Tables from \ref{tab:res1} to \ref{tab:res4} contain the results of the three forecast reconciliation approaches that we consider. From the estimation of the correctly specified MGARCH model without interdependence (Table \ref{tab:res1}), we have a first interesting result: in terms of relative forecast accuracy indexes, forecast reconciliation always improves with respect to the bottom-up case, that is, the forecast from the correctly specified model. In addition, differences across forecast reconciliation methods are minimal, and not affected by the portfolio weights (fixed vs. random). Furthermore, in the DCC-GARCH case, we observe that the distance between the forecast reconciliation cases and the \textit{bu} tends to reduce with increasing sample size, an effect not present for the Scalar BEKK case; again, we link this to the aggregation results available for the latter model. The evidence in favor of forecast reconciliation methods is consistent with the literature \citep[see, e.g. ][]{Wickramasuriya2019}, showing that reconciliation enforces aggregation constraints and generally improves forecast accuracy, with differences in out-of-sample performance relative to the base forecasts reflecting mainly estimation error effects.

If we estimate a misspecified model (without interdependence) when the DGP does not include interdependence, the preference for forecast reconciliation methods is cleare and present under both the Scalar BEKK and DCC-GARCH data generating processes. In both cases, the preference for reconciliation methods slightly worsens with the increase in the sample size and, interesting, remains better than the \textit{base} case when a misspecified Scalar BEKK is estimated on a DCC-GARCH DGP. Similarly to the estimation of a correctly specified model, the three forecast reconciliation approaches are very close to each other.

{Moving to the case of DGPs with interdependence, the results differ according to the data generating model. If we simulate data from an EDCC-GARCH, forecast reconciliation generally improves the bottom-up approach, irrespective of the fitted model and of the portfolio weights; only in the case of a fitted Scalar BEKK and large sample size, reconciliation becomes closer to the univariate case (\textit{bu)}. Moreover, as in the previous cases, reconciliation methods are similar, with no clear preference for one of the approaches. Differently, if we simulate data from a BEKK, forecast reconciliation improves under MSE and MAE losses, with a gain that decreases slightly with sample size. However, under the QLIKE loss when estimating a Scalar BEKK or an EDCC, the forecast reconciliation seems to not improve over the use of a univariate model, even though, we must admit, the values of the loss functions are really close (\textit{bu} vs. forecast reconciliation). 

We consider now the case where we estimate a correctly specified model in the presence of interdependence, namely EDCC-GARCH.\footnote{We highlight that the BEKK model cannot be considered due to its large number of parameters, more than $200$ with $N=9$.} The \textit{bu} approach dominates on the univariate one, as in other correctly specified models (with preference becoming more clear with increasing $T$). Forecast reconciliation improves prediction accuracy, with a decreasing benefit for increasing sample size, and the three methods are extremely close to each other.

For $N=24$, the findings are largely consistent with the $N=9$ case. When the DGP does not present interdependence, even if the estimated multivariate model is correctly specified, forecast reconciliation still improves performance, and differences between reconciliation schemes remain small. With $N=24$ this preference is still strong, although slightly less pronounced than for $N=9$. Under interdependence, if the DGP is EDCC-GARCH and the multivariate estimator is DCC-GARCH, reconciliation yields clear gains; however, when the estimator is Scalar BEKK, the picture changes with respect to $N=9$: under the QLIKE loss, the univariate model is marginally preferred, even relative to reconciled forecasts. Under the Full BEKK DGP, the results for $N=9$ are confirmed: forecast reconciliation improves performance, with a slight advantage for the $shr_A$ method, while under the QLIKE loss, there is no significant gain when the multivariate estimation relies on a Scalar BEKK specification (the \textit{base} method performs comparable to reconciled forecasts). Detailed results are available in the Online Appendix.

\subsubsection{The impact of a noisy proxy}
In the previous sections we contrasted the prediction of the portfolio variance to the true value of the portfolio variance. However, the latter is not observed and is usually replaced by a proxy in empirical analyzes. The early literature on variance forecasting has focused on the use of squared observed returns to proxy for the latent conditional variance. Similarly, when multiple assets are analyzed, the returns cross-product proxies for the conditional covariance. Since early 2000's with the availability of high frequency data, realized variances and covariances, replaced the squared and cross-product of returns, being more precise proxies of the returns conditional variances and covariances. To assess the effect of using a proxy in evaluating the benefits of forecast reconciliation, we rely on the same simulation settings considered above. While the returns cross-product is readily available from our simulation, recovering high frequency data coherent with the existence of a given GARCH-type dynamic on the daily returns is complex and challenging. We chose a simplified approach, mimicking the reduction in the noise provided by the realized covariance estimators using a contaminated proxy. In detail, we generate a noisy proxy of the true covariance by contaminating the true covariance with the returns cross-product. Therefore, we define the proxy as follows:
\begin{equation}
\hat{\bm{\Sigma}}_t = \delta \boldsymbol r_t \boldsymbol r_t^{\prime} + \left(1-\delta\right) \bm{\Sigma}_t,
\end{equation}
with $\delta \in \left(0,1\right]$, and $\bm{\Sigma}_t$ the true covariance (obtained from the simulations), $\boldsymbol r_t$ the simulated daily returns. When $\delta=1$ we are setting the proxy to the cross-product of returns, the most \textit{noisy} proxy. On the other hand, with $\delta <1$ we are assuming the noise in the proxy is smaller, thus mimicking the improvement associated with the use of high frequency data. In the following, we set $\delta \in \left\lbrace 0.25, 0.5, 0.75, 1\right\rbrace$.

Given that we have four different $\delta$ values for each tuple of DGP, fitted model, forecast method and sample size, we graphically represent the average relative MSE; see Figures \ref{fig:plot1} to \ref{fig:plot4}. In all plots, the baseline case is the estimation of the portfolio variance using a univariate GARCH model; notably, this approach, for a given data generating process is not changing across fitted Multivariate GARCH models. Therefore, if it represents the reference forecasting approach, the relative indicators can be contrasted both across forecasting approaches and over models.

In Figures \ref{fig:plot1} and \ref{fig:plot2} the DGPs are without interdependence, and the patterns share some similarity: if the noise in the covariance proxy is increasing, the difference between adopted modeling choices and forecast methods (\textit{base}, \textit{bu} and reconciliation) tends to disappear for increasing sample sizes. In summary, if one believes that the data are not characterized by variance spillovers, then whatever the model and the forecast approach is used, the results will be very close to those of the correctly specified model and close to a univariate GARCH (all relative indicators are close to 1). Moreover, when the noise is smaller, forecast reconciliation benefits are limited, smaller than those observed under the true covariance, and might also disappear, such as in the case of Scalar BEKK. This shows that the improvement of forecast reconciliation might also be hidden by even a limited noise in the covariance proxy.

When moving to the DGPs with interdependence, the results differ. In the case of the EDCC, Figure \ref{fig:plot4}, when fitting the Scalar BEKK and when the noise in the proxy and the sample size both increases, the results are very close to a baseline univariate GARCH and in some cases even worse, with no effect of reconciliation. Unlikely, if a DCC-type model is adopted for portfolio variance forecasting, we do observe a (limited) reduction in the relative MSE compared to the baseline. Moreover, forecast reconciliation becomes comparable to the fit of a misspecified model. This suggests that the noise in the proxy also masks the possible presence of assets interdependence, making it difficult to identify the correct model (we remind that we can perform a comparison across models). In terms of forecasts precision, such a finding is clearly dependent on the strength of the relation across conditional variances. The values we considered in the simulations are in line with values recovered from the observed data. If the DGP is a Full BEKK, Figure \ref{fig:plot3}, the results show one relevant element, the multivariate models, the \textit{bu} forecasts, are the worst even if the noise in the proxy is not large, and the reconciliation improves the forecast quality. However, if the noise in the proxy increases, the fitted models and the forecast approaches tend to converge. Again, this suggests that  if we use a noisy covariance proxy the presence of interdependence in the data might not be detected and thus efficiently exploited in forecast construction; the final outcome seems a suggestion for the use of a simpler model (for instance a DCC) when in reality the data dynamic is much more intricate. Moreover, we confirm previous findings in the literature on the crucial role of using less noisy covariance proxies for model comparison in a forecasting framework. 

In the case $N=24$ ($T=1000$ only), Figure \ref{fig:plot_RV}, results observed under the Scalar BEKK and BEKK data generating processes are confirmed: in the SBEKK DGP case, correctly specified and misspecified models are really close one to the other and reconciliation is not improving over either a univariate fit or a bottom-up approach; for the full BEKK DGP, the noise is making the bottom-up the worst approach (due to misspecification) and forecast reconciliation is not adding much to univariate models. For the DCC DGP, forecast reconciliation is improving, with a more effective outcome in the correctly specified model case. Differently, for the EDCC, miss-specification matters, and both DCC and SBEKK are very close to the univariate case and slightly better than the bottom up, in particular for increasing noise in the proxy. This further confirms the importance of using a less noisy proxy for the covariance matrix when contrasting forecasting models and approaches, the potential advantage of forecast reconciliation, and the risk of not recognizing the existence and relevance of interdependence in multivariate GARCH models.

We close with a note regarding similar plots to those here reported and included in the Appendix for QLIKE and MAE. Interestingly, even if MAE is an inconsistent loss function for model ranking (see \cite{Laurentetal2013}), the results it provides are in line with those of both MSE. QLIKE provides results aligned with the other two indicators only in the case $N=9$. When the cross-sectional dimension increases, QLIKE shows larger preference for the univariate approach.

\begin{figure}[tb]
\centering
\includegraphics[width = \linewidth]{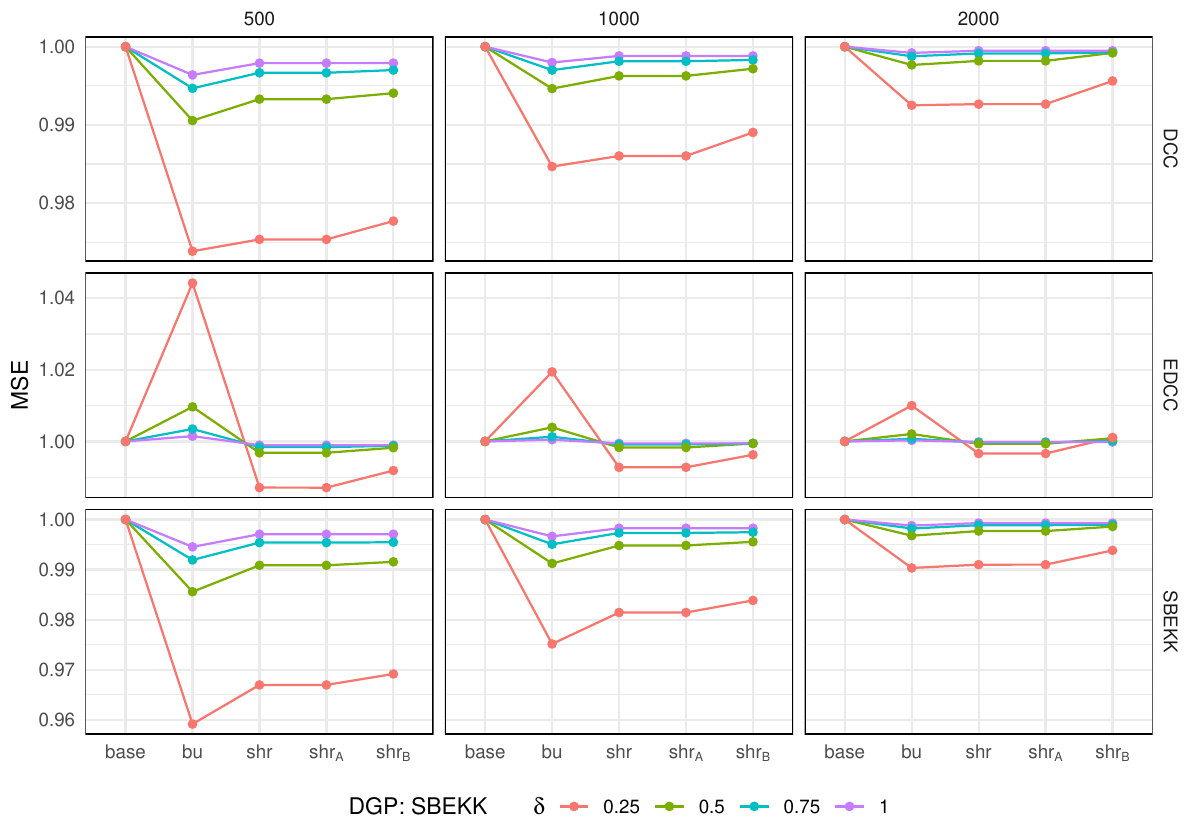}
\vspace*{-1cm}
\caption{Average relative MSE where the reference forecast is the univariate GARCH fitted on simulated portfolio returns with equally weighted portfolios (the $1/N$ case), and the DGP is a Scalar BEKK. The fitted MGARCH models are: the DCC-GARCH (first row, DCC) and the Scalar BEKK (second row, SBEKK). The columns indicates the sample size ($T$): $T=500$ left,  $T=1000$ center and $T=2000$ right column. All values are averages across the 500 experiments.}
\label{fig:plot1}
\end{figure}
\begin{figure}[tb]
\centering
\includegraphics[width = \linewidth]{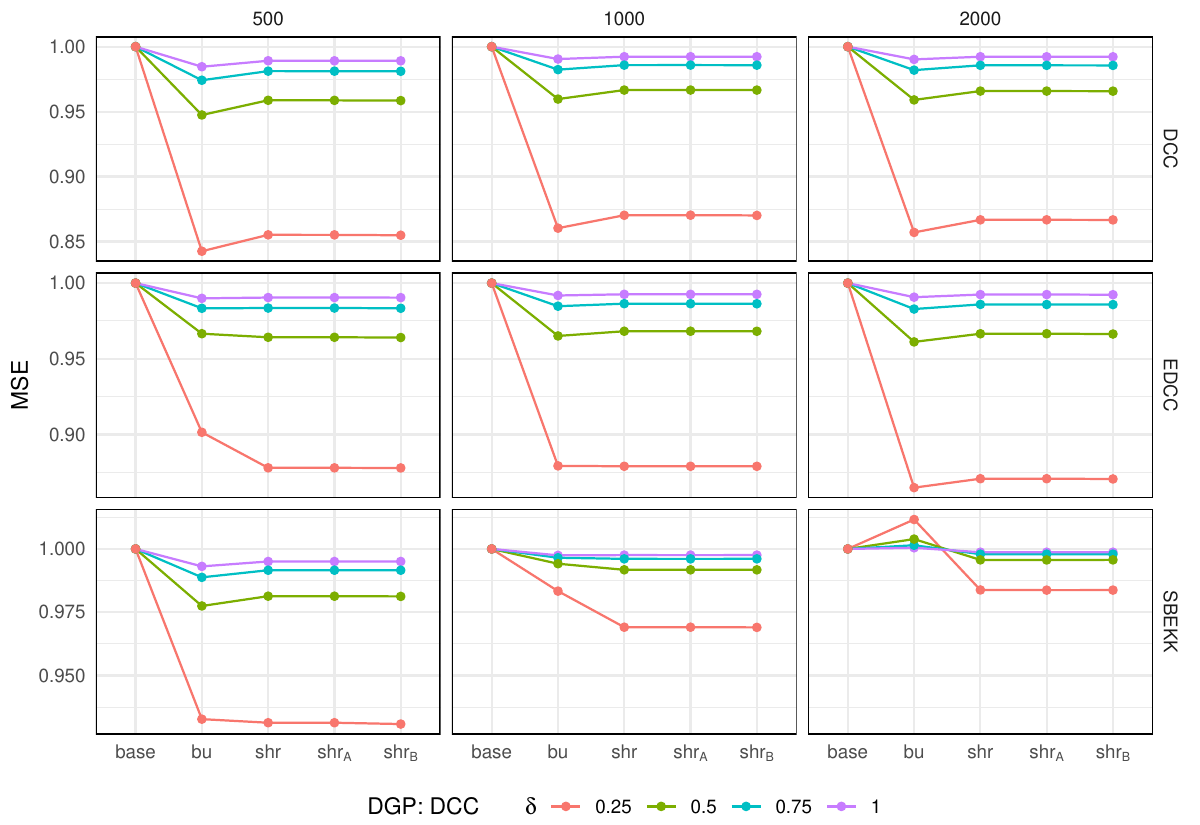}
\vspace*{-1cm}
\caption{Average relative MSE where the reference forecast is the univariate GARCH fitted on simulated portfolio returns with equally weighted portfolios (the $1/N$ case), and the DGP is a DCC-GARCH. The fitted MGARCH models are: the DCC-GARCH (first row, DCC) and the Scalar BEKK (second row, SBEKK). The columns indicates the sample size ($T$): $T=500$ left,  $T=1000$ center and $T=20000$ right column. All values are averages across the 500 experiments.}
\label{fig:plot2}
\end{figure}

\begin{figure}[tb]
\centering
\includegraphics[width = \linewidth]{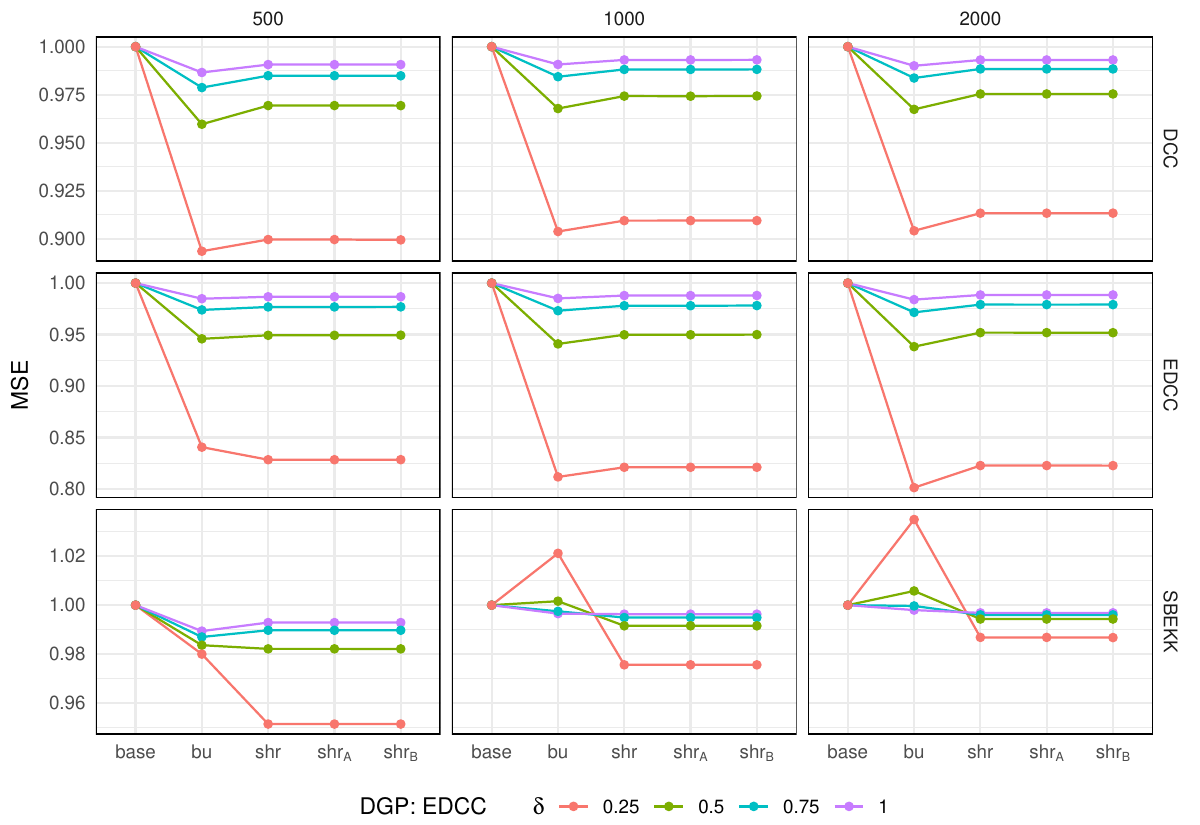}
\vspace*{-1cm}
\caption{Average relative MSE where the reference forecast is the univariate GARCH fitted on simulated portfolio returns with equally weighted portfolios (the $1/N$ case), and the DGP is a EDCC-GARCH. The fitted MGARCH models are: the DCC-GARCH (first row, DCC), the EDCC-GARCH (second row, EDCC) and the Scalar BEKK (third row, SBEKK). The columns indicates the sample size ($T$): $T=500$ left,  $T=1000$ center and $T=2000$ right column. All values are averages across the 500 experiments.}
\label{fig:plot4}
\end{figure}
\begin{figure}[tbp]
\centering
\includegraphics[width = \linewidth]{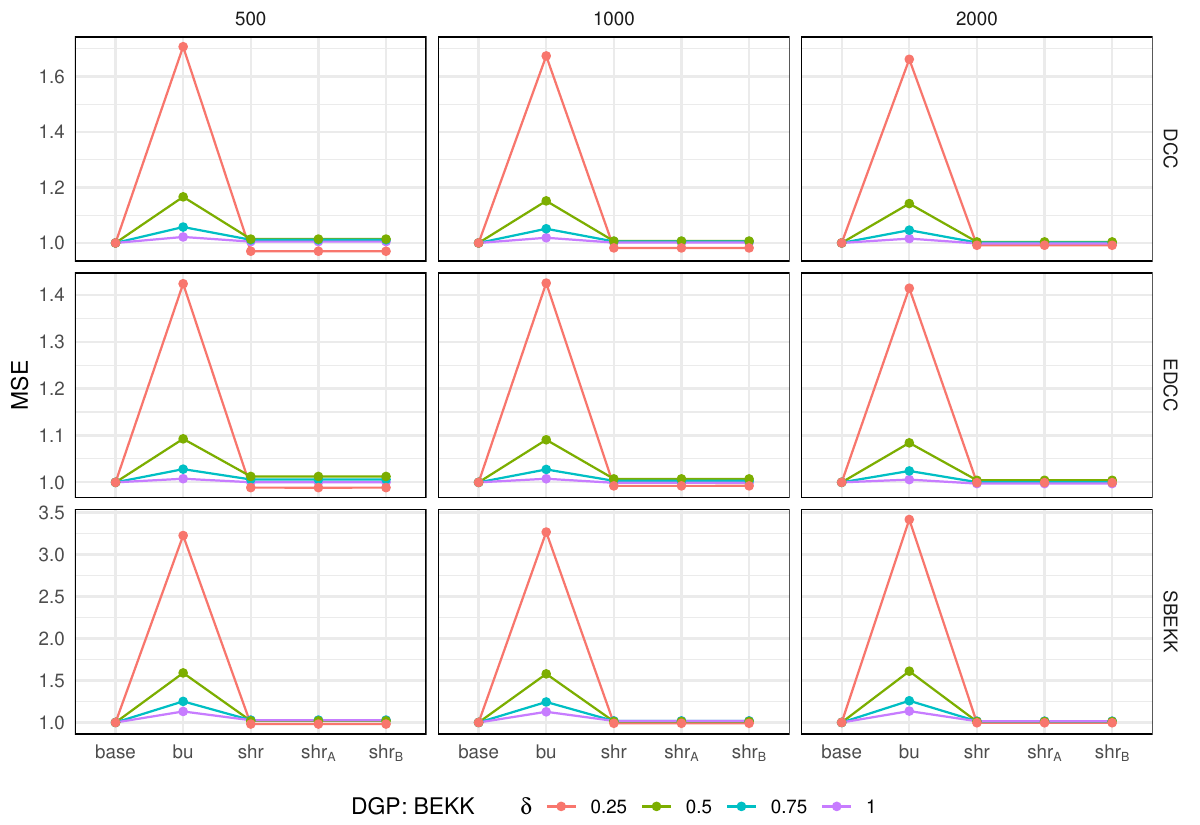}
\vspace*{-1cm}
\caption{Average relative MSE where the reference forecast is the univariate GARCH fitted on simulated portfolio returns with equally weighted portfolios (the $1/N$ case), and the DGP is a Full BEKK. The fitted MGARCH models are: the DCC-GARCH (first row, DCC) and the Scalar BEKK (second row, SBEKK). The columns indicates the sample size ($T$): $T=500$ left,  $T=1000$ center and $T=2000$ right column. All values are averages across the 500 experiments.}
\label{fig:plot3}
\end{figure}

\begin{figure}[tbp]
	\centering
	\includegraphics[width = \linewidth]{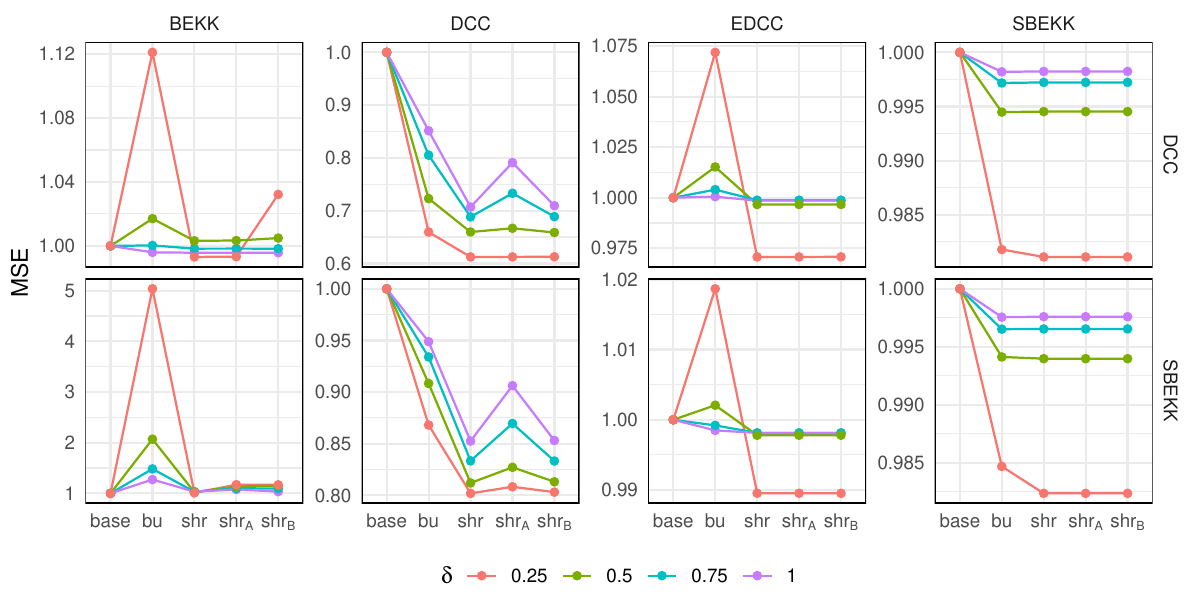}
	\vspace*{-1cm}
	\caption{Average relative MSE where the reference forecast is the univariate GARCH fitted on simulated portfolio returns with equally weighted portfolios (the $1/N$ case) for 24 assets, and the DGP is reported in each column (BEKK, DCC, EDCC and SBEKK). The fitted MGARCH models are: the DCC-GARCH (first row, DCC) and the Scalar BEKK (second row, SBEKK). All values are averages across the 500 experiments.}
	\label{fig:plot_RV}
\end{figure}

\subsection{Pairwise comparison}

We further evaluate differences in predictive accuracy using pairwise Diebold--Mariano tests. The Online Appendix reports summaries in Figures A2--A4, while the corresponding comparisons obtained using noisy covariance proxies are shown in Figures A5--A7.

The figures highlights meaningful differences across data-generating processes. When the DGP follows the Full BEKK specification (BKF), the bottom-up approach (\textit{bu}) tends to outperform the univariate
portfolio-based forecast (\textit{base}) in most pairwise comparisons. In contrast, under the DCC-GARCH generator (DCG) the opposite pattern typically emerges, with the \textit{base} forecast more frequently preferred to \textit{bu}. This advantage of the univariate approach becomes weaker when the multivariate model used for estimation differs from the true DGP, such as when Scalar BEKK (SCB) or DIG specifications are employed, although the \textit{base} forecast generally remains competitive. Under the Scalar BEKK DGP, the relative ranking between \textit{base} and \textit{bu} is less systematic and depends more strongly on the multivariate GARCH used. Across all DGPs, however, forecast reconciliation methods consistently match or outperform the best among the \textit{base} and \textit{bu} forecasts in pairwise comparisons. This result indicates that reconciliation effectively combines the information contained in the univariate and multivariate predictions.

When noisy proxies for the covariance matrix are used in the evaluation stage (Figures A5--A7), the differences between \textit{base} and \textit{bu} become less pronounced. Nevertheless, reconciliation methods remain competitive and often provide improvements relative to the individual forecasts, particularly when less stringent significance thresholds are considered.

\subsection{Model Confidence Set}

To complement the pairwise comparisons, we apply the Model Confidence Set (MCS) procedure of \cite{Hansen2005} with detailed results reported in the Online Appendix.

When the true covariance matrix is used for evaluation (Tables A3--A5), forecast reconciliation methods are included in the MCS with very high frequency across most simulation designs. In many cases the inclusion frequency of the reconciliation approaches exceeds 90\%, and often remains above 80\%. This pattern holds across sample sizes and portfolio weighting schemes. Among the reconciliation strategies, the differences are generally small: the shrinkage-based reconciled forecasts ($shr$ and $shr_A$) typically display very similar inclusion frequencies, while $shr_B$ tends to enter the confidence set slightly less often, although it remains comparable in magnitude.

In contrast, the bottom-up forecast (\textit{bu}) rarely belongs to the MCS in several simulation settings. This is particularly evident in designs based on the Full BEKK generator combined with DCC-type estimation, where the inclusion frequency of the bottom-up approach is close to zero in most cases. The univariate portfolio-based forecast (\textit{base}) performs better than the bottom-up alternative and is often included in the MCS, although it typically remains below the reconciliation methods.

Differences across data-generating processes nevertheless emerge. Under the BEKK-based generator, reconciliation clearly dominates both \textit{base} and \textit{bu}. For DCC-based generators the gap between reconciliation and the univariate approach becomes smaller, and the \textit{base} forecast often enters the confidence set with moderate frequency. Under the Scalar BEKK generator the relative ranking between \textit{base} and \textit{bu} becomes less systematic and depends more strongly on the multivariate model used for estimation.

The same qualitative conclusions hold when the number of assets is increased (Tables A11--A13). Forecast reconciliation remains the class of methods most frequently included in the MCS, confirming that combining univariate and multivariate information provides robust improvements in predictive accuracy.

When noisy covariance proxies are used for evaluation (Tables A7--A9), the evidence becomes less clear-cut. In these settings the inclusion frequencies of the different approaches become more similar, and both the univariate and multivariate forecasts appear more often in the confidence set. Nevertheless, reconciliation methods remain competitive and continue to belong to the MCS in a large fraction of the simulation designs.

\section{An example with real data} \label{sec:app}

We perform an empirical analysis using daily returns from 28 constituents of the Dow Jones Industrial Average index, which are continuously available from January 2, 2003, to December 31, 2024. Excluding weekends and bank holidays, the sample includes a total of 5,537 observations. For the same period, we also collect the daily Realized Covariance, based on 1-minute returns. All data is sourced from the \textit{Capire} database and is freely available at \url{capire.stat.unipd.it}.

Following the approach used in the simulations, we estimate both multivariate models on the set of the 28 assets as well as a univariate model on an equally weighted portfolio. The multivariate models we consider are: the DCC-GARCH, with univariate GARCH(1,1) as marginals; the Extended DCC-GARCH; the Scalar BEKK. In the univariate approach also considers a GARCH(1,1) model.

Estimates are obtained using a rolling window of the most recent 1,500 observations. All the models are re-estimated on the first day of each month, and the estimated coefficients are then used to generate variance forecasts with a horizon up to one month (22 observations), starting from each day of the month. Forecasts are produced on a one-step-ahead basis (i.e., fixed parameters within the month, but moving the information set on a daily basis).

Tables \ref{tab:appDCC} to \ref{tab:appEDC} report forecast evaluation results for a single model, contrasting competing forecast construction approaches, namely, the univariate GARCH on equally weighted portfolio returns (baseline), the bottom-up of the multivariate GARCH models, and the three forecast reconciliation cases. The tables provide comparisons across different forecasting horizons, one day, one week (5 days) and one month (22 days).

When forecasts are obtained by fitting a DCC model, the results differ across loss functions. In fact, for QLIKE the baseline univariate approach has the lowest loss function value and is always included in the confidence set; moreover, the \textit{base} approach is always statistically superior to the \textit{bottom-up} case (see the Diebold-Mariano test outcomes). This evidence is consistent across forecast horizons. Differently, for MAE and MSE the use of forecast reconciliation provides better results, even though the model confidence set always includes the univariate-based forecasts. The \textit{bottom-up} case is part of the confidence set only for the daily horizon. For the Scalar-BEKK model, Table \ref{tab:appSBE}, the results are more favorable to the univariate model (except for the MSE and one day horizon, where $shr_B$ is preferred). Notably, the MCS never includes the \textit{bottom-up} case (i.e., the direct use of the multivariate GARCH). Finally, when fitting the EDCC, the only model partially accounting for interdependence, while the loss function values suggest a preference for forecast reconciliation ($shr$) in most cases (QLIKE is tilted toward the univariate approach for weekly and monthly horizons, while MAE suggest \textit{bottom-up} at the weekly horizon), the MCS includes most forecast approaches, and the Diebold-Mariano tests suggest, with larger frequencies, that the \textit{bottom-up} approach is inferior to the others. These results show some heterogeneity, and are focuses on a single model, across forecasting approaches. Taking a more general viewpoint, and applying Model Confidence Set across models and forecasts, using a 20\% threshold for exclusion from the set, a more clear picture emerges, see Table \ref{tab:appMCS}. The only model which is almost always included in the confidence set is the EDCC (it is excluded under the QLIKE function for weekly and monthly horizons), while the most excluded model is the Scalar BEKK. Forecast reconciliation approaches when they are excluded from the confidence set, they have larger p-values compared to the \textit{bottom-up} cases. Under QLIKE the univariate approach seems the best at the monthly horizon, while for other horizons and for the other loss functions, the confidence set always include the EDCC under forecast reconciliation ($shr$ case). In summary, the use of forecast reconciliation in general improves the forecast ability of MGARCH models, in particular, in situations where we cannot exclude the existence of interdependence and the proxy of the covariance matrix is not too noisy.

\begin{sidewaystable}
	\centering
	\resizebox{\linewidth}{!}{
		
\begin{tabular}[t]{>{}l|cccc>{}c|cccc>{}c|ccccc}
\toprule
\multicolumn{1}{c}{\textbf{ }} & \multicolumn{5}{c}{\textbf{Day}} & \multicolumn{5}{c}{\textbf{Week}} & \multicolumn{5}{c}{\textbf{Month}} \\
\cmidrule(l{3pt}r{3pt}){2-6} \cmidrule(l{3pt}r{3pt}){7-11} \cmidrule(l{3pt}r{3pt}){12-16}
 & $base$ & $bu$ & $shr$ & $shr_{A}$ & $shr_{B}$ & $base$ & $bu$ & $shr$ & $shr_{A}$ & $shr_{B}$ & $base$ & $bu$ & $shr$ & $shr_{A}$ & $shr_{B}$\\
\midrule
\addlinespace[0.3em]
\multicolumn{16}{c}{\textbf{Base forecasts: DCC-GARCH}}\\
\hspace{1em}\textbf{MSE} & 0.416 & \em{0.317} & \textbf{0.307} & \textbf{0.307} & \textbf{0.307} & 1.141 & \em{1.117} & \textbf{1.064} & \textbf{1.064} & \textbf{1.064} & \textbf{1.474} & 1.666 & \em{1.547} & \em{1.547} & \em{1.547}\\
\hspace{1em}AvgRelMSE$_{base}$ & 1.000 & \em{0.761} & \textbf{0.738} & \textbf{0.738} & \textbf{0.738} & 1.000 & \em{0.980} & \textbf{0.933} & \textbf{0.933} & \textbf{0.933} & \textbf{1.000} & 1.130 & \em{1.050} & \em{1.050} & \em{1.050}\\
\hspace{1em}AvgRelMSE$_{bu}$ & 1.314 & \em{1.000} & \textbf{0.969} & \textbf{0.969} & \textbf{0.969} & 1.021 & \em{1.000} & \textbf{0.952} & \textbf{0.952} & \textbf{0.952} & \textbf{0.885} & 1.000 & \em{0.928} & \em{0.928} & \em{0.928}\\
\hspace{1em}$p$-value $dm_{base}$ &  & 0.196 & 0.099 & 0.099 & 0.099 &  & 0.408 & 0.123 & 0.123 & 0.123 &  & 0.992 & 0.934 & 0.934 & 0.934\\
\hspace{1em}$p$-value $dm_{bu}$ & 0.804 &  & 0.401 & 0.401 & 0.401 & 0.592 &  & 0.082 & 0.082 & 0.082 & 0.008 &  & {\footnotesize $<0.001$} & {\footnotesize $<0.001$} & {\footnotesize $<0.001$}\\
\hspace{1em}$p$-value MCS & 0.479 & 0.681 & 1.000 & 1.000 & 1.000 & 0.838 & 0.838 & 1.000 & 1.000 & 1.000 & 1.000 & 0.309 & 0.547 & 0.547 & 0.547\\
\midrule
\hspace{1em}\textbf{MAE} & 0.343 & \em{0.329} & \textbf{0.328} & \textbf{0.328} & \textbf{0.328} & 0.391 & \em{0.381} & \textbf{0.370} & \textbf{0.370} & \textbf{0.370} & \em{0.381} & 0.397 & \textbf{0.377} & \textbf{0.377} & \textbf{0.377}\\
\hspace{1em}AvgRelMAE$_{base}$ & 1.000 & \em{0.958} & \textbf{0.955} & \textbf{0.955} & \textbf{0.955} & 1.000 & \em{0.975} & \textbf{0.944} & \textbf{0.944} & \textbf{0.944} & \em{1.000} & 1.042 & \textbf{0.987} & \textbf{0.987} & \textbf{0.987}\\
\hspace{1em}AvgRelMAE$_{bu}$ & 1.044 & \em{1.000} & \textbf{0.997} & \textbf{0.997} & \textbf{0.997} & 1.026 & \em{1.000} & \textbf{0.969} & \textbf{0.969} & \textbf{0.969} & \em{0.960} & 1.000 & \textbf{0.948} & \textbf{0.948} & \textbf{0.948}\\
\hspace{1em}$p$-value $dm_{base}$ &  & 0.335 & 0.219 & 0.219 & 0.219 &  & 0.258 & 0.009 & 0.009 & 0.009 &  & 0.992 & 0.122 & 0.122 & 0.122\\
\hspace{1em}$p$-value $dm_{bu}$ & 0.665 &  & 0.476 & 0.476 & 0.476 & 0.742 &  & 0.054 & 0.054 & 0.054 & 0.008 &  & {\footnotesize $<0.001$} & {\footnotesize $<0.001$} & {\footnotesize $<0.001$}\\
\hspace{1em}$p$-value MCS & 0.709 & 0.932 & 1.000 & 1.000 & 1.000 & 0.577 & 0.577 & 1.000 & 1.000 & 1.000 & 0.586 & 0.118 & 1.000 & 1.000 & 1.000\\
\midrule
\hspace{1em}\textbf{QLIKE} & \textbf{0.181} & 0.211 & \em{0.182} & \em{0.182} & \em{0.182} & \textbf{0.203} & 0.271 & \em{0.226} & \em{0.226} & \em{0.226} & \textbf{0.202} & 0.275 & \em{0.226} & \em{0.226} & \em{0.226}\\
\hspace{1em}AvgRelQLIKE$_{base}$ & \textbf{1.000} & 1.166 & \em{1.005} & \em{1.005} & \em{1.005} & \textbf{1.000} & 1.337 & \em{1.115} & \em{1.115} & \em{1.115} & \textbf{1.000} & 1.361 & \em{1.122} & \em{1.122} & \em{1.122}\\
\hspace{1em}AvgRelQLIKE$_{bu}$ & \textbf{0.858} & 1.000 & \em{0.862} & \em{0.862} & \em{0.862} & \textbf{0.748} & 1.000 & \em{0.834} & \em{0.834} & \em{0.834} & \textbf{0.735} & 1.000 & \em{0.824} & \em{0.824} & \em{0.824}\\
\hspace{1em}$p$-value $dm_{base}$ &  & 0.968 & 0.542 & 0.542 & 0.542 &  & 1.000 & 0.999 & 0.999 & 0.999 &  & 1.000 & 1.000 & 1.000 & 1.000\\
\hspace{1em}$p$-value $dm_{bu}$ & 0.032 &  & {\footnotesize $<0.001$} & {\footnotesize $<0.001$} & {\footnotesize $<0.001$} & {\footnotesize $<0.001$} &  & {\footnotesize $<0.001$} & {\footnotesize $<0.001$} & {\footnotesize $<0.001$} & {\footnotesize $<0.001$} &  & {\footnotesize $<0.001$} & {\footnotesize $<0.001$} & {\footnotesize $<0.001$}\\
\hspace{1em}$p$-value MCS & 1.000 & 0.003 & 0.907 & 0.907 & 0.907 & 1.000 & {\footnotesize $<0.001$} & 0.025 & 0.025 & 0.025 & 1.000 & {\footnotesize $<0.001$} & {\footnotesize $<0.001$} & {\footnotesize $<0.001$} & {\footnotesize $<0.001$}\\
\bottomrule
\end{tabular}

	}
	\caption{Forecast evaluation of DCC–GARCH base forecasts and reconciliation schemes across forecast horizons (Day, Week, Month). The table reports the Mean Squared Error (MSE), Mean Absolute Error (MAE), and QLIKE loss functions, together with average relative performance measures (AvgRelMSE, AvgRelMAE, AvgRelQLIKE) computed with respect to the base and bottom-up (\textit{bu}) benchmarks. The Diebold–Mariano ($p$-value $dm$) and Model Confidence Set ($p$-value MCS$)$ statistics are reported to examine the statistical significance of forecast differentials. The best result within each row is shown in bold, and the second-best in italics.}
	\label{tab:appDCC}
\end{sidewaystable}

\begin{sidewaystable}
	\centering
	\resizebox{\linewidth}{!}{
	
\begin{tabular}[t]{>{}l|cccc>{}c|cccc>{}c|ccccc}
\toprule
\multicolumn{1}{c}{\textbf{ }} & \multicolumn{5}{c}{\textbf{Day}} & \multicolumn{5}{c}{\textbf{Week}} & \multicolumn{5}{c}{\textbf{Month}} \\
\cmidrule(l{3pt}r{3pt}){2-6} \cmidrule(l{3pt}r{3pt}){7-11} \cmidrule(l{3pt}r{3pt}){12-16}
 & $base$ & $bu$ & $shr$ & $shr_{A}$ & $shr_{B}$ & $base$ & $bu$ & $shr$ & $shr_{A}$ & $shr_{B}$ & $base$ & $bu$ & $shr$ & $shr_{A}$ & $shr_{B}$\\
\midrule
\addlinespace[0.3em]
\multicolumn{16}{c}{\textbf{Base forecasts: SBEKK-GARCH}}\\
\hspace{1em}\textbf{MSE} & \em{0.416} & 0.563 & 0.418 & 0.421 & \textbf{0.416} & \textbf{1.141} & 1.392 & 1.142 & 1.143 & \em{1.141} & \textbf{1.474} & 2.127 & 1.632 & 1.632 & \em{1.631}\\
\hspace{1em}AvgRelMSE$_{base}$ & \em{1.000} & 1.353 & 1.004 & 1.011 & \textbf{0.998} & \textbf{1.000} & 1.220 & 1.001 & 1.002 & \em{1.001} & \textbf{1.000} & 1.443 & 1.107 & 1.107 & \em{1.107}\\
\hspace{1em}AvgRelMSE$_{bu}$ & \em{0.739} & 1.000 & 0.742 & 0.747 & \textbf{0.738} & \textbf{0.820} & 1.000 & 0.821 & 0.821 & \em{0.820} & \textbf{0.693} & 1.000 & 0.767 & 0.767 & \em{0.767}\\
\hspace{1em}$p$-value $dm_{base}$ &  & 0.894 & 0.512 & 0.535 & 0.495 &  & 0.974 & 0.514 & 0.521 & 0.505 &  & 1.000 & 0.998 & 0.998 & 0.998\\
\hspace{1em}$p$-value $dm_{bu}$ & 0.106 &  & 0.038 & 0.041 & 0.036 & 0.026 &  & 0.003 & 0.003 & 0.003 & {\footnotesize $<0.001$} &  & {\footnotesize $<0.001$} & {\footnotesize $<0.001$} & {\footnotesize $<0.001$}\\
\hspace{1em}$p$-value MCS & 0.986 & 0.008 & 0.946 & 0.806 & 1.000 & 1.000 & 0.140 & 0.975 & 0.940 & 0.993 & 1.000 & 0.082 & 0.281 & 0.281 & 0.285\\
\midrule
\hspace{1em}\textbf{MAE} & \textbf{0.343} & 0.492 & \em{0.365} & 0.368 & 0.365 & \textbf{0.391} & 0.559 & \em{0.405} & 0.407 & 0.406 & \textbf{0.381} & 0.571 & 0.408 & \em{0.408} & 0.409\\
\hspace{1em}AvgRelMAE$_{base}$ & \textbf{1.000} & 1.435 & \em{1.062} & 1.073 & 1.063 & \textbf{1.000} & 1.429 & \em{1.035} & 1.039 & 1.038 & \textbf{1.000} & 1.497 & 1.070 & \em{1.070} & 1.073\\
\hspace{1em}AvgRelMAE$_{bu}$ & \textbf{0.697} & 1.000 & \em{0.740} & 0.748 & 0.741 & \textbf{0.700} & 1.000 & \em{0.724} & 0.727 & 0.726 & \textbf{0.668} & 1.000 & 0.715 & \em{0.715} & 0.717\\
\hspace{1em}$p$-value $dm_{base}$ &  & 1.000 & 0.941 & 0.966 & 0.943 &  & 1.000 & 0.981 & 0.989 & 0.987 &  & 1.000 & 1.000 & 1.000 & 1.000\\
\hspace{1em}$p$-value $dm_{bu}$ & {\footnotesize $<0.001$} &  & {\footnotesize $<0.001$} & {\footnotesize $<0.001$} & {\footnotesize $<0.001$} & {\footnotesize $<0.001$} &  & {\footnotesize $<0.001$} & {\footnotesize $<0.001$} & {\footnotesize $<0.001$} & {\footnotesize $<0.001$} &  & {\footnotesize $<0.001$} & {\footnotesize $<0.001$} & {\footnotesize \vphantom{1} $<0.001$}\\
\hspace{1em}$p$-value MCS & 1.000 & {\footnotesize $<0.001$} & 0.047 & 0.008 & 0.044 & 1.000 & {\footnotesize $<0.001$} & 0.198 & 0.139 & 0.151 & 1.000 & {\footnotesize $<0.001$} & 0.001 & 0.001 & {\footnotesize $<0.001$}\\
\midrule
\hspace{1em}\textbf{QLIKE} & \textbf{0.181} & 0.326 & 0.191 & 0.215 & \em{0.187} & \textbf{0.203} & 0.413 & \em{0.232} & 0.247 & 0.232 & \textbf{0.202} & 0.449 & \em{0.238} & 0.247 & 0.238\\
\hspace{1em}AvgRelQLIKE$_{base}$ & \textbf{1.000} & 1.804 & 1.054 & 1.189 & \em{1.036} & \textbf{1.000} & 2.037 & \em{1.143} & 1.218 & 1.143 & \textbf{1.000} & 2.225 & \em{1.177} & 1.223 & 1.182\\
\hspace{1em}AvgRelQLIKE$_{bu}$ & \textbf{0.554} & 1.000 & 0.584 & 0.659 & \em{0.574} & \textbf{0.491} & 1.000 & \em{0.561} & 0.598 & 0.561 & \textbf{0.449} & 1.000 & \em{0.529} & 0.550 & 0.531\\
\hspace{1em}$p$-value $dm_{base}$ &  & 1.000 & 0.837 & 0.984 & 0.731 &  & 1.000 & 1.000 & 1.000 & 1.000 &  & 1.000 & 1.000 & 1.000 & 1.000\\
\hspace{1em}$p$-value $dm_{bu}$ & {\footnotesize $<0.001$} &  & {\footnotesize $<0.001$} & {\footnotesize $<0.001$} & {\footnotesize $<0.001$} & {\footnotesize $<0.001$} &  & {\footnotesize $<0.001$} & {\footnotesize $<0.001$} & {\footnotesize $<0.001$} & {\footnotesize $<0.001$} &  & {\footnotesize $<0.001$} & {\footnotesize $<0.001$} & {\footnotesize $<0.001$}\\
\hspace{1em}$p$-value MCS & 1.000 & {\footnotesize $<0.001$} & 0.199 & 0.083 & 0.554 & 1.000 & {\footnotesize $<0.001$} & 0.011 & 0.006 & 0.011 & 1.000 & 0.001 & 0.001 & 0.001 & 0.001\\
\bottomrule
\end{tabular}

}
	\caption{Forecast evaluation of SBEKK–GARCH base forecasts and reconciliation schemes across forecast horizons (Day, Week, Month). The table reports the Mean Squared Error (MSE), Mean Absolute Error (MAE), and QLIKE loss functions, together with average relative performance measures (AvgRelMSE, AvgRelMAE, AvgRelQLIKE) computed with respect to the base and bottom-up (\textit{bu}) benchmarks. The Diebold–Mariano ($p$-value $dm$) and Model Confidence Set ($p$-value MCS$)$ statistics are reported to examine the statistical significance of forecast differentials. The best result within each row is shown in bold, and the second-best in italics.}
	\label{tab:appSBE}
\end{sidewaystable}

\begin{sidewaystable}
	\centering
	\resizebox{\linewidth}{!}{
		
\begin{tabular}[t]{>{}l|cccc>{}c|cccc>{}c|ccccc}
\toprule
\multicolumn{1}{c}{\textbf{ }} & \multicolumn{5}{c}{\textbf{Day}} & \multicolumn{5}{c}{\textbf{Week}} & \multicolumn{5}{c}{\textbf{Month}} \\
\cmidrule(l{3pt}r{3pt}){2-6} \cmidrule(l{3pt}r{3pt}){7-11} \cmidrule(l{3pt}r{3pt}){12-16}
 & $base$ & $bu$ & $shr$ & $shr_{A}$ & $shr_{B}$ & $base$ & $bu$ & $shr$ & $shr_{A}$ & $shr_{B}$ & $base$ & $bu$ & $shr$ & $shr_{A}$ & $shr_{B}$\\
\midrule
\addlinespace[0.3em]
\multicolumn{16}{c}{\textbf{Base forecasts: EDCC-GARCH}}\\
\hspace{1em}\textbf{MSE} & 0.416 & 0.254 & \textbf{0.243} & 0.246 & \em{0.244} & 1.141 & 0.985 & \textbf{0.983} & 0.984 & \em{0.983} & 1.474 & 1.350 & \textbf{1.349} & 1.350 & \em{1.350}\\
\hspace{1em}AvgRelMSE$_{base}$ & 1.000 & 0.611 & \textbf{0.584} & 0.590 & \em{0.585} & 1.000 & 0.863 & \textbf{0.862} & 0.862 & \em{0.862} & 1.000 & 0.916 & \textbf{0.916} & 0.916 & \em{0.916}\\
\hspace{1em}AvgRelMSE$_{bu}$ & 1.637 & 1.000 & \textbf{0.956} & 0.966 & \em{0.958} & 1.159 & 1.000 & \textbf{0.999} & 0.999 & \em{0.999} & 1.091 & 1.000 & \textbf{0.999} & 0.999 & \em{0.999}\\
\hspace{1em}$p$-value $dm_{base}$ &  & 0.074 & 0.054 & 0.056 & 0.054 &  & 0.017 & 0.009 & 0.010 & 0.009 &  & 0.011 & 0.004 & 0.004 & 0.004\\
\hspace{1em}$p$-value $dm_{bu}$ & 0.926 &  & 0.087 & 0.146 & 0.092 & 0.983 &  & 0.435 & 0.463 & 0.439 & 0.989 &  & 0.448 & 0.455 & 0.449\\
\hspace{1em}$p$-value MCS & 0.250 & 0.325 & 1.000 & 0.383 & 0.383 & 0.306 & 0.964 & 1.000 & 0.964 & 0.964 & 0.156 & 0.935 & 1.000 & 0.935 & 0.935\\
\midrule
\hspace{1em}\textbf{MAE} & 0.343 & 0.306 & \textbf{0.299} & 0.301 & \em{0.299} & 0.391 & \textbf{0.344} & \em{0.345} & 0.345 & 0.345 & 0.381 & 0.360 & \textbf{0.359} & 0.359 & \em{0.359}\\
\hspace{1em}AvgRelMAE$_{base}$ & 1.000 & 0.891 & \textbf{0.870} & 0.877 & \em{0.872} & 1.000 & \textbf{0.879} & \em{0.881} & 0.882 & 0.882 & 1.000 & 0.945 & \textbf{0.941} & 0.942 & \em{0.941}\\
\hspace{1em}AvgRelMAE$_{bu}$ & 1.123 & 1.000 & \textbf{0.977} & 0.985 & \em{0.979} & 1.137 & \textbf{1.000} & \em{1.002} & 1.004 & 1.003 & 1.058 & 1.000 & \textbf{0.996} & 0.996 & \em{0.996}\\
\hspace{1em}$p$-value $dm_{base}$ &  & 0.114 & 0.069 & 0.077 & 0.070 &  & {\footnotesize $<0.001$} & {\footnotesize $<0.001$} & {\footnotesize $<0.001$} & {\footnotesize $<0.001$} &  & {\footnotesize $<0.001$} & {\footnotesize $<0.001$} & {\footnotesize $<0.001$} & {\footnotesize $<0.001$}\\
\hspace{1em}$p$-value $dm_{bu}$ & 0.886 &  & 0.033 & 0.098 & 0.037 & 1.000 &  & 0.658 & 0.734 & 0.674 & 1.000 &  & 0.045 & 0.058 & 0.048\\
\hspace{1em}$p$-value MCS & 0.270 & 0.270 & 1.000 & 0.383 & 0.383 & 0.061 & 1.000 & 0.852 & 0.695 & 0.832 & 0.104 & 0.469 & 1.000 & 0.469 & 0.469\\
\midrule
\hspace{1em}\textbf{QLIKE} & 0.181 & 0.183 & \textbf{0.176} & 0.177 & \em{0.176} & \textbf{0.203} & 0.214 & \em{0.208} & 0.209 & 0.209 & \textbf{0.202} & 0.221 & \em{0.215} & 0.215 & 0.215\\
\hspace{1em}AvgRelQLIKE$_{base}$ & 1.000 & 1.011 & \textbf{0.972} & 0.977 & \em{0.972} & \textbf{1.000} & 1.056 & \em{1.027} & 1.028 & 1.027 & \textbf{1.000} & 1.095 & \em{1.064} & 1.065 & 1.064\\
\hspace{1em}AvgRelQLIKE$_{bu}$ & 0.989 & 1.000 & \textbf{0.961} & 0.966 & \em{0.961} & \textbf{0.947} & 1.000 & \em{0.973} & 0.973 & 0.973 & \textbf{0.913} & 1.000 & \em{0.972} & 0.972 & 0.972\\
\hspace{1em}$p$-value $dm_{base}$ &  & 0.556 & 0.348 & 0.375 & 0.350 &  & 0.942 & 0.812 & 0.820 & 0.813 &  & 1.000 & 1.000 & 1.000 & 1.000\\
\hspace{1em}$p$-value $dm_{bu}$ & 0.444 &  & {\footnotesize $<0.001$} & 0.003 & {\footnotesize $<0.001$} & 0.058 &  & {\footnotesize $<0.001$} & {\footnotesize $<0.001$} & {\footnotesize $<0.001$} & {\footnotesize $<0.001$} &  & {\footnotesize $<0.001$} & {\footnotesize $<0.001$} & {\footnotesize $<0.001$}\\
\hspace{1em}$p$-value MCS & 0.679 & 0.356 & 1.000 & 0.679 & 0.679 & 1.000 & 0.038 & 0.469 & 0.426 & 0.465 & 1.000 & {\footnotesize $<0.001$} & 0.028 & 0.026 & 0.028\\
\bottomrule
\end{tabular}

	}
	\caption{Forecast evaluation of EDCC–GARCH base forecasts and reconciliation schemes across forecast horizons (Day, Week, Month). The table reports the Mean Squared Error (MSE), Mean Absolute Error (MAE), and QLIKE loss functions, together with average relative performance measures (AvgRelMSE, AvgRelMAE, AvgRelQLIKE) computed with respect to the base and bottom-up (\textit{bu}) benchmarks. The Diebold–Mariano ($p$-value $dm$) and Model Confidence Set ($p$-value MCS$)$ statistics are reported to examine the statistical significance of forecast differentials. The best result within each row is shown in bold, and the second-best in italics.}
	\label{tab:appEDC}
\end{sidewaystable}

\begin{table}
	\resizebox{\linewidth}{!}{
	
\begin{tabular}[t]{l>{}l|cc>{}c|cc>{}c|ccc}
\toprule
\multicolumn{2}{c}{\textbf{ }} & \multicolumn{3}{c}{\textbf{MSE}} & \multicolumn{3}{c}{\textbf{MAE}} & \multicolumn{3}{c}{\textbf{QLIKE}} \\
\cmidrule(l{3pt}r{3pt}){3-5} \cmidrule(l{3pt}r{3pt}){6-8} \cmidrule(l{3pt}r{3pt}){9-11}
\textbf{Base forecasts} & \textbf{Approach} & Day & Week & Month & Day & Week & Month & Day & Week & Month\\
\midrule
GARCH & $base$ & 0.423 & 0.306 & 0.325 & 0.300 & \textit{0.061} & \textit{0.127} & 0.832 & 1.000 & 1.000\\
\addlinespace
EDCC-GARCH & $bu$ & 0.423 & 0.964 & 0.935 & 0.300 & 1.000 & 0.469 & 0.832 & \textit{0.085} & \textit{0.006}\\
EDCC-GARCH & $shr$ & 1.000 & 1.000 & 1.000 & 1.000 & 0.852 & 1.000 & 1.000 & 0.469 & \textit{0.028}\\
EDCC-GARCH & $shr_{B}$ & 0.423 & 0.964 & 0.935 & 0.383 & 0.832 & 0.469 & 0.832 & 0.465 & \textit{0.028}\\
EDCC-GARCH & $shr_{A}$ & 0.423 & 0.964 & 0.935 & 0.383 & 0.695 & 0.469 & 0.832 & 0.426 & \textit{0.026}\\
\addlinespace
DCC-GARCH & $bu$ & 0.423 & 0.305 & 0.216 & 0.300 & \textit{0.025} & \textit{0.006} & \textit{0.017} & \textit{0.001} & \textit{0.001}\\
DCC-GARCH & $shr$ & 0.423 & 0.305 & 0.314 & 0.300 & \textit{0.045} & \textit{0.127} & 0.832 & \textit{0.085} & \textit{0.005}\\
DCC-GARCH & $shr_{B}$ & 0.423 & 0.305 & 0.318 & 0.300 & \textit{0.045} & \textit{0.127} & 0.832 & \textit{0.085} & \textit{0.005}\\
DCC-GARCH & $shr_{A}$ & 0.423 & 0.305 & 0.325 & 0.300 & \textit{0.045} & \textit{0.127} & 0.832 & \textit{0.085} & \textit{0.006}\\
\addlinespace
SBEKK-GARCH & $bu$ & \textit{0.017} & \textit{0.066} & \textit{0.138} & \textit{{\footnotesize $<0.001$}} & \textit{{\footnotesize $<0.001$}} & \textit{{\footnotesize $<0.001$}} & \textit{{\footnotesize $<0.001$}} & \textit{{\footnotesize $<0.001$}} & \textit{0.001}\\
SBEKK-GARCH & $shr$ & 0.252 & 0.253 & \textit{0.163} & \textit{0.073} & \textit{0.023} & \textit{0.002} & 0.332 & \textit{0.085} & \textit{0.001}\\
SBEKK-GARCH & $shr_{B}$ & 0.255 & 0.263 & \textit{0.163} & \textit{0.060} & \textit{0.017} & \textit{0.001} & 0.563 & \textit{0.085} & \textit{0.001}\\
SBEKK-GARCH & $shr_{A}$ & 0.218 & 0.230 & \textit{0.162} & \textit{0.073} & \textit{0.020} & \textit{0.001} & \textit{0.155} & \textit{0.007} & \textit{0.001}\\
\bottomrule
\end{tabular}

	}
	\caption{Model Confidence Set ($p$-value MCS$)$ for all the forecasting approach across forecast horizons (Day, Week, Month) and loss functions (MSE, MAE, QLIKE).}
	\label{tab:appMCS}
\end{table}

\section{Conclusion}

Through simulations, we show that forecast reconciliation approaches might improve forecast accuracy of portfolio risk when the underlying model is a Multivariate GARCH, possibly including interdependence. Our results also shed light on the role exerted by noisy proxies in forecast evaluation, highlighting the fundamental impact of the noise that could mask the possible presence of interdependence across conditional variances and covariances. An empirical example illustrates the use of reconciliation with real data and a proxy based on realized covariances.

Our conclusions provide a first look at the role of forecast reconciliation in risk management settings, where the focus moves toward quantiles and conditional expectations. Although current research is moving in that direction, we believe that this topic represents an interesting area for future developments.

\bibliographystyle{elsarticle-harv} 
\bibliography{biblio_CGL}

\end{document}


\title{\textbf{Online appendix:}\\Multivariate GARCH and portfolio variance prediction: a forecast reconciliation perspective}
\bigskip
\bigskip

\author{Massimiliano Caporin\thanks{Corresponding author: Department of Statistical Sciences, University of Padova, Via C. Battisti 241, 35121 Padova, Italy - email: massimiliano.caporin@unipd.it - phone: +39-049-827-4199.}\\
Department of Statistical Sciences\\
University of Padova, Italy\\
massimiliano.caporin@unipd.it\\ 
and\\
Daniele Girolimetto\\
Department of Statistical Sciences\\
University of Padova, Italy\\
daniele.girolimetto@unipd.it\\
and\\
Emanuele Lopetuso\\
Department of Statistical Sciences\\
University of Padova, Italy\\
emanuele.lopetuso@unipd.it}
\maketitle

\tableofcontents
\newpage

\section{Simulation results}

\begin{table}[H]
	\centering
\caption{Notation used in the simulations, including portfolio weighting schemes, MGARCH models, and forecasting approaches.}
	\label{tab:notationModel}
\begin{tabular}{cp{0.75\linewidth}}
		\toprule
		\textbf{Notation} & \textbf{Description} \\
		\midrule
		
		\multicolumn{2}{l}{Portfolio weighting scheme:}\\
		EQ & Portfolio weights equal to $1/N$\\
		RW & Random portfolio weights\\
		
		\addlinespace
		\multicolumn{2}{l}{MGARCH models for the DGP and estimation:}\\
		BKF & Full BEKK model\\
		DCG & Dynamic Conditional Correlation GARCH model\\
		EDCC & Extended Dynamic Conditional Correlation GARCH model\\
		SCB & Scalar BEKK GARCH model\\
		
		\addlinespace
		\multicolumn{2}{l}{Forecasting approaches:}\\
		$base$ & Univariate GARCH fitted on portfolio returns\\
		$bu$ & Bottom-up forecast obtained from the multivariate covariance matrix\\
		$shr$ & Forecast reconciliation using shrinkage estimator\\
		$shr_A$ & Reconciliation with non-linear constrained optimization\\
		$shr_B$ & Reconciliation based on correlation decomposition\\
		
		\bottomrule
	\end{tabular}
\end{table}

\subsection{True portfolio variance and covariance matrix of $N=9$ assets}

\begin{landscape}\centering
\begingroup
\setlength{\LTcapwidth}{\linewidth}
\setlength{\tabcolsep}{4pt}
\footnotesize

\begin{longtable}[t]{>{}c||>{}c>{}c|cc>{}c||>{}c>{}c|cc>{}c||>{}c>{}c|ccc}
\caption{Average accuracy indices across different simulation settings. Each setting is identified by a label composed of three elements: (i) the data-generating process (DGP), (ii) the MGARCH model used for estimation, and (iii) the portfolio weighting scheme (see \autoref{tab:notationModel} for details). The top rows indicate the sample size ($T$) and the variance forecasting method employed: the univariate GARCH on portfolio returns (base), the bottom-up approach using MGARCH models (bu), and the three forecast reconciliation strategies discussed in the paper ($shr$, $shr_{A}$, $shr_{B}$). The reported accuracy measures are grouped into three categories: average accuracy indices (MSE, MAE, QLIKE); average relative measures (AvgRelMSE, AvgRelMAE, AvgRelQLIKE) using the univariate and the multivariate GARCH forecast as the benchmark.}\\
\toprule
\multicolumn{1}{c}{ } & \multicolumn{5}{c}{$T =500$} & \multicolumn{5}{c}{$T =1000$} & \multicolumn{5}{c}{$T =2000$} \\
\cmidrule(l{3pt}r{3pt}){2-6} \cmidrule(l{3pt}r{3pt}){7-11} \cmidrule(l{3pt}r{3pt}){12-16}
Index & $base$ & $bu$ & $shr$ & $shr_{A}$ & $shr_{B}$ & $base$ & $bu$ & $shr$ & $shr_{A}$ & $shr_{B}$ & $base$ & $bu$ & $shr$ & $shr_{A}$ & $shr_{B}$\\
\midrule
\endfirsthead
\caption[]{Average accuracy indices across different simulation settings. Each setting is identified by a label composed of three elements: (i) the data-generating process (DGP), (ii) the MGARCH model used for estimation, and (iii) the portfolio weighting scheme (see \autoref{tab:notationModel} for details). The top rows indicate the sample size ($T$) and the variance forecasting method employed: the univariate GARCH on portfolio returns (base), the bottom-up approach using MGARCH models (bu), and the three forecast reconciliation strategies discussed in the paper ($shr$, $shr_{A}$, $shr_{B}$). The reported accuracy measures are grouped into three categories: average accuracy indices (MSE, MAE, QLIKE); average relative measures (AvgRelMSE, AvgRelMAE, AvgRelQLIKE) using the univariate and the multivariate GARCH forecast as the benchmark. \textit{(continued)}}\\
\toprule
\multicolumn{1}{c}{ } & \multicolumn{5}{c}{$T =500$} & \multicolumn{5}{c}{$T =1000$} & \multicolumn{5}{c}{$T =2000$} \\
\cmidrule(l{3pt}r{3pt}){2-6} \cmidrule(l{3pt}r{3pt}){7-11} \cmidrule(l{3pt}r{3pt}){12-16}
Index & $base$ & $bu$ & $shr$ & $shr_{A}$ & $shr_{B}$ & $base$ & $bu$ & $shr$ & $shr_{A}$ & $shr_{B}$ & $base$ & $bu$ & $shr$ & $shr_{A}$ & $shr_{B}$\\
\midrule
\endhead

\endfoot
\bottomrule
\endlastfoot
\addlinespace[0.5em]
\multicolumn{16}{c}{\textbf{BKF -- DCC -- EQ}}\\
\nopagebreak \addlinespace[0em]
\multicolumn{16}{l}{\textit{Average indexes}}\\
\nopagebreak MSE & 129.132 & 744.274 & \em{52.480} & \textbf{52.458} & 52.510 & 172.623 & 1445.672 & \em{94.363} & \textbf{93.321} & 96.412 & 128.564 & 1408.928 & \em{56.598} & \textbf{56.473} & 56.673\\
\nopagebreak 
MAE & 3.383 & 8.029 & \em{2.502} & \textbf{2.501} & 2.502 & 2.835 & 7.573 & \em{2.333} & \textbf{2.332} & 2.335 & 2.670 & 8.128 & \em{2.245} & \textbf{2.244} & 2.246\\
\nopagebreak 
QLIKE & 0.012 & 0.072 & 0.011 & \em{0.011} & \textbf{0.011} & 0.008 & 0.062 & 0.008 & \em{0.008} & \textbf{0.008} & 0.006 & 0.055 & 0.006 & \textbf{0.006} & \em{0.006}\\
\nopagebreak 
\addlinespace[0em]
\multicolumn{16}{l}{\textit{Relative indexes (base benchmark)}}\\
\nopagebreak AvgRelMSE & 1.000 & 7.054 & \em{0.557} & \textbf{0.557} & 0.558 & 1.000 & 8.168 & \em{0.637} & \textbf{0.637} & 0.638 & 1.000 & 8.968 & \em{0.689} & \textbf{0.689} & 0.690\\
\nopagebreak 
AvgRelMAE & 1.000 & 2.700 & \em{0.813} & \textbf{0.813} & 0.813 & 1.000 & 2.930 & \em{0.858} & \textbf{0.858} & 0.858 & 1.000 & 3.103 & \em{0.885} & \textbf{0.885} & 0.886\\
\end{longtable}

\endgroup
\end{landscape}

\begin{landscape}\centering
	\begingroup
	\setlength{\LTcapwidth}{\linewidth}
	\footnotesize
    
\begin{longtable}[t]{>{}c||>{}c>{}c|cc>{}c||>{}c>{}c|cc>{}c||>{}c>{}c|ccc}
\caption{The number of times (in \%) that model is in the Model Confidence Set with different thresholds ($\gamma \in \{70\%, 75\%, 80\%, 85\%m 90\%, 95\%\}$) across different simulation settings using the MAE as loss function. Each setting is identified by a label composed of three elements: (i) the data-generating process (DGP), (ii) the MGARCH model used for estimation, and (iii) the portfolio weighting scheme (see \autoref{tab:notationModel} for details). The sample size ($T$) is also reported for each setting.  The top rows indicate the noisy level of the proxy $\delta \in \left\lbrace 0.25, 0.5, 0.75, 1\right\rbrace$ and the variance forecasting method employed: the univariate GARCH on portfolio returns (base), the bottom-up approach using MGARCH models (bu), and the three forecast reconciliation strategies discussed in the paper ($shr$, $shr_{A}$, $shr_{B}$).}\\
\toprule
\multicolumn{1}{c}{ } & \multicolumn{5}{c}{$T =500$} & \multicolumn{5}{c}{$T =1000$} & \multicolumn{5}{c}{$T =2000$} \\
\cmidrule(l{3pt}r{3pt}){2-6} \cmidrule(l{3pt}r{3pt}){7-11} \cmidrule(l{3pt}r{3pt}){12-16}
Threshold & $base$ & $bu$ & $shr$ & $shr_{A}$ & $shr_{B}$ & $base$ & $bu$ & $shr$ & $shr_{A}$ & $shr_{B}$ & $base$ & $bu$ & $shr$ & $shr_{A}$ & $shr_{B}$\\
\midrule
\endfirsthead
\caption[]{The number of times (in \%) that model is in the Model Confidence Set with different thresholds ($\gamma \in \{70\%, 75\%, 80\%, 85\%m 90\%, 95\%\}$) across different simulation settings using the MAE as loss function. Each setting is identified by a label composed of three elements: (i) the data-generating process (DGP), (ii) the MGARCH model used for estimation, and (iii) the portfolio weighting scheme (see \autoref{tab:notationModel} for details). The sample size ($T$) is also reported for each setting.  The top rows indicate the noisy level of the proxy $\delta \in \left\lbrace 0.25, 0.5, 0.75, 1\right\rbrace$ and the variance forecasting method employed: the univariate GARCH on portfolio returns (base), the bottom-up approach using MGARCH models (bu), and the three forecast reconciliation strategies discussed in the paper ($shr$, $shr_{A}$, $shr_{B}$). \textit{(continued)}}\\
\toprule
\multicolumn{1}{c}{ } & \multicolumn{5}{c}{$T =500$} & \multicolumn{5}{c}{$T =1000$} & \multicolumn{5}{c}{$T =2000$} \\
\cmidrule(l{3pt}r{3pt}){2-6} \cmidrule(l{3pt}r{3pt}){7-11} \cmidrule(l{3pt}r{3pt}){12-16}
Threshold & $base$ & $bu$ & $shr$ & $shr_{A}$ & $shr_{B}$ & $base$ & $bu$ & $shr$ & $shr_{A}$ & $shr_{B}$ & $base$ & $bu$ & $shr$ & $shr_{A}$ & $shr_{B}$\\
\midrule
\endhead

\endfoot
\bottomrule
\endlastfoot
\addlinespace[0.5em]
\multicolumn{16}{c}{\textbf{BKF -- DCC -- EQ}}\\
\nopagebreak 70 & 45.4 & 0.6 & 83.4 & \em{83.6} & \textbf{84.0} & 44.8 & 0.4 & \em{84.4} & \textbf{85.6} & \em{84.4} & 43.6 & 0.0 & 81.6 & \textbf{83.4} & \em{83.2}\\
\nopagebreak 
75 & 46.8 & 0.6 & \em{84.6} & \em{84.6} & \textbf{85.4} & 47.0 & 0.4 & 85.6 & \textbf{86.6} & \em{86.2} & 47.2 & 0.0 & 84.2 & \em{85.0} & \textbf{85.6}\\
\nopagebreak 
80 & 49.4 & 1.0 & \em{86.6} & 86.2 & \textbf{87.6} & 50.6 & 0.4 & \em{86.8} & \textbf{87.8} & \textbf{87.8} & 49.8 & 0.0 & 85.6 & \em{86.0} & \textbf{87.0}\\
\nopagebreak 
85 & 52.2 & 1.0 & \em{88.8} & 88.4 & \textbf{89.4} & 54.4 & 0.6 & \textbf{90.0} & \em{89.6} & \textbf{90.0} & 53.6 & 0.0 & \em{87.6} & 87.4 & \textbf{88.6}\\
\nopagebreak 
90 & 56.6 & 1.4 & \em{91.0} & \textbf{91.2} & \em{91.0} & \em{59.0} & 0.6 & \textbf{91.4} & \textbf{91.4} & \textbf{91.4} & 57.8 & 0.0 & 89.6 & \em{89.8} & \textbf{90.6}\\
\nopagebreak 
95 & 64.0 & 3.0 & \em{92.2} & \em{92.2} & \textbf{92.4} & 65.4 & 1.6 & \em{93.0} & \em{93.0} & \textbf{93.2} & 64.6 & 1.0 & \textbf{92.2} & \textbf{92.2} & \em{92.0}\\
 
\addlinespace[0.5em]
\multicolumn{16}{c}{\textbf{BKF -- DCC -- RW}}\\
\end{longtable}

\endgroup
\end{landscape}

\begin{landscape}\centering
	\begingroup
	\setlength{\LTcapwidth}{\linewidth}
	\footnotesize
    
\begin{longtable}[t]{>{}c||>{}c>{}c|cc>{}c||>{}c>{}c|cc>{}c||>{}c>{}c|ccc}
\caption{The number of times (in \%) that model is in the Model Confidence Set with different thresholds ($\gamma \in \{70\%, 75\%, 80\%, 85\%m 90\%, 95\%\}$) across different simulation settings using the MSE as loss function. Each setting is identified by a label composed of three elements: (i) the data-generating process (DGP), (ii) the MGARCH model used for estimation, and (iii) the portfolio weighting scheme (see \autoref{tab:notationModel} for details). The sample size ($T$) is also reported for each setting.  The top rows indicate the noisy level of the proxy $\delta \in \left\lbrace 0.25, 0.5, 0.75, 1\right\rbrace$ and the variance forecasting method employed: the univariate GARCH on portfolio returns (base), the bottom-up approach using MGARCH models (bu), and the three forecast reconciliation strategies discussed in the paper ($shr$, $shr_{A}$, $shr_{B}$).}\\
\toprule
\multicolumn{1}{c}{ } & \multicolumn{5}{c}{$T =500$} & \multicolumn{5}{c}{$T =1000$} & \multicolumn{5}{c}{$T =2000$} \\
\cmidrule(l{3pt}r{3pt}){2-6} \cmidrule(l{3pt}r{3pt}){7-11} \cmidrule(l{3pt}r{3pt}){12-16}
Threshold & $base$ & $bu$ & $shr$ & $shr_{A}$ & $shr_{B}$ & $base$ & $bu$ & $shr$ & $shr_{A}$ & $shr_{B}$ & $base$ & $bu$ & $shr$ & $shr_{A}$ & $shr_{B}$\\
\midrule
\endfirsthead
\caption[]{The number of times (in \%) that model is in the Model Confidence Set with different thresholds ($\gamma \in \{70\%, 75\%, 80\%, 85\%m 90\%, 95\%\}$) across different simulation settings using the MSE as loss function. Each setting is identified by a label composed of three elements: (i) the data-generating process (DGP), (ii) the MGARCH model used for estimation, and (iii) the portfolio weighting scheme (see \autoref{tab:notationModel} for details). The sample size ($T$) is also reported for each setting.  The top rows indicate the noisy level of the proxy $\delta \in \left\lbrace 0.25, 0.5, 0.75, 1\right\rbrace$ and the variance forecasting method employed: the univariate GARCH on portfolio returns (base), the bottom-up approach using MGARCH models (bu), and the three forecast reconciliation strategies discussed in the paper ($shr$, $shr_{A}$, $shr_{B}$). \textit{(continued)}}\\
\toprule
\multicolumn{1}{c}{ } & \multicolumn{5}{c}{$T =500$} & \multicolumn{5}{c}{$T =1000$} & \multicolumn{5}{c}{$T =2000$} \\
\cmidrule(l{3pt}r{3pt}){2-6} \cmidrule(l{3pt}r{3pt}){7-11} \cmidrule(l{3pt}r{3pt}){12-16}
Threshold & $base$ & $bu$ & $shr$ & $shr_{A}$ & $shr_{B}$ & $base$ & $bu$ & $shr$ & $shr_{A}$ & $shr_{B}$ & $base$ & $bu$ & $shr$ & $shr_{A}$ & $shr_{B}$\\
\midrule
\endhead

\endfoot
\bottomrule
\endlastfoot
\addlinespace[0.5em]
\multicolumn{16}{c}{\textbf{BKF -- DCC -- EQ}}\\
\nopagebreak 70 & 33.4 & 0.8 & \em{87.4} & 87.0 & \textbf{87.8} & 35.6 & 0.6 & 87.0 & \em{88.2} & \textbf{89.0} & 33.4 & 0.6 & \em{88.0} & \textbf{88.4} & \textbf{88.4}\\
\nopagebreak 
75 & 35.6 & 0.8 & \em{88.2} & 87.8 & \textbf{88.4} & 38.8 & 0.8 & \em{89.2} & \em{89.2} & \textbf{90.4} & 37.4 & 0.6 & 89.0 & \em{89.4} & \textbf{90.4}\\
\nopagebreak 
80 & 39.4 & 1.2 & \em{90.2} & 89.0 & \textbf{90.6} & 41.4 & 0.8 & \em{90.6} & 90.4 & \textbf{92.2} & 40.8 & 0.6 & \em{90.2} & 90.0 & \textbf{90.8}\\
\nopagebreak 
85 & 43.2 & 1.6 & \textbf{91.4} & \em{90.0} & \textbf{91.4} & 44.8 & 1.0 & \em{91.8} & 91.6 & \textbf{93.0} & 45.8 & 1.0 & 91.8 & \em{92.0} & \textbf{92.2}\\
\nopagebreak 
90 & 52.8 & 4.0 & \textbf{93.4} & 92.8 & \em{93.0} & 52.6 & 4.8 & \em{93.6} & 93.2 & \textbf{93.8} & 54.6 & 2.8 & \textbf{93.2} & \em{92.8} & \textbf{93.2}\\
\nopagebreak 
95 & \em{65.8} & 22.0 & \textbf{94.6} & \textbf{94.6} & \textbf{94.6} & \em{66.8} & 20.6 & \textbf{95.0} & \textbf{95.0} & \textbf{95.0} & 68.2 & 22.2 & \textbf{94.4} & \textbf{94.4} & \em{94.2}\\
 
\addlinespace[0.5em]
\multicolumn{16}{c}{\textbf{BKF -- DCC -- RW}}\\
\end{longtable}

	\endgroup
\end{landscape}

\begin{landscape}\centering
	\begingroup
	\setlength{\LTcapwidth}{\linewidth}
	\footnotesize
    
\begin{longtable}[t]{>{}c||>{}c>{}c|cc>{}c||>{}c>{}c|cc>{}c||>{}c>{}c|ccc}
\caption{The number of times (in \%) that model is in the Model Confidence Set with different thresholds ($\gamma \in \{70\%, 75\%, 80\%, 85\%m 90\%, 95\%\}$) across different simulation settings using the QLIKE as loss function. Each setting is identified by a label composed of three elements: (i) the data-generating process (DGP), (ii) the MGARCH model used for estimation, and (iii) the portfolio weighting scheme (see \autoref{tab:notationModel} for details). The sample size ($T$) is also reported for each setting.  The top rows indicate the noisy level of the proxy $\delta \in \left\lbrace 0.25, 0.5, 0.75, 1\right\rbrace$ and the variance forecasting method employed: the univariate GARCH on portfolio returns (base), the bottom-up approach using MGARCH models (bu), and the three forecast reconciliation strategies discussed in the paper ($shr$, $shr_{A}$, $shr_{B}$).}\\
\toprule
\multicolumn{1}{c}{ } & \multicolumn{5}{c}{$T =500$} & \multicolumn{5}{c}{$T =1000$} & \multicolumn{5}{c}{$T =2000$} \\
\cmidrule(l{3pt}r{3pt}){2-6} \cmidrule(l{3pt}r{3pt}){7-11} \cmidrule(l{3pt}r{3pt}){12-16}
Threshold & $base$ & $bu$ & $shr$ & $shr_{A}$ & $shr_{B}$ & $base$ & $bu$ & $shr$ & $shr_{A}$ & $shr_{B}$ & $base$ & $bu$ & $shr$ & $shr_{A}$ & $shr_{B}$\\
\midrule
\endfirsthead
\caption[]{The number of times (in \%) that model is in the Model Confidence Set with different thresholds ($\gamma \in \{70\%, 75\%, 80\%, 85\%m 90\%, 95\%\}$) across different simulation settings using the QLIKE as loss function. Each setting is identified by a label composed of three elements: (i) the data-generating process (DGP), (ii) the MGARCH model used for estimation, and (iii) the portfolio weighting scheme (see \autoref{tab:notationModel} for details). The sample size ($T$) is also reported for each setting.  The top rows indicate the noisy level of the proxy $\delta \in \left\lbrace 0.25, 0.5, 0.75, 1\right\rbrace$ and the variance forecasting method employed: the univariate GARCH on portfolio returns (base), the bottom-up approach using MGARCH models (bu), and the three forecast reconciliation strategies discussed in the paper ($shr$, $shr_{A}$, $shr_{B}$). \textit{(continued)}}\\
\toprule
\multicolumn{1}{c}{ } & \multicolumn{5}{c}{$T =500$} & \multicolumn{5}{c}{$T =1000$} & \multicolumn{5}{c}{$T =2000$} \\
\cmidrule(l{3pt}r{3pt}){2-6} \cmidrule(l{3pt}r{3pt}){7-11} \cmidrule(l{3pt}r{3pt}){12-16}
Threshold & $base$ & $bu$ & $shr$ & $shr_{A}$ & $shr_{B}$ & $base$ & $bu$ & $shr$ & $shr_{A}$ & $shr_{B}$ & $base$ & $bu$ & $shr$ & $shr_{A}$ & $shr_{B}$\\
\midrule
\endhead

\endfoot
\bottomrule
\endlastfoot
\addlinespace[0.5em]
\multicolumn{16}{c}{\textbf{BKF -- DCC -- EQ}}\\
\nopagebreak 70 & 56.2 & 0.4 & 65.2 & \textbf{66.8} & \em{66.2} & 59.0 & 0.4 & 68.4 & \em{69.0} & \textbf{69.2} & 55.2 & 0.0 & \em{72.8} & \em{72.8} & \textbf{73.2}\\
\nopagebreak 
75 & 57.8 & 0.8 & 67.8 & \textbf{69.2} & \em{68.4} & 59.6 & 0.4 & 71.2 & \em{71.6} & \textbf{72.2} & 56.8 & 0.0 & 74.2 & \em{74.6} & \textbf{75.0}\\
\nopagebreak 
80 & 59.8 & 1.0 & 71.0 & \textbf{72.0} & \em{71.4} & 61.6 & 0.6 & 73.2 & \em{73.6} & \textbf{74.2} & 58.6 & 0.0 & 77.0 & \em{77.4} & \textbf{77.6}\\
\nopagebreak 
85 & 61.0 & 1.2 & 74.4 & \textbf{75.0} & \em{74.6} & 63.8 & 0.8 & 76.6 & \em{77.0} & \textbf{77.2} & 60.2 & 0.0 & 79.4 & \em{79.6} & \textbf{79.8}\\
\nopagebreak 
90 & 63.2 & 1.8 & \em{77.6} & \textbf{77.8} & \em{77.6} & 65.8 & 0.8 & 79.0 & \textbf{79.4} & \em{79.2} & 62.4 & 0.2 & \em{81.2} & \textbf{81.4} & \textbf{81.4}\\
\nopagebreak 
95 & \em{66.4} & 2.8 & \textbf{82.8} & \textbf{82.8} & \textbf{82.8} & \em{69.0} & 1.0 & \textbf{83.2} & \textbf{83.2} & \textbf{83.2} & \em{66.0} & 0.2 & \textbf{85.8} & \textbf{85.8} & \textbf{85.8}\\
 
\addlinespace[0.5em]
\multicolumn{16}{c}{\textbf{BKF -- DCC -- RW}}\\
\end{longtable}

	\endgroup
\end{landscape}

\begin{figure}[H]
	\centering
	\includegraphics[width=0.9\linewidth]{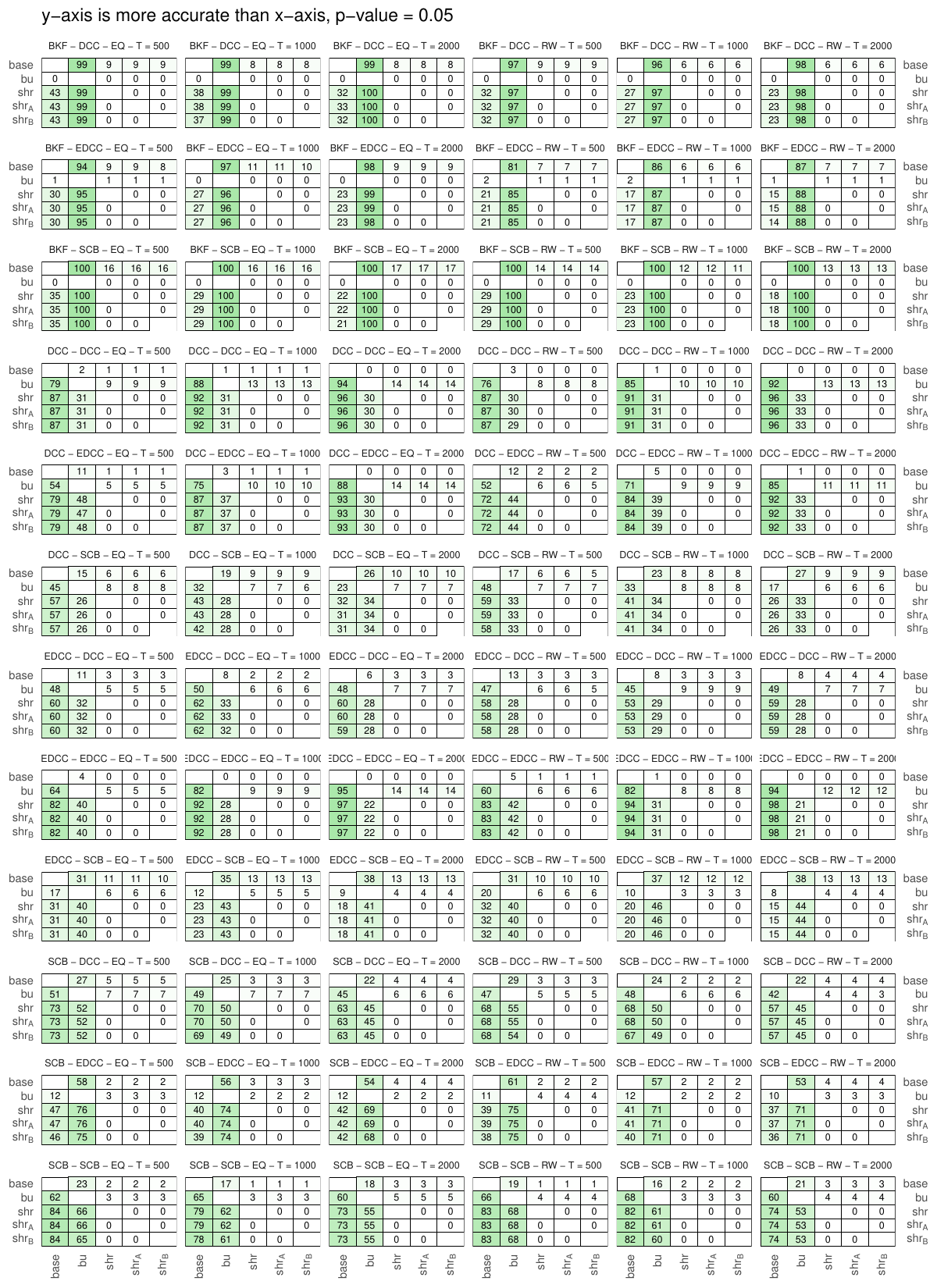}
	\caption{Qualitative evaluation using the Diebold-Mariano across different simulation settings using an absolute loss function. Each setting is identified by a label composed of three elements: (i) the data-generating process (DGP), (ii) the MGARCH model used for estimation, and (iii) the portfolio weighting scheme (see \autoref{tab:notationModel} for details). The sample size ($T$) is also reported. The rows and columns indicate the variance forecasting method employed: the univariate GARCH on portfolio returns (base), the bottom-up approach using MGARCH models (bu), and the three forecast reconciliation strategies discussed in the paper ($shr$, $shr_{A}$, $shr_{B}$). Each cell reports the number of times (in \%) the forecasting model in the row statistically outperforms (p-values $< 0.05$ and using Bonferroni correction) the model in the column.}
\end{figure}
	
\begin{figure}[H]
	\centering
	\includegraphics[width=0.9\linewidth]{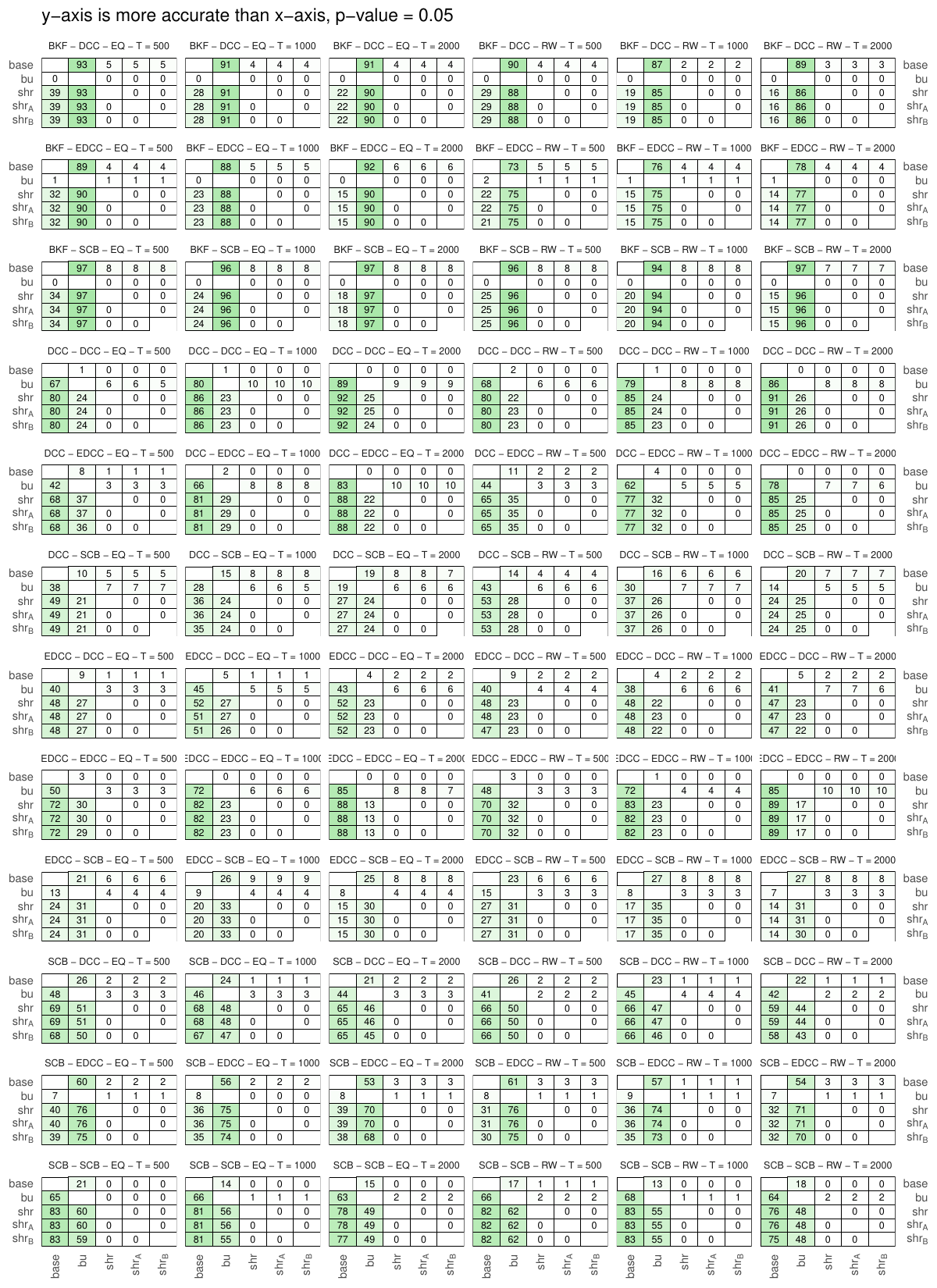}
	\caption{Qualitative evaluation using the Diebold-Mariano across different simulation settings using a square loss function. Each setting is identified by a label composed of three elements: (i) the data-generating process (DGP), (ii) the MGARCH model used for estimation, and (iii) the portfolio weighting scheme (see \autoref{tab:notationModel} for details). The sample size ($T$) is also reported. The rows and columns indicate the variance forecasting method employed: the univariate GARCH on portfolio returns (base), the bottom-up approach using MGARCH models (bu), and the three forecast reconciliation strategies discussed in the paper ($shr$, $shr_{A}$, $shr_{B}$). Each cell reports the number of times (in \%) the forecasting model in the row statistically outperforms (p-values $< 0.05$ and using Bonferroni correction) the model in the column.}
\end{figure}

\begin{figure}[H]
	\centering
	\includegraphics[width=0.9\linewidth]{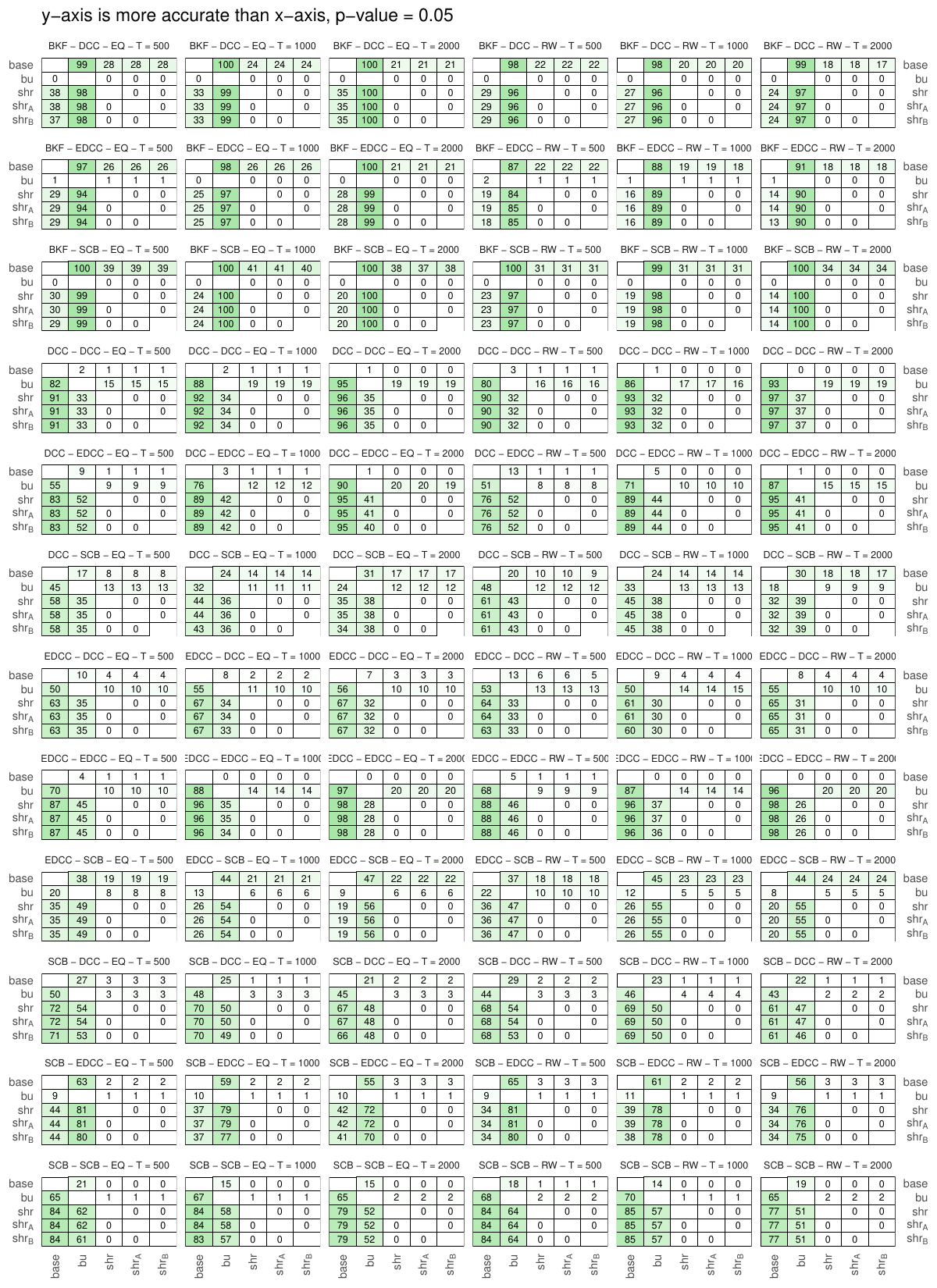}
	\caption{Qualitative evaluation using the Diebold-Mariano across different simulation settings using a QLIKE loss function. Each setting is identified by a label composed of three elements: (i) the data-generating process (DGP), (ii) the MGARCH model used for estimation, and (iii) the portfolio weighting scheme (see \autoref{tab:notationModel} for details). The sample size ($T$) is also reported. The rows and columns indicate the variance forecasting method employed: the univariate GARCH on portfolio returns (base), the bottom-up approach using MGARCH models (bu), and the three forecast reconciliation strategies discussed in the paper ($shr$, $shr_{A}$, $shr_{B}$). Each cell reports the number of times (in \%) the forecasting model in the row statistically outperforms (p-values $< 0.05$ and using Bonferroni correction) the model in the column.}
\end{figure}

\newpage
\subsection{Proxy portfolio variance and covariance matrix of $N=9$ assets}

\begin{landscape}\centering
	\begingroup
	\setlength{\LTcapwidth}{\linewidth}
	\setlength{\tabcolsep}{2pt}
	\scriptsize
    
\begin{longtable}[t]{>{}c||>{}c>{}c|cc>{}c||>{}c>{}c|cc>{}c||>{}c>{}c|cc>{}c||>{}c>{}c|ccc}
\caption{Average accuracy indices across different simulation settings. Each setting is identified by a label composed of three elements: (i) the data-generating process (DGP), (ii) the MGARCH model used for estimation, and (iii) the portfolio weighting scheme (see \autoref{tab:notationModel} for details). The sample size ($T$) is also reported for each setting.  The top rows indicate the noisy level of the proxy $\delta \in \left\lbrace 0.25, 0.5, 0.75, 1\right\rbrace$ and the variance forecasting method employed: the univariate GARCH on portfolio returns (base), the bottom-up approach using MGARCH models (bu), and the three forecast reconciliation strategies discussed in the paper ($shr$, $shr_{A}$, $shr_{B}$). The reported accuracy measures are grouped into three categories: average accuracy indices (MSE, MAE, QLIKE); average relative measures (AvgRelMSE, AvgRelMAE, AvgRelQLIKE) using the univariate and the multivariate GARCH forecast as the benchmark.}\\
\toprule
\multicolumn{1}{c}{ } & \multicolumn{5}{c}{$\delta = 0.25$} & \multicolumn{5}{c}{$\delta = 0.5$} & \multicolumn{5}{c}{$\delta = 0.75$} & \multicolumn{5}{c}{$\delta = 1$} \\
\cmidrule(l{3pt}r{3pt}){2-6} \cmidrule(l{3pt}r{3pt}){7-11} \cmidrule(l{3pt}r{3pt}){12-16} \cmidrule(l{3pt}r{3pt}){17-21}
Index & $base$ & $bu$ & $shr$ & $shr_{A}$ & $shr_{B}$ & $base$ & $bu$ & $shr$ & $shr_{A}$ & $shr_{B}$ & $base$ & $bu$ & $shr$ & $shr_{A}$ & $shr_{B}$ & $base$ & $bu$ & $shr$ & $shr_{A}$ & $shr_{B}$\\
\midrule
\endfirsthead
\caption[]{Average accuracy indices across different simulation settings. Each setting is identified by a label composed of three elements: (i) the data-generating process (DGP), (ii) the MGARCH model used for estimation, and (iii) the portfolio weighting scheme (see \autoref{tab:notationModel} for details). The sample size ($T$) is also reported for each setting.  The top rows indicate the noisy level of the proxy $\delta \in \left\lbrace 0.25, 0.5, 0.75, 1\right\rbrace$ and the variance forecasting method employed: the univariate GARCH on portfolio returns (base), the bottom-up approach using MGARCH models (bu), and the three forecast reconciliation strategies discussed in the paper ($shr$, $shr_{A}$, $shr_{B}$). The reported accuracy measures are grouped into three categories: average accuracy indices (MSE, MAE, QLIKE); average relative measures (AvgRelMSE, AvgRelMAE, AvgRelQLIKE) using the univariate and the multivariate GARCH forecast as the benchmark. \textit{(continued)}}\\
\toprule
\multicolumn{1}{c}{ } & \multicolumn{5}{c}{$\delta = 0.25$} & \multicolumn{5}{c}{$\delta = 0.5$} & \multicolumn{5}{c}{$\delta = 0.75$} & \multicolumn{5}{c}{$\delta = 1$} \\
\cmidrule(l{3pt}r{3pt}){2-6} \cmidrule(l{3pt}r{3pt}){7-11} \cmidrule(l{3pt}r{3pt}){12-16} \cmidrule(l{3pt}r{3pt}){17-21}
Index & $base$ & $bu$ & $shr$ & $shr_{A}$ & $shr_{B}$ & $base$ & $bu$ & $shr$ & $shr_{A}$ & $shr_{B}$ & $base$ & $bu$ & $shr$ & $shr_{A}$ & $shr_{B}$ & $base$ & $bu$ & $shr$ & $shr_{A}$ & $shr_{B}$\\
\midrule
\endhead

\endfoot
\bottomrule
\endlastfoot
\addlinespace[0.5em]
\multicolumn{21}{c}{\textbf{BKF -- DCC -- EQ ($T=500$)}}\\
\nopagebreak \addlinespace[0em]
\multicolumn{21}{l}{\textit{Average indexes}}\\
\nopagebreak MSE & 817.0 & 1352.0 & 792.0 & \textbf{791.5} & \em{792.0} & \textbf{2971.7} & 3426.5 & 3028.4 & \em{3028.3} & 3028.4 & \textbf{6593.1} & 6967.7 & 6686.4 & \em{6686.3} & 6686.4 & \textbf{11681.3} & 11975.7 & 11770.6 & \em{11770.5} & 11770.6\\
\nopagebreak 
MAE & 7.531 & 10.118 & 7.449 & \textbf{7.448} & \em{7.449} & 14.240 & 14.870 & 14.156 & \textbf{14.156} & \em{14.156} & 21.284 & 21.112 & 21.005 & \textbf{21.005} & \em{21.005} & 28.397 & \textbf{27.913} & 27.953 & \em{27.952} & 27.953\\
\nopagebreak 
QLIKE & \textbf{0.056} & 0.114 & 0.057 & 0.057 & \em{0.057} & \textbf{0.173} & 0.228 & 0.183 & 0.183 & \em{0.183} & \textbf{0.395} & 0.447 & 0.413 & \em{0.413} & 0.413 & \textbf{1.295} & 1.344 & 1.319 & \em{1.319} & 1.319\\
\nopagebreak 
\addlinespace[0em]
\multicolumn{21}{l}{\textit{Relative indexes (base benchmark)}}\\
\nopagebreak AvgRelMSE & 1.000 & 1.708 & 0.970 & \em{0.970} & \textbf{0.970} & \textbf{1.000} & 1.166 & 1.014 & 1.014 & \em{1.014} & \textbf{1.000} & 1.057 & 1.011 & \em{1.011} & 1.011 & \textbf{1.000} & 1.022 & 1.005 & \em{1.005} & 1.005\\
\nopagebreak 
AvgRelMAE & 1.000 & 1.408 & 0.994 & \em{0.994} & \textbf{0.994} & \textbf{1.000} & 1.085 & 1.009 & 1.009 & \em{1.009} & \textbf{1.000} & 1.021 & 1.004 & \em{1.004} & 1.004 & \textbf{1.000} & 1.005 & 1.001 & \em{1.000} & 1.001\\
\end{longtable}

	\endgroup
\end{landscape}

\begin{landscape}\centering
	\begingroup
	\setlength{\LTcapwidth}{\linewidth}
	\setlength{\tabcolsep}{4pt}
	\scriptsize
    
\begin{longtable}[t]{>{}c||>{}c>{}c|cc>{}c||>{}c>{}c|cc>{}c||>{}c>{}c|cc>{}c||>{}c>{}c|ccc}
\caption{The number of times (in \%) that model is in the Model Confidence Set with different thresholds ($\gamma \in \{70\%, 75\%, 80\%, 85\%m 90\%, 95\%\}$) across different simulation settings using the MAE as loss function. Each setting is identified by a label composed of three elements: (i) the data-generating process (DGP), (ii) the MGARCH model used for estimation, and (iii) the portfolio weighting scheme (see \autoref{tab:notationModel} for details). The sample size ($T$) is also reported for each setting.  The top rows indicate the noisy level of the proxy $\delta \in \left\lbrace 0.25, 0.5, 0.75, 1\right\rbrace$ and the variance forecasting method employed: the univariate GARCH on portfolio returns (base), the bottom-up approach using MGARCH models (bu), and the three forecast reconciliation strategies discussed in the paper ($shr$, $shr_{A}$, $shr_{B}$).}\\
\toprule
\multicolumn{1}{c}{ } & \multicolumn{5}{c}{$\delta = 0.25$} & \multicolumn{5}{c}{$\delta = 0.5$} & \multicolumn{5}{c}{$\delta = 0.75$} & \multicolumn{5}{c}{$\delta = 1$} \\
\cmidrule(l{3pt}r{3pt}){2-6} \cmidrule(l{3pt}r{3pt}){7-11} \cmidrule(l{3pt}r{3pt}){12-16} \cmidrule(l{3pt}r{3pt}){17-21}
Threshold & $base$ & $bu$ & $shr$ & $shr_{A}$ & $shr_{B}$ & $base$ & $bu$ & $shr$ & $shr_{A}$ & $shr_{B}$ & $base$ & $bu$ & $shr$ & $shr_{A}$ & $shr_{B}$ & $base$ & $bu$ & $shr$ & $shr_{A}$ & $shr_{B}$\\
\midrule
\endfirsthead
\caption[]{The number of times (in \%) that model is in the Model Confidence Set with different thresholds ($\gamma \in \{70\%, 75\%, 80\%, 85\%m 90\%, 95\%\}$) across different simulation settings using the MAE as loss function. Each setting is identified by a label composed of three elements: (i) the data-generating process (DGP), (ii) the MGARCH model used for estimation, and (iii) the portfolio weighting scheme (see \autoref{tab:notationModel} for details). The sample size ($T$) is also reported for each setting.  The top rows indicate the noisy level of the proxy $\delta \in \left\lbrace 0.25, 0.5, 0.75, 1\right\rbrace$ and the variance forecasting method employed: the univariate GARCH on portfolio returns (base), the bottom-up approach using MGARCH models (bu), and the three forecast reconciliation strategies discussed in the paper ($shr$, $shr_{A}$, $shr_{B}$). \textit{(continued)}}\\
\toprule
\multicolumn{1}{c}{ } & \multicolumn{5}{c}{$\delta = 0.25$} & \multicolumn{5}{c}{$\delta = 0.5$} & \multicolumn{5}{c}{$\delta = 0.75$} & \multicolumn{5}{c}{$\delta = 1$} \\
\cmidrule(l{3pt}r{3pt}){2-6} \cmidrule(l{3pt}r{3pt}){7-11} \cmidrule(l{3pt}r{3pt}){12-16} \cmidrule(l{3pt}r{3pt}){17-21}
Threshold & $base$ & $bu$ & $shr$ & $shr_{A}$ & $shr_{B}$ & $base$ & $bu$ & $shr$ & $shr_{A}$ & $shr_{B}$ & $base$ & $bu$ & $shr$ & $shr_{A}$ & $shr_{B}$ & $base$ & $bu$ & $shr$ & $shr_{A}$ & $shr_{B}$\\
\midrule
\endhead

\endfoot
\bottomrule
\endlastfoot
\addlinespace[0.5em]
\multicolumn{21}{c}{\textbf{BKF -- DCC -- EQ ($T=500$)}}\\
\nopagebreak 70 & 70.4 & 6.2 & \em{71.0} & 70.0 & \textbf{72.2} & \textbf{90.0} & 37.2 & \em{62.8} & 62.6 & \em{62.8} & \textbf{83.6} & \em{66.6} & 65.0 & 65.0 & 65.0 & \em{74.8} & \textbf{76.4} & 61.0 & 61.0 & 61.0\\
\nopagebreak 
75 & \em{73.8} & 6.8 & 73.2 & 72.2 & \textbf{74.4} & \textbf{91.0} & 39.6 & 66.0 & 65.8 & \em{66.2} & \textbf{85.2} & \em{69.8} & 68.0 & 68.0 & 68.0 & \textbf{78.4} & \textbf{78.4} & \em{65.2} & \em{65.2} & \em{65.2}\\
\nopagebreak 
80 & 75.4 & 7.2 & \em{76.6} & 75.6 & \textbf{78.0} & \textbf{91.8} & 43.0 & 70.4 & 70.2 & \em{70.6} & \textbf{86.4} & \em{73.8} & 71.2 & 71.2 & 71.2 & \textbf{81.2} & \em{79.6} & 68.2 & 68.2 & 68.2\\
\nopagebreak 
85 & 79.6 & 7.8 & \em{79.8} & 78.8 & \textbf{80.6} & \textbf{93.2} & 47.6 & 73.8 & 74.0 & \em{74.2} & \textbf{87.8} & \em{75.4} & 75.0 & 75.0 & 75.0 & \textbf{83.4} & \em{81.4} & 72.4 & 72.4 & 72.4\\
\nopagebreak 
90 & 82.4 & 9.8 & \em{82.8} & \em{82.8} & \textbf{83.2} & \textbf{95.0} & 53.6 & \em{77.0} & 76.8 & \em{77.0} & \textbf{90.8} & \em{78.6} & 78.0 & 78.0 & 78.0 & \textbf{86.8} & \em{82.6} & 76.4 & 76.4 & 76.4\\
\nopagebreak 
95 & 86.6 & 15.0 & 88.4 & \em{88.6} & \textbf{88.8} & \textbf{97.4} & 62.8 & \em{82.4} & \em{82.4} & \em{82.4} & \textbf{94.4} & 82.6 & \em{83.4} & \em{83.4} & \em{83.4} & \textbf{91.8} & \em{86.8} & 82.8 & 82.8 & 82.8\\
 
\addlinespace[0.5em]
\multicolumn{21}{c}{\textbf{BKF -- DCC -- EQ ($T=1000$)}}\\
\end{longtable}

	\endgroup
\end{landscape}

\begin{landscape}\centering
\begingroup
\setlength{\LTcapwidth}{\linewidth}
\setlength{\tabcolsep}{4pt}
\scriptsize
    
\begin{longtable}[t]{>{}c||>{}c>{}c|cc>{}c||>{}c>{}c|cc>{}c||>{}c>{}c|cc>{}c||>{}c>{}c|ccc}
\caption{The number of times (in \%) that model is in the Model Confidence Set with different thresholds ($\gamma \in \{70\%, 75\%, 80\%, 85\%m 90\%, 95\%\}$) across different simulation settings using the MSE as loss function. Each setting is identified by a label composed of three elements: (i) the data-generating process (DGP), (ii) the MGARCH model used for estimation, and (iii) the portfolio weighting scheme (see \autoref{tab:notationModel} for details). The sample size ($T$) is also reported for each setting.  The top rows indicate the noisy level of the proxy $\delta \in \left\lbrace 0.25, 0.5, 0.75, 1\right\rbrace$ and the variance forecasting method employed: the univariate GARCH on portfolio returns (base), the bottom-up approach using MGARCH models (bu), and the three forecast reconciliation strategies discussed in the paper ($shr$, $shr_{A}$, $shr_{B}$).}\\
\toprule
\multicolumn{1}{c}{ } & \multicolumn{5}{c}{$\delta = 0.25$} & \multicolumn{5}{c}{$\delta = 0.5$} & \multicolumn{5}{c}{$\delta = 0.75$} & \multicolumn{5}{c}{$\delta = 1$} \\
\cmidrule(l{3pt}r{3pt}){2-6} \cmidrule(l{3pt}r{3pt}){7-11} \cmidrule(l{3pt}r{3pt}){12-16} \cmidrule(l{3pt}r{3pt}){17-21}
Threshold & $base$ & $bu$ & $shr$ & $shr_{A}$ & $shr_{B}$ & $base$ & $bu$ & $shr$ & $shr_{A}$ & $shr_{B}$ & $base$ & $bu$ & $shr$ & $shr_{A}$ & $shr_{B}$ & $base$ & $bu$ & $shr$ & $shr_{A}$ & $shr_{B}$\\
\midrule
\endfirsthead
\caption[]{The number of times (in \%) that model is in the Model Confidence Set with different thresholds ($\gamma \in \{70\%, 75\%, 80\%, 85\%m 90\%, 95\%\}$) across different simulation settings using the MSE as loss function. Each setting is identified by a label composed of three elements: (i) the data-generating process (DGP), (ii) the MGARCH model used for estimation, and (iii) the portfolio weighting scheme (see \autoref{tab:notationModel} for details). The sample size ($T$) is also reported for each setting.  The top rows indicate the noisy level of the proxy $\delta \in \left\lbrace 0.25, 0.5, 0.75, 1\right\rbrace$ and the variance forecasting method employed: the univariate GARCH on portfolio returns (base), the bottom-up approach using MGARCH models (bu), and the three forecast reconciliation strategies discussed in the paper ($shr$, $shr_{A}$, $shr_{B}$). \textit{(continued)}}\\
\toprule
\multicolumn{1}{c}{ } & \multicolumn{5}{c}{$\delta = 0.25$} & \multicolumn{5}{c}{$\delta = 0.5$} & \multicolumn{5}{c}{$\delta = 0.75$} & \multicolumn{5}{c}{$\delta = 1$} \\
\cmidrule(l{3pt}r{3pt}){2-6} \cmidrule(l{3pt}r{3pt}){7-11} \cmidrule(l{3pt}r{3pt}){12-16} \cmidrule(l{3pt}r{3pt}){17-21}
Threshold & $base$ & $bu$ & $shr$ & $shr_{A}$ & $shr_{B}$ & $base$ & $bu$ & $shr$ & $shr_{A}$ & $shr_{B}$ & $base$ & $bu$ & $shr$ & $shr_{A}$ & $shr_{B}$ & $base$ & $bu$ & $shr$ & $shr_{A}$ & $shr_{B}$\\
\midrule
\endhead

\endfoot
\bottomrule
\endlastfoot
\addlinespace[0.5em]
\multicolumn{21}{c}{\textbf{BKF -- DCC -- EQ ($T=500$)}}\\
\nopagebreak 70 & 62.2 & 5.2 & \em{80.6} & \em{80.6} & \textbf{81.6} & \textbf{93.0} & 18.2 & 59.6 & 59.6 & \em{59.8} & \textbf{93.0} & 46.4 & \em{67.6} & \em{67.6} & \em{67.6} & \textbf{85.8} & 66.2 & \em{76.4} & \em{76.4} & \em{76.4}\\
\nopagebreak 
75 & 65.8 & 5.8 & \em{82.2} & \em{82.2} & \textbf{82.8} & \textbf{94.0} & 21.6 & 63.6 & 63.6 & \em{63.8} & \textbf{94.0} & 49.6 & \em{72.4} & \em{72.4} & \em{72.4} & \textbf{88.0} & 70.2 & \em{79.4} & \em{79.4} & \em{79.4}\\
\nopagebreak 
80 & 70.4 & 6.8 & \em{85.6} & \em{85.6} & \textbf{86.2} & \textbf{94.6} & 25.2 & 68.8 & 68.8 & \em{69.0} & \textbf{94.6} & 53.6 & \em{77.0} & \em{77.0} & \em{77.0} & \textbf{90.6} & 75.4 & \em{83.0} & \em{83.0} & \em{83.0}\\
\nopagebreak 
85 & 76.8 & 8.6 & \em{88.6} & 88.4 & \textbf{89.0} & \textbf{96.6} & 31.4 & 77.0 & 77.0 & \em{77.2} & \textbf{94.8} & 60.8 & \em{83.0} & \em{83.0} & \em{83.0} & \textbf{91.8} & 81.4 & \em{88.0} & \em{88.0} & \em{88.0}\\
\nopagebreak 
90 & 83.0 & 12.0 & \textbf{91.8} & \em{91.4} & \textbf{91.8} & \textbf{97.8} & 42.6 & \em{84.0} & \em{84.0} & \em{84.0} & \textbf{96.0} & 71.0 & \em{87.2} & \em{87.2} & \em{87.2} & \textbf{94.2} & 87.4 & \em{92.4} & \em{92.4} & \em{92.4}\\
\nopagebreak 
95 & 88.0 & 33.6 & \textbf{94.4} & \textbf{94.4} & \em{94.2} & \textbf{98.8} & 60.6 & \em{92.0} & \em{92.0} & \em{92.0} & \textbf{98.2} & 84.2 & \em{95.0} & \em{95.0} & \em{95.0} & \em{96.4} & 94.0 & \textbf{96.8} & \textbf{96.8} & \textbf{96.8}\\
 
\addlinespace[0.5em]
\multicolumn{21}{c}{\textbf{BKF -- DCC -- EQ ($T=1000$)}}\\
\end{longtable}

\endgroup
\end{landscape}

\begin{landscape}\centering
\begingroup
\setlength{\LTcapwidth}{\linewidth}
\setlength{\tabcolsep}{4pt}
\scriptsize
    
\begin{longtable}[t]{>{}c||>{}c>{}c|cc>{}c||>{}c>{}c|cc>{}c||>{}c>{}c|cc>{}c||>{}c>{}c|ccc}
\caption{The number of times (in \%) that model is in the Model Confidence Set with different thresholds ($\gamma \in \{70\%, 75\%, 80\%, 85\%m 90\%, 95\%\}$) across different simulation settings using the QLIKE as loss function. Each setting is identified by a label composed of three elements: (i) the data-generating process (DGP), (ii) the MGARCH model used for estimation, and (iii) the portfolio weighting scheme (see \autoref{tab:notationModel} for details). The sample size ($T$) is also reported for each setting.  The top rows indicate the noisy level of the proxy $\delta \in \left\lbrace 0.25, 0.5, 0.75, 1\right\rbrace$ and the variance forecasting method employed: the univariate GARCH on portfolio returns (base), the bottom-up approach using MGARCH models (bu), and the three forecast reconciliation strategies discussed in the paper ($shr$, $shr_{A}$, $shr_{B}$).}\\
\toprule
\multicolumn{1}{c}{ } & \multicolumn{5}{c}{$\delta = 0.25$} & \multicolumn{5}{c}{$\delta = 0.5$} & \multicolumn{5}{c}{$\delta = 0.75$} & \multicolumn{5}{c}{$\delta = 1$} \\
\cmidrule(l{3pt}r{3pt}){2-6} \cmidrule(l{3pt}r{3pt}){7-11} \cmidrule(l{3pt}r{3pt}){12-16} \cmidrule(l{3pt}r{3pt}){17-21}
Threshold & $base$ & $bu$ & $shr$ & $shr_{A}$ & $shr_{B}$ & $base$ & $bu$ & $shr$ & $shr_{A}$ & $shr_{B}$ & $base$ & $bu$ & $shr$ & $shr_{A}$ & $shr_{B}$ & $base$ & $bu$ & $shr$ & $shr_{A}$ & $shr_{B}$\\
\midrule
\endfirsthead
\caption[]{The number of times (in \%) that model is in the Model Confidence Set with different thresholds ($\gamma \in \{70\%, 75\%, 80\%, 85\%m 90\%, 95\%\}$) across different simulation settings using the QLIKE as loss function. Each setting is identified by a label composed of three elements: (i) the data-generating process (DGP), (ii) the MGARCH model used for estimation, and (iii) the portfolio weighting scheme (see \autoref{tab:notationModel} for details). The sample size ($T$) is also reported for each setting.  The top rows indicate the noisy level of the proxy $\delta \in \left\lbrace 0.25, 0.5, 0.75, 1\right\rbrace$ and the variance forecasting method employed: the univariate GARCH on portfolio returns (base), the bottom-up approach using MGARCH models (bu), and the three forecast reconciliation strategies discussed in the paper ($shr$, $shr_{A}$, $shr_{B}$). \textit{(continued)}}\\
\toprule
\multicolumn{1}{c}{ } & \multicolumn{5}{c}{$\delta = 0.25$} & \multicolumn{5}{c}{$\delta = 0.5$} & \multicolumn{5}{c}{$\delta = 0.75$} & \multicolumn{5}{c}{$\delta = 1$} \\
\cmidrule(l{3pt}r{3pt}){2-6} \cmidrule(l{3pt}r{3pt}){7-11} \cmidrule(l{3pt}r{3pt}){12-16} \cmidrule(l{3pt}r{3pt}){17-21}
Threshold & $base$ & $bu$ & $shr$ & $shr_{A}$ & $shr_{B}$ & $base$ & $bu$ & $shr$ & $shr_{A}$ & $shr_{B}$ & $base$ & $bu$ & $shr$ & $shr_{A}$ & $shr_{B}$ & $base$ & $bu$ & $shr$ & $shr_{A}$ & $shr_{B}$\\
\midrule
\endhead

\endfoot
\bottomrule
\endlastfoot
\addlinespace[0.5em]
\multicolumn{21}{c}{\textbf{BKF -- DCC -- EQ ($T=500$)}}\\
\nopagebreak 70 & \textbf{70.6} & 1.8 & 65.0 & 64.4 & \em{65.8} & \textbf{97.0} & 6.8 & \em{38.0} & \em{38.0} & \em{38.0} & \textbf{96.8} & 21.2 & \em{40.4} & \em{40.4} & \em{40.4} & \textbf{95.4} & 35.4 & \em{46.0} & \em{46.0} & \em{46.0}\\
\nopagebreak 
75 & \textbf{72.0} & 2.0 & 66.2 & 65.8 & \em{66.8} & \textbf{98.0} & 8.2 & \em{41.8} & \em{41.8} & \em{41.8} & \textbf{97.2} & 24.4 & \em{43.8} & \em{43.8} & \em{43.8} & \textbf{96.0} & 39.6 & \em{51.6} & \em{51.6} & \em{51.6}\\
\nopagebreak 
80 & \textbf{73.4} & 2.4 & 68.8 & 68.6 & \em{69.2} & \textbf{98.6} & 10.6 & \em{47.4} & \em{47.4} & \em{47.4} & \textbf{97.8} & 28.2 & \em{49.8} & \em{49.8} & \em{49.8} & \textbf{96.0} & 43.6 & \em{56.4} & \em{56.4} & \em{56.4}\\
\nopagebreak 
85 & \textbf{76.4} & 2.6 & 72.4 & 72.0 & \em{72.8} & \textbf{99.0} & 12.6 & \em{53.6} & \em{53.6} & \em{53.6} & \textbf{98.0} & 33.8 & \em{56.6} & \em{56.6} & \em{56.6} & \textbf{96.8} & 49.4 & \em{61.8} & \em{61.8} & \em{61.8}\\
\nopagebreak 
90 & \textbf{80.4} & 3.4 & 76.0 & 75.8 & \em{76.6} & \textbf{99.2} & 16.8 & \em{60.8} & \em{60.8} & \em{60.8} & \textbf{98.6} & 39.4 & \em{63.8} & \em{63.8} & \em{63.8} & \textbf{97.6} & 58.6 & \em{70.6} & \em{70.6} & \em{70.6}\\
\nopagebreak 
95 & \textbf{86.8} & 5.6 & \em{81.2} & \em{81.2} & \em{81.2} & \textbf{99.4} & 27.4 & \em{70.8} & \em{70.8} & \em{70.8} & \textbf{98.8} & 53.4 & \em{74.2} & \em{74.2} & \em{74.2} & \textbf{98.6} & 71.0 & \em{81.0} & \em{81.0} & \em{81.0}\\
 
\addlinespace[0.5em]
\multicolumn{21}{c}{\textbf{BKF -- DCC -- EQ ($T=1000$)}}\\
\end{longtable}

\endgroup
\end{landscape}

\begin{figure}[H]
	\centering
	\includegraphics[width=0.9\linewidth]{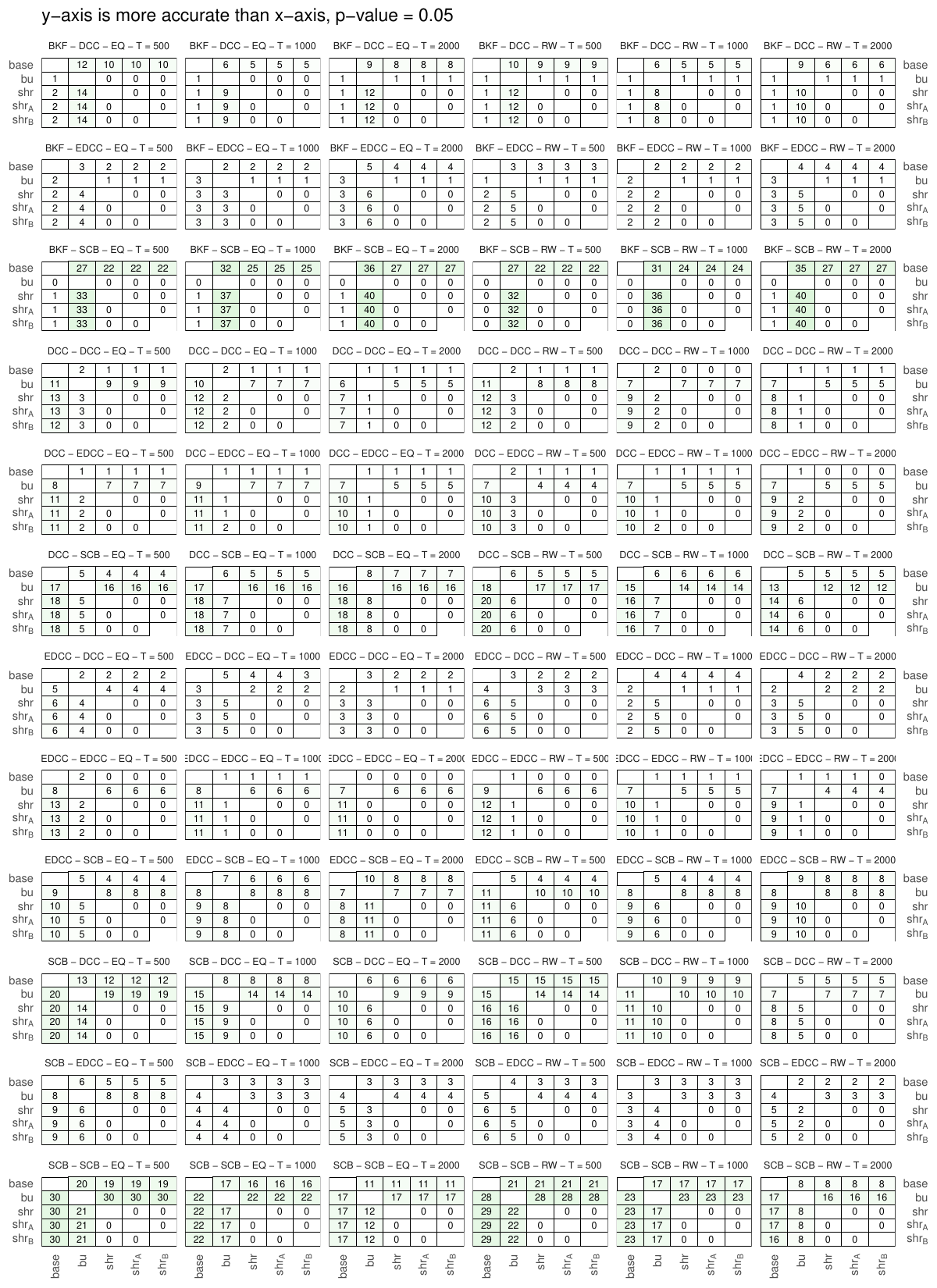}
	\caption{Qualitative evaluation using the Diebold-Mariano across different simulation settings using an absolute loss function and noisy level $\delta = 1$. Each setting is identified by a label composed of three elements: (i) the data-generating process (DGP), (ii) the MGARCH model used for estimation, and (iii) the portfolio weighting scheme (see \autoref{tab:notationModel} for details). The sample size ($T$) is also reported. The rows and columns indicate the variance forecasting method employed: the univariate GARCH on portfolio returns (base), the bottom-up approach using MGARCH models (bu), and the three forecast reconciliation strategies discussed in the paper ($shr$, $shr_{A}$, $shr_{B}$). Each cell reports the number of times (in \%) the forecasting model in the row statistically outperforms (p-values $< 0.05$ and using Bonferroni correction) the model in the column.}
\end{figure}
\begin{figure}[H]
\centering
	\includegraphics[width=0.9\linewidth]{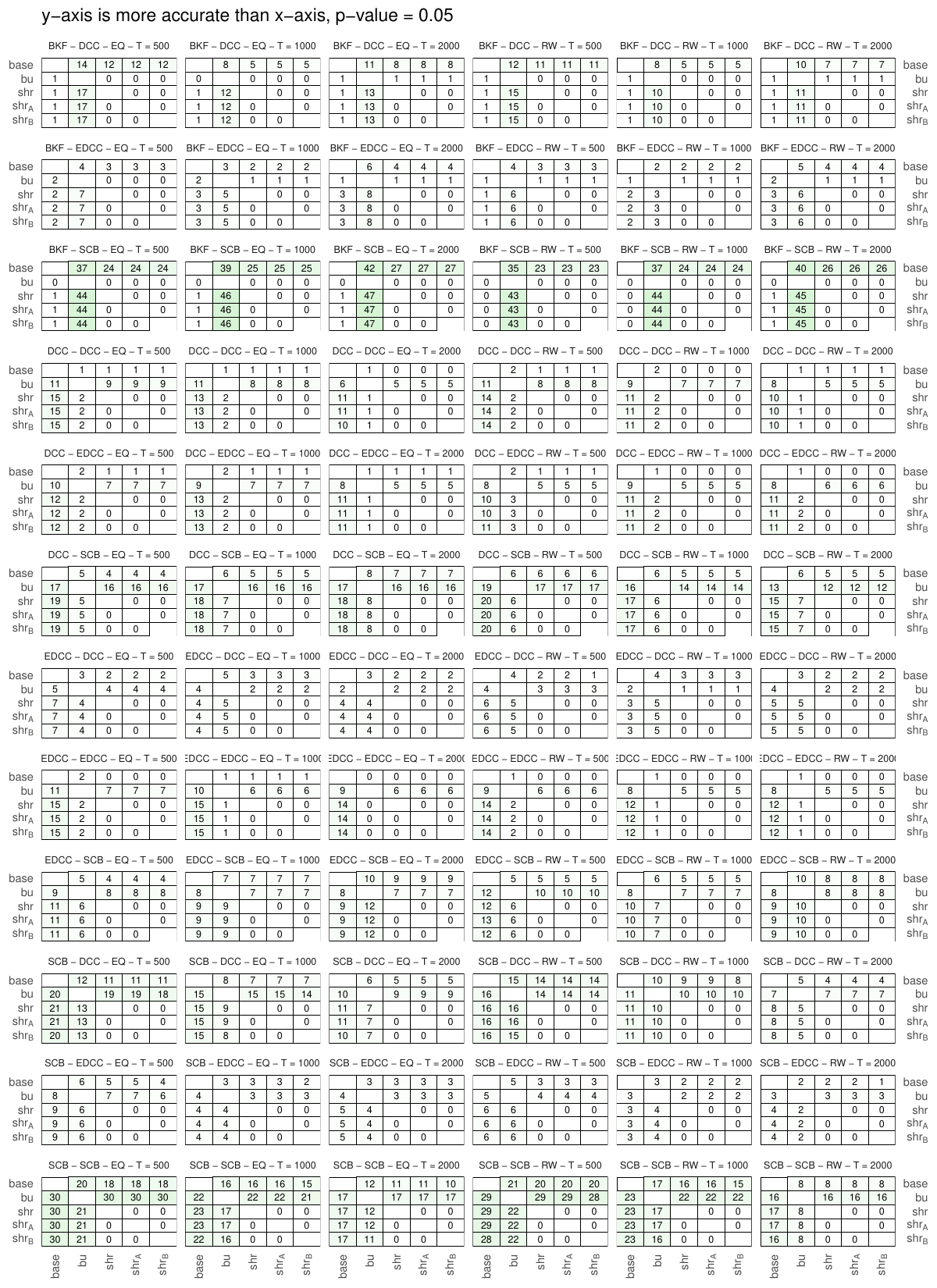}
\caption{Qualitative evaluation using the Diebold-Mariano across different simulation settings using an absolute loss function and noisy level $\delta = 0.75$. Each setting is identified by a label composed of three elements: (i) the data-generating process (DGP), (ii) the MGARCH model used for estimation, and (iii) the portfolio weighting scheme (see \autoref{tab:notationModel} for details). The sample size ($T$) is also reported. The rows and columns indicate the variance forecasting method employed: the univariate GARCH on portfolio returns (base), the bottom-up approach using MGARCH models (bu), and the three forecast reconciliation strategies discussed in the paper ($shr$, $shr_{A}$, $shr_{B}$). Each cell reports the number of times (in \%) the forecasting model in the row statistically outperforms (p-values $< 0.05$ and using Bonferroni correction) the model in the column.}
\end{figure}
\begin{figure}[H]
\centering
	\includegraphics[width=0.9\linewidth]{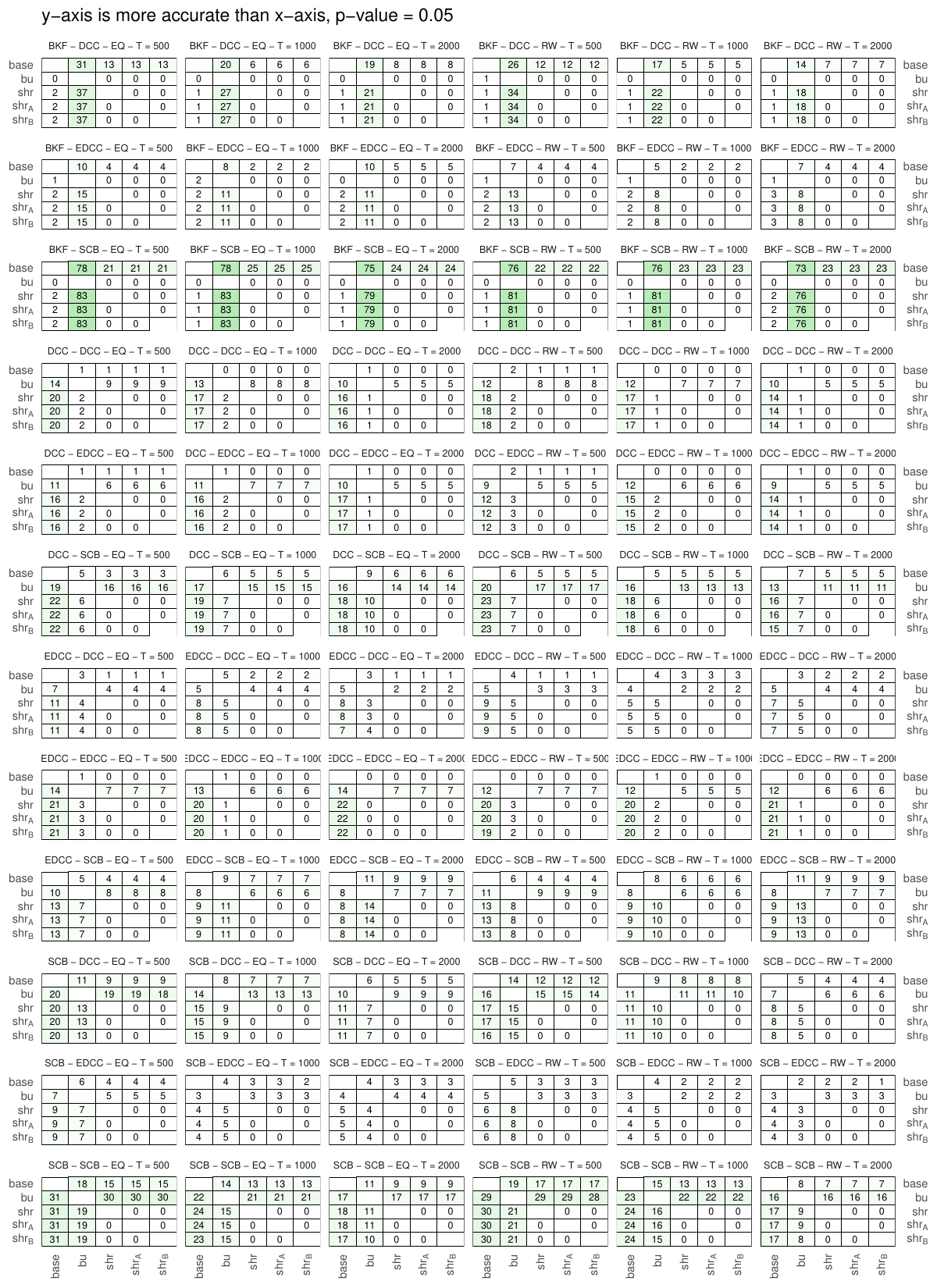}
\caption{Qualitative evaluation using the Diebold-Mariano across different simulation settings using an absolute loss function and noisy level $\delta = 0.5$. Each setting is identified by a label composed of three elements: (i) the data-generating process (DGP), (ii) the MGARCH model used for estimation, and (iii) the portfolio weighting scheme (see \autoref{tab:notationModel} for details). The sample size ($T$) is also reported. The rows and columns indicate the variance forecasting method employed: the univariate GARCH on portfolio returns (base), the bottom-up approach using MGARCH models (bu), and the three forecast reconciliation strategies discussed in the paper ($shr$, $shr_{A}$, $shr_{B}$). Each cell reports the number of times (in \%) the forecasting model in the row statistically outperforms (p-values $< 0.05$ and using Bonferroni correction) the model in the column.}
\end{figure}
\begin{figure}[H]
\centering
	\includegraphics[width=0.9\linewidth]{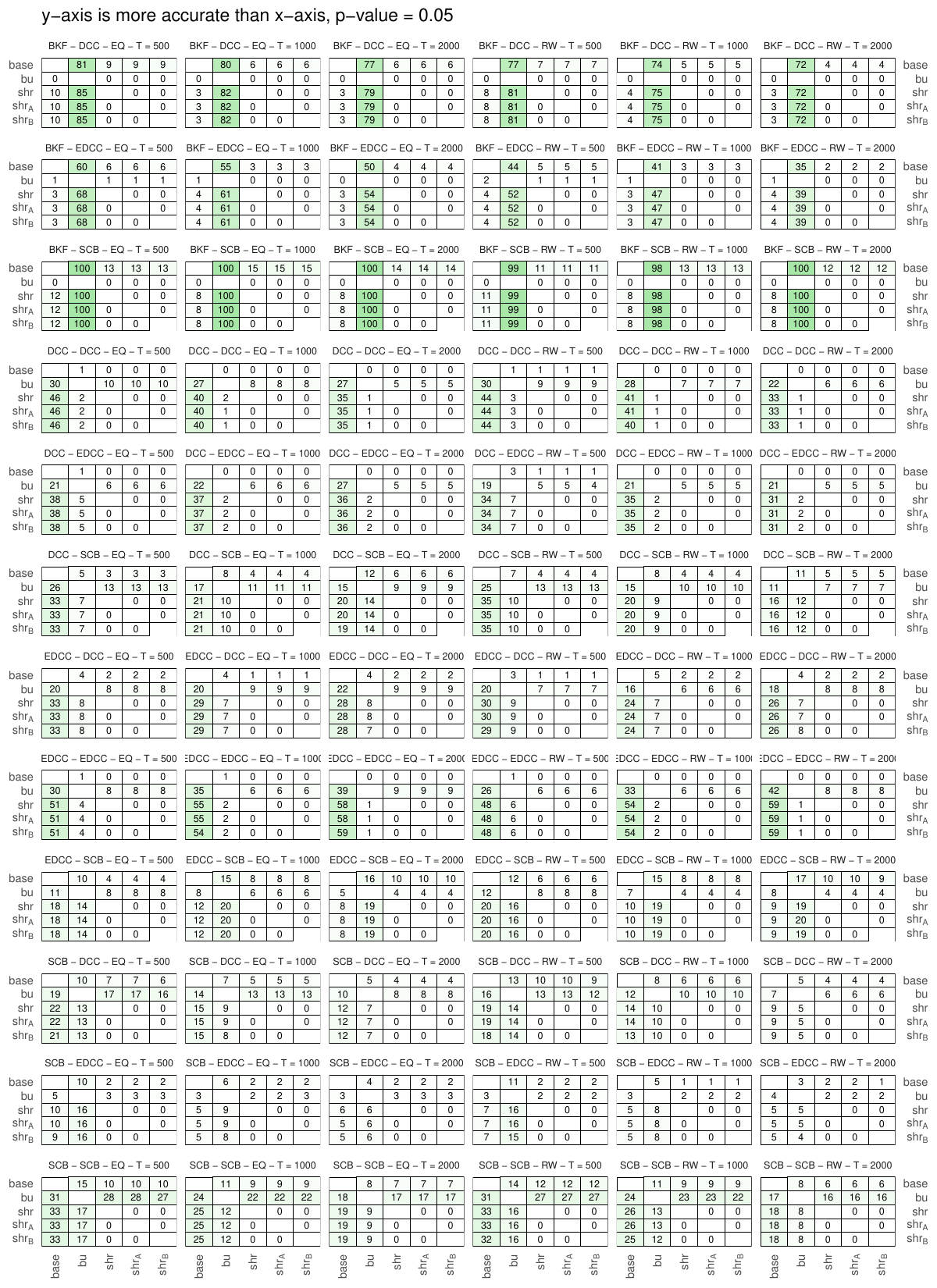}
\caption{Qualitative evaluation using the Diebold-Mariano across different simulation settings using an absolute loss function and noisy level $\delta = 0.25$. Each setting is identified by a label composed of three elements: (i) the data-generating process (DGP), (ii) the MGARCH model used for estimation, and (iii) the portfolio weighting scheme (see \autoref{tab:notationModel} for details). The sample size ($T$) is also reported. The rows and columns indicate the variance forecasting method employed: the univariate GARCH on portfolio returns (base), the bottom-up approach using MGARCH models (bu), and the three forecast reconciliation strategies discussed in the paper ($shr$, $shr_{A}$, $shr_{B}$). Each cell reports the number of times (in \%) the forecasting model in the row statistically outperforms (p-values $< 0.05$ and using Bonferroni correction) the model in the column.}
\end{figure}

\begin{figure}[H]
	\centering
	\includegraphics[width=0.9\linewidth]{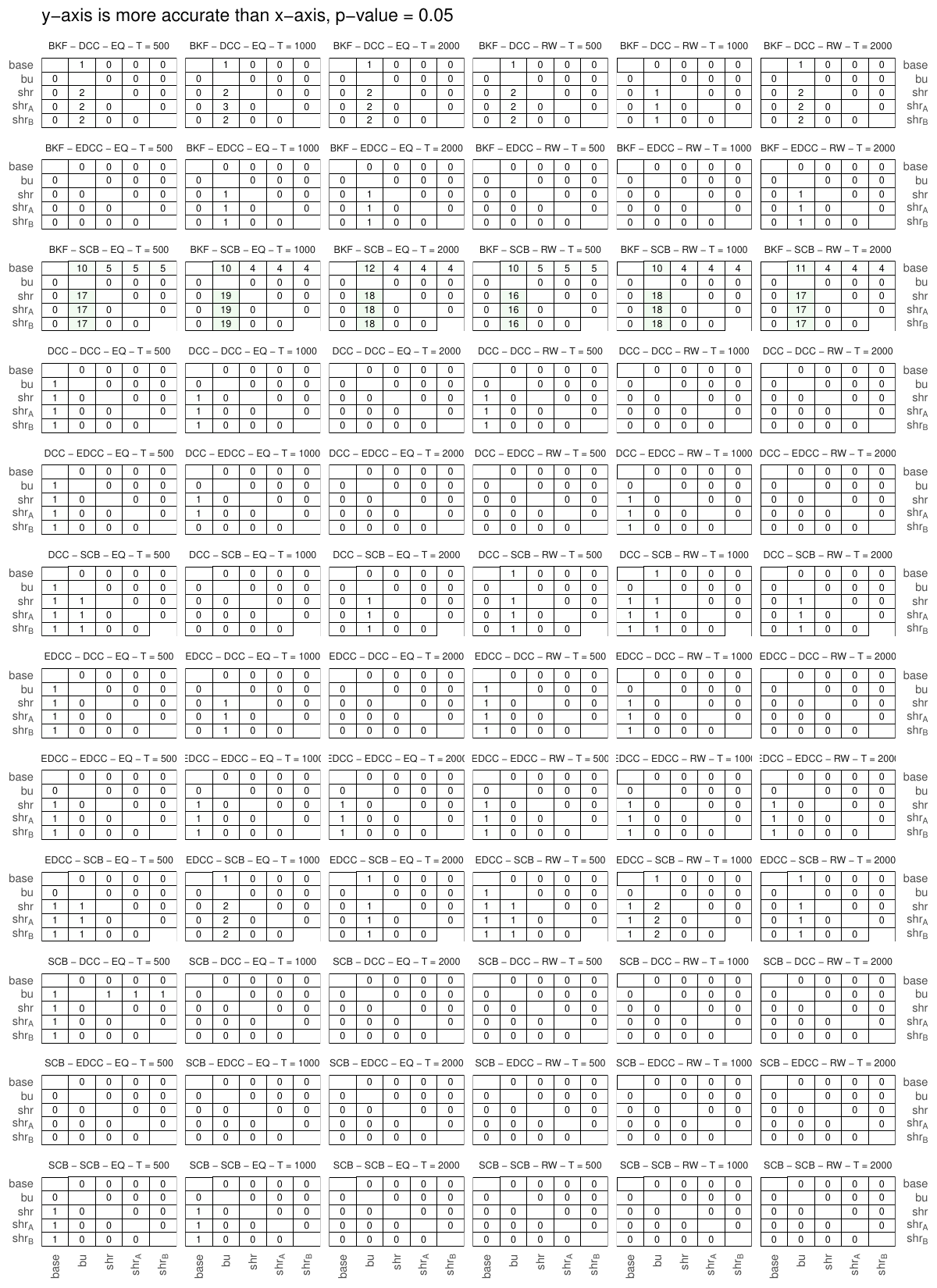}
	\caption{Qualitative evaluation using the Diebold-Mariano across different simulation settings using a square loss function and noisy level $\delta = 1$. Each setting is identified by a label composed of three elements: (i) the data-generating process (DGP), (ii) the MGARCH model used for estimation, and (iii) the portfolio weighting scheme (see \autoref{tab:notationModel} for details). The sample size ($T$) is also reported. The rows and columns indicate the variance forecasting method employed: the univariate GARCH on portfolio returns (base), the bottom-up approach using MGARCH models (bu), and the three forecast reconciliation strategies discussed in the paper ($shr$, $shr_{A}$, $shr_{B}$). Each cell reports the number of times (in \%) the forecasting model in the row statistically outperforms (p-values $< 0.05$ and using Bonferroni correction) the model in the column.}
\end{figure}
\begin{figure}[H]
\centering
	\includegraphics[width=0.9\linewidth]{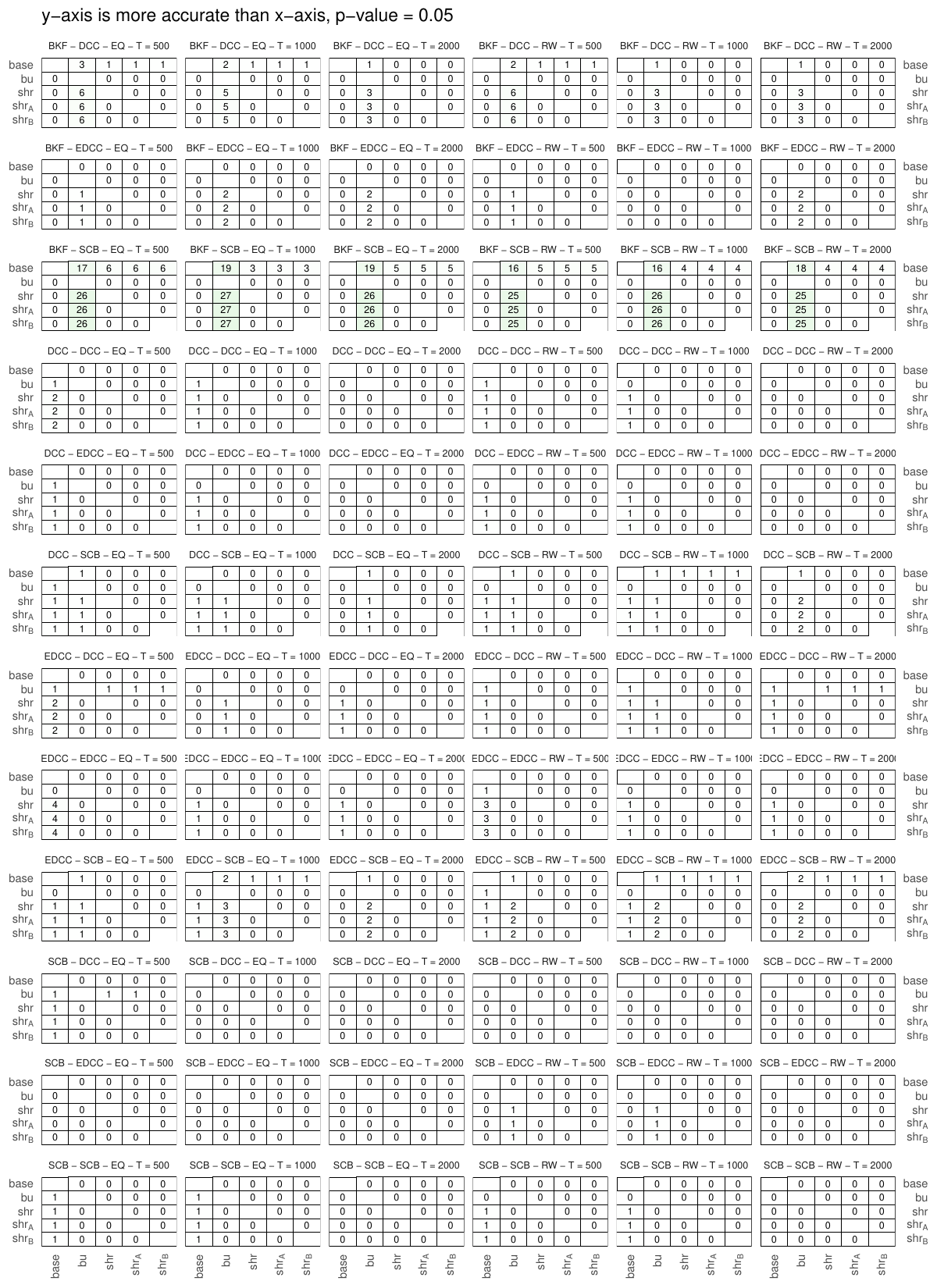}
\caption{Qualitative evaluation using the Diebold-Mariano across different simulation settings using a square loss function and noisy level $\delta = 0.75$. Each setting is identified by a label composed of three elements: (i) the data-generating process (DGP), (ii) the MGARCH model used for estimation, and (iii) the portfolio weighting scheme (see \autoref{tab:notationModel} for details). The sample size ($T$) is also reported. The rows and columns indicate the variance forecasting method employed: the univariate GARCH on portfolio returns (base), the bottom-up approach using MGARCH models (bu), and the three forecast reconciliation strategies discussed in the paper ($shr$, $shr_{A}$, $shr_{B}$). Each cell reports the number of times (in \%) the forecasting model in the row statistically outperforms (p-values $< 0.05$ and using Bonferroni correction) the model in the column.}
\end{figure}
\begin{figure}[H]
\centering
	\includegraphics[width=0.9\linewidth]{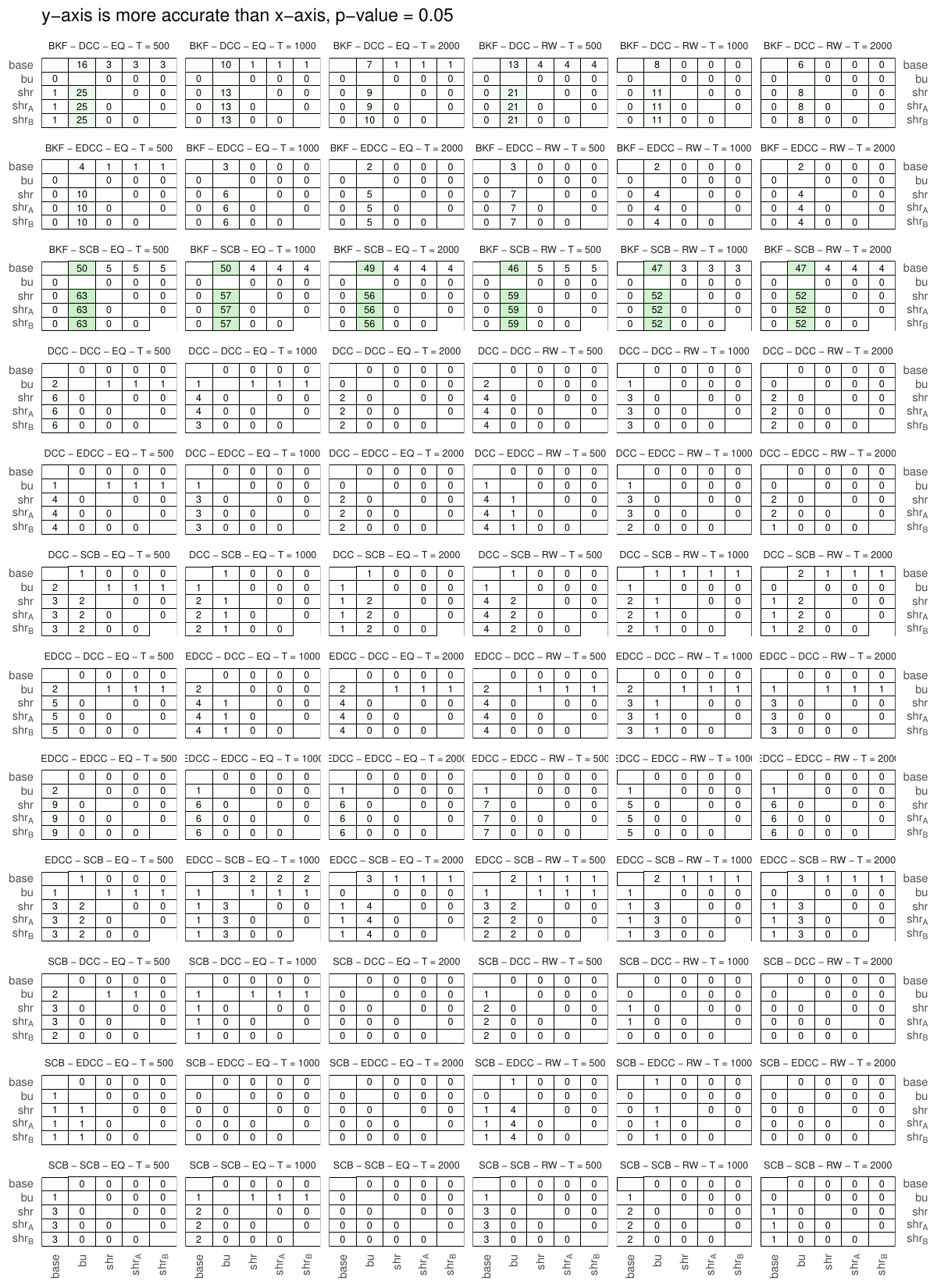}
\caption{Qualitative evaluation using the Diebold-Mariano across different simulation settings using a square loss function and noisy level $\delta = 0.5$. Each setting is identified by a label composed of three elements: (i) the data-generating process (DGP), (ii) the MGARCH model used for estimation, and (iii) the portfolio weighting scheme (see \autoref{tab:notationModel} for details). The sample size ($T$) is also reported. The rows and columns indicate the variance forecasting method employed: the univariate GARCH on portfolio returns (base), the bottom-up approach using MGARCH models (bu), and the three forecast reconciliation strategies discussed in the paper ($shr$, $shr_{A}$, $shr_{B}$). Each cell reports the number of times (in \%) the forecasting model in the row statistically outperforms (p-values $< 0.05$ and using Bonferroni correction) the model in the column.}
\end{figure}
\begin{figure}[H]
\centering
	\includegraphics[width=0.9\linewidth]{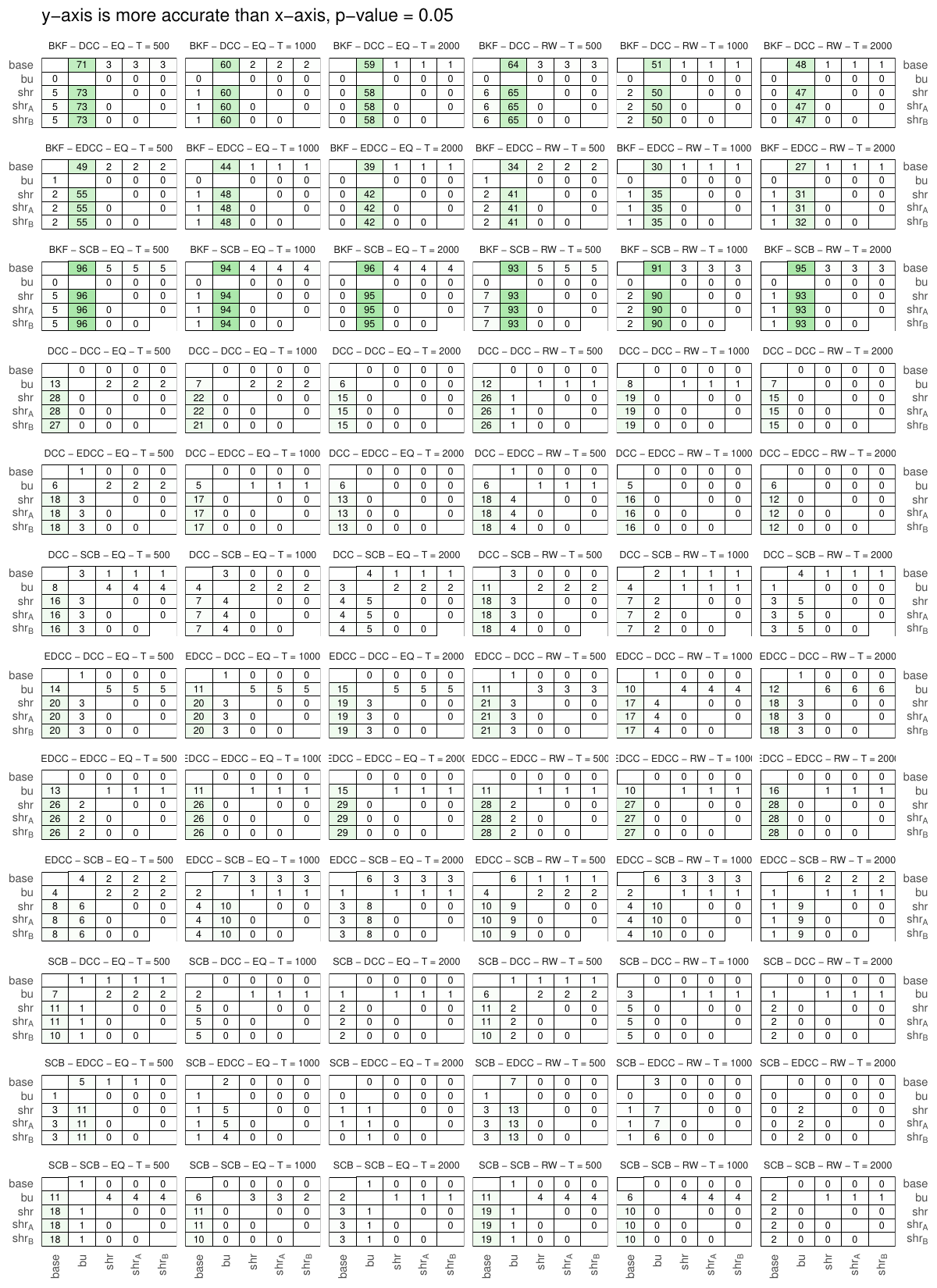}
\caption{Qualitative evaluation using the Diebold-Mariano across different simulation settings using a square loss function and noisy level $\delta = 0.25$. Each setting is identified by a label composed of three elements: (i) the data-generating process (DGP), (ii) the MGARCH model used for estimation, and (iii) the portfolio weighting scheme (see \autoref{tab:notationModel} for details). The sample size ($T$) is also reported. The rows and columns indicate the variance forecasting method employed: the univariate GARCH on portfolio returns (base), the bottom-up approach using MGARCH models (bu), and the three forecast reconciliation strategies discussed in the paper ($shr$, $shr_{A}$, $shr_{B}$). Each cell reports the number of times (in \%) the forecasting model in the row statistically outperforms (p-values $< 0.05$ and using Bonferroni correction) the model in the column.}
\end{figure}

\begin{figure}[H]
	\centering
	\includegraphics[width=0.9\linewidth]{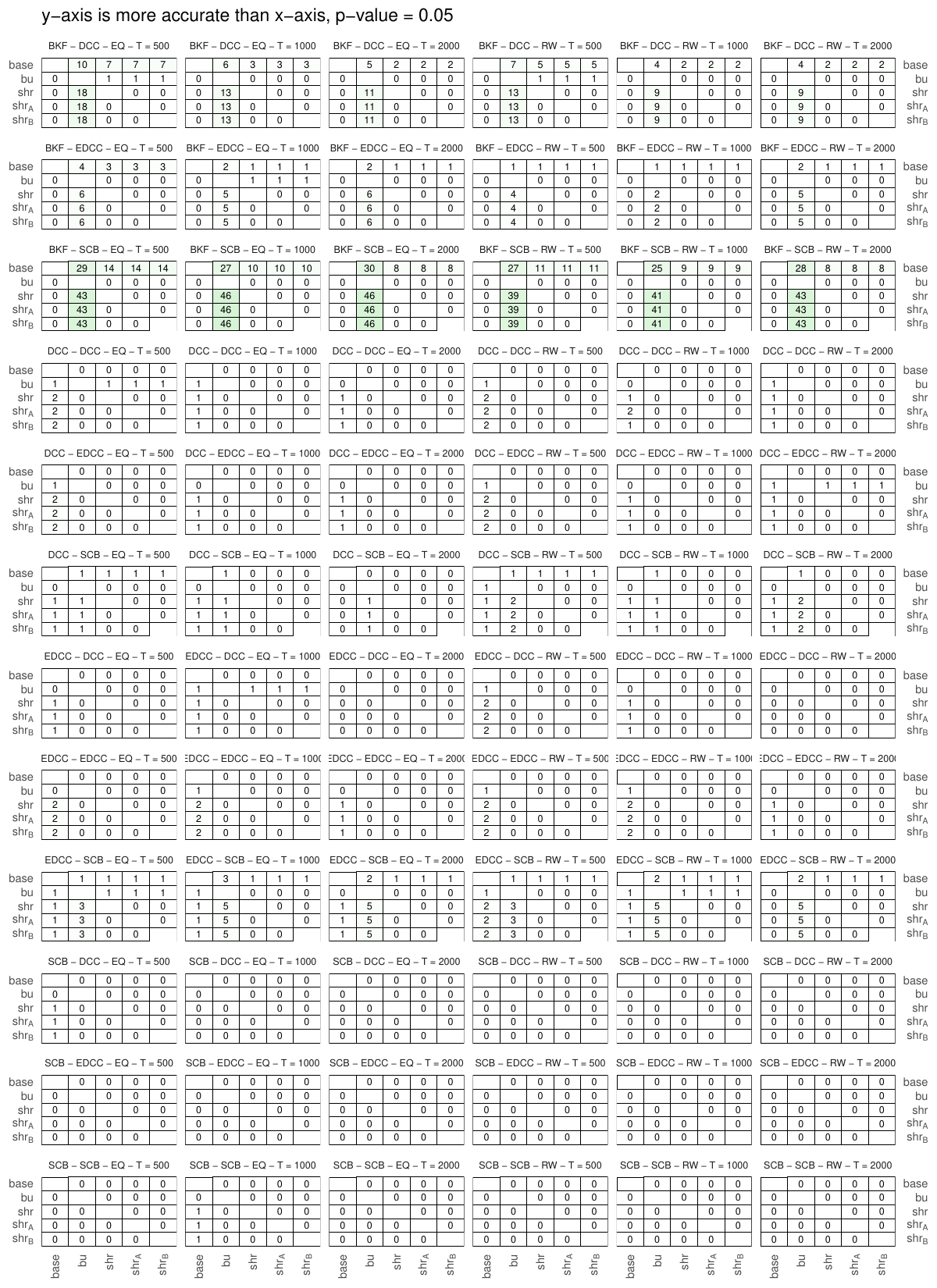}
	\caption{Qualitative evaluation using the Diebold-Mariano across different simulation settings using a QLIKE loss function and noisy level $\delta = 1$. Each setting is identified by a label composed of three elements: (i) the data-generating process (DGP), (ii) the MGARCH model used for estimation, and (iii) the portfolio weighting scheme (see \autoref{tab:notationModel} for details). The sample size ($T$) is also reported. The rows and columns indicate the variance forecasting method employed: the univariate GARCH on portfolio returns (base), the bottom-up approach using MGARCH models (bu), and the three forecast reconciliation strategies discussed in the paper ($shr$, $shr_{A}$, $shr_{B}$). Each cell reports the number of times (in \%) the forecasting model in the row statistically outperforms (p-values $< 0.05$ and using Bonferroni correction) the model in the column.}
\end{figure}
\begin{figure}[H]
\centering
	\includegraphics[width=0.9\linewidth]{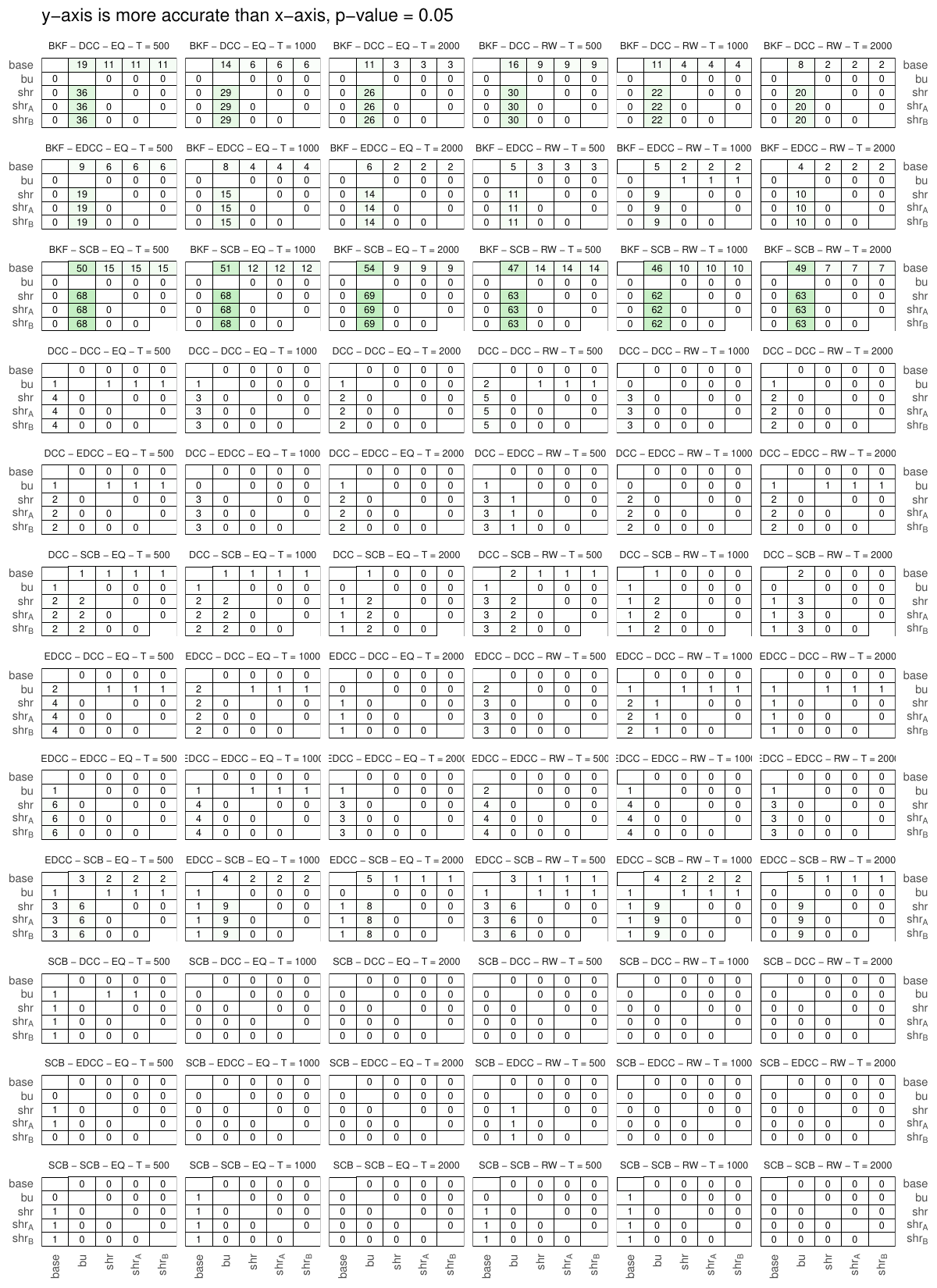}
\caption{Qualitative evaluation using the Diebold-Mariano across different simulation settings using a QLIKE loss function and noisy level $\delta = 0.75$. Each setting is identified by a label composed of three elements: (i) the data-generating process (DGP), (ii) the MGARCH model used for estimation, and (iii) the portfolio weighting scheme (see \autoref{tab:notationModel} for details). The sample size ($T$) is also reported. The rows and columns indicate the variance forecasting method employed: the univariate GARCH on portfolio returns (base), the bottom-up approach using MGARCH models (bu), and the three forecast reconciliation strategies discussed in the paper ($shr$, $shr_{A}$, $shr_{B}$). Each cell reports the number of times (in \%) the forecasting model in the row statistically outperforms (p-values $< 0.05$ and using Bonferroni correction) the model in the column.}
\end{figure}
\begin{figure}[H]
\centering
	\includegraphics[width=0.9\linewidth]{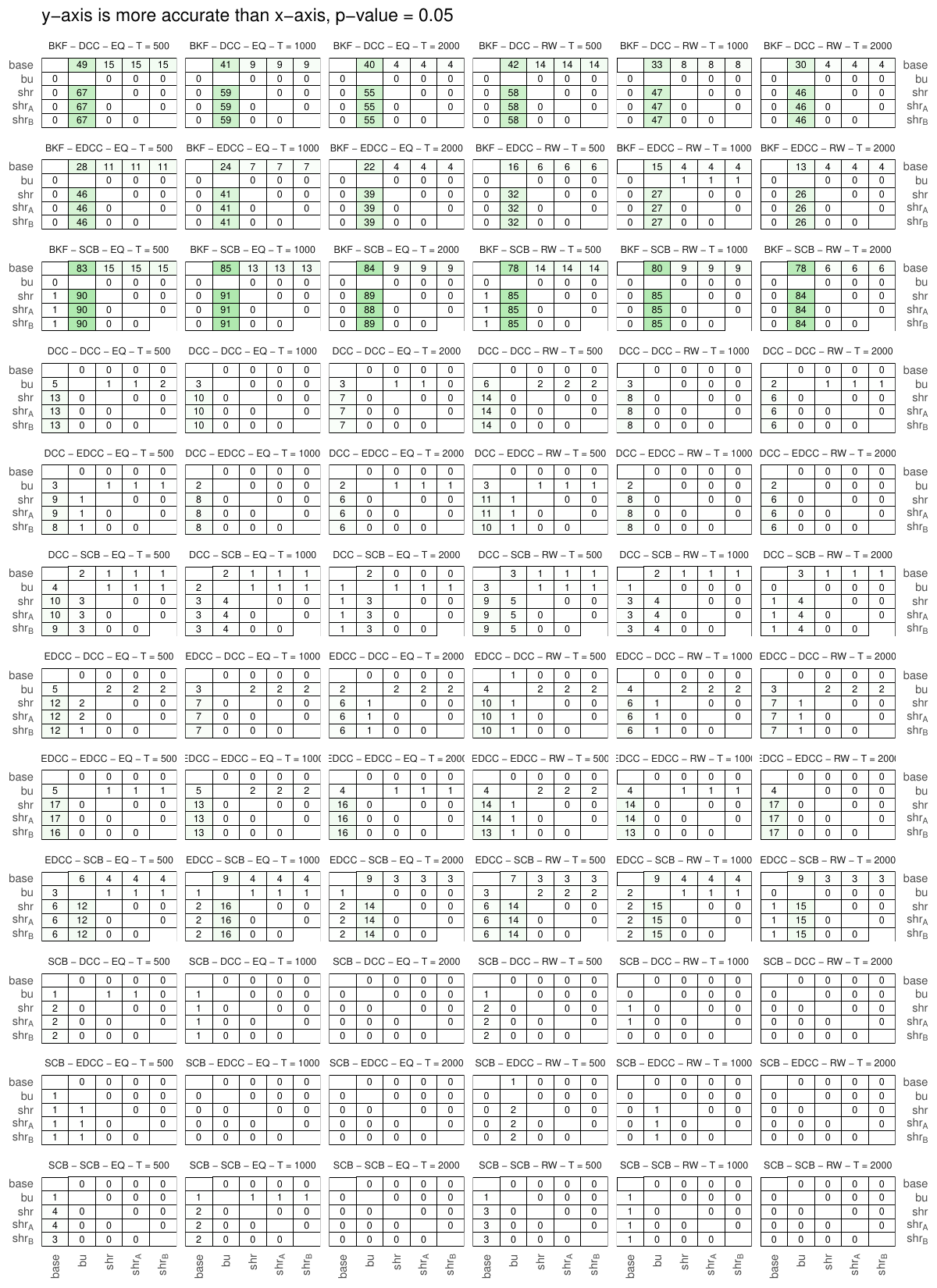}
\caption{Qualitative evaluation using the Diebold-Mariano across different simulation settings using a QLIKE loss function and noisy level $\delta = 0.5$. Each setting is identified by a label composed of three elements: (i) the data-generating process (DGP), (ii) the MGARCH model used for estimation, and (iii) the portfolio weighting scheme (see \autoref{tab:notationModel} for details). The sample size ($T$) is also reported. The rows and columns indicate the variance forecasting method employed: the univariate GARCH on portfolio returns (base), the bottom-up approach using MGARCH models (bu), and the three forecast reconciliation strategies discussed in the paper ($shr$, $shr_{A}$, $shr_{B}$). Each cell reports the number of times (in \%) the forecasting model in the row statistically outperforms (p-values $< 0.05$ and using Bonferroni correction) the model in the column.}
\end{figure}
\begin{figure}[H]
\centering
	\includegraphics[width=0.9\linewidth]{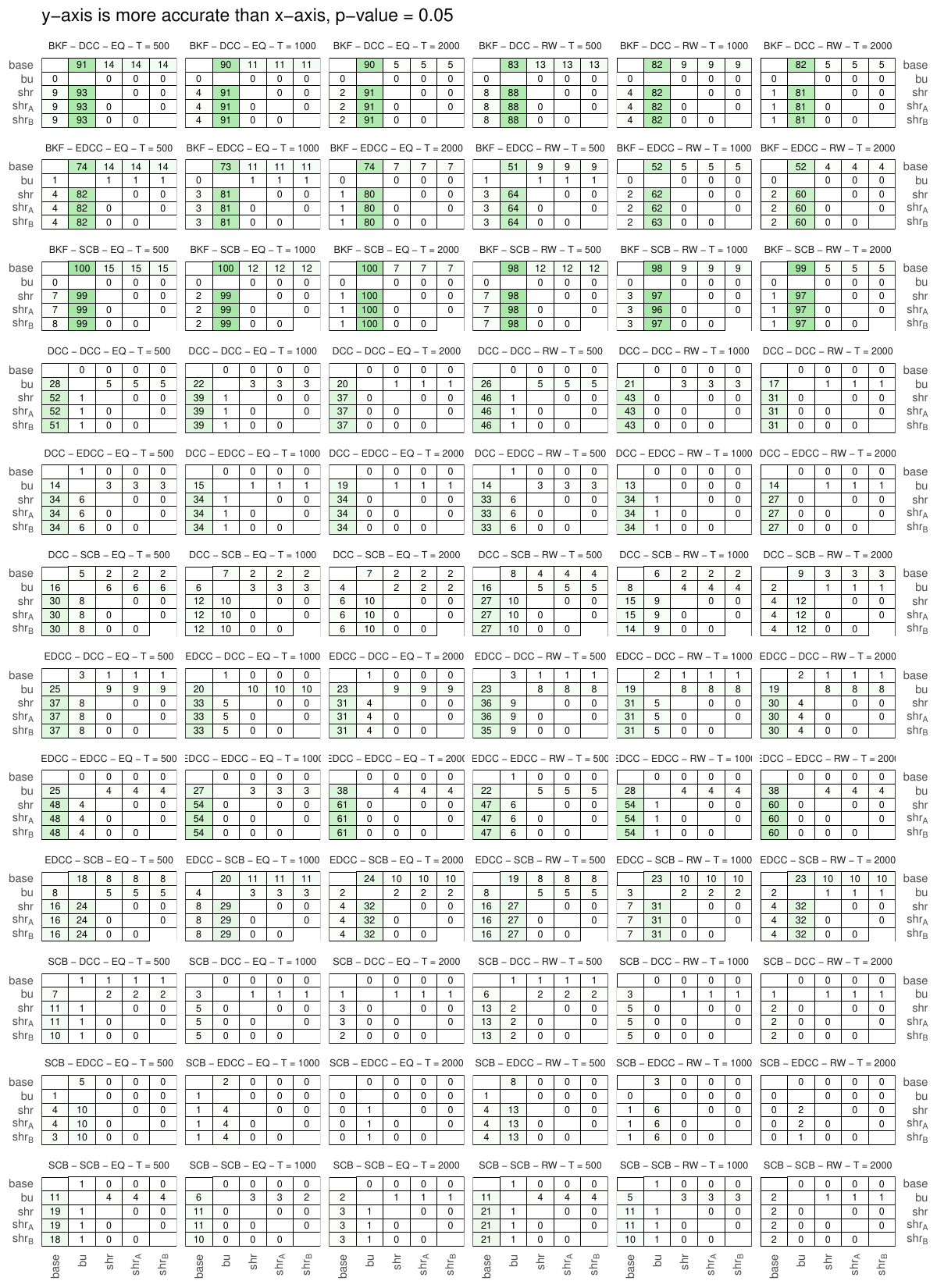}
\caption{Qualitative evaluation using the Diebold-Mariano across different simulation settings using a QLIKE loss function and noisy level $\delta = 0.25$. Each setting is identified by a label composed of three elements: (i) the data-generating process (DGP), (ii) the MGARCH model used for estimation, and (iii) the portfolio weighting scheme (see \autoref{tab:notationModel} for details). The sample size ($T$) is also reported. The rows and columns indicate the variance forecasting method employed: the univariate GARCH on portfolio returns (base), the bottom-up approach using MGARCH models (bu), and the three forecast reconciliation strategies discussed in the paper ($shr$, $shr_{A}$, $shr_{B}$). Each cell reports the number of times (in \%) the forecasting model in the row statistically outperforms (p-values $< 0.05$ and using Bonferroni correction) the model in the column.}
\end{figure}

\subsubsection{Visual results}

\begin{figure}[H]
\centering
\includegraphics[width = 0.8\linewidth]{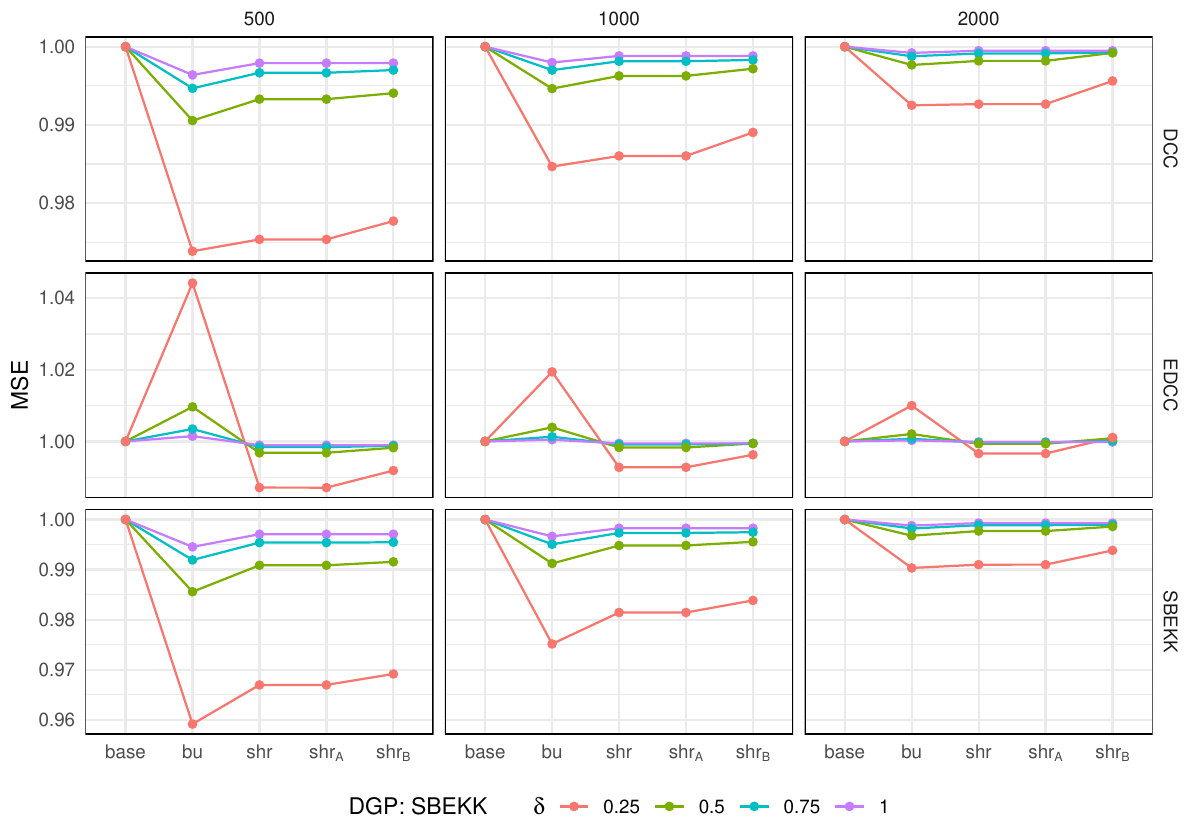}
\caption{Average relative MSE where the reference forecast is the univariate GARCH fitted on simulated portfolio returns with equally weighted portfolios (the $1/N$ case), and the DGP is a Scalar BEKK. The fitted MGARCH models are: the DCC-GARCH (first row, DCC) and the Scalar BEKK (second row, SBEKK). The columns indicates the sample size ($T$): $T=500$ left,  $T=1000$ center and $T=2000$ right column. All values are averages across the 500 experiments.}
\vspace*{1em}
\includegraphics[width = 0.8\linewidth]{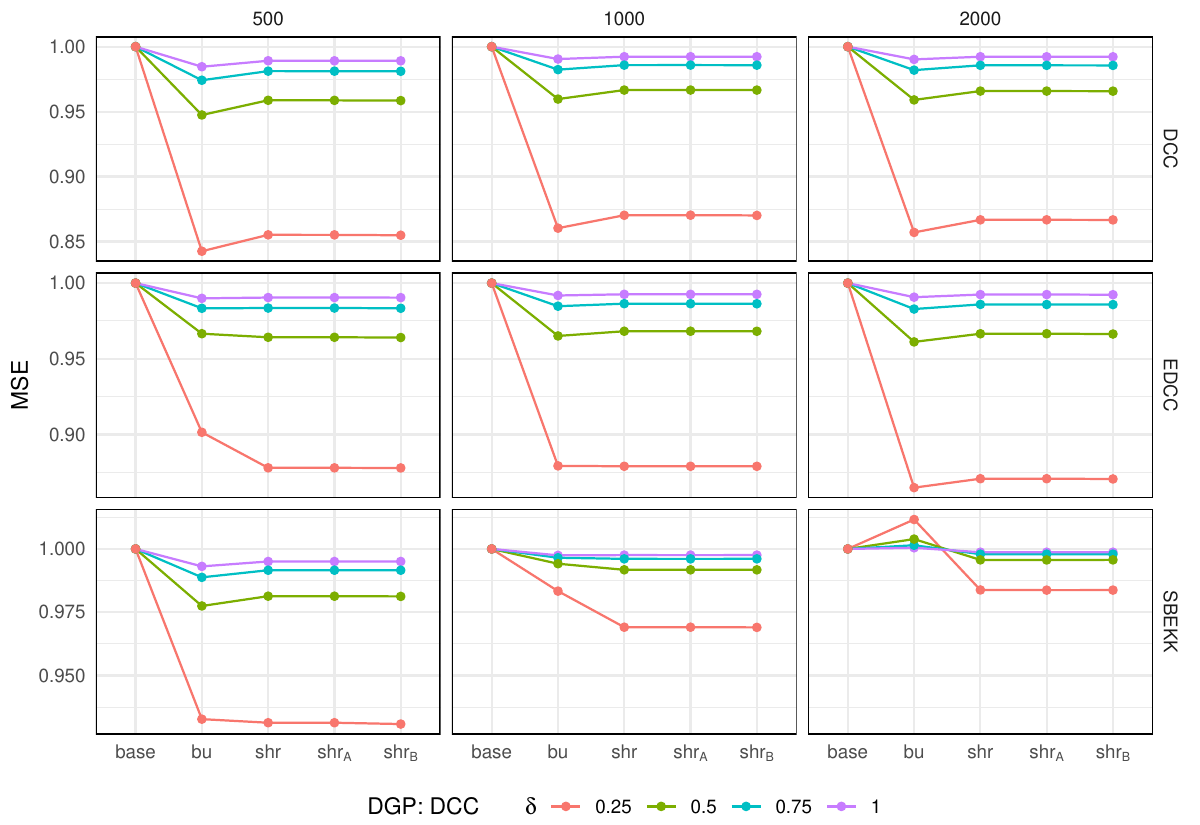}
\caption{Average relative MSE where the reference forecast is the univariate GARCH fitted on simulated portfolio returns with equally weighted portfolios (the $1/N$ case), and the DGP is a DCC-GARCH. The fitted MGARCH models are: the DCC-GARCH (first row, DCC) and the Scalar BEKK (second row, SBEKK). The columns indicates the sample size ($T$): $T=500$ left,  $T=1000$ center and $T=20000$ right column. All values are averages across the 500 experiments.}
\end{figure}

\begin{figure}[H]
\centering
\includegraphics[width = 0.8\linewidth]{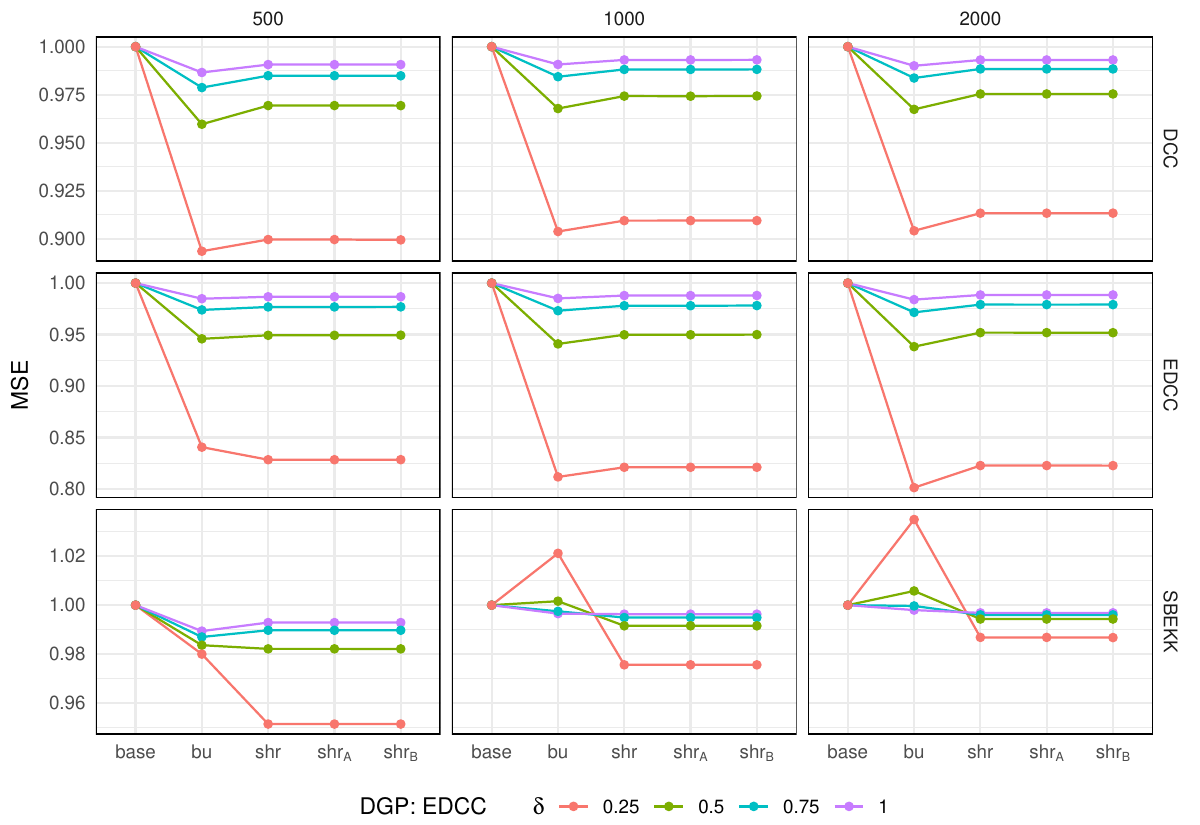}
\caption{Average relative MSE where the reference forecast is the univariate GARCH fitted on simulated portfolio returns with equally weighted portfolios (the $1/N$ case), and the DGP is a EDCC-GARCH. The fitted MGARCH models are: the DCC-GARCH (first row, DCC), the EDCC-GARCH (second row, EDCC) and the Scalar BEKK (third row, SBEKK). The columns indicates the sample size ($T$): $T=500$ left,  $T=1000$ center and $T=2000$ right column. All values are averages across the 500 experiments.}
\vspace*{1em}
\includegraphics[width = 0.8\linewidth]{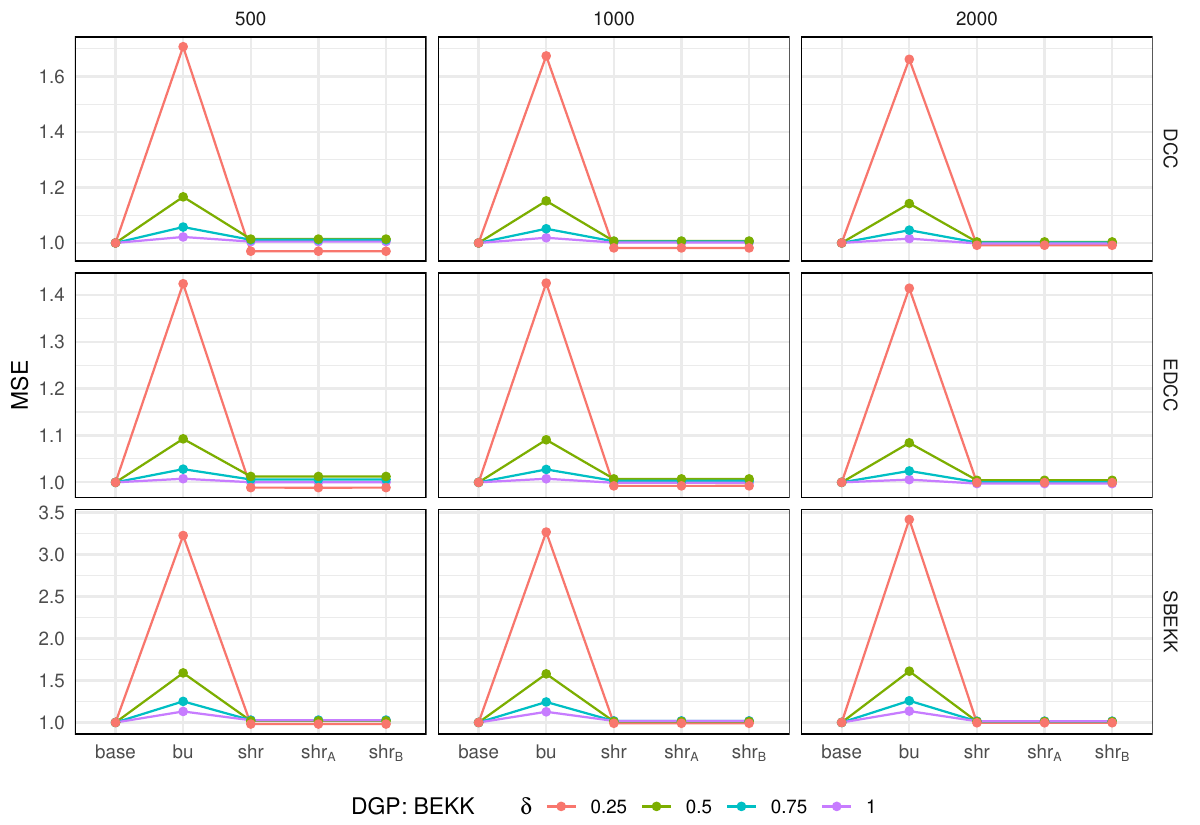}
\caption{Average relative MSE where the reference forecast is the univariate GARCH fitted on simulated portfolio returns with equally weighted portfolios (the $1/N$ case), and the DGP is a Full BEKK. The fitted MGARCH models are: the DCC-GARCH (first row, DCC) and the Scalar BEKK (second row, SBEKK). The columns indicates the sample size ($T$): $T=500$ left,  $T=1000$ center and $T=2000$ right column. All values are averages across the 500 experiments.}
\end{figure}

\begin{figure}[H]
\centering
\includegraphics[width = 0.8\linewidth]{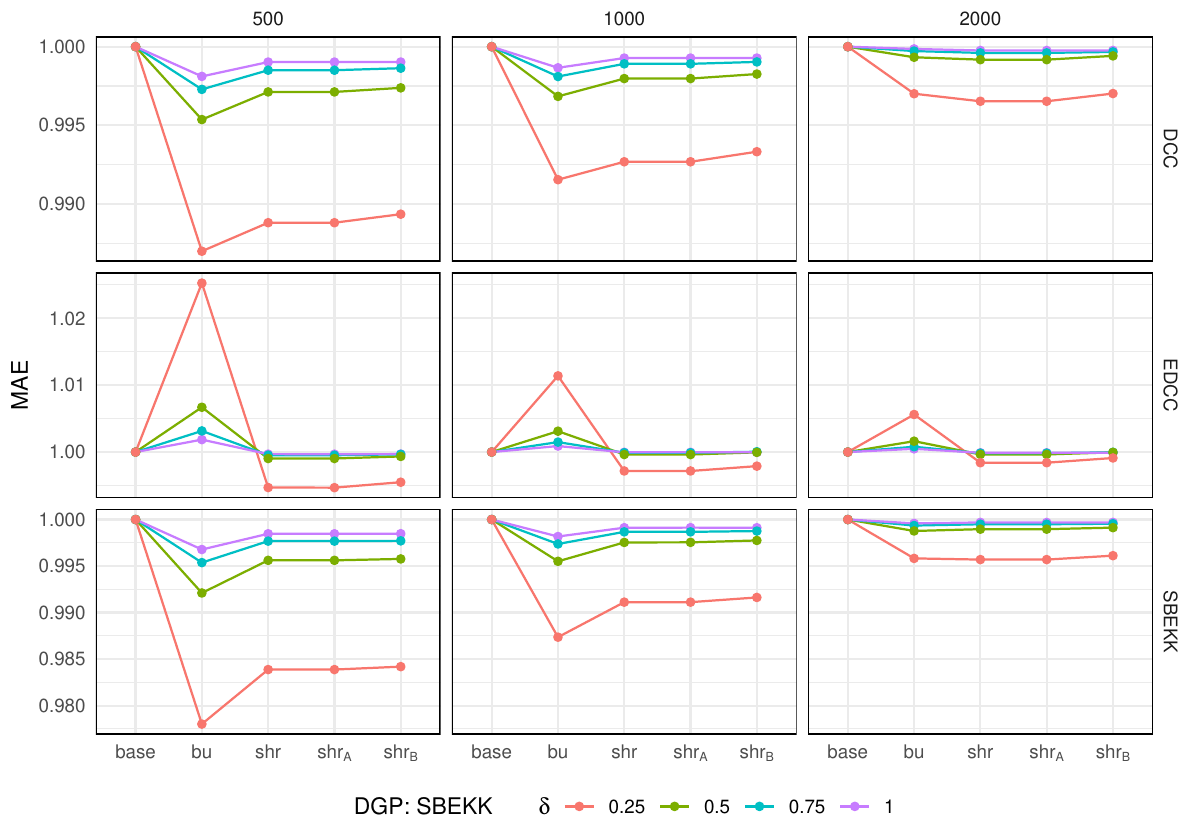}
\caption{Average relative MAE where the reference forecast is the univariate GARCH fitted on simulated portfolio returns with equally weighted portfolios (the $1/N$ case), and the DGP is a Scalar BEKK. The fitted MGARCH models are: the DCC-GARCH (first row, DCC) and the Scalar BEKK (second row, SBEKK). The columns indicates the sample size ($T$): $T=500$ left,  $T=1000$ center and $T=2000$ right column. All values are averages across the 500 experiments.}
\vspace*{1em}
\includegraphics[width = 0.8\linewidth]{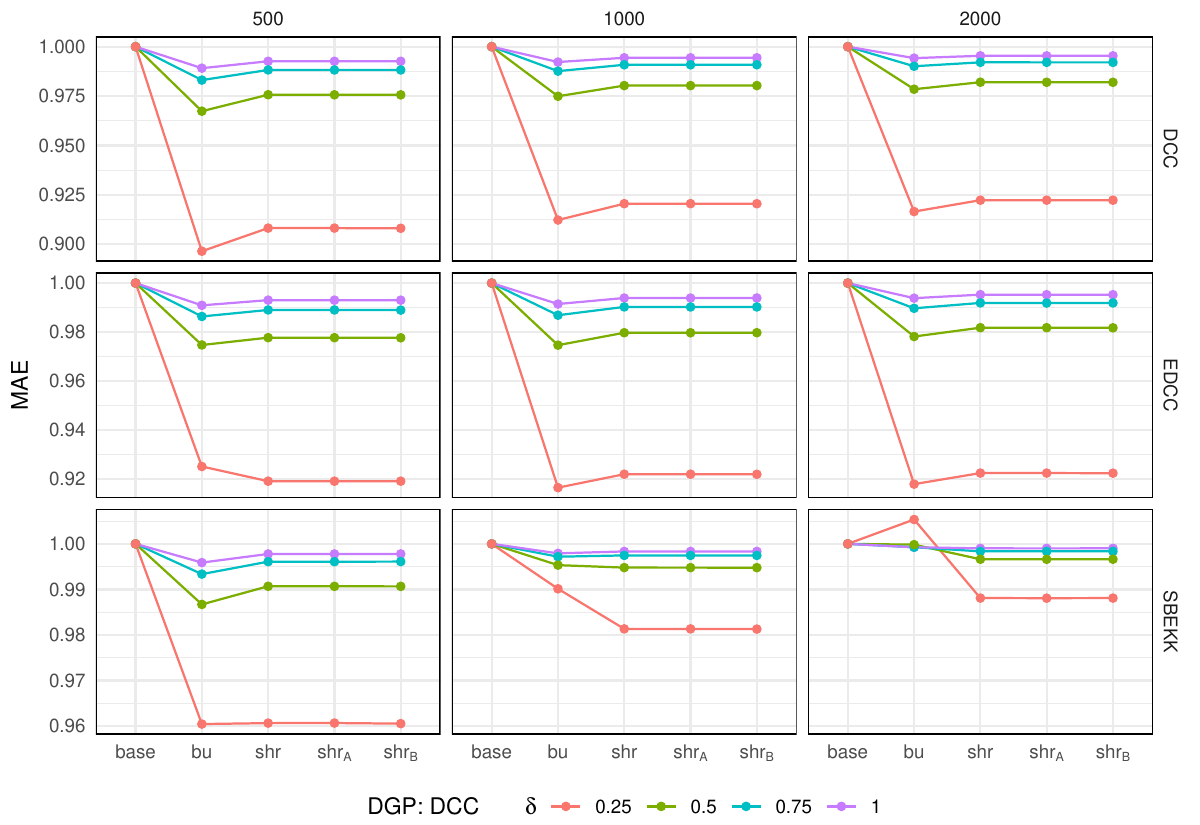}
\caption{Average relative MAE where the reference forecast is the univariate GARCH fitted on simulated portfolio returns with equally weighted portfolios (the $1/N$ case), and the DGP is a DCC-GARCH. The fitted MGARCH models are: the DCC-GARCH (first row, DCC) and the Scalar BEKK (second row, SBEKK). The columns indicates the sample size ($T$): $T=500$ left,  $T=1000$ center and $T=20000$ right column. All values are averages across the 500 experiments.}
\end{figure}

\begin{figure}[H]
\centering
\includegraphics[width = 0.8\linewidth]{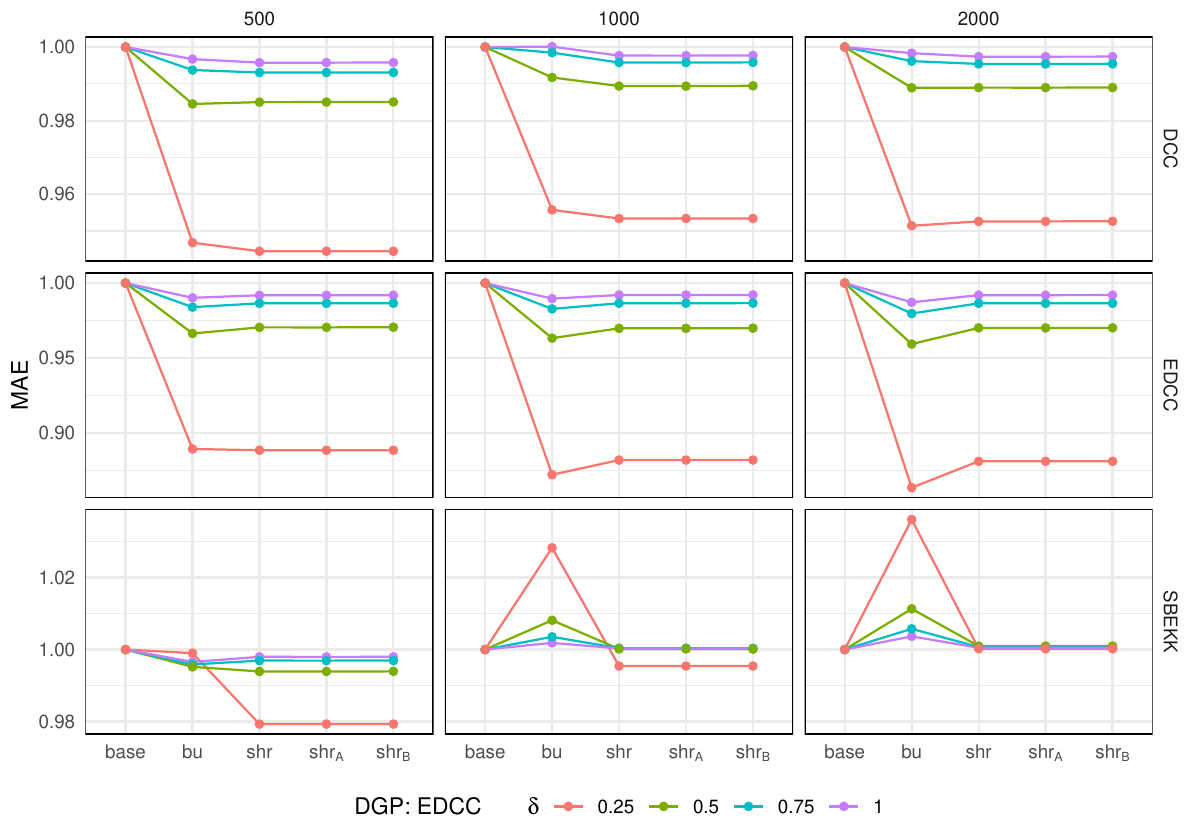}
\caption{Average relative MAE where the reference forecast is the univariate GARCH fitted on simulated portfolio returns with equally weighted portfolios (the $1/N$ case), and the DGP is a EDCC-GARCH. The fitted MGARCH models are: the DCC-GARCH (first row, DCC), the EDCC-GARCH (second row, EDCC) and the Scalar BEKK (third row, SBEKK). The columns indicates the sample size ($T$): $T=500$ left,  $T=1000$ center and $T=2000$ right column. All values are averages across the 500 experiments.}
\vspace*{1em}
\includegraphics[width = 0.8\linewidth]{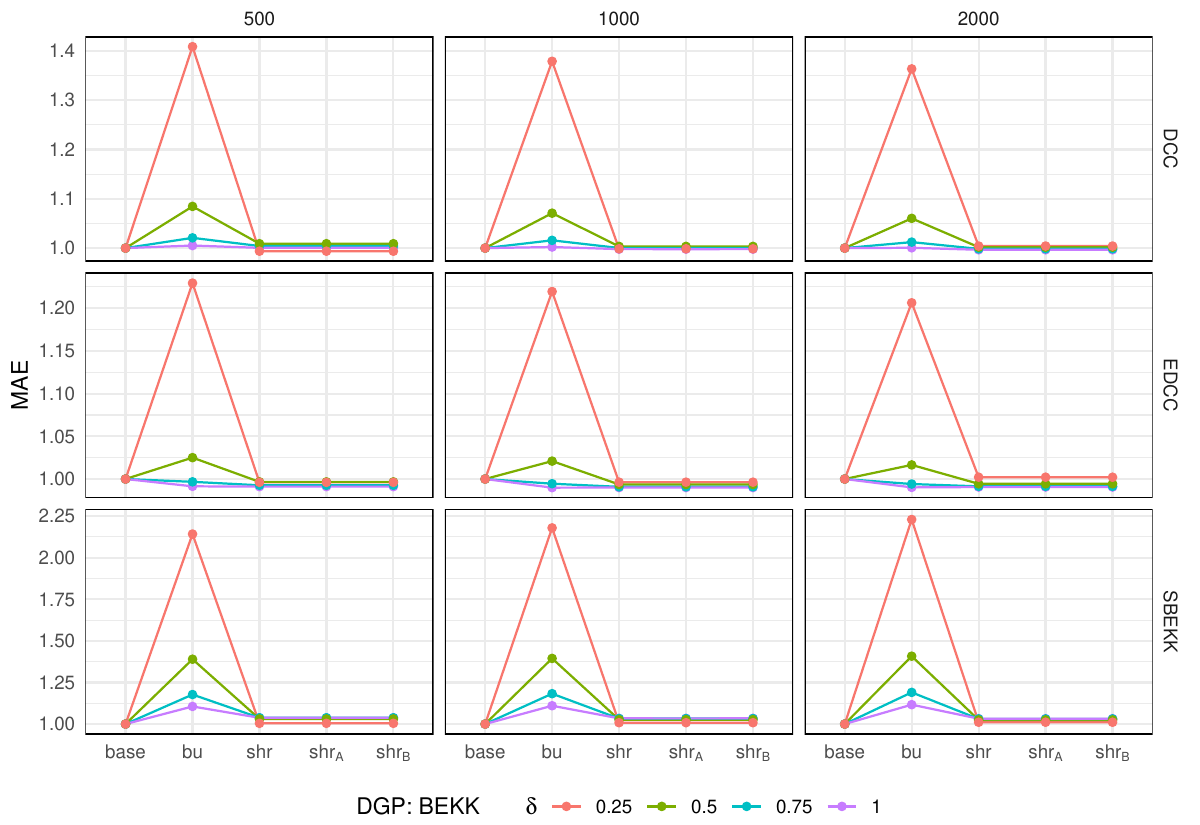}
\caption{Average relative MAE where the reference forecast is the univariate GARCH fitted on simulated portfolio returns with equally weighted portfolios (the $1/N$ case), and the DGP is a Full BEKK. The fitted MGARCH models are: the DCC-GARCH (first row, DCC) and the Scalar BEKK (second row, SBEKK). The columns indicates the sample size ($T$): $T=500$ left,  $T=1000$ center and $T=2000$ right column. All values are averages across the 500 experiments.}
\end{figure}

\begin{figure}[H]
\centering
\includegraphics[width = 0.8\linewidth]{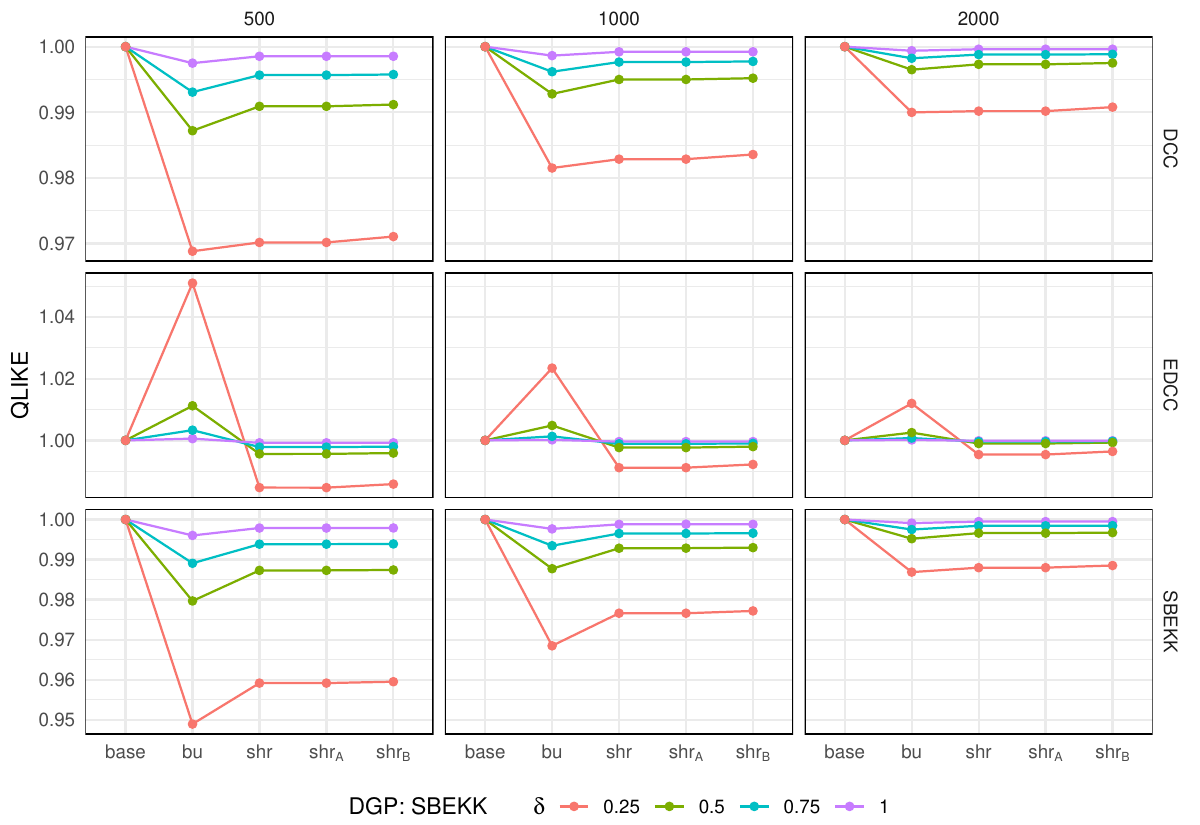}
\caption{Average relative QLIKE where the reference forecast is the univariate GARCH fitted on simulated portfolio returns with equally weighted portfolios (the $1/N$ case), and the DGP is a Scalar BEKK. The fitted MGARCH models are: the DCC-GARCH (first row, DCC) and the Scalar BEKK (second row, SBEKK). The columns indicates the sample size ($T$): $T=500$ left,  $T=1000$ center and $T=2000$ right column. All values are averages across the 500 experiments.}
\vspace*{1em}
\includegraphics[width = 0.8\linewidth]{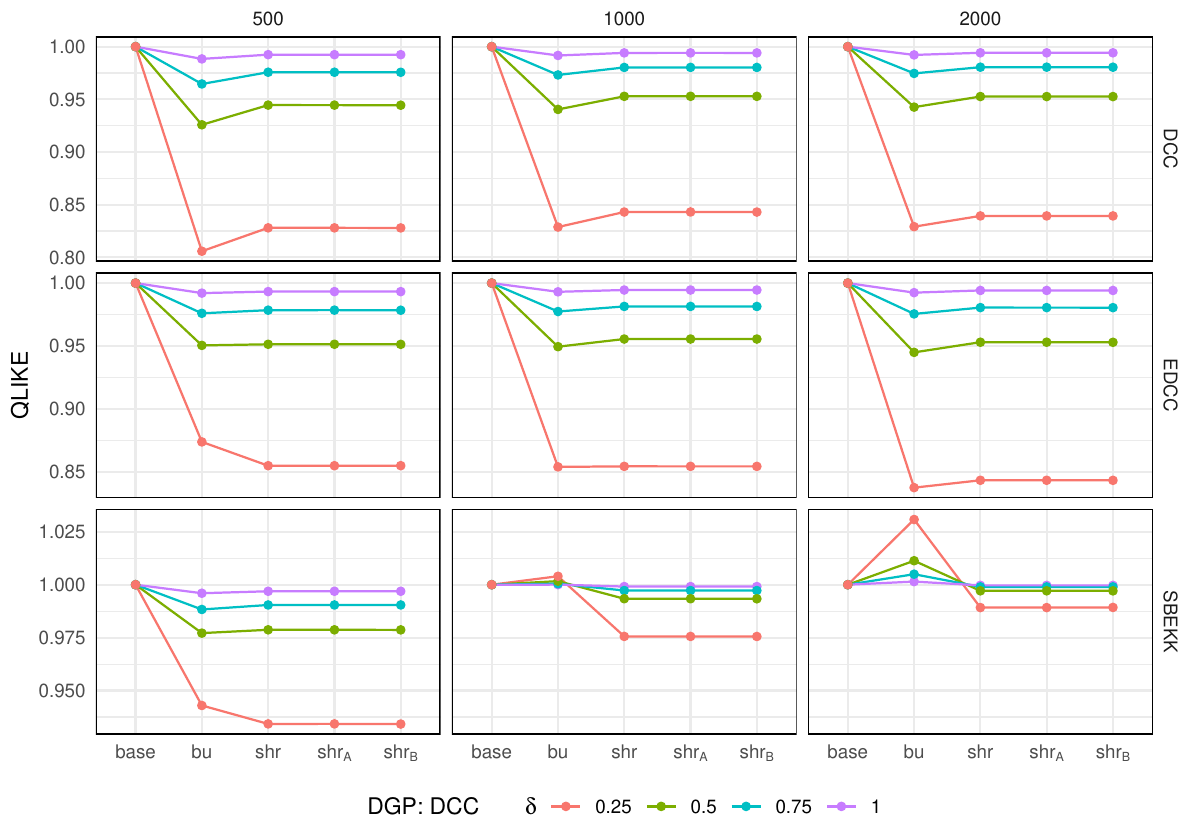}
\caption{Average relative QLIKE where the reference forecast is the univariate GARCH fitted on simulated portfolio returns with equally weighted portfolios (the $1/N$ case), and the DGP is a DCC-GARCH. The fitted MGARCH models are: the DCC-GARCH (first row, DCC) and the Scalar BEKK (second row, SBEKK). The columns indicates the sample size ($T$): $T=500$ left,  $T=1000$ center and $T=20000$ right column. All values are averages across the 500 experiments.}
\end{figure}

\begin{figure}[H]
\centering
\includegraphics[width = 0.8\linewidth]{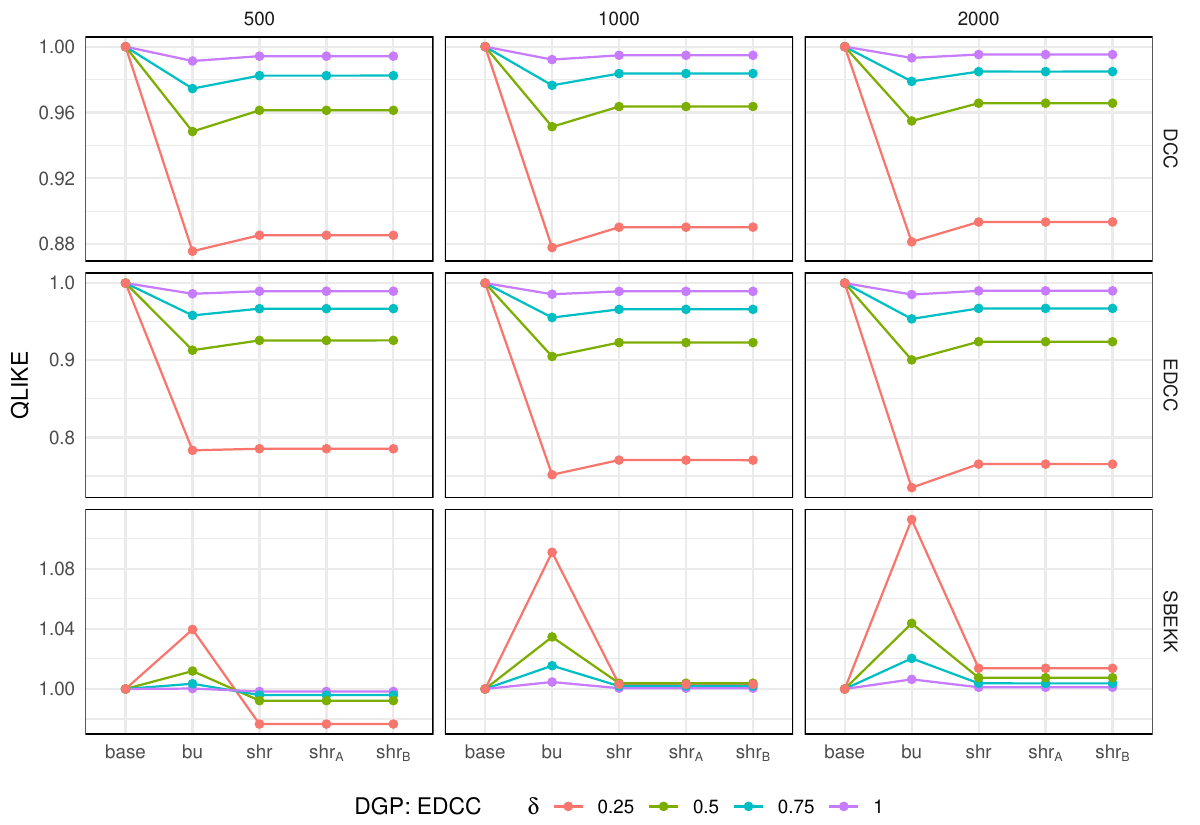}
\caption{Average relative QLIKE where the reference forecast is the univariate GARCH fitted on simulated portfolio returns with equally weighted portfolios (the $1/N$ case), and the DGP is a EDCC-GARCH. The fitted MGARCH models are: the DCC-GARCH (first row, DCC), the EDCC-GARCH (second row, EDCC) and the Scalar BEKK (third row, SBEKK). The columns indicates the sample size ($T$): $T=500$ left,  $T=1000$ center and $T=2000$ right column. All values are averages across the 500 experiments.}
\vspace*{1em}
\includegraphics[width = 0.8\linewidth]{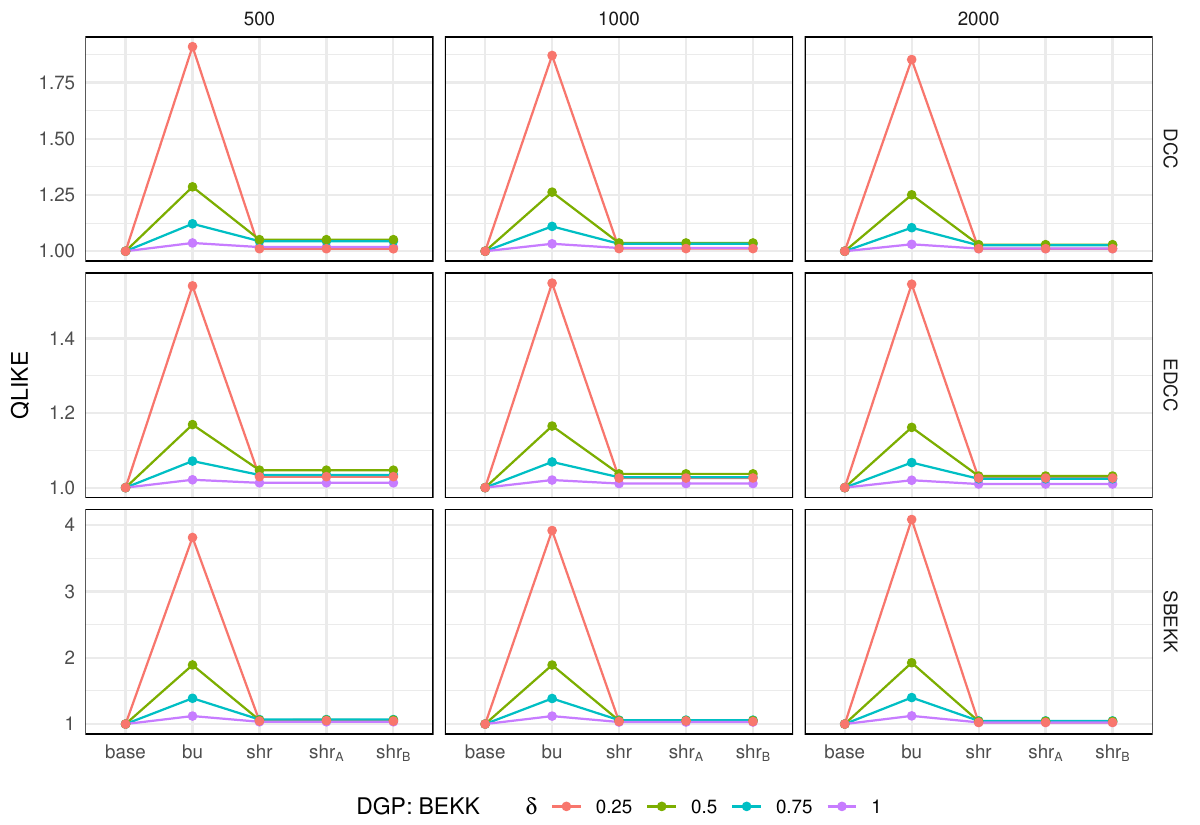}
\caption{Average relative QLIKE where the reference forecast is the univariate GARCH fitted on simulated portfolio returns with equally weighted portfolios (the $1/N$ case), and the DGP is a Full BEKK. The fitted MGARCH models are: the DCC-GARCH (first row, DCC) and the Scalar BEKK (second row, SBEKK). The columns indicates the sample size ($T$): $T=500$ left,  $T=1000$ center and $T=2000$ right column. All values are averages across the 500 experiments.}
\end{figure}

\begin{figure}[H]
\centering
\includegraphics[width = 0.8\linewidth]{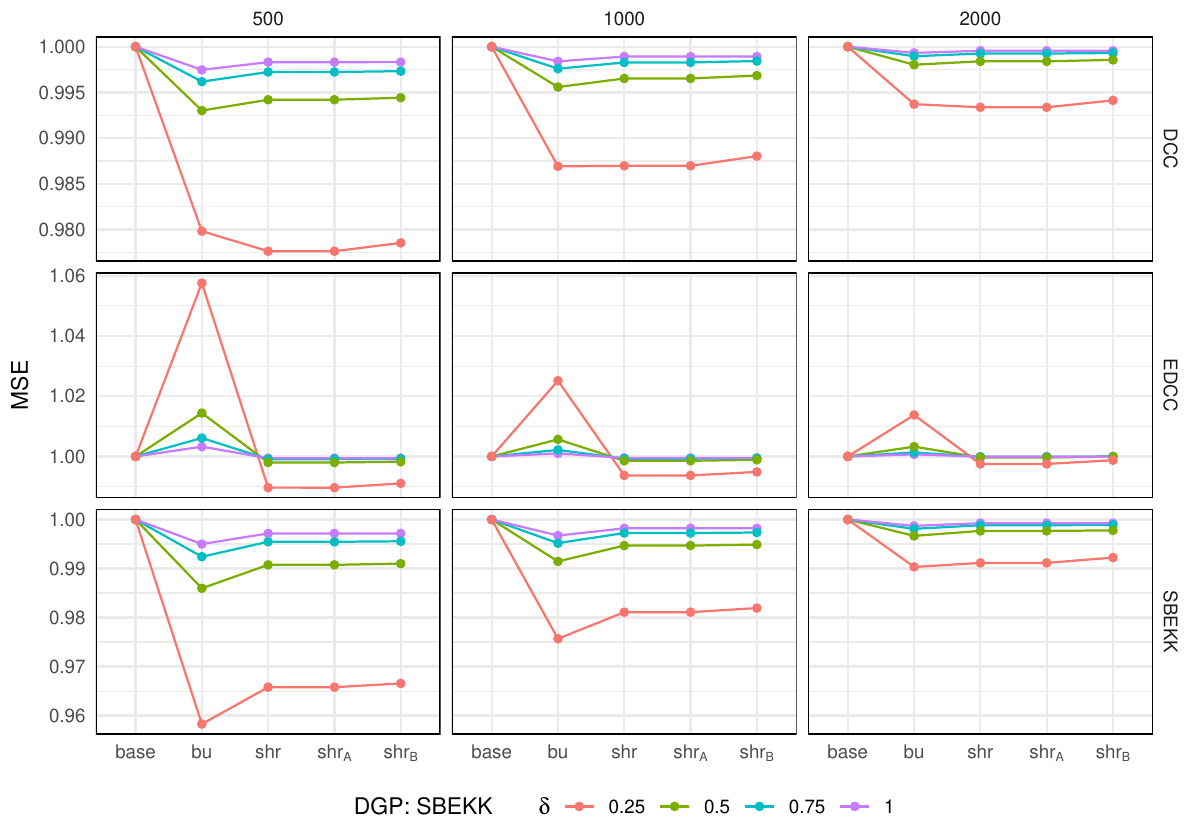}
\caption{Average relative MSE where the reference forecast is the univariate GARCH fitted on simulated portfolio returns with random weighted portfolios, and the DGP is a Scalar BEKK. The fitted MGARCH models are: the DCC-GARCH (first row, DCC) and the Scalar BEKK (second row, SBEKK). The columns indicates the sample size ($T$): $T=500$ left,  $T=1000$ center and $T=2000$ right column. All values are averages across the 500 experiments.}
\vspace*{1em}
\includegraphics[width = 0.8\linewidth]{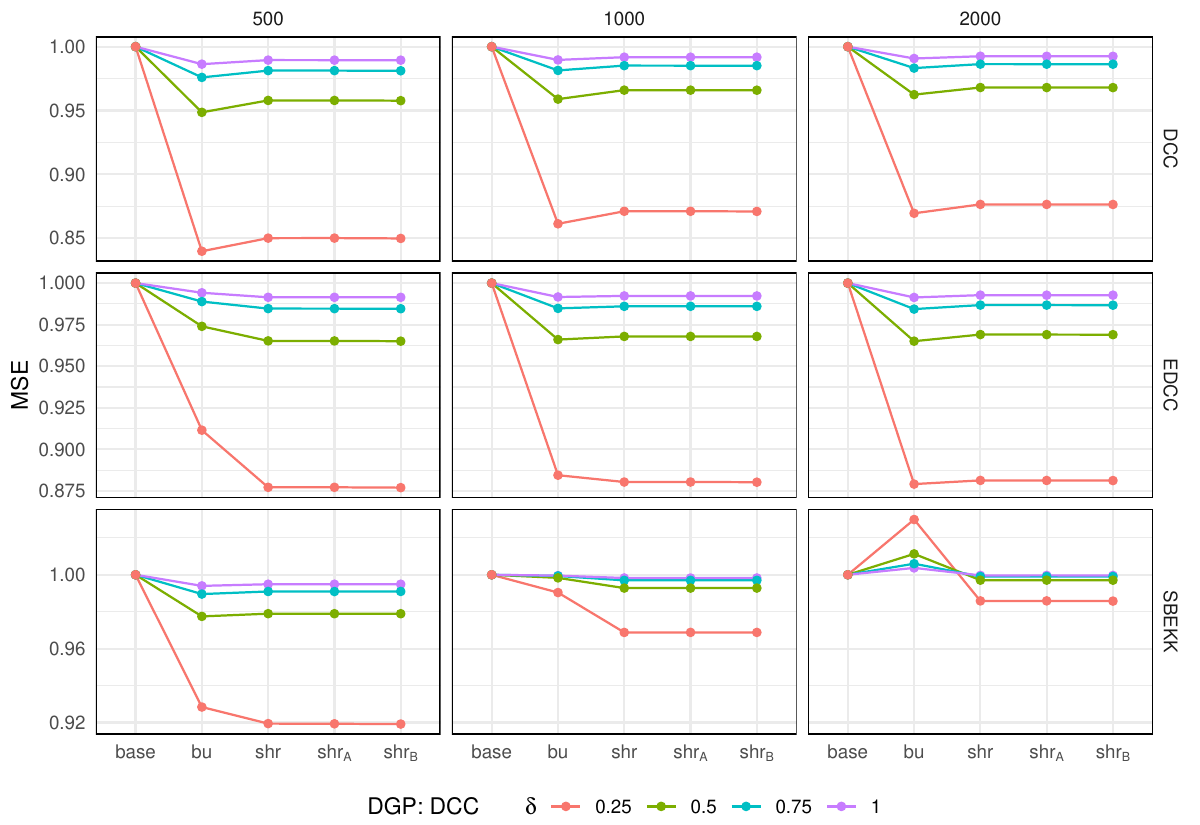}
\caption{Average relative MSE where the reference forecast is the univariate GARCH fitted on simulated portfolio returns with random weighted portfolios, and the DGP is a DCC-GARCH. The fitted MGARCH models are: the DCC-GARCH (first row, DCC) and the Scalar BEKK (second row, SBEKK). The columns indicates the sample size ($T$): $T=500$ left,  $T=1000$ center and $T=20000$ right column. All values are averages across the 500 experiments.}
\end{figure}

\begin{figure}[H]
\centering
\includegraphics[width = 0.8\linewidth]{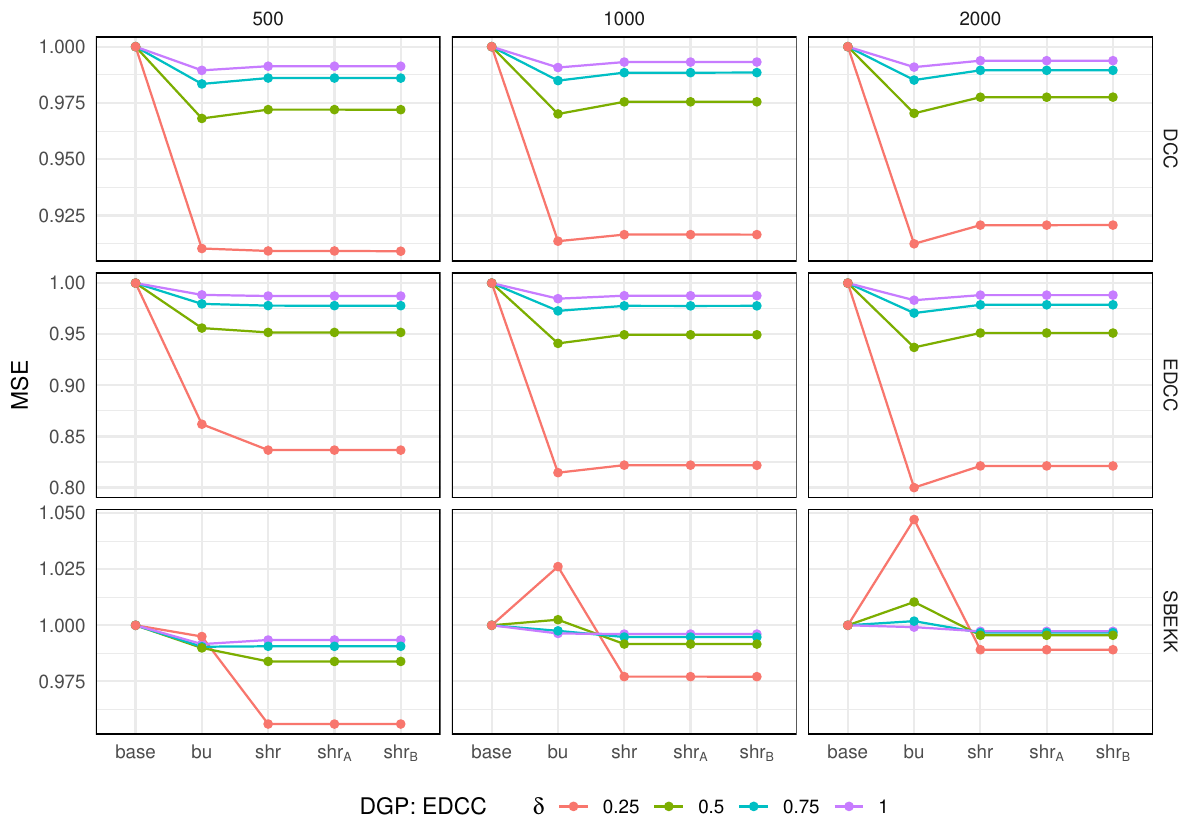}
\caption{Average relative MSE where the reference forecast is the univariate GARCH fitted on simulated portfolio returns with random weighted portfolios, and the DGP is a EDCC-GARCH. The fitted MGARCH models are: the DCC-GARCH (first row, DCC), the EDCC-GARCH (second row, EDCC) and the Scalar BEKK (third row, SBEKK). The columns indicates the sample size ($T$): $T=500$ left,  $T=1000$ center and $T=2000$ right column. All values are averages across the 500 experiments.}
\vspace*{1em}
\includegraphics[width = 0.8\linewidth]{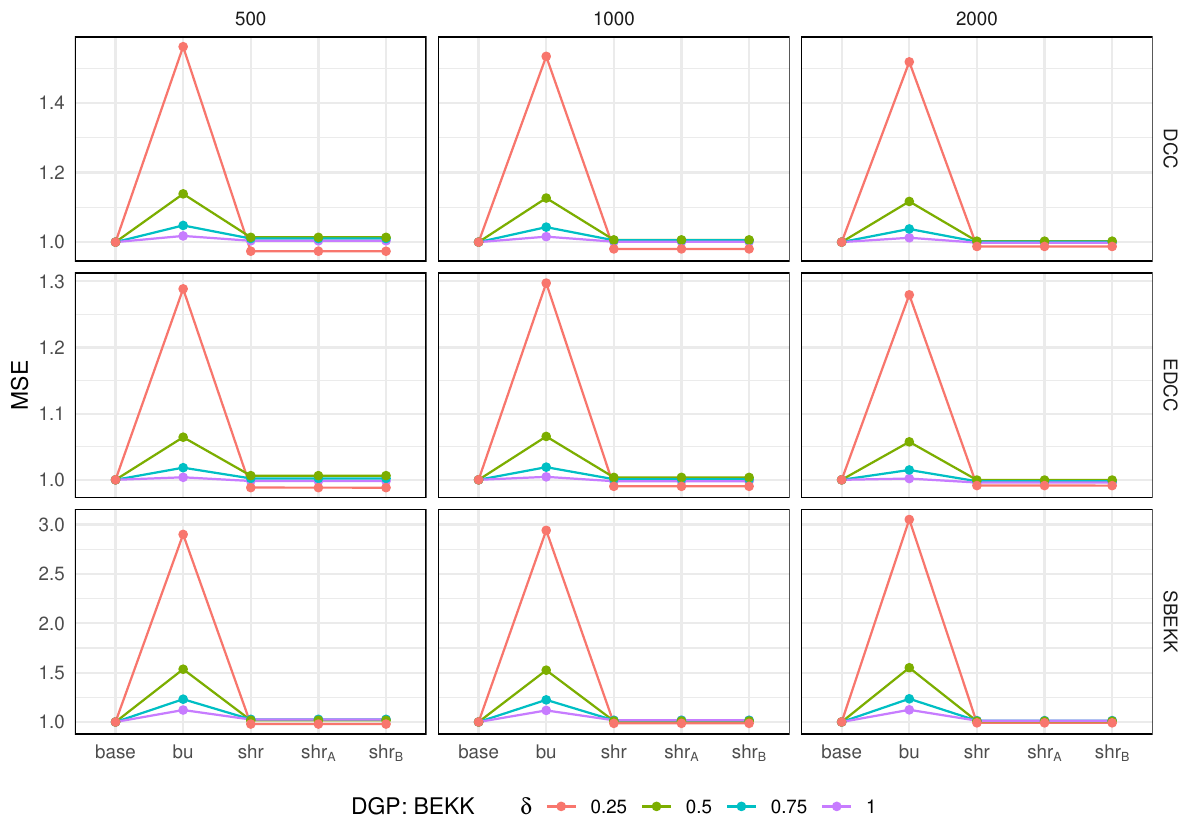}
\caption{Average relative MSE where the reference forecast is the univariate GARCH fitted on simulated portfolio returns with random weighted portfolios, and the DGP is a Full BEKK. The fitted MGARCH models are: the DCC-GARCH (first row, DCC) and the Scalar BEKK (second row, SBEKK). The columns indicates the sample size ($T$): $T=500$ left,  $T=1000$ center and $T=2000$ right column. All values are averages across the 500 experiments.}
\end{figure}

\begin{figure}[H]
\centering
\includegraphics[width = 0.8\linewidth]{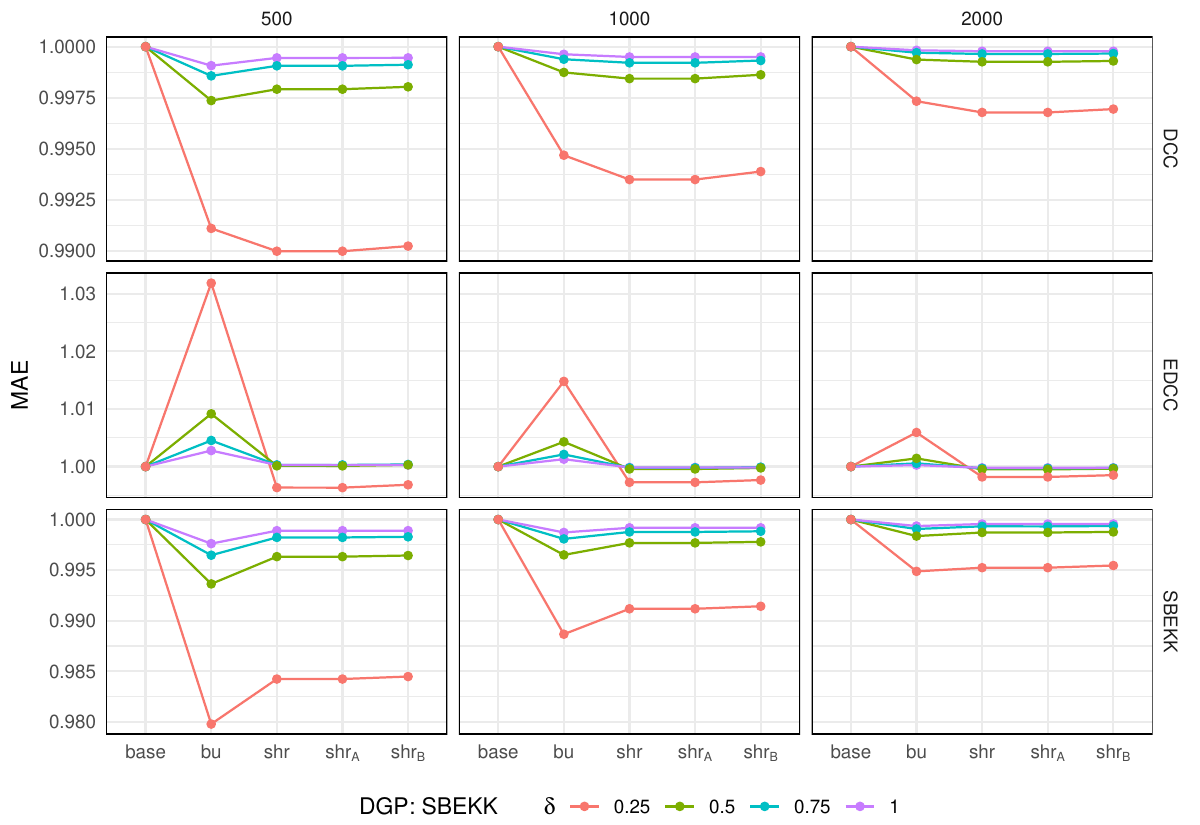}
\caption{Average relative MAE where the reference forecast is the univariate GARCH fitted on simulated portfolio returns with random weighted portfolios, and the DGP is a Scalar BEKK. The fitted MGARCH models are: the DCC-GARCH (first row, DCC) and the Scalar BEKK (second row, SBEKK). The columns indicates the sample size ($T$): $T=500$ left,  $T=1000$ center and $T=2000$ right column. All values are averages across the 500 experiments.}
\vspace*{1em}
\includegraphics[width = 0.8\linewidth]{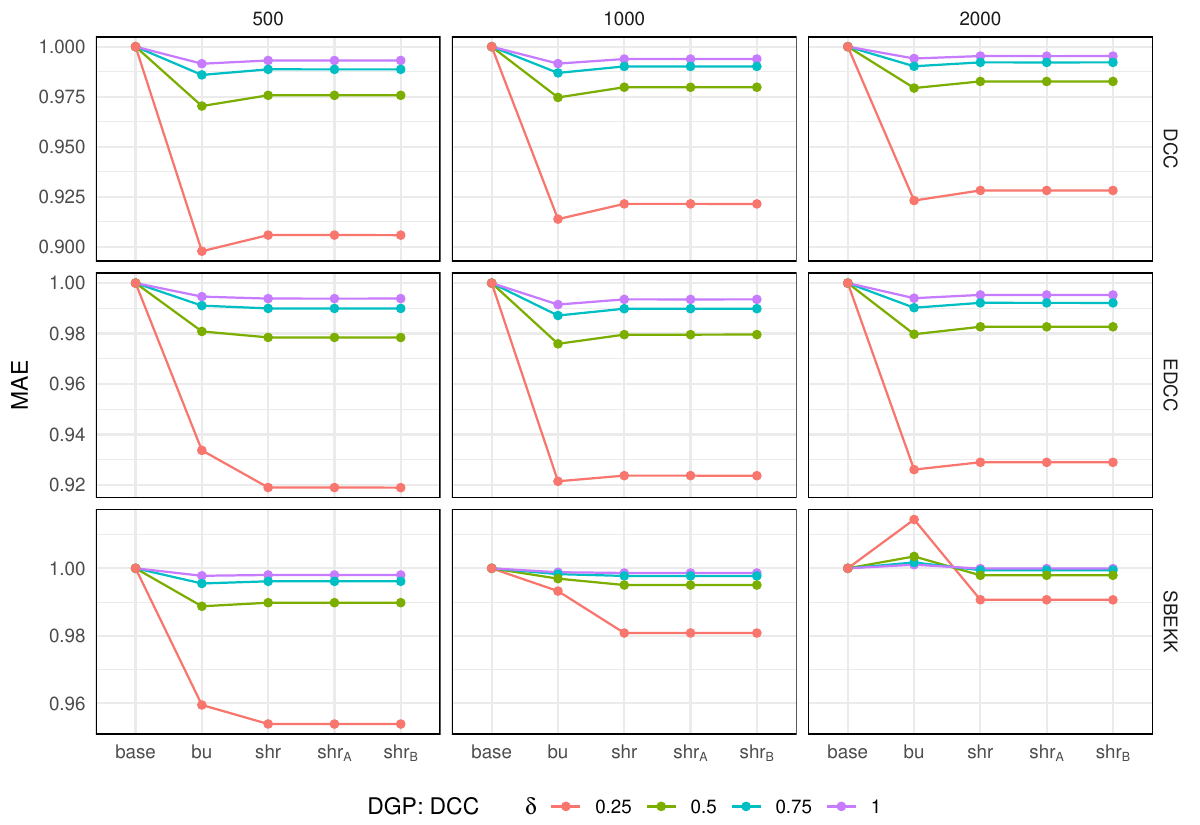}
\caption{Average relative MAE where the reference forecast is the univariate GARCH fitted on simulated portfolio returns with random weighted portfolios, and the DGP is a DCC-GARCH. The fitted MGARCH models are: the DCC-GARCH (first row, DCC) and the Scalar BEKK (second row, SBEKK). The columns indicates the sample size ($T$): $T=500$ left,  $T=1000$ center and $T=20000$ right column. All values are averages across the 500 experiments.}
\end{figure}

\begin{figure}[H]
\centering
\includegraphics[width = 0.8\linewidth]{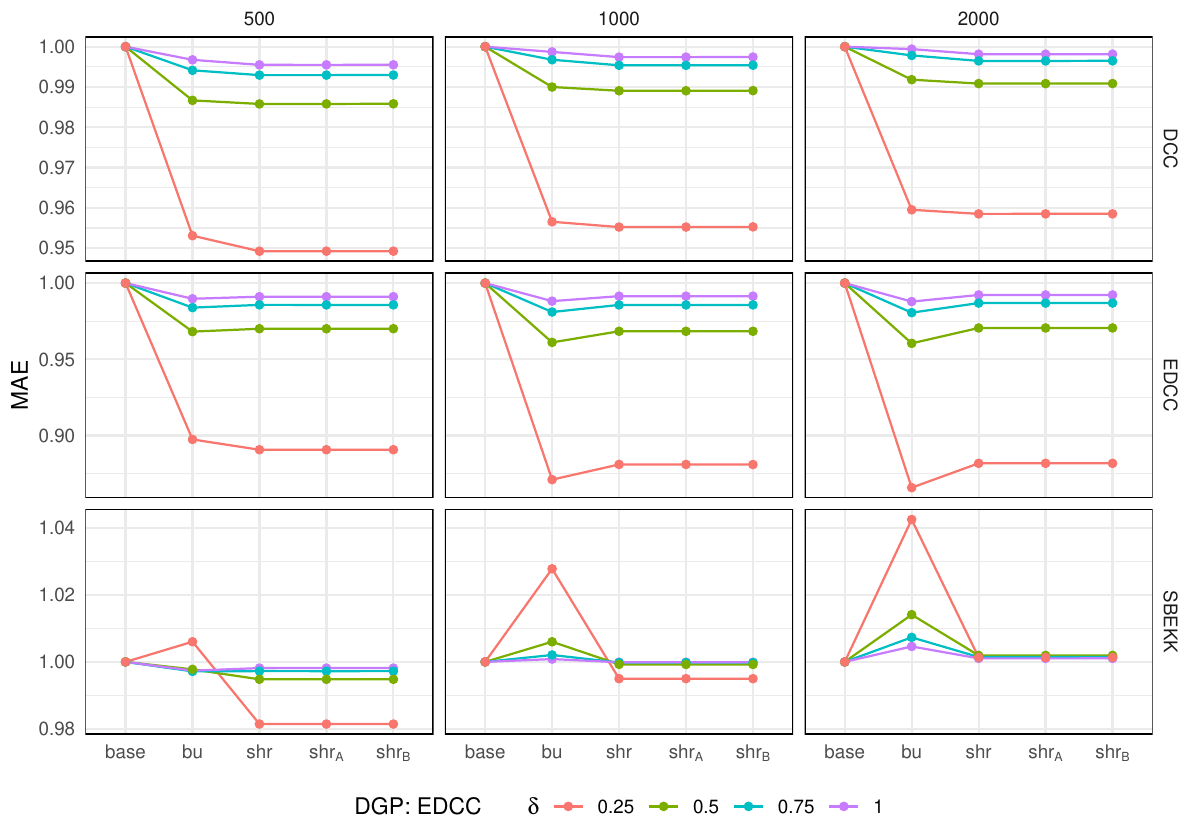}
\caption{Average relative MAE where the reference forecast is the univariate GARCH fitted on simulated portfolio returns with random weighted portfolios, and the DGP is a EDCC-GARCH. The fitted MGARCH models are: the DCC-GARCH (first row, DCC), the EDCC-GARCH (second row, EDCC) and the Scalar BEKK (third row, SBEKK). The columns indicates the sample size ($T$): $T=500$ left,  $T=1000$ center and $T=2000$ right column. All values are averages across the 500 experiments.}
\vspace*{1em}
\includegraphics[width = 0.8\linewidth]{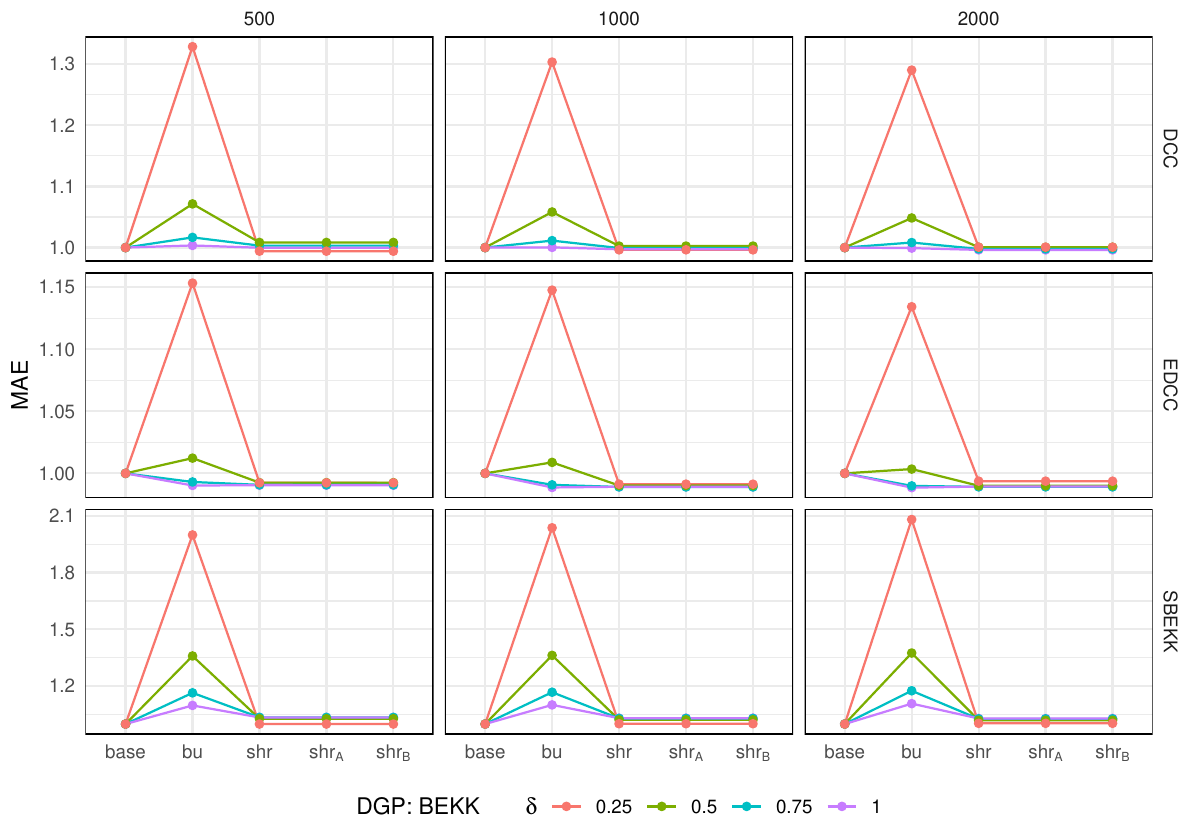}
\caption{Average relative MAE where the reference forecast is the univariate GARCH fitted on simulated portfolio returns with random weighted portfolios, and the DGP is a Full BEKK. The fitted MGARCH models are: the DCC-GARCH (first row, DCC) and the Scalar BEKK (second row, SBEKK). The columns indicates the sample size ($T$): $T=500$ left,  $T=1000$ center and $T=2000$ right column. All values are averages across the 500 experiments.}
\end{figure}

\begin{figure}[H]
\centering
\includegraphics[width = 0.8\linewidth]{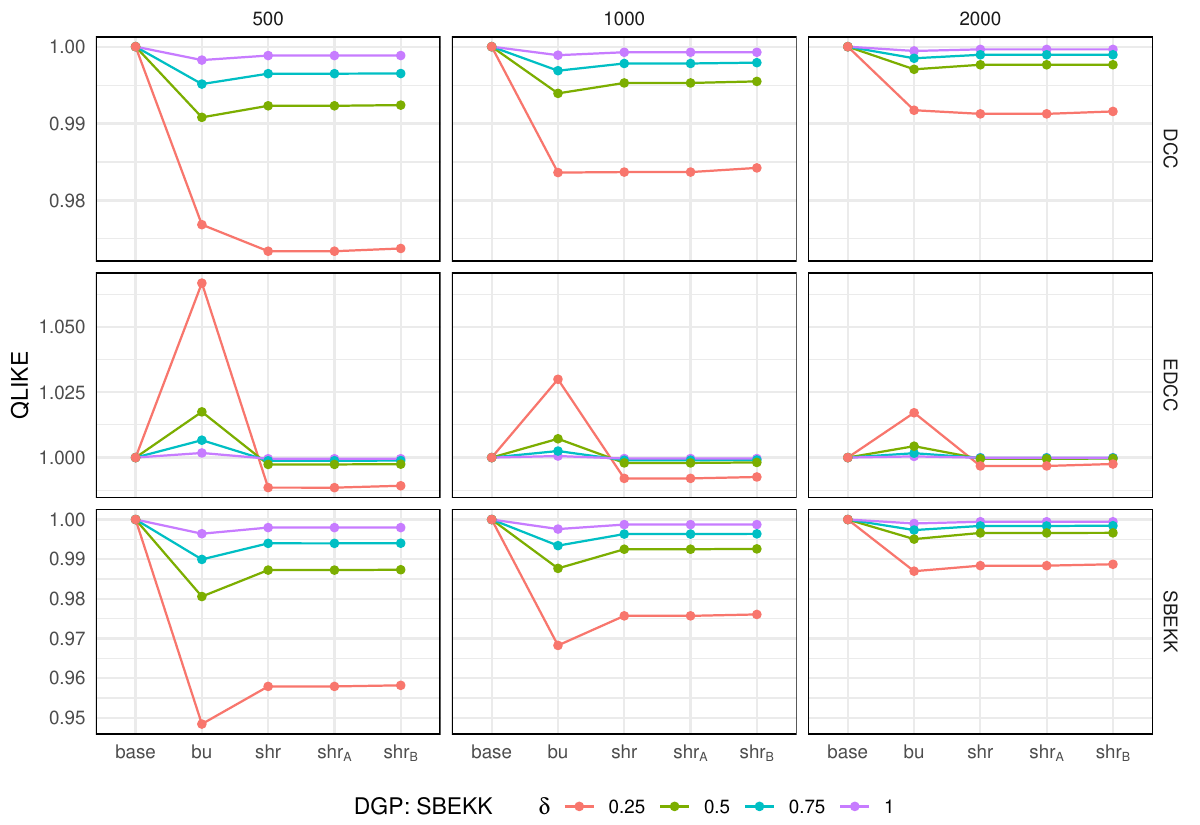}
\caption{Average relative QLIKE where the reference forecast is the univariate GARCH fitted on simulated portfolio returns with random weighted portfolios, and the DGP is a Scalar BEKK. The fitted MGARCH models are: the DCC-GARCH (first row, DCC) and the Scalar BEKK (second row, SBEKK). The columns indicates the sample size ($T$): $T=500$ left,  $T=1000$ center and $T=2000$ right column. All values are averages across the 500 experiments.}
\vspace*{1em}
\includegraphics[width = 0.8\linewidth]{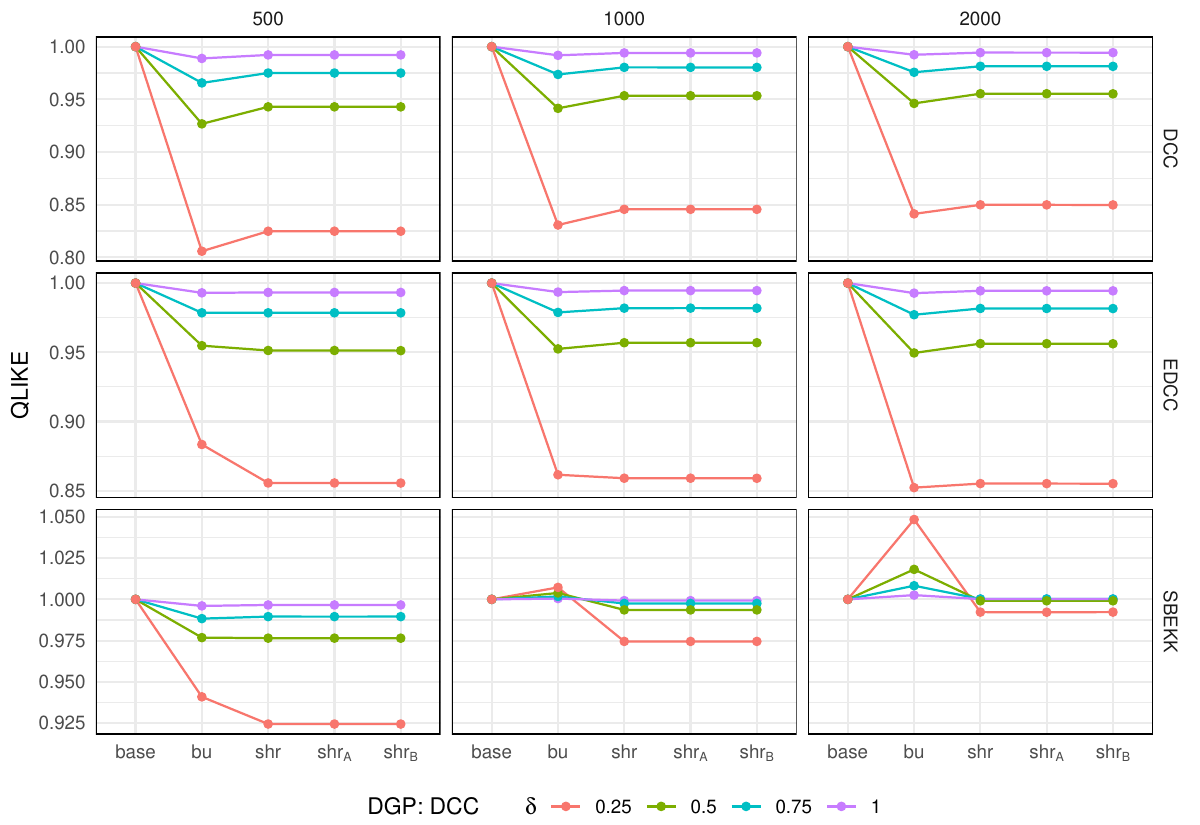}
\caption{Average relative QLIKE where the reference forecast is the univariate GARCH fitted on simulated portfolio returns with random weighted portfolios, and the DGP is a DCC-GARCH. The fitted MGARCH models are: the DCC-GARCH (first row, DCC) and the Scalar BEKK (second row, SBEKK). The columns indicates the sample size ($T$): $T=500$ left,  $T=1000$ center and $T=20000$ right column. All values are averages across the 500 experiments.}
\end{figure}

\begin{figure}[H]
\centering
\includegraphics[width = 0.8\linewidth]{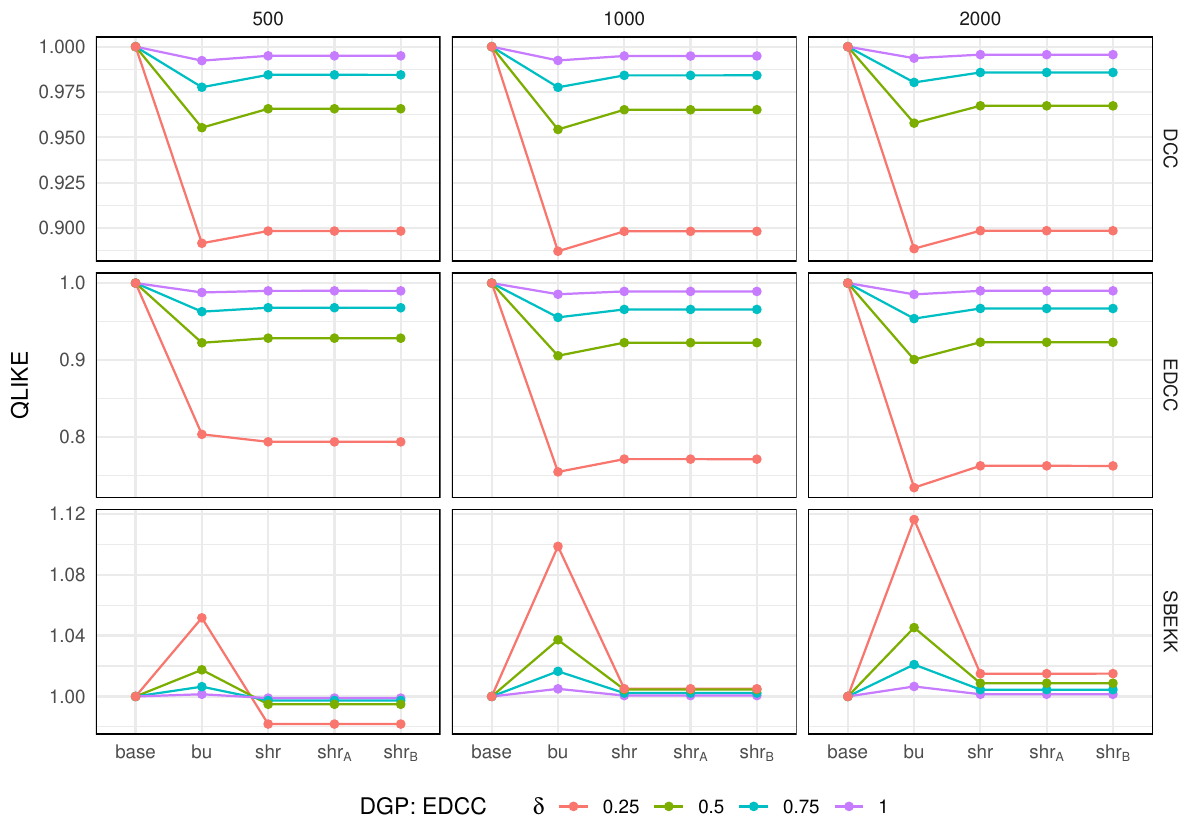}
\caption{Average relative QLIKE where the reference forecast is the univariate GARCH fitted on simulated portfolio returns with random weighted portfolios, and the DGP is a EDCC-GARCH. The fitted MGARCH models are: the DCC-GARCH (first row, DCC), the EDCC-GARCH (second row, EDCC) and the Scalar BEKK (third row, SBEKK). The columns indicates the sample size ($T$): $T=500$ left,  $T=1000$ center and $T=2000$ right column. All values are averages across the 500 experiments.}
\vspace*{1em}
\includegraphics[width = 0.8\linewidth]{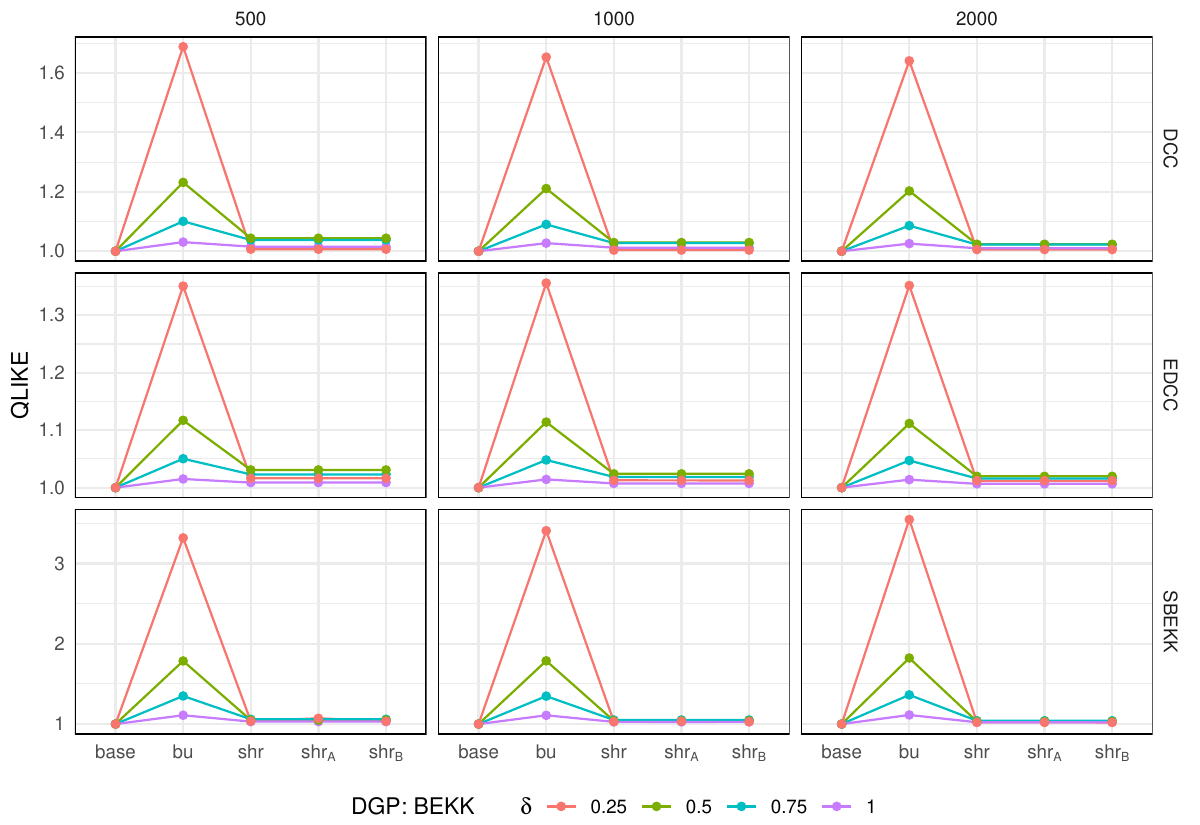}
\caption{Average relative QLIKE where the reference forecast is the univariate GARCH fitted on simulated portfolio returns with random weighted portfolios, and the DGP is a Full BEKK. The fitted MGARCH models are: the DCC-GARCH (first row, DCC) and the Scalar BEKK (second row, SBEKK). The columns indicates the sample size ($T$): $T=500$ left,  $T=1000$ center and $T=2000$ right column. All values are averages across the 500 experiments.}
\end{figure}

\newpage
\subsection{True portfolio variance and covariance matrix of $N=24$ assets}

\begin{landscape}\centering
	\begingroup
	\setlength{\LTcapwidth}{\linewidth}
	\setlength{\tabcolsep}{4pt}
	\footnotesize
	
\begin{longtable}[t]{>{}c||>{}c>{}c|cc>{}c||>{}c>{}c|ccc}
\caption{Average accuracy indices across different simulation settings ($N=24$ and $T=1000$). Each setting is identified by a label composed of three elements: (i) the data-generating process (DGP), (ii) the MGARCH model used for estimation, and (iii) the portfolio weighting scheme (see \autoref{tab:notationModel} for details). The top rows indicate the variance forecasting method employed: the univariate GARCH on portfolio returns (base), the bottom-up approach using MGARCH models (bu), and the three forecast reconciliation strategies discussed in the paper ($shr$, $shr_{A}$, $shr_{B}$). The reported accuracy measures are grouped into three categories: average accuracy indices (MSE, MAE, QLIKE); average relative measures (AvgRelMSE, AvgRelMAE, AvgRelQLIKE) using the univariate and the multivariate GARCH forecast as the benchmark.}\\
\toprule
Index & $base$ & $bu$ & $shr$ & $shr_{A}$ & $shr_{B}$ & $base$ & $bu$ & $shr$ & $shr_{A}$ & $shr_{B}$\\
\midrule
\endfirsthead
\caption[]{Average accuracy indices across different simulation settings ($N=24$ and $T=1000$). Each setting is identified by a label composed of three elements: (i) the data-generating process (DGP), (ii) the MGARCH model used for estimation, and (iii) the portfolio weighting scheme (see \autoref{tab:notationModel} for details). The top rows indicate the variance forecasting method employed: the univariate GARCH on portfolio returns (base), the bottom-up approach using MGARCH models (bu), and the three forecast reconciliation strategies discussed in the paper ($shr$, $shr_{A}$, $shr_{B}$). The reported accuracy measures are grouped into three categories: average accuracy indices (MSE, MAE, QLIKE); average relative measures (AvgRelMSE, AvgRelMAE, AvgRelQLIKE) using the univariate and the multivariate GARCH forecast as the benchmark. \textit{(continued)}}\\
\toprule
Index & $base$ & $bu$ & $shr$ & $shr_{A}$ & $shr_{B}$ & $base$ & $bu$ & $shr$ & $shr_{A}$ & $shr_{B}$\\
\midrule
\endhead

\endfoot
\bottomrule
\endlastfoot
\addlinespace[0.5em]
\multicolumn{1}{c}{} & \multicolumn{5}{c}{\textbf{BKF -- DCC -- EQ}} & \multicolumn{5}{c}{\textbf{BKF -- DCC -- RW}}\\
\nopagebreak \addlinespace[0em]
\multicolumn{11}{l}{\textit{Average indexes}}\\
\nopagebreak MSE & 7586.552 & 19925.876 & \em{4682.709} & \textbf{4660.824} & 5132.878 & 8424.214 & 21083.215 & \em{5066.222} & \textbf{5044.954} & 5428.564\\
\nopagebreak 
MAE & 20.132 & 32.816 & \em{15.178} & \textbf{15.122} & 16.173 & 20.766 & 32.905 & \em{15.811} & \textbf{15.752} & 16.689\\
\nopagebreak 
QLIKE & 0.007 & 0.017 & \em{0.006} & \textbf{0.006} & 0.006 & 0.007 & 0.018 & \em{0.006} & \textbf{0.006} & 0.006\\
\nopagebreak 
\addlinespace[0em]
\multicolumn{11}{l}{\textit{Relative indexes (base benchmark)}}\\
\nopagebreak AvgRelMSE & 1.000 & 3.170 & \em{0.538} & \textbf{0.535} & 0.725 & 1.000 & 2.917 & \em{0.575} & \textbf{0.573} & 0.729\\
\nopagebreak 
AvgRelMAE & 1.000 & 1.773 & \em{0.784} & \textbf{0.783} & 0.848 & 1.000 & 1.707 & \em{0.809} & \textbf{0.808} & 0.862\\
\nopagebreak 
AvgRelQLIKE & 1.000 & 3.326 & \em{0.840} & \textbf{0.839} & 0.912 & 1.000 & 3.032 & \em{0.868} & \textbf{0.867} & 0.925\\
\nopagebreak 
\addlinespace[0em]
\multicolumn{11}{l}{\textit{Relative indexes (bu benchmark)}}\\
\nopagebreak AvgRelMSE & 0.315 & 1.000 & \em{0.170} & \textbf{0.169} & 0.229 & 0.343 & 1.000 & \em{0.197} & \textbf{0.196} & 0.250\\
\nopagebreak 
AvgRelMAE & 0.564 & 1.000 & \em{0.442} & \textbf{0.442} & 0.478 & 0.586 & 1.000 & \em{0.474} & \textbf{0.473} & 0.505\\
\nopagebreak 
AvgRelQLIKE & 0.301 & 1.000 & \em{0.252} & \textbf{0.252} & 0.274 & 0.330 & 1.000 & \em{0.286} & \textbf{0.286} & 0.305\\
 
\addlinespace[0.5em]
\multicolumn{1}{c}{} & \multicolumn{5}{c}{\textbf{BKF -- SCB -- EQ}} & \multicolumn{5}{c}{\textbf{BKF -- SCB -- RW}}\\
\nopagebreak \addlinespace[0em]
\multicolumn{11}{l}{\textit{Average indexes}}\\
\nopagebreak MSE & 7586.552 & 308846.471 & \em{5375.408} & 24567.577 & \textbf{4874.462} & 8424.214 & 296383.158 & \em{5734.310} & 7212.052 & \textbf{5226.377}\\
\nopagebreak 
MAE & 20.132 & 156.357 & \em{18.424} & 19.076 & \textbf{18.396} & 20.766 & 156.164 & \em{18.970} & 19.052 & \textbf{18.936}\\
\nopagebreak 
QLIKE & \textbf{0.007} & 0.332 & 0.010 & 0.010 & \em{0.010} & \textbf{0.007} & 0.329 & 0.010 & 0.010 & \em{0.010}\\
\nopagebreak 
\addlinespace[0em]
\multicolumn{11}{l}{\textit{Relative indexes (base benchmark)}}\\
\nopagebreak AvgRelMSE & 1.000 & 54.010 & \textbf{0.721} & \em{0.723} & 0.735 & 1.000 & 49.202 & \textbf{0.742} & \em{0.742} & 0.750\\
\nopagebreak 
AvgRelMAE & 1.000 & 8.344 & \textbf{0.957} & \em{0.958} & 0.959 & 1.000 & 7.986 & \textbf{0.965} & \em{0.965} & 0.966\\
\nopagebreak 
AvgRelQLIKE & \textbf{1.000} & 59.773 & 1.301 & 1.302 & \em{1.286} & \textbf{1.000} & 53.922 & 1.311 & 1.311 & \em{1.297}\\
\nopagebreak 
\addlinespace[0em]
\multicolumn{11}{l}{\textit{Relative indexes (bu benchmark)}}\\
\nopagebreak AvgRelMSE & 0.019 & 1.000 & \textbf{0.013} & \em{0.013} & 0.014 & 0.020 & 1.000 & \textbf{0.015} & \em{0.015} & 0.015\\
\nopagebreak 
AvgRelMAE & 0.120 & 1.000 & \textbf{0.115} & \em{0.115} & 0.115 & 0.125 & 1.000 & \textbf{0.121} & \em{0.121} & 0.121\\
\nopagebreak 
AvgRelQLIKE & \textbf{0.017} & 1.000 & 0.022 & 0.022 & \em{0.022} & \textbf{0.019} & 1.000 & 0.024 & 0.024 & \em{0.024}\\
 
\addlinespace[0.5em]
\multicolumn{1}{c}{} & \multicolumn{5}{c}{\textbf{DCC -- DCC -- EQ}} & \multicolumn{5}{c}{\textbf{DCC -- DCC -- RW}}\\
\nopagebreak \addlinespace[0em]
\multicolumn{11}{l}{\textit{Average indexes}}\\
\nopagebreak MSE & 0.210 & 0.401 & \em{0.124} & \textbf{0.124} & 0.124 & 0.227 & 0.431 & \em{0.147} & \textbf{0.147} & \vphantom{1} 0.147\\
\nopagebreak 
MAE & 0.248 & 0.347 & \em{0.221} & \textbf{0.221} & 0.221 & 0.266 & 0.362 & \em{0.240} & \textbf{0.240} & \vphantom{1} 0.240\\
\nopagebreak 
QLIKE & 0.010 & 0.018 & \em{0.009} & \textbf{0.009} & 0.009 & 0.011 & 0.018 & \em{0.010} & \textbf{0.010} & \vphantom{1} 0.010\\
\nopagebreak 
\addlinespace[0em]
\multicolumn{11}{l}{\textit{Relative indexes (base benchmark)}}\\
\nopagebreak AvgRelMSE & 1.000 & 1.688 & \em{0.745} & \textbf{0.745} & 0.746 & 1.000 & 1.578 & \em{0.766} & \textbf{0.766} & \vphantom{1} 0.767\\
\nopagebreak 
AvgRelMAE & 1.000 & 1.344 & \em{0.908} & \textbf{0.907} & 0.908 & 1.000 & 1.294 & \em{0.914} & \textbf{0.913} & \vphantom{1} 0.914\\
\nopagebreak 
AvgRelQLIKE & 1.000 & 1.669 & \em{0.871} & \textbf{0.871} & 0.872 & 1.000 & 1.545 & \em{0.871} & \textbf{0.870} & \vphantom{1} 0.872\\
\nopagebreak 
\addlinespace[0em]
\multicolumn{11}{l}{\textit{Relative indexes (bu benchmark)}}\\
\nopagebreak AvgRelMSE & 0.592 & 1.000 & \em{0.441} & \textbf{0.441} & 0.442 & 0.634 & 1.000 & \em{0.485} & \textbf{0.485} & \vphantom{1} 0.486\\
\nopagebreak 
AvgRelMAE & 0.744 & 1.000 & \em{0.675} & \textbf{0.675} & 0.676 & 0.773 & 1.000 & \em{0.706} & \textbf{0.706} & \vphantom{1} 0.706\\
\nopagebreak 
AvgRelQLIKE & 0.599 & 1.000 & \em{0.522} & \textbf{0.521} & 0.522 & 0.647 & 1.000 & \em{0.564} & \textbf{0.563} & \vphantom{1} 0.564\\
 
\addlinespace[0.5em]
\multicolumn{1}{c}{} & \multicolumn{5}{c}{\textbf{DCC -- SCB -- EQ}} & \multicolumn{5}{c}{\textbf{DCC -- SCB -- RW}}\\
\nopagebreak \addlinespace[0em]
\multicolumn{11}{l}{\textit{Average indexes}}\\
\nopagebreak MSE & 0.210 & 0.205 & \em{0.141} & 0.141 & \textbf{0.141} & 0.227 & 0.230 & \em{0.162} & 0.162 & \vphantom{1} \textbf{0.162}\\
\nopagebreak 
MAE & 0.248 & 0.289 & \em{0.235} & 0.235 & \textbf{0.235} & 0.266 & 0.307 & 0.253 & \em{0.253} & \vphantom{1} \textbf{0.253}\\
\nopagebreak 
QLIKE & \textbf{0.010} & 0.015 & 0.010 & 0.010 & \em{0.010} & 0.011 & 0.016 & 0.011 & \em{0.011} & \vphantom{1} \textbf{0.011}\\
\nopagebreak 
\addlinespace[0em]
\multicolumn{11}{l}{\textit{Relative indexes (base benchmark)}}\\
\nopagebreak AvgRelMSE & 1.000 & 1.278 & \em{0.886} & 0.886 & \textbf{0.886} & 1.000 & 1.257 & \em{0.897} & 0.897 & \vphantom{1} \textbf{0.897}\\
\nopagebreak 
AvgRelMAE & 1.000 & 1.199 & \em{0.972} & 0.972 & \textbf{0.972} & 1.000 & 1.180 & 0.976 & \em{0.976} & \vphantom{1} \textbf{0.976}\\
\nopagebreak 
AvgRelQLIKE & \textbf{1.000} & 1.456 & 1.004 & 1.004 & \em{1.004} & \textbf{1.000} & 1.403 & 1.002 & 1.002 & \vphantom{1} \em{1.002}\\
\nopagebreak 
\addlinespace[0em]
\multicolumn{11}{l}{\textit{Relative indexes (bu benchmark)}}\\
\nopagebreak AvgRelMSE & 0.782 & 1.000 & \em{0.693} & 0.694 & \textbf{0.693} & 0.795 & 1.000 & \em{0.713} & 0.714 & \vphantom{1} \textbf{0.713}\\
\nopagebreak 
AvgRelMAE & 0.834 & 1.000 & \em{0.811} & 0.811 & \textbf{0.811} & 0.847 & 1.000 & 0.827 & \em{0.827} & \vphantom{1} \textbf{0.827}\\
\nopagebreak 
AvgRelQLIKE & \textbf{0.687} & 1.000 & 0.690 & 0.690 & \em{0.690} & \textbf{0.713} & 1.000 & 0.714 & 0.714 & \vphantom{1} \em{0.714}\\
 
\addlinespace[0.5em]
\multicolumn{1}{c}{} & \multicolumn{5}{c}{\textbf{EDCC -- DCC -- EQ}} & \multicolumn{5}{c}{\textbf{EDCC -- DCC -- RW}}\\
\nopagebreak \addlinespace[0em]
\multicolumn{11}{l}{\textit{Average indexes}}\\
\nopagebreak MSE & 0.210 & 0.401 & \em{0.124} & \textbf{0.124} & 0.124 & 0.227 & 0.431 & \em{0.147} & \textbf{0.147} & 0.147\\
\nopagebreak 
MAE & 0.248 & 0.347 & \em{0.221} & \textbf{0.221} & 0.221 & 0.266 & 0.362 & \em{0.240} & \textbf{0.240} & 0.240\\
\nopagebreak 
QLIKE & 0.010 & 0.018 & \em{0.009} & \textbf{0.009} & 0.009 & 0.011 & 0.018 & \em{0.010} & \textbf{0.010} & 0.010\\
\nopagebreak 
\addlinespace[0em]
\multicolumn{11}{l}{\textit{Relative indexes (base benchmark)}}\\
\nopagebreak AvgRelMSE & 1.000 & 1.688 & \em{0.745} & \textbf{0.745} & 0.746 & 1.000 & 1.578 & \em{0.766} & \textbf{0.766} & 0.767\\
\nopagebreak 
AvgRelMAE & 1.000 & 1.344 & \em{0.908} & \textbf{0.907} & 0.908 & 1.000 & 1.294 & \em{0.914} & \textbf{0.913} & 0.914\\
\nopagebreak 
AvgRelQLIKE & 1.000 & 1.669 & \em{0.871} & \textbf{0.871} & 0.872 & 1.000 & 1.545 & \em{0.871} & \textbf{0.870} & 0.872\\
\nopagebreak 
\addlinespace[0em]
\multicolumn{11}{l}{\textit{Relative indexes (bu benchmark)}}\\
\nopagebreak AvgRelMSE & 0.592 & 1.000 & \em{0.441} & \textbf{0.441} & 0.442 & 0.634 & 1.000 & \em{0.485} & \textbf{0.485} & 0.486\\
\nopagebreak 
AvgRelMAE & 0.744 & 1.000 & \em{0.675} & \textbf{0.675} & 0.676 & 0.773 & 1.000 & \em{0.706} & \textbf{0.706} & 0.706\\
\nopagebreak 
AvgRelQLIKE & 0.599 & 1.000 & \em{0.522} & \textbf{0.521} & 0.522 & 0.647 & 1.000 & \em{0.564} & \textbf{0.563} & 0.564\\
 
\addlinespace[0.5em]
\multicolumn{1}{c}{} & \multicolumn{5}{c}{\textbf{EDCC -- SCB -- EQ}} & \multicolumn{5}{c}{\textbf{EDCC -- SCB -- RW}}\\
\nopagebreak \addlinespace[0em]
\multicolumn{11}{l}{\textit{Average indexes}}\\
\nopagebreak MSE & 0.210 & 0.205 & \em{0.141} & 0.141 & \textbf{0.141} & 0.227 & 0.230 & \em{0.162} & 0.162 & \textbf{0.162}\\
\nopagebreak 
MAE & 0.248 & 0.289 & \em{0.235} & 0.235 & \textbf{0.235} & 0.266 & 0.307 & 0.253 & \em{0.253} & \textbf{0.253}\\
\nopagebreak 
QLIKE & \textbf{0.010} & 0.015 & 0.010 & 0.010 & \em{0.010} & 0.011 & 0.016 & 0.011 & \em{0.011} & \textbf{0.011}\\
\nopagebreak 
\addlinespace[0em]
\multicolumn{11}{l}{\textit{Relative indexes (base benchmark)}}\\
\nopagebreak AvgRelMSE & 1.000 & 1.278 & \em{0.886} & 0.886 & \textbf{0.886} & 1.000 & 1.257 & \em{0.897} & 0.897 & \textbf{0.897}\\
\nopagebreak 
AvgRelMAE & 1.000 & 1.199 & \em{0.972} & 0.972 & \textbf{0.972} & 1.000 & 1.180 & 0.976 & \em{0.976} & \textbf{0.976}\\
\nopagebreak 
AvgRelQLIKE & \textbf{1.000} & 1.456 & 1.004 & 1.004 & \em{1.004} & \textbf{1.000} & 1.403 & 1.002 & 1.002 & \em{1.002}\\
\nopagebreak 
\addlinespace[0em]
\multicolumn{11}{l}{\textit{Relative indexes (bu benchmark)}}\\
\nopagebreak AvgRelMSE & 0.782 & 1.000 & \em{0.693} & 0.694 & \textbf{0.693} & 0.795 & 1.000 & \em{0.713} & 0.714 & \textbf{0.713}\\
\nopagebreak 
AvgRelMAE & 0.834 & 1.000 & \em{0.811} & 0.811 & \textbf{0.811} & 0.847 & 1.000 & 0.827 & \em{0.827} & \textbf{0.827}\\
\nopagebreak 
AvgRelQLIKE & \textbf{0.687} & 1.000 & 0.690 & 0.690 & \em{0.690} & \textbf{0.713} & 1.000 & 0.714 & 0.714 & \em{0.714}\\
 
\addlinespace[0.5em]
\multicolumn{1}{c}{} & \multicolumn{5}{c}{\textbf{SCB -- DCC -- EQ}} & \multicolumn{5}{c}{\textbf{SCB -- DCC -- RW}}\\
\nopagebreak \addlinespace[0em]
\multicolumn{11}{l}{\textit{Average indexes}}\\
\nopagebreak MSE & 0.616 & 0.305 & \em{0.255} & \textbf{0.255} & 0.255 & 0.665 & 0.331 & 0.276 & \em{0.276} & \textbf{0.276}\\
\nopagebreak 
MAE & 0.602 & 0.440 & \em{0.394} & \textbf{0.394} & 0.394 & 0.621 & 0.454 & 0.406 & \em{0.406} & \textbf{0.406}\\
\nopagebreak 
QLIKE & 0.0019 & 0.0010 & \em{0.0008} & \textbf{0.0008} & 0.0008 & 0.0019 & 0.0010 & 0.0008 & \em{0.0008} & \textbf{0.0008}\\
\nopagebreak 
\addlinespace[0em]
\multicolumn{11}{l}{\textit{Relative indexes (base benchmark)}}\\
\nopagebreak AvgRelMSE & 1.000 & 0.414 & \em{0.330} & \textbf{0.330} & 0.330 & 1.000 & 0.428 & 0.341 & \em{0.341} & \textbf{0.341}\\
\nopagebreak 
AvgRelMAE & 1.000 & 0.685 & \em{0.607} & \textbf{0.607} & 0.607 & 1.000 & 0.696 & 0.616 & \em{0.616} & \textbf{0.616}\\
\nopagebreak 
AvgRelQLIKE & 1.000 & 0.425 & \em{0.339} & \textbf{0.339} & 0.339 & 1.000 & 0.439 & 0.350 & \em{0.350} & \textbf{0.350}\\
\nopagebreak 
\addlinespace[0em]
\multicolumn{11}{l}{\textit{Relative indexes (bu benchmark)}}\\
\nopagebreak AvgRelMSE & 2.418 & 1.000 & \em{0.799} & \textbf{0.799} & 0.799 & 2.336 & 1.000 & 0.797 & \em{0.797} & \textbf{0.797}\\
\nopagebreak 
AvgRelMAE & 1.459 & 1.000 & \em{0.886} & \textbf{0.886} & 0.886 & 1.437 & 1.000 & 0.885 & \em{0.885} & \textbf{0.885}\\
\nopagebreak 
AvgRelQLIKE & 2.353 & 1.000 & \em{0.799} & \textbf{0.799} & 0.799 & 2.279 & 1.000 & 0.797 & \em{0.797} & \textbf{0.797}\\
 
\addlinespace[0.5em]
\multicolumn{1}{c}{} & \multicolumn{5}{c}{\textbf{SCB -- SCB -- EQ}} & \multicolumn{5}{c}{\textbf{SCB -- SCB -- RW}}\\
\nopagebreak \addlinespace[0em]
\multicolumn{11}{l}{\textit{Average indexes}}\\
\nopagebreak MSE & 0.616 & \em{0.457} & \textbf{0.280} & \textbf{0.280} & \textbf{0.280} & 0.665 & \em{0.490} & \textbf{0.295} & \textbf{0.295} & \textbf{0.295}\\
\nopagebreak 
MAE & 0.602 & \em{0.525} & \textbf{0.397} & \textbf{0.397} & \textbf{0.397} & 0.621 & \em{0.538} & \textbf{0.402} & \textbf{0.402} & \textbf{0.402}\\
\nopagebreak 
QLIKE & 0.0019 & \em{0.0014} & \textbf{0.0009} & \textbf{0.0009} & \textbf{0.0009} & 0.0019 & \em{0.0014} & \textbf{0.0009} & \textbf{0.0009} & \textbf{0.0009}\\
\nopagebreak 
\addlinespace[0em]
\multicolumn{11}{l}{\textit{Relative indexes (base benchmark)}}\\
\nopagebreak AvgRelMSE & 1.000 & \em{0.776} & \textbf{0.399} & \textbf{0.399} & \textbf{0.399} & 1.000 & \em{0.772} & \textbf{0.385} & \textbf{0.385} & \textbf{0.385}\\
\nopagebreak 
AvgRelMAE & 1.000 & \em{0.879} & \textbf{0.631} & \textbf{0.631} & \textbf{0.631} & 1.000 & \em{0.873} & \textbf{0.618} & \textbf{0.618} & \textbf{0.618}\\
\nopagebreak 
AvgRelQLIKE & 1.000 & \em{0.792} & \textbf{0.406} & \textbf{0.406} & \textbf{0.406} & 1.000 & \em{0.785} & \textbf{0.390} & \textbf{0.390} & \textbf{0.390}\\
\nopagebreak 
\addlinespace[0em]
\multicolumn{11}{l}{\textit{Relative indexes (bu benchmark)}}\\
\nopagebreak AvgRelMSE & 1.289 & \em{1.000} & \textbf{0.514} & \textbf{0.514} & \textbf{0.514} & 1.296 & \em{1.000} & \textbf{0.499} & \textbf{0.499} & \textbf{0.499}\\
\nopagebreak 
AvgRelMAE & 1.138 & \em{1.000} & \textbf{0.718} & \textbf{0.718} & \textbf{0.718} & 1.146 & \em{1.000} & \textbf{0.708} & \textbf{0.708} & \textbf{0.708}\\
\nopagebreak 
AvgRelQLIKE & 1.263 & \em{1.000} & \textbf{0.513} & \textbf{0.513} & \textbf{0.513} & 1.274 & \em{1.000} & \textbf{0.497} & \textbf{0.497} & \textbf{0.497}\\*
\end{longtable}

	\endgroup
\end{landscape}

\begingroup
\setlength{\LTcapwidth}{\linewidth}
\footnotesize

\begin{longtable}[t]{>{}c||>{}c>{}c|cc>{}c||>{}c>{}c|ccc}
\caption{The number of times (in \%) that model is in the Model Confidence Set with different thresholds ($\gamma \in \{70\%, 75\%, 80\%, 85\%m 90\%, 95\%\}$) across different simulation settings ($N=24$ and $T=1000$) using the MAE as loss function. Each setting is identified by a label composed of three elements: (i) the data-generating process (DGP), (ii) the MGARCH model used for estimation, and (iii) the portfolio weighting scheme (see \autoref{tab:notationModel} for details). The top rows indicate the noisy level of the proxy $\delta \in \left\lbrace 0.25, 0.5, 0.75, 1\right\rbrace$ and the variance forecasting method employed: the univariate GARCH on portfolio returns (base), the bottom-up approach using MGARCH models (bu), and the three forecast reconciliation strategies discussed in the paper ($shr$, $shr_{A}$, $shr_{B}$).}\\
\toprule
Threshold & $base$ & $bu$ & $shr$ & $shr_{A}$ & $shr_{B}$ & $base$ & $bu$ & $shr$ & $shr_{A}$ & $shr_{B}$\\
\midrule
\endfirsthead
\caption[]{The number of times (in \%) that model is in the Model Confidence Set with different thresholds ($\gamma \in \{70\%, 75\%, 80\%, 85\%m 90\%, 95\%\}$) across different simulation settings ($N=24$ and $T=1000$) using the MAE as loss function. Each setting is identified by a label composed of three elements: (i) the data-generating process (DGP), (ii) the MGARCH model used for estimation, and (iii) the portfolio weighting scheme (see \autoref{tab:notationModel} for details). The top rows indicate the noisy level of the proxy $\delta \in \left\lbrace 0.25, 0.5, 0.75, 1\right\rbrace$ and the variance forecasting method employed: the univariate GARCH on portfolio returns (base), the bottom-up approach using MGARCH models (bu), and the three forecast reconciliation strategies discussed in the paper ($shr$, $shr_{A}$, $shr_{B}$). \textit{(continued)}}\\
\toprule
Threshold & $base$ & $bu$ & $shr$ & $shr_{A}$ & $shr_{B}$ & $base$ & $bu$ & $shr$ & $shr_{A}$ & $shr_{B}$\\
\midrule
\endhead

\endfoot
\bottomrule
\endlastfoot
\addlinespace[0.5em]
\multicolumn{1}{c}{} & \multicolumn{5}{c}{\textbf{BKF -- DCC -- EQ}} & \multicolumn{5}{c}{\textbf{BKF -- DCC -- RW}}\\
\nopagebreak 70 & 40.6 & 3.4 & \em{77.6} & \textbf{81.2} & 48.0 & 44.8 & 3.8 & \em{78.0} & \textbf{79.6} & 50.0\\
\nopagebreak 
75 & 43.8 & 4.0 & \em{79.8} & \textbf{82.8} & 52.0 & 47.0 & 4.4 & \textbf{80.4} & \em{80.2} & 53.4\\
\nopagebreak 
80 & 46.6 & 4.6 & \em{84.2} & \textbf{85.4} & 57.8 & 50.8 & 4.8 & \textbf{83.2} & \textbf{83.2} & \em{60.6}\\
\nopagebreak 
85 & 51.8 & 5.2 & \em{86.0} & \textbf{88.6} & 67.6 & 54.2 & 6.4 & \textbf{86.6} & \em{86.4} & 68.2\\
\nopagebreak 
90 & 55.8 & 7.0 & \em{89.8} & \textbf{90.0} & 74.8 & 58.0 & 7.8 & \textbf{89.2} & \em{88.4} & 75.4\\
\nopagebreak 
95 & 61.2 & 10.0 & \em{93.4} & \textbf{93.8} & 85.0 & 64.6 & 11.0 & \textbf{93.6} & \textbf{93.6} & \em{86.4}\\
 
\addlinespace[0.5em]
\multicolumn{1}{c}{} & \multicolumn{5}{c}{\textbf{BKF -- SCB -- EQ}} & \multicolumn{5}{c}{\textbf{BKF -- SCB -- RW}}\\
\nopagebreak 70 & 60.8 & 0.0 & 69.6 & \em{70.2} & \textbf{73.8} & 64.0 & 0.0 & 67.4 & \em{67.8} & \textbf{71.8}\\
\nopagebreak 
75 & 63.6 & 0.0 & 71.4 & \em{71.8} & \textbf{75.2} & 66.4 & 0.0 & \em{69.2} & 69.0 & \textbf{72.4}\\
\nopagebreak 
80 & 66.2 & 0.0 & 72.6 & \em{72.8} & \textbf{75.8} & 68.6 & 0.0 & \em{71.0} & \em{71.0} & \textbf{74.2}\\
\nopagebreak 
85 & 69.2 & 0.0 & \em{76.4} & \em{76.4} & \textbf{78.0} & 72.2 & 0.0 & \em{73.2} & \em{73.2} & \textbf{75.2}\\
\nopagebreak 
90 & 72.8 & 0.0 & \em{79.0} & \textbf{79.2} & \textbf{79.2} & 75.4 & 0.0 & 78.4 & \em{78.6} & \textbf{78.8}\\
\nopagebreak 
95 & \em{79.6} & 0.2 & \textbf{82.4} & \textbf{82.4} & \textbf{82.4} & \textbf{81.8} & 0.2 & \em{81.0} & \em{81.0} & \em{81.0}\\
 
\addlinespace[0.5em]
\multicolumn{1}{c}{} & \multicolumn{5}{c}{\textbf{DCC -- DCC -- EQ}} & \multicolumn{5}{c}{\textbf{DCC -- DCC -- RW}}\\
\nopagebreak 70 & 64.2 & 26.0 & \em{84.0} & \textbf{85.0} & \em{84.0} & 66.6 & 28.8 & \em{85.4} & \textbf{87.0} & \vphantom{1} \textbf{87.0}\\
\nopagebreak 
75 & 68.2 & 29.0 & \textbf{86.8} & \textbf{86.8} & \em{86.2} & 71.0 & 32.6 & \em{87.8} & \textbf{89.0} & \vphantom{1} \textbf{89.0}\\
\nopagebreak 
80 & 72.0 & 31.6 & \textbf{88.2} & \em{88.0} & 87.6 & 74.6 & 35.6 & 88.8 & \em{89.4} & \vphantom{1} \textbf{90.0}\\
\nopagebreak 
85 & 75.6 & 36.0 & \em{90.2} & \textbf{90.6} & 89.6 & 77.8 & 40.2 & 91.2 & \em{91.4} & \vphantom{1} \textbf{91.8}\\
\nopagebreak 
90 & 81.4 & 41.8 & \em{92.4} & \textbf{92.6} & \textbf{92.6} & 83.0 & 46.6 & \em{94.0} & \em{94.0} & \vphantom{1} \textbf{94.6}\\
\nopagebreak 
95 & \em{89.0} & 51.8 & \textbf{94.6} & \textbf{94.6} & \textbf{94.6} & 89.2 & 55.2 & \em{95.4} & \em{95.4} & \vphantom{1} \textbf{95.6}\\
 
\addlinespace[0.5em]
\multicolumn{1}{c}{} & \multicolumn{5}{c}{\textbf{DCC -- SCB -- EQ}} & \multicolumn{5}{c}{\textbf{DCC -- SCB -- RW}}\\
\nopagebreak 70 & 68.4 & 34.2 & \em{72.2} & \em{72.2} & \textbf{72.4} & 69.2 & 31.6 & 72.4 & \em{72.6} & \vphantom{1} \textbf{73.2}\\
\nopagebreak 
75 & 70.8 & 37.2 & \textbf{74.4} & \em{74.2} & \textbf{74.4} & 72.6 & 36.0 & \em{75.0} & 74.8 & \vphantom{1} \textbf{75.6}\\
\nopagebreak 
80 & 74.0 & 40.2 & \textbf{76.8} & \em{76.6} & \textbf{76.8} & 75.8 & 40.2 & \em{78.0} & 77.8 & \vphantom{1} \textbf{78.4}\\
\nopagebreak 
85 & 77.0 & 42.8 & \em{79.2} & \textbf{79.4} & \textbf{79.4} & 77.6 & 44.4 & \em{80.4} & \em{80.4} & \vphantom{1} \textbf{80.8}\\
\nopagebreak 
90 & 81.8 & 47.8 & 82.0 & \em{82.2} & \textbf{82.4} & 79.8 & 48.8 & \em{82.6} & \em{82.6} & \vphantom{1} \textbf{82.8}\\
\nopagebreak 
95 & \textbf{86.2} & 55.6 & 85.0 & 85.0 & \em{85.2} & \em{85.0} & 55.4 & \em{85.0} & \textbf{85.2} & \vphantom{1} \textbf{85.2}\\
 
\addlinespace[0.5em]
\multicolumn{1}{c}{} & \multicolumn{5}{c}{\textbf{EDCC -- DCC -- EQ}} & \multicolumn{5}{c}{\textbf{EDCC -- DCC -- RW}}\\
\nopagebreak 70 & 64.2 & 26.0 & \em{84.0} & \textbf{85.0} & \em{84.0} & 66.6 & 28.8 & \em{85.4} & \textbf{87.0} & \textbf{87.0}\\
\nopagebreak 
75 & 68.2 & 29.0 & \textbf{86.8} & \textbf{86.8} & \em{86.2} & 71.0 & 32.6 & \em{87.8} & \textbf{89.0} & \textbf{89.0}\\
\nopagebreak 
80 & 72.0 & 31.6 & \textbf{88.2} & \em{88.0} & 87.6 & 74.6 & 35.6 & 88.8 & \em{89.4} & \textbf{90.0}\\
\nopagebreak 
85 & 75.6 & 36.0 & \em{90.2} & \textbf{90.6} & 89.6 & 77.8 & 40.2 & 91.2 & \em{91.4} & \textbf{91.8}\\
\nopagebreak 
90 & 81.4 & 41.8 & \em{92.4} & \textbf{92.6} & \textbf{92.6} & 83.0 & 46.6 & \em{94.0} & \em{94.0} & \textbf{94.6}\\
\nopagebreak 
95 & \em{89.0} & 51.8 & \textbf{94.6} & \textbf{94.6} & \textbf{94.6} & 89.2 & 55.2 & \em{95.4} & \em{95.4} & \textbf{95.6}\\
 
\addlinespace[0.5em]
\multicolumn{1}{c}{} & \multicolumn{5}{c}{\textbf{EDCC -- SCB -- EQ}} & \multicolumn{5}{c}{\textbf{EDCC -- SCB -- RW}}\\
\nopagebreak 70 & 68.4 & 34.2 & \em{72.2} & \em{72.2} & \textbf{72.4} & 69.2 & 31.6 & 72.4 & \em{72.6} & \textbf{73.2}\\
\nopagebreak 
75 & 70.8 & 37.2 & \textbf{74.4} & \em{74.2} & \textbf{74.4} & 72.6 & 36.0 & \em{75.0} & 74.8 & \textbf{75.6}\\
\nopagebreak 
80 & 74.0 & 40.2 & \textbf{76.8} & \em{76.6} & \textbf{76.8} & 75.8 & 40.2 & \em{78.0} & 77.8 & \textbf{78.4}\\
\nopagebreak 
85 & 77.0 & 42.8 & \em{79.2} & \textbf{79.4} & \textbf{79.4} & 77.6 & 44.4 & \em{80.4} & \em{80.4} & \textbf{80.8}\\
\nopagebreak 
90 & 81.8 & 47.8 & 82.0 & \em{82.2} & \textbf{82.4} & 79.8 & 48.8 & \em{82.6} & \em{82.6} & \textbf{82.8}\\
\nopagebreak 
95 & \textbf{86.2} & 55.6 & 85.0 & 85.0 & \em{85.2} & \em{85.0} & 55.4 & \em{85.0} & \textbf{85.2} & \textbf{85.2}\\
 
\addlinespace[0.5em]
\multicolumn{1}{c}{} & \multicolumn{5}{c}{\textbf{SCB -- DCC -- EQ}} & \multicolumn{5}{c}{\textbf{SCB -- DCC -- RW}}\\
\nopagebreak 70 & 22.6 & \em{41.2} & \textbf{84.0} & \textbf{84.0} & \textbf{84.0} & 21.2 & 41.4 & \em{82.6} & \em{82.6} & \textbf{82.8}\\
\nopagebreak 
75 & 23.0 & \em{42.8} & \textbf{85.2} & \textbf{85.2} & \textbf{85.2} & 22.4 & 43.2 & \em{83.8} & \em{83.8} & \textbf{84.0}\\
\nopagebreak 
80 & 24.6 & \em{46.6} & \textbf{86.6} & \textbf{86.6} & \textbf{86.6} & 23.2 & 46.0 & \em{85.8} & \em{85.8} & \textbf{86.0}\\
\nopagebreak 
85 & 26.2 & \em{49.0} & \textbf{87.4} & \textbf{87.4} & \textbf{87.4} & 24.8 & 48.6 & \em{87.6} & \em{87.6} & \textbf{87.8}\\
\nopagebreak 
90 & 28.4 & \em{52.2} & \textbf{88.8} & \textbf{88.8} & \textbf{88.8} & 28.6 & \em{52.8} & \textbf{90.0} & \textbf{90.0} & \textbf{90.0}\\
\nopagebreak 
95 & 30.6 & \em{56.2} & \textbf{90.2} & \textbf{90.2} & \textbf{90.2} & 30.8 & \em{58.0} & \textbf{92.0} & \textbf{92.0} & \textbf{92.0}\\
 
\addlinespace[0.5em]
\multicolumn{1}{c}{} & \multicolumn{5}{c}{\textbf{SCB -- SCB -- EQ}} & \multicolumn{5}{c}{\textbf{SCB -- SCB -- RW}}\\
\nopagebreak 70 & 22.8 & \em{26.6} & \textbf{88.8} & \textbf{88.8} & \textbf{88.8} & 20.2 & \em{28.8} & \textbf{88.4} & \textbf{88.4} & \textbf{88.4}\\
\nopagebreak 
75 & 24.2 & \em{28.8} & \textbf{89.2} & \textbf{89.2} & \textbf{89.2} & 21.4 & \em{30.6} & \textbf{88.8} & \textbf{88.8} & \textbf{88.8}\\
\nopagebreak 
80 & 26.0 & \em{29.6} & \textbf{89.4} & \textbf{89.4} & \textbf{89.4} & 22.6 & \em{32.6} & \textbf{89.2} & \textbf{89.2} & \textbf{89.2}\\
\nopagebreak 
85 & 27.6 & \em{31.4} & \textbf{90.2} & \textbf{90.2} & \textbf{90.2} & 23.8 & \em{33.8} & \textbf{89.8} & \textbf{89.8} & \textbf{89.8}\\
\nopagebreak 
90 & 31.0 & \em{33.2} & \textbf{91.2} & \textbf{91.2} & \textbf{91.2} & 26.6 & \em{36.0} & \textbf{91.0} & \textbf{91.0} & \textbf{91.0}\\
\nopagebreak 
95 & 33.0 & \em{38.0} & \textbf{93.8} & \textbf{93.8} & \textbf{93.8} & 30.0 & \em{39.4} & \textbf{92.6} & \textbf{92.6} & \textbf{92.6}\\*
\end{longtable}

\endgroup

\begingroup
\setlength{\LTcapwidth}{\linewidth}
\footnotesize

\begin{longtable}[t]{>{}c||>{}c>{}c|cc>{}c||>{}c>{}c|ccc}
\caption{The number of times (in \%) that model is in the Model Confidence Set with different thresholds ($\gamma \in \{70\%, 75\%, 80\%, 85\%m 90\%, 95\%\}$) across different simulation settings ($N=24$ and $T=1000$) using the MSE as loss function. Each setting is identified by a label composed of three elements: (i) the data-generating process (DGP), (ii) the MGARCH model used for estimation, and (iii) the portfolio weighting scheme (see \autoref{tab:notationModel} for details). The top rows indicate the noisy level of the proxy $\delta \in \left\lbrace 0.25, 0.5, 0.75, 1\right\rbrace$ and the variance forecasting method employed: the univariate GARCH on portfolio returns (base), the bottom-up approach using MGARCH models (bu), and the three forecast reconciliation strategies discussed in the paper ($shr$, $shr_{A}$, $shr_{B}$).}\\
\toprule
Threshold & $base$ & $bu$ & $shr$ & $shr_{A}$ & $shr_{B}$ & $base$ & $bu$ & $shr$ & $shr_{A}$ & $shr_{B}$\\
\midrule
\endfirsthead
\caption[]{The number of times (in \%) that model is in the Model Confidence Set with different thresholds ($\gamma \in \{70\%, 75\%, 80\%, 85\%m 90\%, 95\%\}$) across different simulation settings ($N=24$ and $T=1000$) using the MSE as loss function. Each setting is identified by a label composed of three elements: (i) the data-generating process (DGP), (ii) the MGARCH model used for estimation, and (iii) the portfolio weighting scheme (see \autoref{tab:notationModel} for details). The top rows indicate the noisy level of the proxy $\delta \in \left\lbrace 0.25, 0.5, 0.75, 1\right\rbrace$ and the variance forecasting method employed: the univariate GARCH on portfolio returns (base), the bottom-up approach using MGARCH models (bu), and the three forecast reconciliation strategies discussed in the paper ($shr$, $shr_{A}$, $shr_{B}$). \textit{(continued)}}\\
\toprule
Threshold & $base$ & $bu$ & $shr$ & $shr_{A}$ & $shr_{B}$ & $base$ & $bu$ & $shr$ & $shr_{A}$ & $shr_{B}$\\
\midrule
\endhead

\endfoot
\bottomrule
\endlastfoot
\addlinespace[0.5em]
\multicolumn{1}{c}{} & \multicolumn{5}{c}{\textbf{BKF -- DCC -- EQ}} & \multicolumn{5}{c}{\textbf{BKF -- DCC -- RW}}\\
\nopagebreak 70 & 40.6 & 3.4 & \em{77.6} & \textbf{81.2} & 48.0 & 44.8 & 3.8 & \em{78.0} & \textbf{79.6} & 50.0\\
\nopagebreak 
75 & 43.8 & 4.0 & \em{79.8} & \textbf{82.8} & 52.0 & 47.0 & 4.4 & \textbf{80.4} & \em{80.2} & 53.4\\
\nopagebreak 
80 & 46.6 & 4.6 & \em{84.2} & \textbf{85.4} & 57.8 & 50.8 & 4.8 & \textbf{83.2} & \textbf{83.2} & \em{60.6}\\
\nopagebreak 
85 & 51.8 & 5.2 & \em{86.0} & \textbf{88.6} & 67.6 & 54.2 & 6.4 & \textbf{86.6} & \em{86.4} & 68.2\\
\nopagebreak 
90 & 55.8 & 7.0 & \em{89.8} & \textbf{90.0} & 74.8 & 58.0 & 7.8 & \textbf{89.2} & \em{88.4} & 75.4\\
\nopagebreak 
95 & 61.2 & 10.0 & \em{93.4} & \textbf{93.8} & 85.0 & 64.6 & 11.0 & \textbf{93.6} & \textbf{93.6} & \em{86.4}\\
 
\addlinespace[0.5em]
\multicolumn{1}{c}{} & \multicolumn{5}{c}{\textbf{BKF -- SCB -- EQ}} & \multicolumn{5}{c}{\textbf{BKF -- SCB -- RW}}\\
\nopagebreak 70 & 60.8 & 0.0 & 69.6 & \em{70.2} & \textbf{73.8} & 64.0 & 0.0 & 67.4 & \em{67.8} & \textbf{71.8}\\
\nopagebreak 
75 & 63.6 & 0.0 & 71.4 & \em{71.8} & \textbf{75.2} & 66.4 & 0.0 & \em{69.2} & 69.0 & \textbf{72.4}\\
\nopagebreak 
80 & 66.2 & 0.0 & 72.6 & \em{72.8} & \textbf{75.8} & 68.6 & 0.0 & \em{71.0} & \em{71.0} & \textbf{74.2}\\
\nopagebreak 
85 & 69.2 & 0.0 & \em{76.4} & \em{76.4} & \textbf{78.0} & 72.2 & 0.0 & \em{73.2} & \em{73.2} & \textbf{75.2}\\
\nopagebreak 
90 & 72.8 & 0.0 & \em{79.0} & \textbf{79.2} & \textbf{79.2} & 75.4 & 0.0 & 78.4 & \em{78.6} & \textbf{78.8}\\
\nopagebreak 
95 & \em{79.6} & 0.2 & \textbf{82.4} & \textbf{82.4} & \textbf{82.4} & \textbf{81.8} & 0.2 & \em{81.0} & \em{81.0} & \em{81.0}\\
 
\addlinespace[0.5em]
\multicolumn{1}{c}{} & \multicolumn{5}{c}{\textbf{DCC -- DCC -- EQ}} & \multicolumn{5}{c}{\textbf{DCC -- DCC -- RW}}\\
\nopagebreak 70 & 64.2 & 26.0 & \em{84.0} & \textbf{85.0} & \em{84.0} & 66.6 & 28.8 & \em{85.4} & \textbf{87.0} & \vphantom{1} \textbf{87.0}\\
\nopagebreak 
75 & 68.2 & 29.0 & \textbf{86.8} & \textbf{86.8} & \em{86.2} & 71.0 & 32.6 & \em{87.8} & \textbf{89.0} & \vphantom{1} \textbf{89.0}\\
\nopagebreak 
80 & 72.0 & 31.6 & \textbf{88.2} & \em{88.0} & 87.6 & 74.6 & 35.6 & 88.8 & \em{89.4} & \vphantom{1} \textbf{90.0}\\
\nopagebreak 
85 & 75.6 & 36.0 & \em{90.2} & \textbf{90.6} & 89.6 & 77.8 & 40.2 & 91.2 & \em{91.4} & \vphantom{1} \textbf{91.8}\\
\nopagebreak 
90 & 81.4 & 41.8 & \em{92.4} & \textbf{92.6} & \textbf{92.6} & 83.0 & 46.6 & \em{94.0} & \em{94.0} & \vphantom{1} \textbf{94.6}\\
\nopagebreak 
95 & \em{89.0} & 51.8 & \textbf{94.6} & \textbf{94.6} & \textbf{94.6} & 89.2 & 55.2 & \em{95.4} & \em{95.4} & \vphantom{1} \textbf{95.6}\\
 
\addlinespace[0.5em]
\multicolumn{1}{c}{} & \multicolumn{5}{c}{\textbf{DCC -- SCB -- EQ}} & \multicolumn{5}{c}{\textbf{DCC -- SCB -- RW}}\\
\nopagebreak 70 & 68.4 & 34.2 & \em{72.2} & \em{72.2} & \textbf{72.4} & 69.2 & 31.6 & 72.4 & \em{72.6} & \vphantom{1} \textbf{73.2}\\
\nopagebreak 
75 & 70.8 & 37.2 & \textbf{74.4} & \em{74.2} & \textbf{74.4} & 72.6 & 36.0 & \em{75.0} & 74.8 & \vphantom{1} \textbf{75.6}\\
\nopagebreak 
80 & 74.0 & 40.2 & \textbf{76.8} & \em{76.6} & \textbf{76.8} & 75.8 & 40.2 & \em{78.0} & 77.8 & \vphantom{1} \textbf{78.4}\\
\nopagebreak 
85 & 77.0 & 42.8 & \em{79.2} & \textbf{79.4} & \textbf{79.4} & 77.6 & 44.4 & \em{80.4} & \em{80.4} & \vphantom{1} \textbf{80.8}\\
\nopagebreak 
90 & 81.8 & 47.8 & 82.0 & \em{82.2} & \textbf{82.4} & 79.8 & 48.8 & \em{82.6} & \em{82.6} & \vphantom{1} \textbf{82.8}\\
\nopagebreak 
95 & \textbf{86.2} & 55.6 & 85.0 & 85.0 & \em{85.2} & \em{85.0} & 55.4 & \em{85.0} & \textbf{85.2} & \vphantom{1} \textbf{85.2}\\
 
\addlinespace[0.5em]
\multicolumn{1}{c}{} & \multicolumn{5}{c}{\textbf{EDCC -- DCC -- EQ}} & \multicolumn{5}{c}{\textbf{EDCC -- DCC -- RW}}\\
\nopagebreak 70 & 64.2 & 26.0 & \em{84.0} & \textbf{85.0} & \em{84.0} & 66.6 & 28.8 & \em{85.4} & \textbf{87.0} & \textbf{87.0}\\
\nopagebreak 
75 & 68.2 & 29.0 & \textbf{86.8} & \textbf{86.8} & \em{86.2} & 71.0 & 32.6 & \em{87.8} & \textbf{89.0} & \textbf{89.0}\\
\nopagebreak 
80 & 72.0 & 31.6 & \textbf{88.2} & \em{88.0} & 87.6 & 74.6 & 35.6 & 88.8 & \em{89.4} & \textbf{90.0}\\
\nopagebreak 
85 & 75.6 & 36.0 & \em{90.2} & \textbf{90.6} & 89.6 & 77.8 & 40.2 & 91.2 & \em{91.4} & \textbf{91.8}\\
\nopagebreak 
90 & 81.4 & 41.8 & \em{92.4} & \textbf{92.6} & \textbf{92.6} & 83.0 & 46.6 & \em{94.0} & \em{94.0} & \textbf{94.6}\\
\nopagebreak 
95 & \em{89.0} & 51.8 & \textbf{94.6} & \textbf{94.6} & \textbf{94.6} & 89.2 & 55.2 & \em{95.4} & \em{95.4} & \textbf{95.6}\\
 
\addlinespace[0.5em]
\multicolumn{1}{c}{} & \multicolumn{5}{c}{\textbf{EDCC -- SCB -- EQ}} & \multicolumn{5}{c}{\textbf{EDCC -- SCB -- RW}}\\
\nopagebreak 70 & 68.4 & 34.2 & \em{72.2} & \em{72.2} & \textbf{72.4} & 69.2 & 31.6 & 72.4 & \em{72.6} & \textbf{73.2}\\
\nopagebreak 
75 & 70.8 & 37.2 & \textbf{74.4} & \em{74.2} & \textbf{74.4} & 72.6 & 36.0 & \em{75.0} & 74.8 & \textbf{75.6}\\
\nopagebreak 
80 & 74.0 & 40.2 & \textbf{76.8} & \em{76.6} & \textbf{76.8} & 75.8 & 40.2 & \em{78.0} & 77.8 & \textbf{78.4}\\
\nopagebreak 
85 & 77.0 & 42.8 & \em{79.2} & \textbf{79.4} & \textbf{79.4} & 77.6 & 44.4 & \em{80.4} & \em{80.4} & \textbf{80.8}\\
\nopagebreak 
90 & 81.8 & 47.8 & 82.0 & \em{82.2} & \textbf{82.4} & 79.8 & 48.8 & \em{82.6} & \em{82.6} & \textbf{82.8}\\
\nopagebreak 
95 & \textbf{86.2} & 55.6 & 85.0 & 85.0 & \em{85.2} & \em{85.0} & 55.4 & \em{85.0} & \textbf{85.2} & \textbf{85.2}\\
 
\addlinespace[0.5em]
\multicolumn{1}{c}{} & \multicolumn{5}{c}{\textbf{SCB -- DCC -- EQ}} & \multicolumn{5}{c}{\textbf{SCB -- DCC -- RW}}\\
\nopagebreak 70 & 22.6 & \em{41.2} & \textbf{84.0} & \textbf{84.0} & \textbf{84.0} & 21.2 & 41.4 & \em{82.6} & \em{82.6} & \textbf{82.8}\\
\nopagebreak 
75 & 23.0 & \em{42.8} & \textbf{85.2} & \textbf{85.2} & \textbf{85.2} & 22.4 & 43.2 & \em{83.8} & \em{83.8} & \textbf{84.0}\\
\nopagebreak 
80 & 24.6 & \em{46.6} & \textbf{86.6} & \textbf{86.6} & \textbf{86.6} & 23.2 & 46.0 & \em{85.8} & \em{85.8} & \textbf{86.0}\\
\nopagebreak 
85 & 26.2 & \em{49.0} & \textbf{87.4} & \textbf{87.4} & \textbf{87.4} & 24.8 & 48.6 & \em{87.6} & \em{87.6} & \textbf{87.8}\\
\nopagebreak 
90 & 28.4 & \em{52.2} & \textbf{88.8} & \textbf{88.8} & \textbf{88.8} & 28.6 & \em{52.8} & \textbf{90.0} & \textbf{90.0} & \textbf{90.0}\\
\nopagebreak 
95 & 30.6 & \em{56.2} & \textbf{90.2} & \textbf{90.2} & \textbf{90.2} & 30.8 & \em{58.0} & \textbf{92.0} & \textbf{92.0} & \textbf{92.0}\\
 
\addlinespace[0.5em]
\multicolumn{1}{c}{} & \multicolumn{5}{c}{\textbf{SCB -- SCB -- EQ}} & \multicolumn{5}{c}{\textbf{SCB -- SCB -- RW}}\\
\nopagebreak 70 & 22.8 & \em{26.6} & \textbf{88.8} & \textbf{88.8} & \textbf{88.8} & 20.2 & \em{28.8} & \textbf{88.4} & \textbf{88.4} & \textbf{88.4}\\
\nopagebreak 
75 & 24.2 & \em{28.8} & \textbf{89.2} & \textbf{89.2} & \textbf{89.2} & 21.4 & \em{30.6} & \textbf{88.8} & \textbf{88.8} & \textbf{88.8}\\
\nopagebreak 
80 & 26.0 & \em{29.6} & \textbf{89.4} & \textbf{89.4} & \textbf{89.4} & 22.6 & \em{32.6} & \textbf{89.2} & \textbf{89.2} & \textbf{89.2}\\
\nopagebreak 
85 & 27.6 & \em{31.4} & \textbf{90.2} & \textbf{90.2} & \textbf{90.2} & 23.8 & \em{33.8} & \textbf{89.8} & \textbf{89.8} & \textbf{89.8}\\
\nopagebreak 
90 & 31.0 & \em{33.2} & \textbf{91.2} & \textbf{91.2} & \textbf{91.2} & 26.6 & \em{36.0} & \textbf{91.0} & \textbf{91.0} & \textbf{91.0}\\
\nopagebreak 
95 & 33.0 & \em{38.0} & \textbf{93.8} & \textbf{93.8} & \textbf{93.8} & 30.0 & \em{39.4} & \textbf{92.6} & \textbf{92.6} & \textbf{92.6}\\*
\end{longtable}

\endgroup

\begingroup
\setlength{\LTcapwidth}{\linewidth}
\footnotesize

\begin{longtable}[t]{>{}c||>{}c>{}c|cc>{}c||>{}c>{}c|ccc}
\caption{The number of times (in \%) that model is in the Model Confidence Set with different thresholds ($\gamma \in \{70\%, 75\%, 80\%, 85\%m 90\%, 95\%\}$) across different simulation settings ($N=24$ and $T=1000$) using the QLIKE as loss function. Each setting is identified by a label composed of three elements: (i) the data-generating process (DGP), (ii) the MGARCH model used for estimation, and (iii) the portfolio weighting scheme (see \autoref{tab:notationModel} for details). The top rows indicate the noisy level of the proxy $\delta \in \left\lbrace 0.25, 0.5, 0.75, 1\right\rbrace$ and the variance forecasting method employed: the univariate GARCH on portfolio returns (base), the bottom-up approach using MGARCH models (bu), and the three forecast reconciliation strategies discussed in the paper ($shr$, $shr_{A}$, $shr_{B}$).}\\
\toprule
Threshold & $base$ & $bu$ & $shr$ & $shr_{A}$ & $shr_{B}$ & $base$ & $bu$ & $shr$ & $shr_{A}$ & $shr_{B}$\\
\midrule
\endfirsthead
\caption[]{The number of times (in \%) that model is in the Model Confidence Set with different thresholds ($\gamma \in \{70\%, 75\%, 80\%, 85\%m 90\%, 95\%\}$) across different simulation settings ($N=24$ and $T=1000$) using the QLIKE as loss function. Each setting is identified by a label composed of three elements: (i) the data-generating process (DGP), (ii) the MGARCH model used for estimation, and (iii) the portfolio weighting scheme (see \autoref{tab:notationModel} for details). The top rows indicate the noisy level of the proxy $\delta \in \left\lbrace 0.25, 0.5, 0.75, 1\right\rbrace$ and the variance forecasting method employed: the univariate GARCH on portfolio returns (base), the bottom-up approach using MGARCH models (bu), and the three forecast reconciliation strategies discussed in the paper ($shr$, $shr_{A}$, $shr_{B}$). \textit{(continued)}}\\
\toprule
Threshold & $base$ & $bu$ & $shr$ & $shr_{A}$ & $shr_{B}$ & $base$ & $bu$ & $shr$ & $shr_{A}$ & $shr_{B}$\\
\midrule
\endhead

\endfoot
\bottomrule
\endlastfoot
\addlinespace[0.5em]
\multicolumn{1}{c}{} & \multicolumn{5}{c}{\textbf{BKF -- DCC -- EQ}} & \multicolumn{5}{c}{\textbf{BKF -- DCC -- RW}}\\
\nopagebreak 70 & 40.6 & 3.4 & \em{77.6} & \textbf{81.2} & 48.0 & 44.8 & 3.8 & \em{78.0} & \textbf{79.6} & 50.0\\
\nopagebreak 
75 & 43.8 & 4.0 & \em{79.8} & \textbf{82.8} & 52.0 & 47.0 & 4.4 & \textbf{80.4} & \em{80.2} & 53.4\\
\nopagebreak 
80 & 46.6 & 4.6 & \em{84.2} & \textbf{85.4} & 57.8 & 50.8 & 4.8 & \textbf{83.2} & \textbf{83.2} & \em{60.6}\\
\nopagebreak 
85 & 51.8 & 5.2 & \em{86.0} & \textbf{88.6} & 67.6 & 54.2 & 6.4 & \textbf{86.6} & \em{86.4} & 68.2\\
\nopagebreak 
90 & 55.8 & 7.0 & \em{89.8} & \textbf{90.0} & 74.8 & 58.0 & 7.8 & \textbf{89.2} & \em{88.4} & 75.4\\
\nopagebreak 
95 & 61.2 & 10.0 & \em{93.4} & \textbf{93.8} & 85.0 & 64.6 & 11.0 & \textbf{93.6} & \textbf{93.6} & \em{86.4}\\
 
\addlinespace[0.5em]
\multicolumn{1}{c}{} & \multicolumn{5}{c}{\textbf{BKF -- SCB -- EQ}} & \multicolumn{5}{c}{\textbf{BKF -- SCB -- RW}}\\
\nopagebreak 70 & 60.8 & 0.0 & 69.6 & \em{70.2} & \textbf{73.8} & 64.0 & 0.0 & 67.4 & \em{67.8} & \textbf{71.8}\\
\nopagebreak 
75 & 63.6 & 0.0 & 71.4 & \em{71.8} & \textbf{75.2} & 66.4 & 0.0 & \em{69.2} & 69.0 & \textbf{72.4}\\
\nopagebreak 
80 & 66.2 & 0.0 & 72.6 & \em{72.8} & \textbf{75.8} & 68.6 & 0.0 & \em{71.0} & \em{71.0} & \textbf{74.2}\\
\nopagebreak 
85 & 69.2 & 0.0 & \em{76.4} & \em{76.4} & \textbf{78.0} & 72.2 & 0.0 & \em{73.2} & \em{73.2} & \textbf{75.2}\\
\nopagebreak 
90 & 72.8 & 0.0 & \em{79.0} & \textbf{79.2} & \textbf{79.2} & 75.4 & 0.0 & 78.4 & \em{78.6} & \textbf{78.8}\\
\nopagebreak 
95 & \em{79.6} & 0.2 & \textbf{82.4} & \textbf{82.4} & \textbf{82.4} & \textbf{81.8} & 0.2 & \em{81.0} & \em{81.0} & \em{81.0}\\
 
\addlinespace[0.5em]
\multicolumn{1}{c}{} & \multicolumn{5}{c}{\textbf{DCC -- DCC -- EQ}} & \multicolumn{5}{c}{\textbf{DCC -- DCC -- RW}}\\
\nopagebreak 70 & 64.2 & 26.0 & \em{84.0} & \textbf{85.0} & \em{84.0} & 66.6 & 28.8 & \em{85.4} & \textbf{87.0} & \vphantom{1} \textbf{87.0}\\
\nopagebreak 
75 & 68.2 & 29.0 & \textbf{86.8} & \textbf{86.8} & \em{86.2} & 71.0 & 32.6 & \em{87.8} & \textbf{89.0} & \vphantom{1} \textbf{89.0}\\
\nopagebreak 
80 & 72.0 & 31.6 & \textbf{88.2} & \em{88.0} & 87.6 & 74.6 & 35.6 & 88.8 & \em{89.4} & \vphantom{1} \textbf{90.0}\\
\nopagebreak 
85 & 75.6 & 36.0 & \em{90.2} & \textbf{90.6} & 89.6 & 77.8 & 40.2 & 91.2 & \em{91.4} & \vphantom{1} \textbf{91.8}\\
\nopagebreak 
90 & 81.4 & 41.8 & \em{92.4} & \textbf{92.6} & \textbf{92.6} & 83.0 & 46.6 & \em{94.0} & \em{94.0} & \vphantom{1} \textbf{94.6}\\
\nopagebreak 
95 & \em{89.0} & 51.8 & \textbf{94.6} & \textbf{94.6} & \textbf{94.6} & 89.2 & 55.2 & \em{95.4} & \em{95.4} & \vphantom{1} \textbf{95.6}\\
 
\addlinespace[0.5em]
\multicolumn{1}{c}{} & \multicolumn{5}{c}{\textbf{DCC -- SCB -- EQ}} & \multicolumn{5}{c}{\textbf{DCC -- SCB -- RW}}\\
\nopagebreak 70 & 68.4 & 34.2 & \em{72.2} & \em{72.2} & \textbf{72.4} & 69.2 & 31.6 & 72.4 & \em{72.6} & \vphantom{1} \textbf{73.2}\\
\nopagebreak 
75 & 70.8 & 37.2 & \textbf{74.4} & \em{74.2} & \textbf{74.4} & 72.6 & 36.0 & \em{75.0} & 74.8 & \vphantom{1} \textbf{75.6}\\
\nopagebreak 
80 & 74.0 & 40.2 & \textbf{76.8} & \em{76.6} & \textbf{76.8} & 75.8 & 40.2 & \em{78.0} & 77.8 & \vphantom{1} \textbf{78.4}\\
\nopagebreak 
85 & 77.0 & 42.8 & \em{79.2} & \textbf{79.4} & \textbf{79.4} & 77.6 & 44.4 & \em{80.4} & \em{80.4} & \vphantom{1} \textbf{80.8}\\
\nopagebreak 
90 & 81.8 & 47.8 & 82.0 & \em{82.2} & \textbf{82.4} & 79.8 & 48.8 & \em{82.6} & \em{82.6} & \vphantom{1} \textbf{82.8}\\
\nopagebreak 
95 & \textbf{86.2} & 55.6 & 85.0 & 85.0 & \em{85.2} & \em{85.0} & 55.4 & \em{85.0} & \textbf{85.2} & \vphantom{1} \textbf{85.2}\\
 
\addlinespace[0.5em]
\multicolumn{1}{c}{} & \multicolumn{5}{c}{\textbf{EDCC -- DCC -- EQ}} & \multicolumn{5}{c}{\textbf{EDCC -- DCC -- RW}}\\
\nopagebreak 70 & 64.2 & 26.0 & \em{84.0} & \textbf{85.0} & \em{84.0} & 66.6 & 28.8 & \em{85.4} & \textbf{87.0} & \textbf{87.0}\\
\nopagebreak 
75 & 68.2 & 29.0 & \textbf{86.8} & \textbf{86.8} & \em{86.2} & 71.0 & 32.6 & \em{87.8} & \textbf{89.0} & \textbf{89.0}\\
\nopagebreak 
80 & 72.0 & 31.6 & \textbf{88.2} & \em{88.0} & 87.6 & 74.6 & 35.6 & 88.8 & \em{89.4} & \textbf{90.0}\\
\nopagebreak 
85 & 75.6 & 36.0 & \em{90.2} & \textbf{90.6} & 89.6 & 77.8 & 40.2 & 91.2 & \em{91.4} & \textbf{91.8}\\
\nopagebreak 
90 & 81.4 & 41.8 & \em{92.4} & \textbf{92.6} & \textbf{92.6} & 83.0 & 46.6 & \em{94.0} & \em{94.0} & \textbf{94.6}\\
\nopagebreak 
95 & \em{89.0} & 51.8 & \textbf{94.6} & \textbf{94.6} & \textbf{94.6} & 89.2 & 55.2 & \em{95.4} & \em{95.4} & \textbf{95.6}\\
 
\addlinespace[0.5em]
\multicolumn{1}{c}{} & \multicolumn{5}{c}{\textbf{EDCC -- SCB -- EQ}} & \multicolumn{5}{c}{\textbf{EDCC -- SCB -- RW}}\\
\nopagebreak 70 & 68.4 & 34.2 & \em{72.2} & \em{72.2} & \textbf{72.4} & 69.2 & 31.6 & 72.4 & \em{72.6} & \textbf{73.2}\\
\nopagebreak 
75 & 70.8 & 37.2 & \textbf{74.4} & \em{74.2} & \textbf{74.4} & 72.6 & 36.0 & \em{75.0} & 74.8 & \textbf{75.6}\\
\nopagebreak 
80 & 74.0 & 40.2 & \textbf{76.8} & \em{76.6} & \textbf{76.8} & 75.8 & 40.2 & \em{78.0} & 77.8 & \textbf{78.4}\\
\nopagebreak 
85 & 77.0 & 42.8 & \em{79.2} & \textbf{79.4} & \textbf{79.4} & 77.6 & 44.4 & \em{80.4} & \em{80.4} & \textbf{80.8}\\
\nopagebreak 
90 & 81.8 & 47.8 & 82.0 & \em{82.2} & \textbf{82.4} & 79.8 & 48.8 & \em{82.6} & \em{82.6} & \textbf{82.8}\\
\nopagebreak 
95 & \textbf{86.2} & 55.6 & 85.0 & 85.0 & \em{85.2} & \em{85.0} & 55.4 & \em{85.0} & \textbf{85.2} & \textbf{85.2}\\
 
\addlinespace[0.5em]
\multicolumn{1}{c}{} & \multicolumn{5}{c}{\textbf{SCB -- DCC -- EQ}} & \multicolumn{5}{c}{\textbf{SCB -- DCC -- RW}}\\
\nopagebreak 70 & 22.6 & \em{41.2} & \textbf{84.0} & \textbf{84.0} & \textbf{84.0} & 21.2 & 41.4 & \em{82.6} & \em{82.6} & \textbf{82.8}\\
\nopagebreak 
75 & 23.0 & \em{42.8} & \textbf{85.2} & \textbf{85.2} & \textbf{85.2} & 22.4 & 43.2 & \em{83.8} & \em{83.8} & \textbf{84.0}\\
\nopagebreak 
80 & 24.6 & \em{46.6} & \textbf{86.6} & \textbf{86.6} & \textbf{86.6} & 23.2 & 46.0 & \em{85.8} & \em{85.8} & \textbf{86.0}\\
\nopagebreak 
85 & 26.2 & \em{49.0} & \textbf{87.4} & \textbf{87.4} & \textbf{87.4} & 24.8 & 48.6 & \em{87.6} & \em{87.6} & \textbf{87.8}\\
\nopagebreak 
90 & 28.4 & \em{52.2} & \textbf{88.8} & \textbf{88.8} & \textbf{88.8} & 28.6 & \em{52.8} & \textbf{90.0} & \textbf{90.0} & \textbf{90.0}\\
\nopagebreak 
95 & 30.6 & \em{56.2} & \textbf{90.2} & \textbf{90.2} & \textbf{90.2} & 30.8 & \em{58.0} & \textbf{92.0} & \textbf{92.0} & \textbf{92.0}\\
 
\addlinespace[0.5em]
\multicolumn{1}{c}{} & \multicolumn{5}{c}{\textbf{SCB -- SCB -- EQ}} & \multicolumn{5}{c}{\textbf{SCB -- SCB -- RW}}\\
\nopagebreak 70 & 22.8 & \em{26.6} & \textbf{88.8} & \textbf{88.8} & \textbf{88.8} & 20.2 & \em{28.8} & \textbf{88.4} & \textbf{88.4} & \textbf{88.4}\\
\nopagebreak 
75 & 24.2 & \em{28.8} & \textbf{89.2} & \textbf{89.2} & \textbf{89.2} & 21.4 & \em{30.6} & \textbf{88.8} & \textbf{88.8} & \textbf{88.8}\\
\nopagebreak 
80 & 26.0 & \em{29.6} & \textbf{89.4} & \textbf{89.4} & \textbf{89.4} & 22.6 & \em{32.6} & \textbf{89.2} & \textbf{89.2} & \textbf{89.2}\\
\nopagebreak 
85 & 27.6 & \em{31.4} & \textbf{90.2} & \textbf{90.2} & \textbf{90.2} & 23.8 & \em{33.8} & \textbf{89.8} & \textbf{89.8} & \textbf{89.8}\\
\nopagebreak 
90 & 31.0 & \em{33.2} & \textbf{91.2} & \textbf{91.2} & \textbf{91.2} & 26.6 & \em{36.0} & \textbf{91.0} & \textbf{91.0} & \textbf{91.0}\\
\nopagebreak 
95 & 33.0 & \em{38.0} & \textbf{93.8} & \textbf{93.8} & \textbf{93.8} & 30.0 & \em{39.4} & \textbf{92.6} & \textbf{92.6} & \textbf{92.6}\\*
\end{longtable}

\endgroup

\begin{figure}[H]
	\centering
	\includegraphics[width=0.9\linewidth]{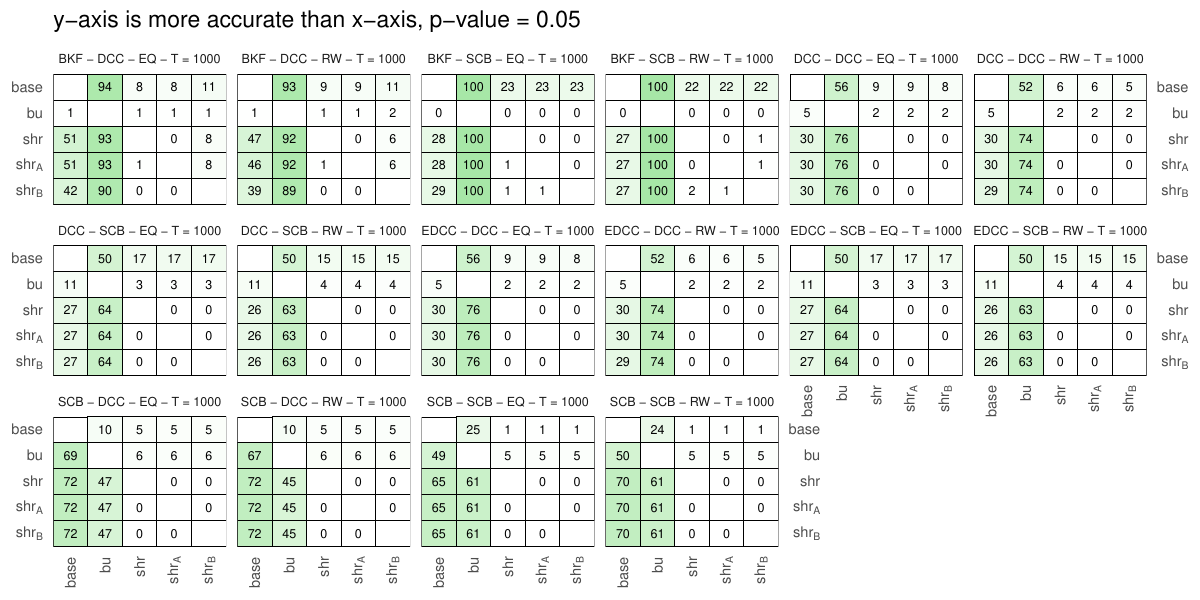}
	\caption{Qualitative evaluation using the Diebold-Mariano across different simulation settings ($N = 24$ and $T = 1000$) using an absolute loss function. Each setting is identified by a label composed of three elements: (i) the data-generating process (DGP), (ii) the MGARCH model used for estimation, and (iii) the portfolio weighting scheme (see \autoref{tab:notationModel} for details). The rows and columns indicate the variance forecasting method employed: the univariate GARCH on portfolio returns (base), the bottom-up approach using MGARCH models (bu), and the three forecast reconciliation strategies discussed in the paper ($shr$, $shr_{A}$, $shr_{B}$). Each cell reports the number of times (in \%) the forecasting model in the row statistically outperforms (p-values $< 0.05$ and using Bonferroni correction) the model in the column.}
\end{figure}

\begin{figure}[H]
	\centering
	\includegraphics[width=0.9\linewidth]{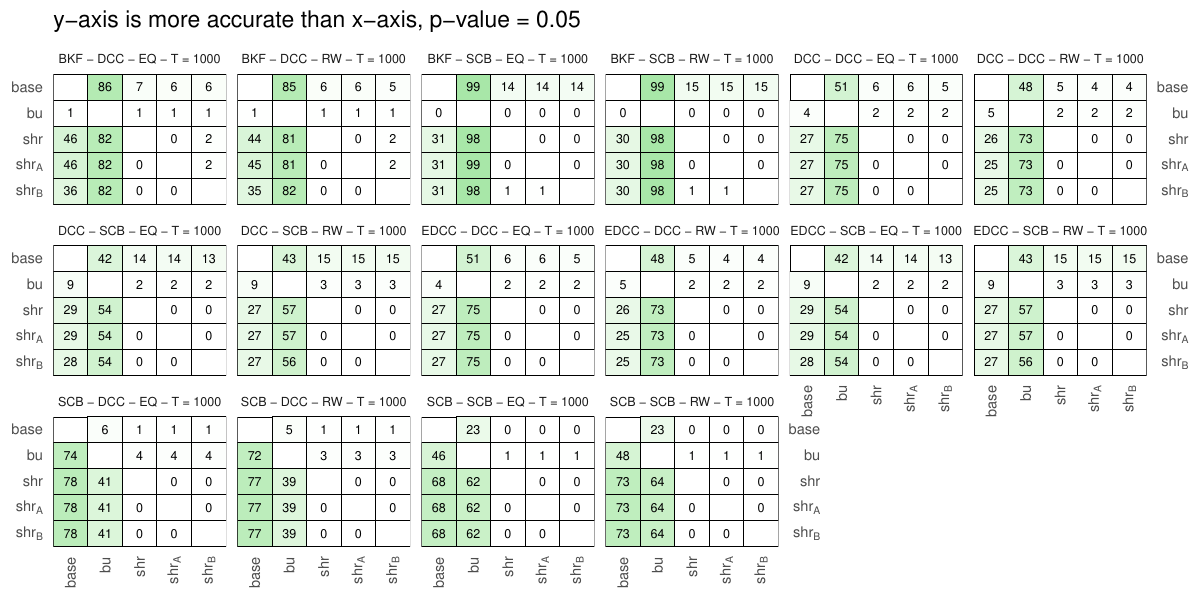}
	\caption{Qualitative evaluation using the Diebold-Mariano across different simulation settings ($N = 24$ and $T = 1000$) using a square loss function. Each setting is identified by a label composed of three elements: (i) the data-generating process (DGP), (ii) the MGARCH model used for estimation, and (iii) the portfolio weighting scheme (see \autoref{tab:notationModel} for details). The rows and columns indicate the variance forecasting method employed: the univariate GARCH on portfolio returns (base), the bottom-up approach using MGARCH models (bu), and the three forecast reconciliation strategies discussed in the paper ($shr$, $shr_{A}$, $shr_{B}$). Each cell reports the number of times (in \%) the forecasting model in the row statistically outperforms (p-values $< 0.05$ and using Bonferroni correction) the model in the column.}
\end{figure}

\begin{figure}[H]
	\centering
	\includegraphics[width=0.9\linewidth]{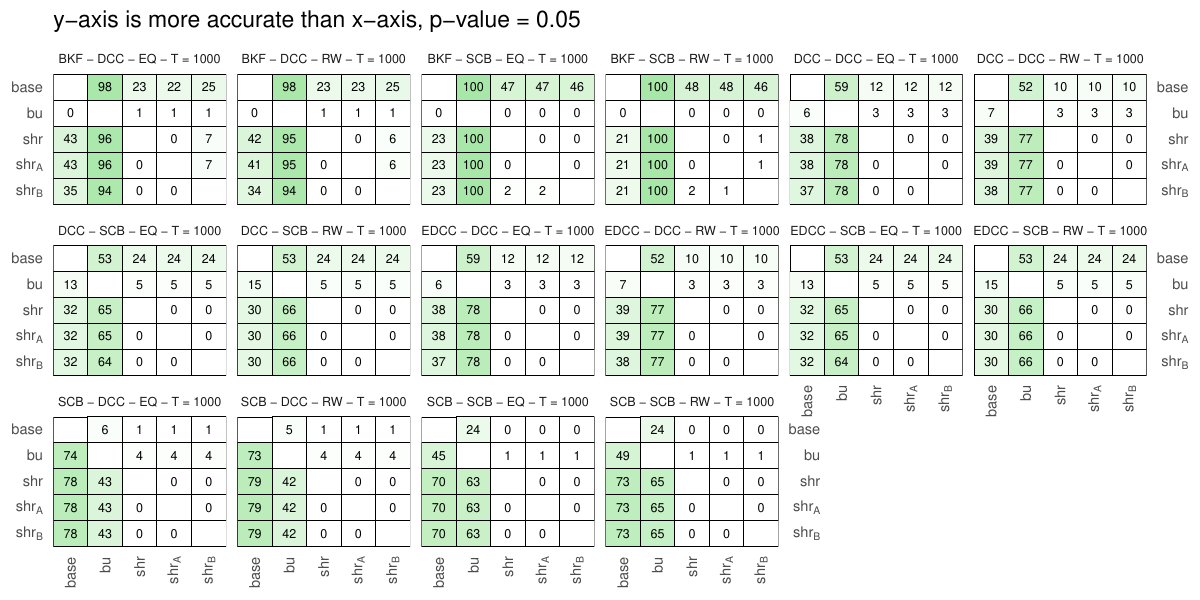}
	\caption{Qualitative evaluation using the Diebold-Mariano across different simulation settings ($N = 24$ and $T = 1000$) using a QLIKE loss function. Each setting is identified by a label composed of three elements: (i) the data-generating process (DGP), (ii) the MGARCH model used for estimation, and (iii) the portfolio weighting scheme (see \autoref{tab:notationModel} for details). The rows and columns indicate the variance forecasting method employed: the univariate GARCH on portfolio returns (base), the bottom-up approach using MGARCH models (bu), and the three forecast reconciliation strategies discussed in the paper ($shr$, $shr_{A}$, $shr_{B}$). Each cell reports the number of times (in \%) the forecasting model in the row statistically outperforms (p-values $< 0.05$ and using Bonferroni correction) the model in the column.}
\end{figure}

\newpage
\subsection{Proxy portfolio variance and covariance matrix of $N=24$ assets}

\begin{landscape}\centering
	\begingroup
	\setlength{\LTcapwidth}{\linewidth}
	\setlength{\tabcolsep}{2pt}
	\scriptsize
	
\begin{longtable}[t]{>{}c||>{}c>{}c|cc>{}c||>{}c>{}c|cc>{}c||>{}c>{}c|cc>{}c||>{}c>{}c|ccc}
\caption{Average accuracy indices across different simulation settings. Each setting is identified by a label composed of three elements: (i) the data-generating process (DGP), (ii) the MGARCH model used for estimation, and (iii) the portfolio weighting scheme (see \autoref{tab:notationModel} for details). The sample size ($T$) is also reported for each setting.  The top rows indicate the noisy level of the proxy $\delta \in \left\lbrace 0.25, 0.5, 0.75, 1\right\rbrace$ and the variance forecasting method employed: the univariate GARCH on portfolio returns (base), the bottom-up approach using MGARCH models (bu), and the three forecast reconciliation strategies discussed in the paper ($shr$, $shr_{A}$, $shr_{B}$). The reported accuracy measures are grouped into three categories: average accuracy indices (MSE, MAE, QLIKE); average relative measures (AvgRelMSE, AvgRelMAE, AvgRelQLIKE) using the univariate and the multivariate GARCH forecast as the benchmark.}\\
\toprule
\multicolumn{1}{c}{ } & \multicolumn{5}{c}{$\delta = 0.25$} & \multicolumn{5}{c}{$\delta = 0.5$} & \multicolumn{5}{c}{$\delta = 0.75$} & \multicolumn{5}{c}{$\delta = 1$} \\
\cmidrule(l{3pt}r{3pt}){2-6} \cmidrule(l{3pt}r{3pt}){7-11} \cmidrule(l{3pt}r{3pt}){12-16} \cmidrule(l{3pt}r{3pt}){17-21}
Index & $base$ & $bu$ & $shr$ & $shr_{A}$ & $shr_{B}$ & $base$ & $bu$ & $shr$ & $shr_{A}$ & $shr_{B}$ & $base$ & $bu$ & $shr$ & $shr_{A}$ & $shr_{B}$ & $base$ & $bu$ & $shr$ & $shr_{A}$ & $shr_{B}$\\
\midrule
\endfirsthead
\caption[]{Average accuracy indices across different simulation settings. Each setting is identified by a label composed of three elements: (i) the data-generating process (DGP), (ii) the MGARCH model used for estimation, and (iii) the portfolio weighting scheme (see \autoref{tab:notationModel} for details). The sample size ($T$) is also reported for each setting.  The top rows indicate the noisy level of the proxy $\delta \in \left\lbrace 0.25, 0.5, 0.75, 1\right\rbrace$ and the variance forecasting method employed: the univariate GARCH on portfolio returns (base), the bottom-up approach using MGARCH models (bu), and the three forecast reconciliation strategies discussed in the paper ($shr$, $shr_{A}$, $shr_{B}$). The reported accuracy measures are grouped into three categories: average accuracy indices (MSE, MAE, QLIKE); average relative measures (AvgRelMSE, AvgRelMAE, AvgRelQLIKE) using the univariate and the multivariate GARCH forecast as the benchmark. \textit{(continued)}}\\
\toprule
\multicolumn{1}{c}{ } & \multicolumn{5}{c}{$\delta = 0.25$} & \multicolumn{5}{c}{$\delta = 0.5$} & \multicolumn{5}{c}{$\delta = 0.75$} & \multicolumn{5}{c}{$\delta = 1$} \\
\cmidrule(l{3pt}r{3pt}){2-6} \cmidrule(l{3pt}r{3pt}){7-11} \cmidrule(l{3pt}r{3pt}){12-16} \cmidrule(l{3pt}r{3pt}){17-21}
Index & $base$ & $bu$ & $shr$ & $shr_{A}$ & $shr_{B}$ & $base$ & $bu$ & $shr$ & $shr_{A}$ & $shr_{B}$ & $base$ & $bu$ & $shr$ & $shr_{A}$ & $shr_{B}$ & $base$ & $bu$ & $shr$ & $shr_{A}$ & $shr_{B}$\\
\midrule
\endhead

\endfoot
\bottomrule
\endlastfoot
\addlinespace[0.5em]
\multicolumn{21}{c}{\textbf{BKF -- DCC -- EQ ($T=1000$)}}\\
\nopagebreak \addlinespace[0em]
\multicolumn{21}{l}{\textit{Average indexes}}\\
\nopagebreak MSE & \textbf{5.4}\scalebox{.5}{$\times 10^4$} & 6.5\scalebox{.5}{$\times 10^4$} & \em{5.6}\scalebox{.5}{$\times 10^4$} & 5.6\scalebox{.5}{$\times 10^4$} & 5.6\scalebox{.5}{$\times 10^4$} & \textbf{19.5}\scalebox{.5}{$\times 10^4$} & 20.4\scalebox{.5}{$\times 10^4$} & \em{19.9}\scalebox{.5}{$\times 10^4$} & 19.9\scalebox{.5}{$\times 10^4$} & 20.0\scalebox{.5}{$\times 10^4$} & \textbf{43.0}\scalebox{.5}{$\times 10^4$} & 43.7\scalebox{.5}{$\times 10^4$} & \em{43.4}\scalebox{.5}{$\times 10^4$} & 43.4\scalebox{.5}{$\times 10^4$} & 43.4\scalebox{.5}{$\times 10^4$} & \textbf{75.8}\scalebox{.5}{$\times 10^4$} & 76.5\scalebox{.5}{$\times 10^4$} & \em{76.2}\scalebox{.5}{$\times 10^4$} & 76.2\scalebox{.5}{$\times 10^4$} & 76.2\scalebox{.5}{$\times 10^4$}\\
\nopagebreak 
MAE & 54.641 & 56.593 & \em{53.702} & \textbf{53.688} & 54.645 & 105.9 & \textbf{103.7} & \em{104.0} & 104.0 & 104.1 & 158.6 & \textbf{155.4} & 155.8 & 155.8 & \em{155.8} & 211.5 & \textbf{208.0} & 208.2 & 208.2 & \em{208.2}\\
\nopagebreak 
QLIKE & \textbf{0.052} & 0.062 & \em{0.052} & 0.052 & 0.057 & \textbf{0.169} & 0.179 & \em{0.174} & 0.174 & 0.175 & \textbf{0.392} & 0.400 & \em{0.398} & 0.398 & 0.398 & \textbf{1.295} & 1.302 & 1.302 & 1.302 & \em{1.302}\\
\nopagebreak 
\addlinespace[0em]
\multicolumn{21}{l}{\textit{Relative indexes (base benchmark)}}\\
\nopagebreak AvgRelMSE & 1.000 & 1.121 & \textbf{0.993} & \em{0.993} & 1.032 & \textbf{1.000} & 1.017 & \em{1.003} & 1.003 & 1.005 & 1.000 & 1.000 & \em{0.998} & 0.998 & \textbf{0.998} & 1.000 & 0.996 & \em{0.996} & 0.996 & \textbf{0.996}\\
\nopagebreak 
AvgRelMAE & 1.000 & 1.052 & \em{0.995} & \textbf{0.995} & 1.011 & 1.000 & 0.995 & \textbf{0.994} & \em{0.994} & 0.995 & 1.000 & \textbf{0.991} & 0.992 & 0.992 & \em{0.992} & 1.000 & \textbf{0.992} & 0.992 & 0.992 & \em{0.992}\\
\end{longtable}

	\endgroup
\end{landscape}

\begin{landscape}\centering
	\begingroup
	\setlength{\LTcapwidth}{\linewidth}
	\setlength{\tabcolsep}{4pt}
	\scriptsize
	
\begin{longtable}[t]{>{}c||>{}c>{}c|cc>{}c||>{}c>{}c|cc>{}c||>{}c>{}c|cc>{}c||>{}c>{}c|ccc}
\caption{The number of times (in \%) that model is in the Model Confidence Set with different thresholds ($\gamma \in \{70\%, 75\%, 80\%, 85\%m 90\%, 95\%\}$) across different simulation settings using the MAE as loss function. Each setting is identified by a label composed of three elements: (i) the data-generating process (DGP), (ii) the MGARCH model used for estimation, and (iii) the portfolio weighting scheme (see \autoref{tab:notationModel} for details). The sample size ($T$) is also reported for each setting.  The top rows indicate the noisy level of the proxy $\delta \in \left\lbrace 0.25, 0.5, 0.75, 1\right\rbrace$ and the variance forecasting method employed: the univariate GARCH on portfolio returns (base), the bottom-up approach using MGARCH models (bu), and the three forecast reconciliation strategies discussed in the paper ($shr$, $shr_{A}$, $shr_{B}$).}\\
\toprule
\multicolumn{1}{c}{ } & \multicolumn{5}{c}{$\delta = 0.25$} & \multicolumn{5}{c}{$\delta = 0.5$} & \multicolumn{5}{c}{$\delta = 0.75$} & \multicolumn{5}{c}{$\delta = 1$} \\
\cmidrule(l{3pt}r{3pt}){2-6} \cmidrule(l{3pt}r{3pt}){7-11} \cmidrule(l{3pt}r{3pt}){12-16} \cmidrule(l{3pt}r{3pt}){17-21}
Threshold & $base$ & $bu$ & $shr$ & $shr_{A}$ & $shr_{B}$ & $base$ & $bu$ & $shr$ & $shr_{A}$ & $shr_{B}$ & $base$ & $bu$ & $shr$ & $shr_{A}$ & $shr_{B}$ & $base$ & $bu$ & $shr$ & $shr_{A}$ & $shr_{B}$\\
\midrule
\endfirsthead
\caption[]{The number of times (in \%) that model is in the Model Confidence Set with different thresholds ($\gamma \in \{70\%, 75\%, 80\%, 85\%m 90\%, 95\%\}$) across different simulation settings using the MAE as loss function. Each setting is identified by a label composed of three elements: (i) the data-generating process (DGP), (ii) the MGARCH model used for estimation, and (iii) the portfolio weighting scheme (see \autoref{tab:notationModel} for details). The sample size ($T$) is also reported for each setting.  The top rows indicate the noisy level of the proxy $\delta \in \left\lbrace 0.25, 0.5, 0.75, 1\right\rbrace$ and the variance forecasting method employed: the univariate GARCH on portfolio returns (base), the bottom-up approach using MGARCH models (bu), and the three forecast reconciliation strategies discussed in the paper ($shr$, $shr_{A}$, $shr_{B}$). \textit{(continued)}}\\
\toprule
\multicolumn{1}{c}{ } & \multicolumn{5}{c}{$\delta = 0.25$} & \multicolumn{5}{c}{$\delta = 0.5$} & \multicolumn{5}{c}{$\delta = 0.75$} & \multicolumn{5}{c}{$\delta = 1$} \\
\cmidrule(l{3pt}r{3pt}){2-6} \cmidrule(l{3pt}r{3pt}){7-11} \cmidrule(l{3pt}r{3pt}){12-16} \cmidrule(l{3pt}r{3pt}){17-21}
Threshold & $base$ & $bu$ & $shr$ & $shr_{A}$ & $shr_{B}$ & $base$ & $bu$ & $shr$ & $shr_{A}$ & $shr_{B}$ & $base$ & $bu$ & $shr$ & $shr_{A}$ & $shr_{B}$ & $base$ & $bu$ & $shr$ & $shr_{A}$ & $shr_{B}$\\
\midrule
\endhead

\endfoot
\bottomrule
\endlastfoot
\addlinespace[0.5em]
\multicolumn{21}{c}{\textbf{BKF -- DCC -- EQ ($T=1000$)}}\\
\nopagebreak 70 & \textbf{79.8} & 48.0 & 73.6 & \em{74.0} & 64.6 & \em{71.4} & \textbf{81.2} & 62.8 & 62.8 & 63.8 & \em{59.0} & \textbf{85.4} & 54.2 & 54.2 & 56.4 & 52.6 & \textbf{83.6} & 57.6 & 57.2 & \em{58.2}\\
\nopagebreak 
75 & \textbf{81.8} & 50.4 & 76.4 & \em{76.6} & 69.6 & \em{75.8} & \textbf{83.0} & 68.0 & 68.2 & 69.2 & \em{63.6} & \textbf{87.0} & 60.2 & 60.0 & 62.0 & 56.6 & \textbf{84.8} & 62.2 & 62.0 & \em{62.8}\\
\nopagebreak 
80 & \textbf{84.0} & 54.0 & 79.2 & \em{79.4} & 75.2 & \em{77.8} & \textbf{84.6} & 71.0 & 71.2 & 72.2 & 65.6 & \textbf{88.4} & 64.8 & 64.6 & \em{66.0} & 61.2 & \textbf{86.4} & \em{66.2} & \em{66.2} & \em{66.2}\\
\nopagebreak 
85 & \textbf{85.6} & 57.8 & \em{82.6} & \em{82.6} & 79.0 & \em{80.0} & \textbf{87.6} & 76.8 & 77.0 & 77.4 & \em{72.4} & \textbf{89.8} & 71.2 & 71.0 & 72.2 & 68.2 & \textbf{89.8} & \em{73.0} & 72.8 & \em{73.0}\\
\nopagebreak 
90 & \textbf{89.4} & 61.6 & \em{86.8} & \em{86.8} & 85.4 & \em{83.6} & \textbf{90.8} & 81.8 & 82.0 & 82.4 & 77.0 & \textbf{91.4} & 78.0 & 77.6 & \em{78.4} & 71.2 & \textbf{91.0} & \em{77.6} & \em{77.6} & \em{77.6}\\
\nopagebreak 
95 & \textbf{92.8} & 68.0 & \em{92.6} & 92.4 & 92.4 & \em{89.8} & \textbf{92.6} & 88.0 & 87.8 & 87.8 & 84.4 & \textbf{94.0} & \em{85.6} & 85.4 & \em{85.6} & 81.0 & \textbf{93.2} & \em{86.0} & \em{86.0} & \em{86.0}\\
 
\addlinespace[0.5em]
\multicolumn{21}{c}{\textbf{BKF -- DCC -- RW ($T=1000$)}}\\
\end{longtable}%

	\endgroup
\end{landscape}

\begin{figure}[H]
	\centering
	\includegraphics[width=0.9\linewidth]{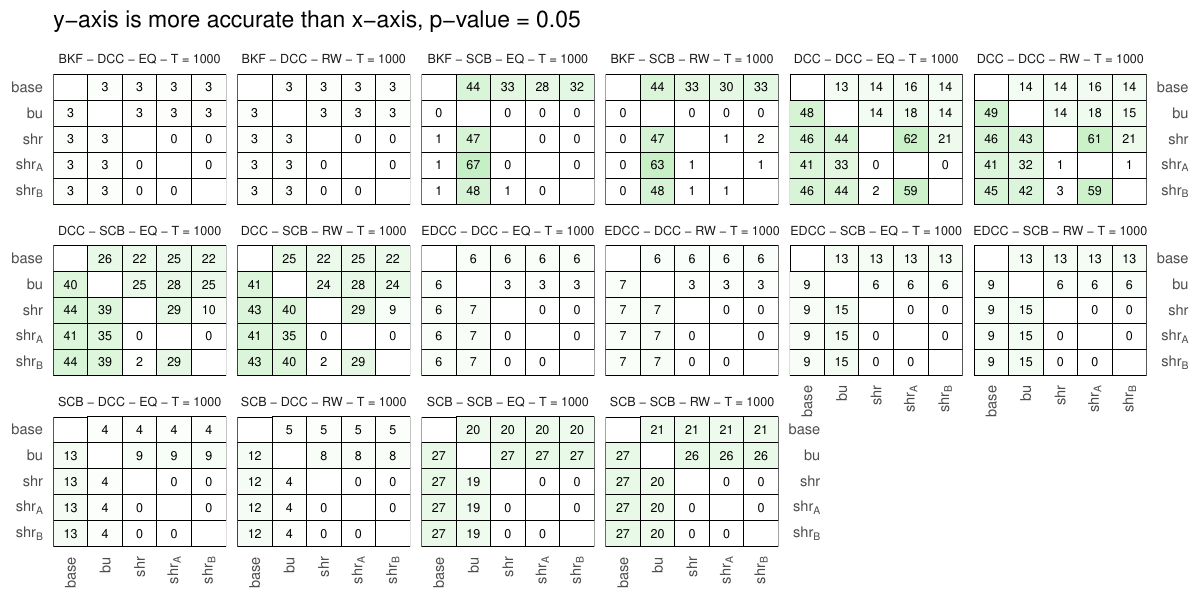}
	\caption{Qualitative evaluation using the Diebold-Mariano across different simulation settings using an absolute loss function and noisy level $\delta = 1$. Each setting is identified by a label composed of three elements: (i) the data-generating process (DGP), (ii) the MGARCH model used for estimation, and (iii) the portfolio weighting scheme (see \autoref{tab:notationModel} for details). The sample size ($T$) is also reported. The rows and columns indicate the variance forecasting method employed: the univariate GARCH on portfolio returns (base), the bottom-up approach using MGARCH models (bu), and the three forecast reconciliation strategies discussed in the paper ($shr$, $shr_{A}$, $shr_{B}$). Each cell reports the number of times (in \%) the forecasting model in the row statistically outperforms (p-values $< 0.05$ and using Bonferroni correction) the model in the column.}
\end{figure}
\begin{figure}[H]
	\centering
	\includegraphics[width=0.9\linewidth]{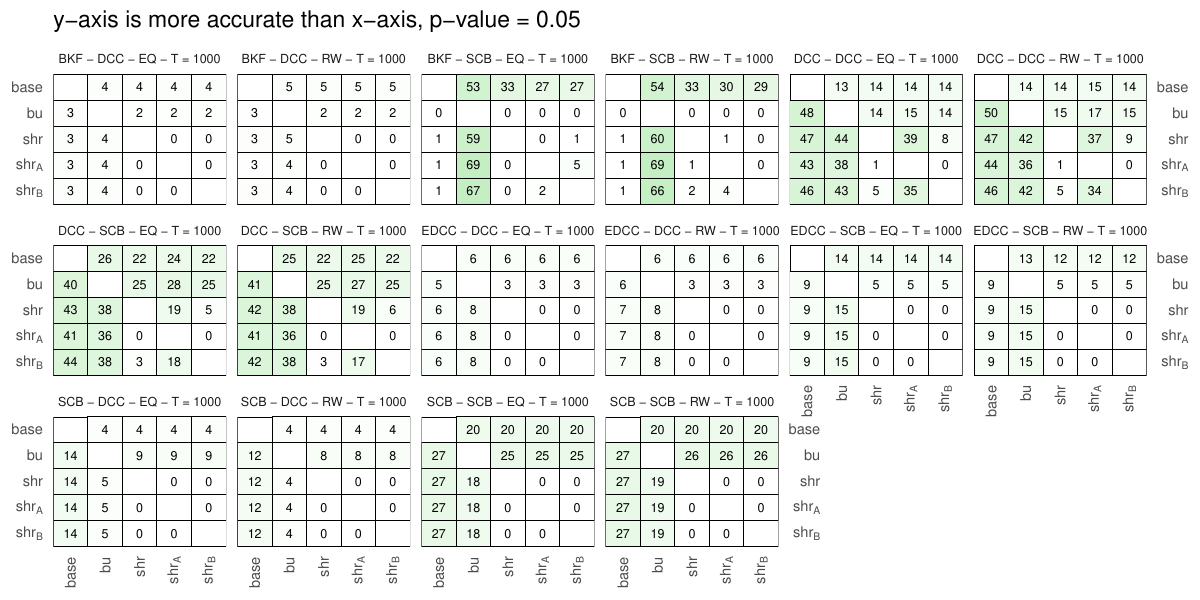}
	\caption{Qualitative evaluation using the Diebold-Mariano across different simulation settings using an absolute loss function and noisy level $\delta = 0.75$. Each setting is identified by a label composed of three elements: (i) the data-generating process (DGP), (ii) the MGARCH model used for estimation, and (iii) the portfolio weighting scheme (see \autoref{tab:notationModel} for details). The sample size ($T$) is also reported. The rows and columns indicate the variance forecasting method employed: the univariate GARCH on portfolio returns (base), the bottom-up approach using MGARCH models (bu), and the three forecast reconciliation strategies discussed in the paper ($shr$, $shr_{A}$, $shr_{B}$). Each cell reports the number of times (in \%) the forecasting model in the row statistically outperforms (p-values $< 0.05$ and using Bonferroni correction) the model in the column.}
\end{figure}
\begin{figure}[H]
	\centering
	\includegraphics[width=0.9\linewidth]{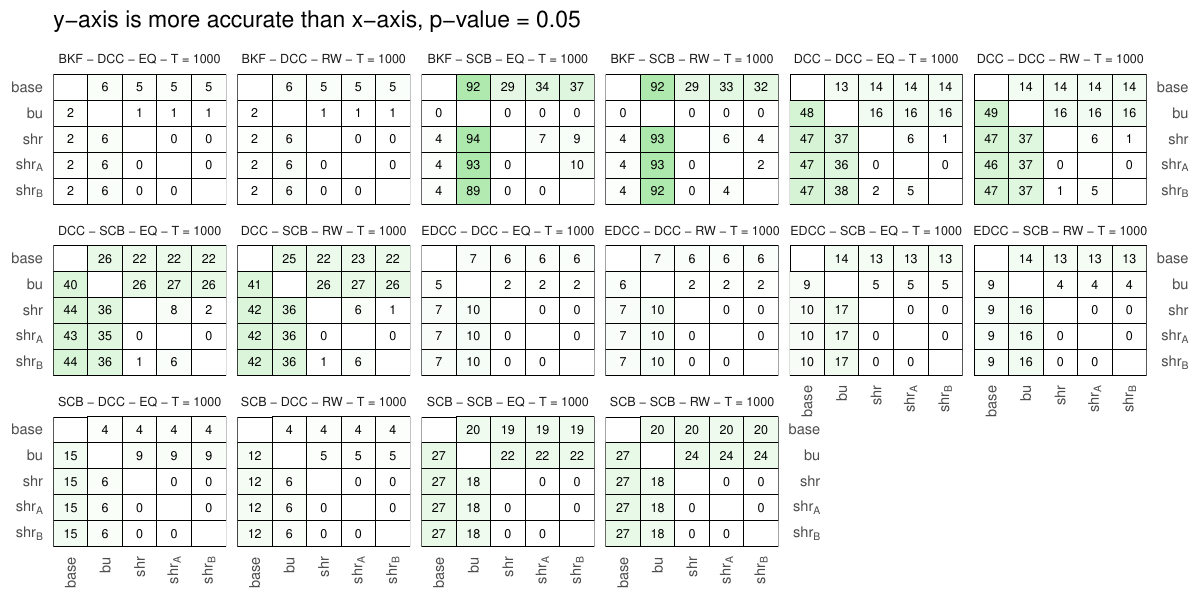}
	\caption{Qualitative evaluation using the Diebold-Mariano across different simulation settings using an absolute loss function and noisy level $\delta = 0.5$. Each setting is identified by a label composed of three elements: (i) the data-generating process (DGP), (ii) the MGARCH model used for estimation, and (iii) the portfolio weighting scheme (see \autoref{tab:notationModel} for details). The sample size ($T$) is also reported. The rows and columns indicate the variance forecasting method employed: the univariate GARCH on portfolio returns (base), the bottom-up approach using MGARCH models (bu), and the three forecast reconciliation strategies discussed in the paper ($shr$, $shr_{A}$, $shr_{B}$). Each cell reports the number of times (in \%) the forecasting model in the row statistically outperforms (p-values $< 0.05$ and using Bonferroni correction) the model in the column.}
\end{figure}
\begin{figure}[H]
	\centering
	\includegraphics[width=0.9\linewidth]{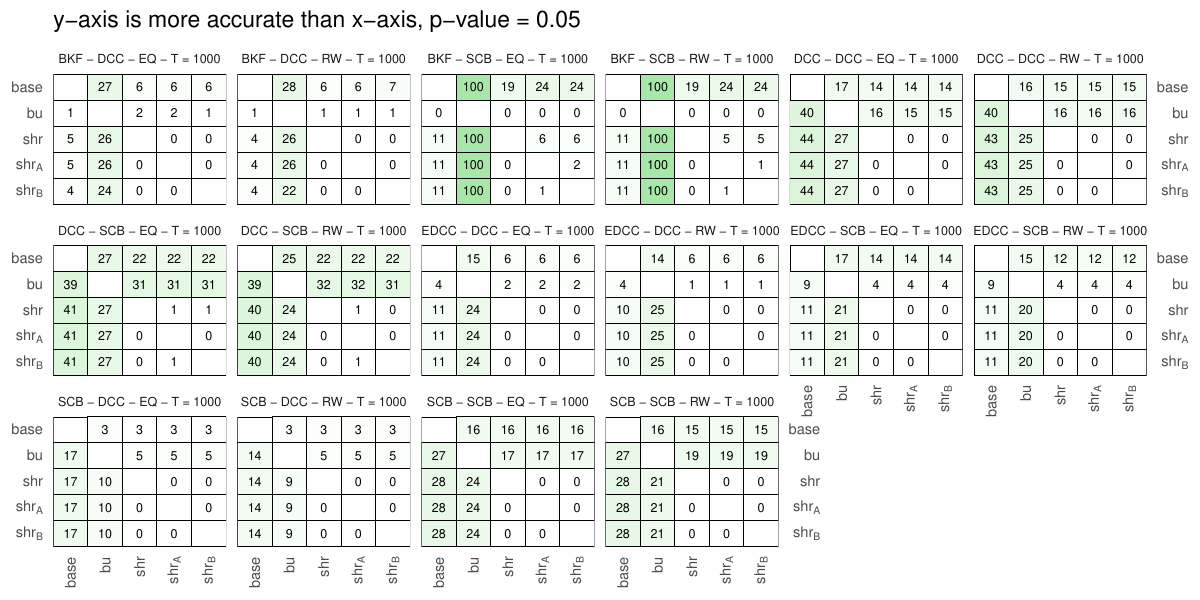}
	\caption{Qualitative evaluation using the Diebold-Mariano across different simulation settings using an absolute loss function and noisy level $\delta = 0.25$. Each setting is identified by a label composed of three elements: (i) the data-generating process (DGP), (ii) the MGARCH model used for estimation, and (iii) the portfolio weighting scheme (see \autoref{tab:notationModel} for details). The sample size ($T$) is also reported. The rows and columns indicate the variance forecasting method employed: the univariate GARCH on portfolio returns (base), the bottom-up approach using MGARCH models (bu), and the three forecast reconciliation strategies discussed in the paper ($shr$, $shr_{A}$, $shr_{B}$). Each cell reports the number of times (in \%) the forecasting model in the row statistically outperforms (p-values $< 0.05$ and using Bonferroni correction) the model in the column.}
\end{figure}

\begin{figure}[H]
	\centering
	\includegraphics[width=0.9\linewidth]{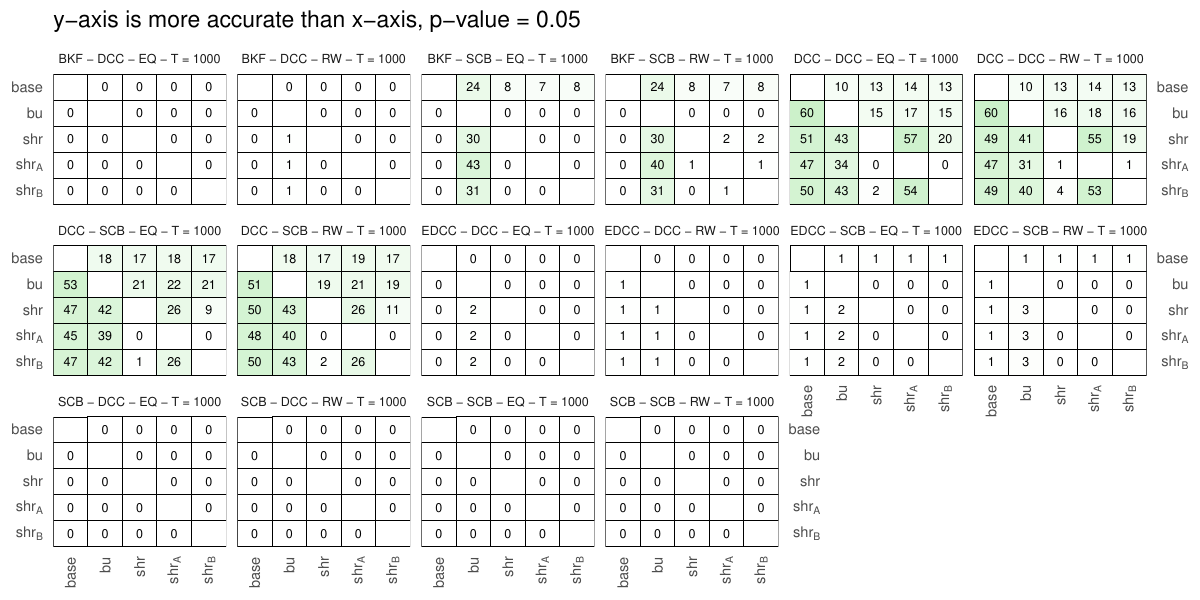}
	\caption{Qualitative evaluation using the Diebold-Mariano across different simulation settings using a square loss function and noisy level $\delta = 1$. Each setting is identified by a label composed of three elements: (i) the data-generating process (DGP), (ii) the MGARCH model used for estimation, and (iii) the portfolio weighting scheme (see \autoref{tab:notationModel} for details). The sample size ($T$) is also reported. The rows and columns indicate the variance forecasting method employed: the univariate GARCH on portfolio returns (base), the bottom-up approach using MGARCH models (bu), and the three forecast reconciliation strategies discussed in the paper ($shr$, $shr_{A}$, $shr_{B}$). Each cell reports the number of times (in \%) the forecasting model in the row statistically outperforms (p-values $< 0.05$ and using Bonferroni correction) the model in the column.}
\end{figure}
\begin{figure}[H]
	\centering
	\includegraphics[width=0.9\linewidth]{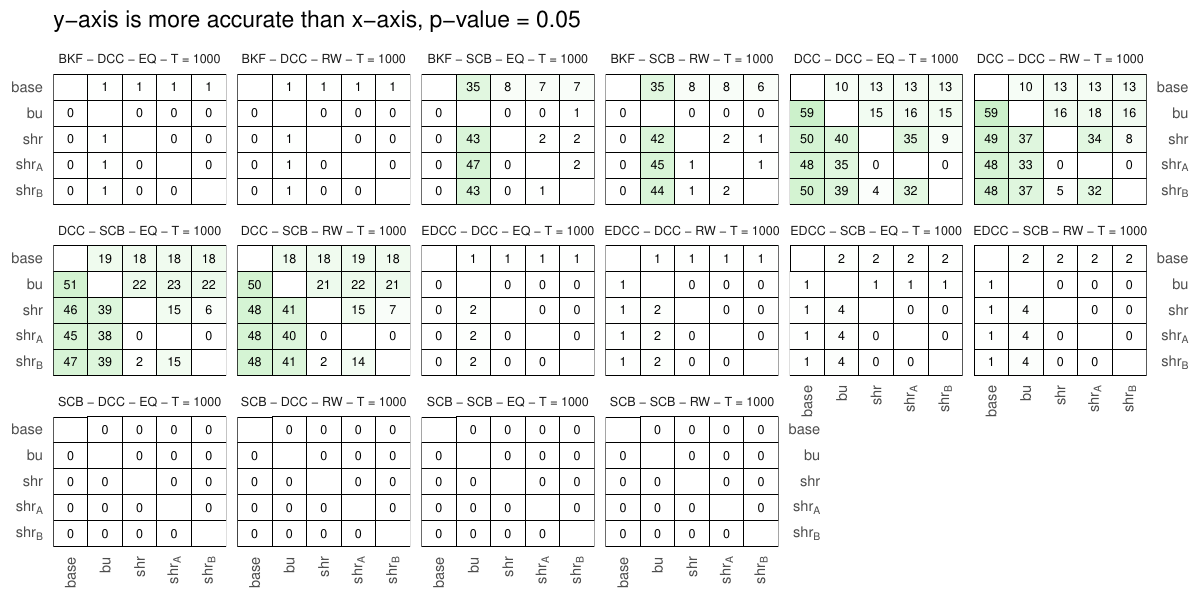}
	\caption{Qualitative evaluation using the Diebold-Mariano across different simulation settings using a square loss function and noisy level $\delta = 0.75$. Each setting is identified by a label composed of three elements: (i) the data-generating process (DGP), (ii) the MGARCH model used for estimation, and (iii) the portfolio weighting scheme (see \autoref{tab:notationModel} for details). The sample size ($T$) is also reported. The rows and columns indicate the variance forecasting method employed: the univariate GARCH on portfolio returns (base), the bottom-up approach using MGARCH models (bu), and the three forecast reconciliation strategies discussed in the paper ($shr$, $shr_{A}$, $shr_{B}$). Each cell reports the number of times (in \%) the forecasting model in the row statistically outperforms (p-values $< 0.05$ and using Bonferroni correction) the model in the column.}
\end{figure}
\begin{figure}[H]
	\centering
	\includegraphics[width=0.9\linewidth]{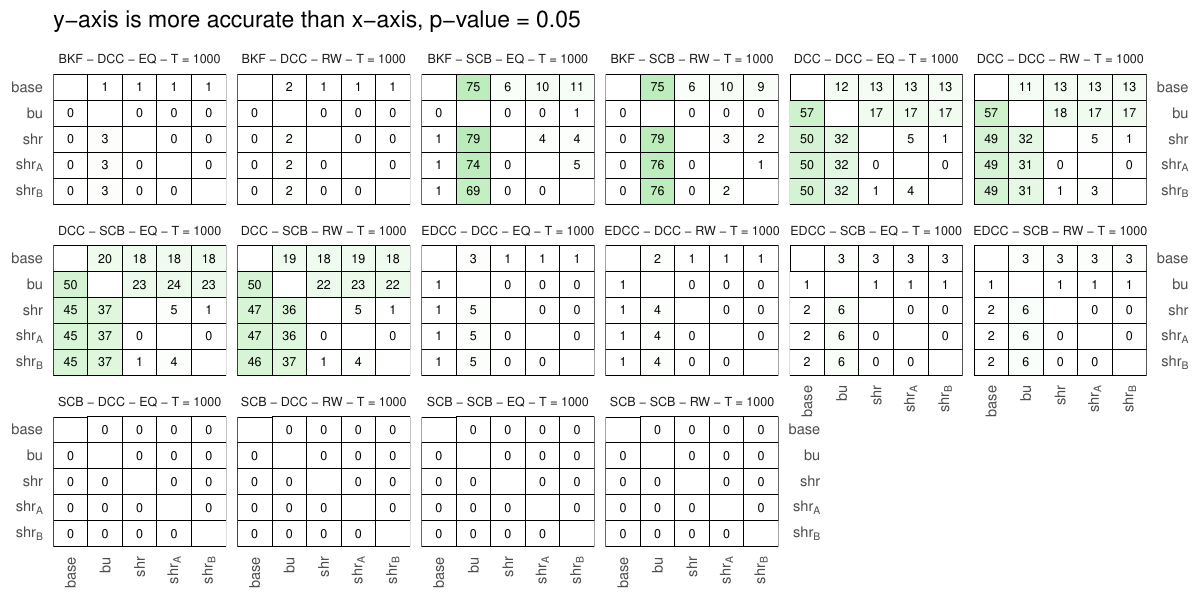}
	\caption{Qualitative evaluation using the Diebold-Mariano across different simulation settings using a square loss function and noisy level $\delta = 0.5$. Each setting is identified by a label composed of three elements: (i) the data-generating process (DGP), (ii) the MGARCH model used for estimation, and (iii) the portfolio weighting scheme (see \autoref{tab:notationModel} for details). The sample size ($T$) is also reported. The rows and columns indicate the variance forecasting method employed: the univariate GARCH on portfolio returns (base), the bottom-up approach using MGARCH models (bu), and the three forecast reconciliation strategies discussed in the paper ($shr$, $shr_{A}$, $shr_{B}$). Each cell reports the number of times (in \%) the forecasting model in the row statistically outperforms (p-values $< 0.05$ and using Bonferroni correction) the model in the column.}
\end{figure}
\begin{figure}[H]
	\centering
	\includegraphics[width=0.9\linewidth]{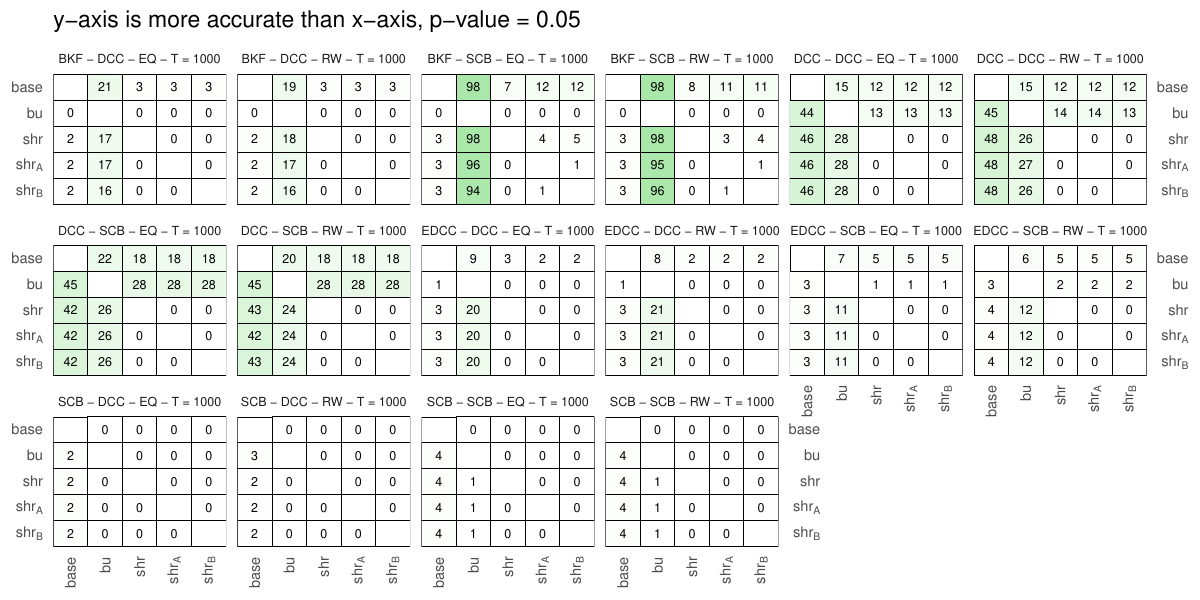}
	\caption{Qualitative evaluation using the Diebold-Mariano across different simulation settings using a square loss function and noisy level $\delta = 0.25$. Each setting is identified by a label composed of three elements: (i) the data-generating process (DGP), (ii) the MGARCH model used for estimation, and (iii) the portfolio weighting scheme (see \autoref{tab:notationModel} for details). The sample size ($T$) is also reported. The rows and columns indicate the variance forecasting method employed: the univariate GARCH on portfolio returns (base), the bottom-up approach using MGARCH models (bu), and the three forecast reconciliation strategies discussed in the paper ($shr$, $shr_{A}$, $shr_{B}$). Each cell reports the number of times (in \%) the forecasting model in the row statistically outperforms (p-values $< 0.05$ and using Bonferroni correction) the model in the column.}
\end{figure}

\begin{figure}[H]
	\centering
	\includegraphics[width=0.9\linewidth]{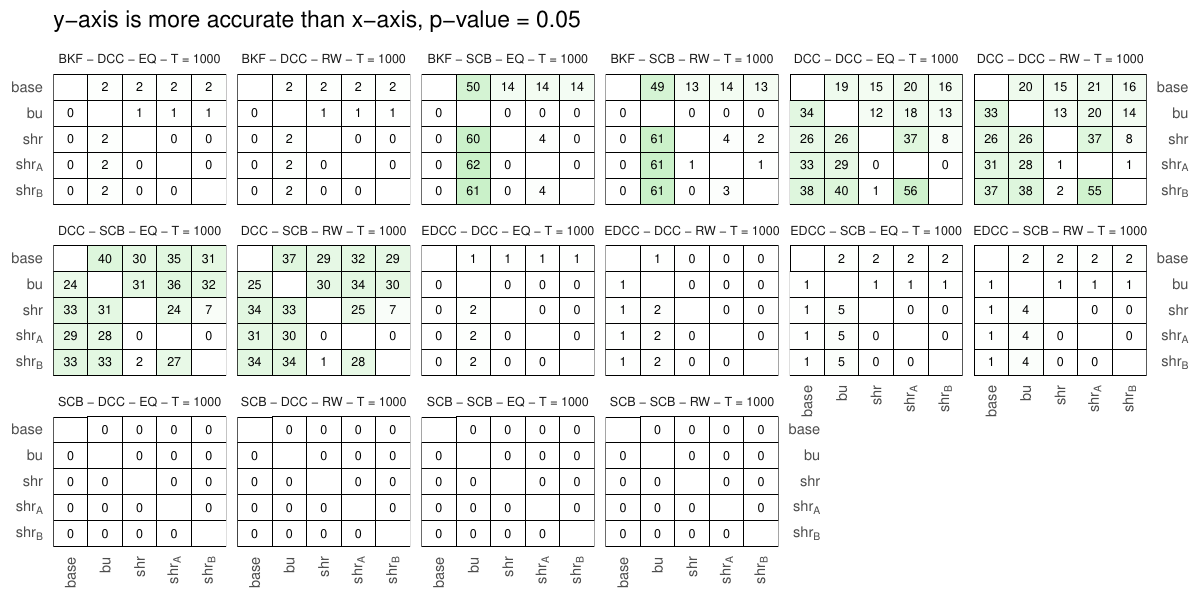}
	\caption{Qualitative evaluation using the Diebold-Mariano across different simulation settings using a QLIKE loss function and noisy level $\delta = 1$. Each setting is identified by a label composed of three elements: (i) the data-generating process (DGP), (ii) the MGARCH model used for estimation, and (iii) the portfolio weighting scheme (see \autoref{tab:notationModel} for details). The sample size ($T$) is also reported. The rows and columns indicate the variance forecasting method employed: the univariate GARCH on portfolio returns (base), the bottom-up approach using MGARCH models (bu), and the three forecast reconciliation strategies discussed in the paper ($shr$, $shr_{A}$, $shr_{B}$). Each cell reports the number of times (in \%) the forecasting model in the row statistically outperforms (p-values $< 0.05$ and using Bonferroni correction) the model in the column.}
\end{figure}
\begin{figure}[H]
	\centering
	\includegraphics[width=0.9\linewidth]{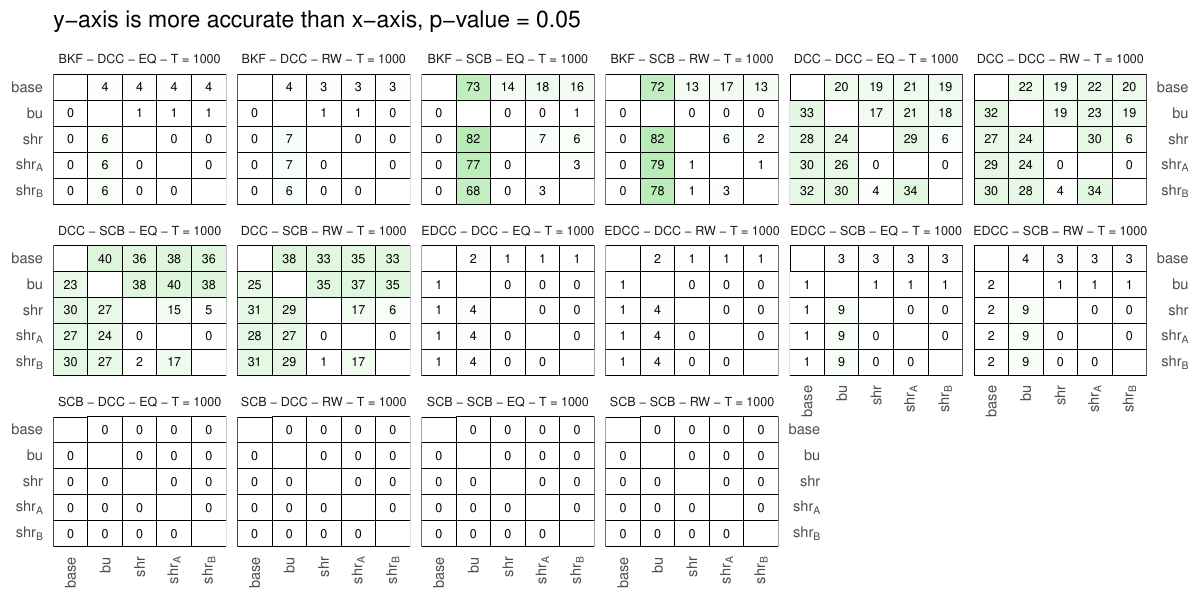}
	\caption{Qualitative evaluation using the Diebold-Mariano across different simulation settings using a QLIKE loss function and noisy level $\delta = 0.75$. Each setting is identified by a label composed of three elements: (i) the data-generating process (DGP), (ii) the MGARCH model used for estimation, and (iii) the portfolio weighting scheme (see \autoref{tab:notationModel} for details). The sample size ($T$) is also reported. The rows and columns indicate the variance forecasting method employed: the univariate GARCH on portfolio returns (base), the bottom-up approach using MGARCH models (bu), and the three forecast reconciliation strategies discussed in the paper ($shr$, $shr_{A}$, $shr_{B}$). Each cell reports the number of times (in \%) the forecasting model in the row statistically outperforms (p-values $< 0.05$ and using Bonferroni correction) the model in the column.}
\end{figure}
\begin{figure}[H]
	\centering
	\includegraphics[width=0.9\linewidth]{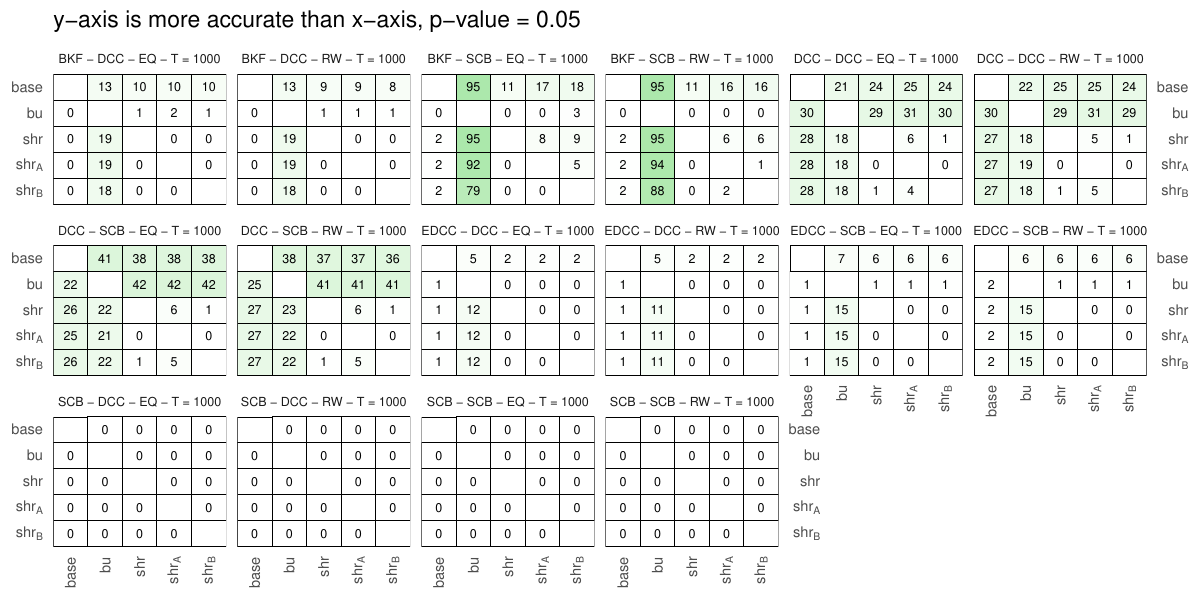}
	\caption{Qualitative evaluation using the Diebold-Mariano across different simulation settings using a QLIKE loss function and noisy level $\delta = 0.5$. Each setting is identified by a label composed of three elements: (i) the data-generating process (DGP), (ii) the MGARCH model used for estimation, and (iii) the portfolio weighting scheme (see \autoref{tab:notationModel} for details). The sample size ($T$) is also reported. The rows and columns indicate the variance forecasting method employed: the univariate GARCH on portfolio returns (base), the bottom-up approach using MGARCH models (bu), and the three forecast reconciliation strategies discussed in the paper ($shr$, $shr_{A}$, $shr_{B}$). Each cell reports the number of times (in \%) the forecasting model in the row statistically outperforms (p-values $< 0.05$ and using Bonferroni correction) the model in the column.}
\end{figure}
\begin{figure}[H]
	\centering
	\includegraphics[width=0.9\linewidth]{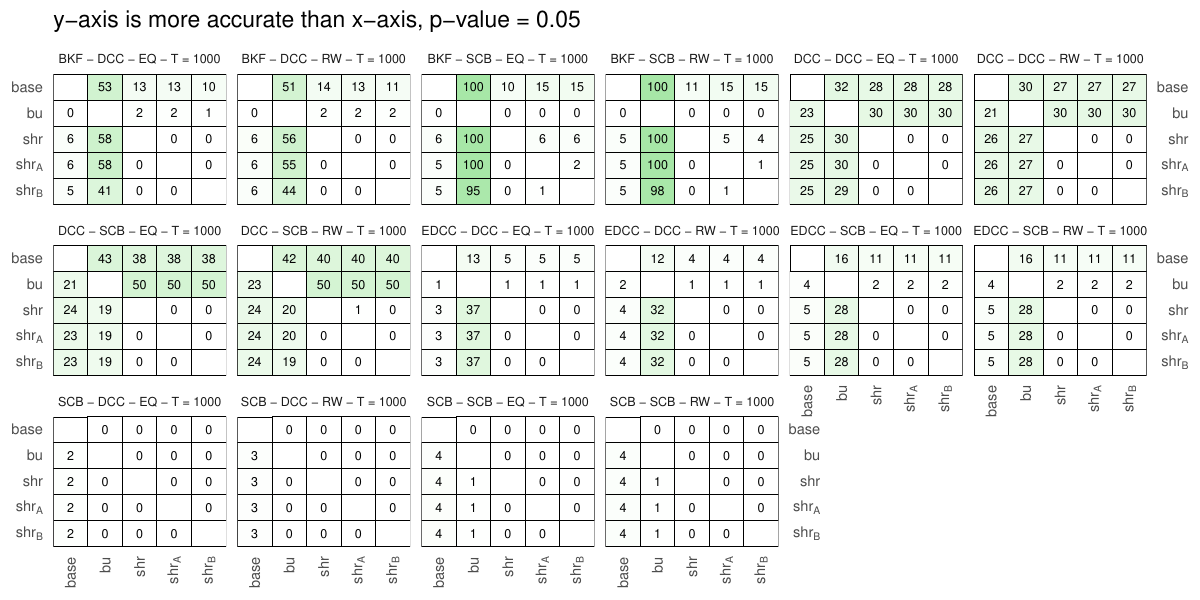}
	\caption{Qualitative evaluation using the Diebold-Mariano across different simulation settings using a QLIKE loss function and noisy level $\delta = 0.25$. Each setting is identified by a label composed of three elements: (i) the data-generating process (DGP), (ii) the MGARCH model used for estimation, and (iii) the portfolio weighting scheme (see \autoref{tab:notationModel} for details). The sample size ($T$) is also reported. The rows and columns indicate the variance forecasting method employed: the univariate GARCH on portfolio returns (base), the bottom-up approach using MGARCH models (bu), and the three forecast reconciliation strategies discussed in the paper ($shr$, $shr_{A}$, $shr_{B}$). Each cell reports the number of times (in \%) the forecasting model in the row statistically outperforms (p-values $< 0.05$ and using Bonferroni correction) the model in the column.}
\end{figure}

\subsubsection{Visual results}

\begin{figure}[H]
	\centering
	\includegraphics[width = \linewidth]{Sim/N24/proxy/plot24_th_MSE_EQ.pdf}
		\vspace*{-2em}
		\caption{Average relative MSE where the reference forecast is the univariate GARCH fitted on simulated portfolio returns with equally weighted portfolios (the $1/N$ case) for 24 assets, and the DGP is reported in each column (BEKK, DCC, EDCC and SBEKK). The fitted MGARCH models are: the DCC-GARCH (first row, DCC) and the Scalar BEKK (second row, SBEKK). All values are averages across the 500 experiments.}
			\vspace{1em}
		\includegraphics[width = \linewidth]{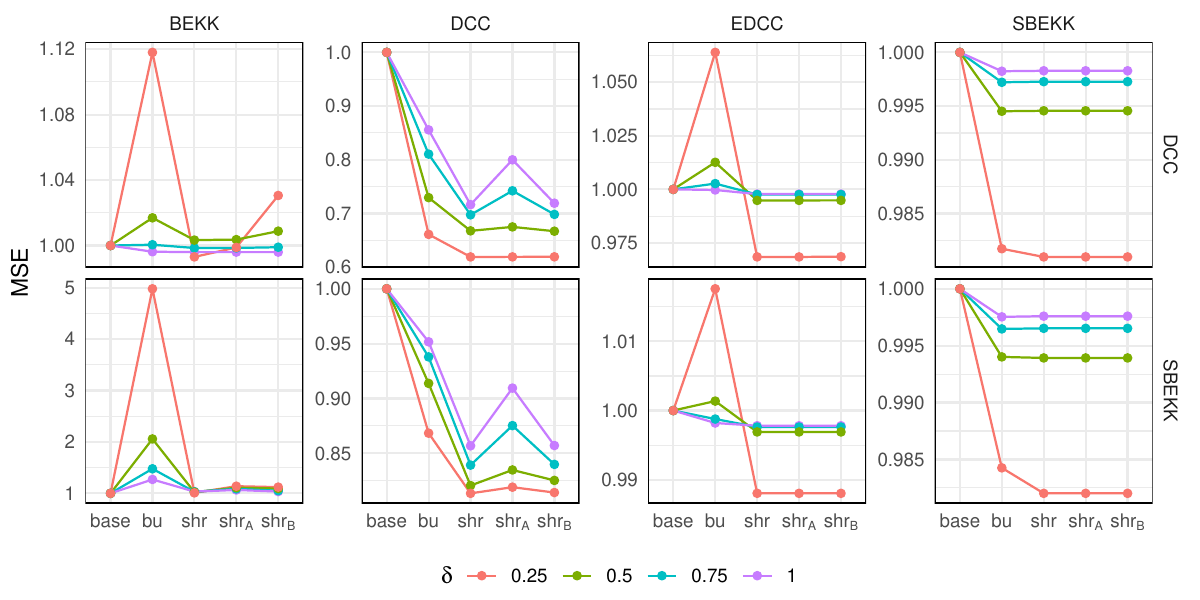}
	\vspace*{-2em}
	\caption{Average relative MSE where the reference forecast is the univariate GARCH fitted on simulated portfolio returns with random weighted portfolios for 24 assets, and the DGP is reported in each column (BEKK, DCC, EDCC and SBEKK). The fitted MGARCH models are: the DCC-GARCH (first row, DCC) and the Scalar BEKK (second row, SBEKK). All values are averages across the 500 experiments.}
\end{figure}

\begin{figure}[H]
	\centering
	\includegraphics[width = \linewidth]{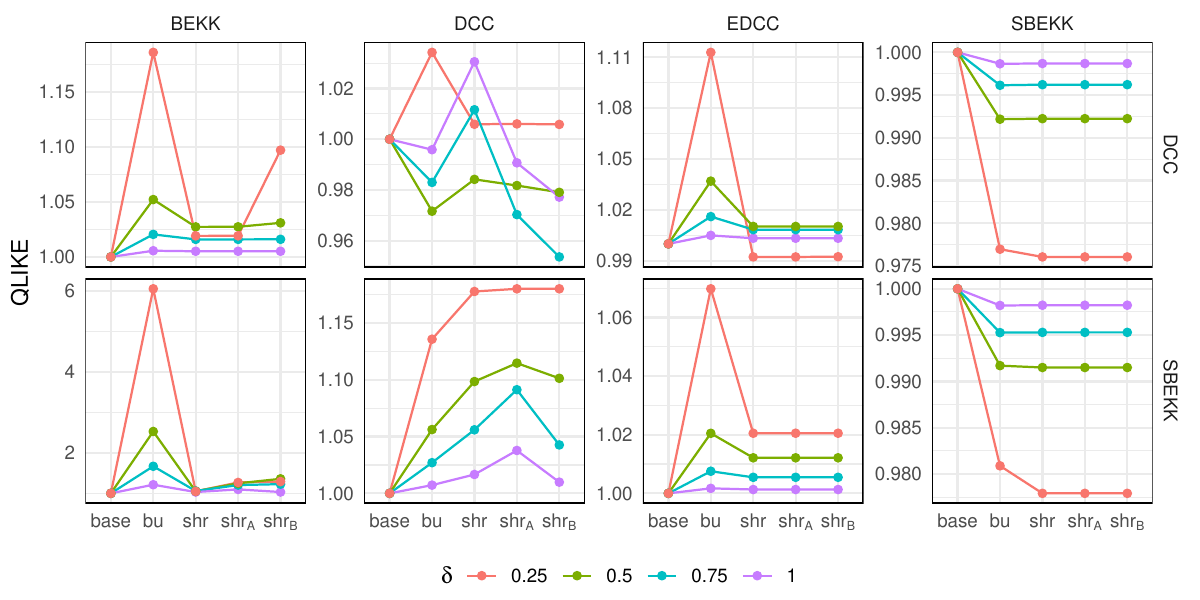}
		\vspace*{-2em}
		\caption{Average relative QLIKE where the reference forecast is the univariate GARCH fitted on simulated portfolio returns with equally weighted portfolios (the $1/N$ case) for 24 assets, and the DGP is reported in each column (BEKK, DCC, EDCC and SBEKK). The fitted MGARCH models are: the DCC-GARCH (first row, DCC) and the Scalar BEKK (second row, SBEKK). All values are averages across the 500 experiments.}
			\vspace{1em}
		\includegraphics[width = \linewidth]{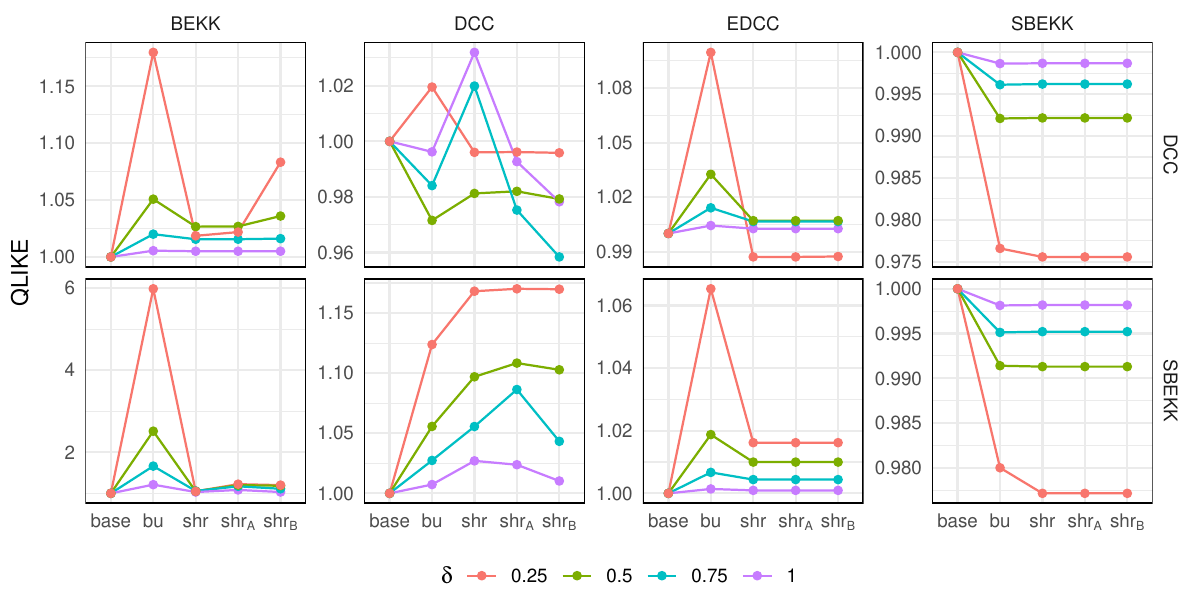}
	\vspace*{-2em}
	\caption{Average relative QLIKE where the reference forecast is the univariate GARCH fitted on simulated portfolio returns with random weighted portfolios for 24 assets, and the DGP is reported in each column (BEKK, DCC, EDCC and SBEKK). The fitted MGARCH models are: the DCC-GARCH (first row, DCC) and the Scalar BEKK (second row, SBEKK). All values are averages across the 500 experiments.}
\end{figure}

\begin{figure}[H]
	\centering
		\includegraphics[width = \linewidth]{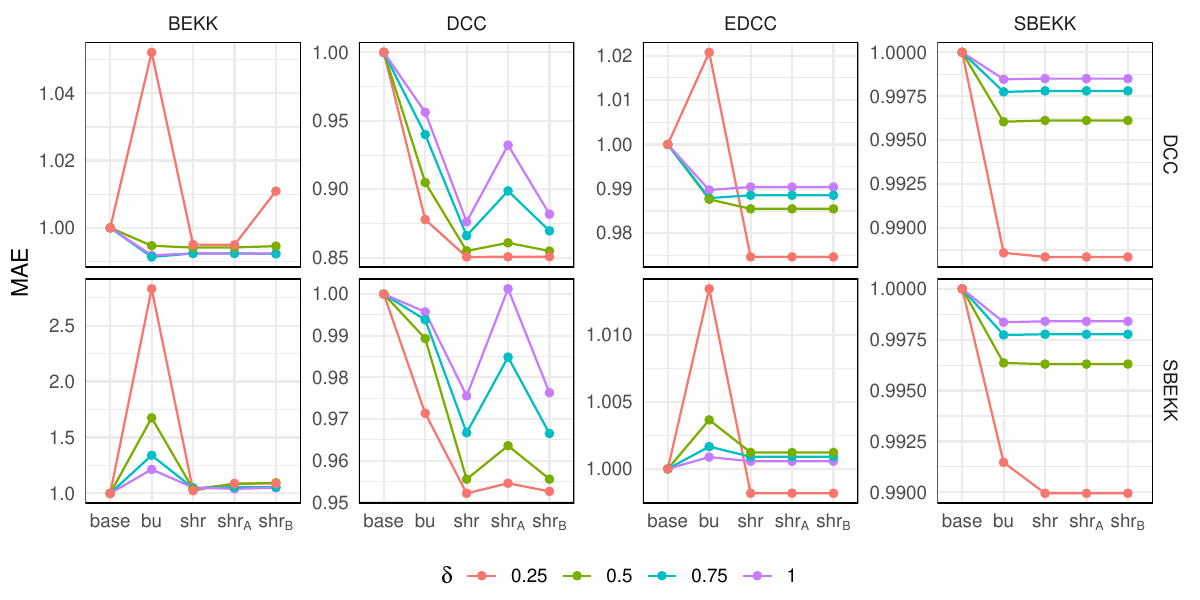}
			\vspace*{-2em}
			\caption{Average relative MAE where the reference forecast is the univariate GARCH fitted on simulated portfolio returns with equally weighted portfolios (the $1/N$ case) for 24 assets, and the DGP is reported in each column (BEKK, DCC, EDCC and SBEKK). The fitted MGARCH models are: the DCC-GARCH (first row, DCC) and the Scalar BEKK (second row, SBEKK). All values are averages across the 500 experiments.}
		\includegraphics[width = \linewidth]{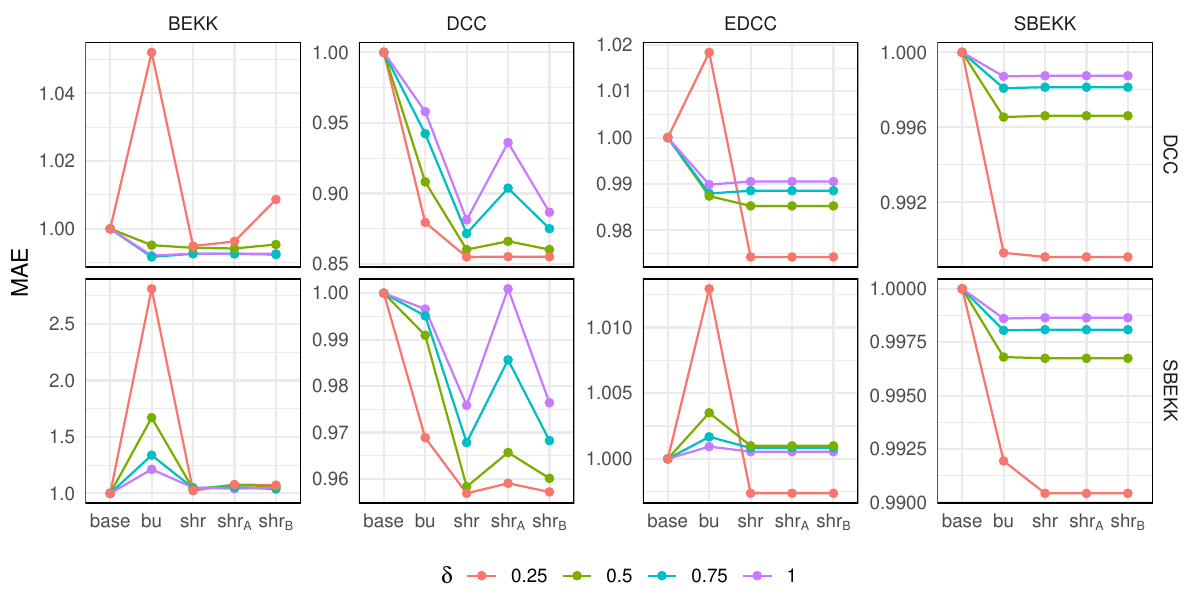}
	\vspace*{-2em}
	\caption{Average relative MAE where the reference forecast is the univariate GARCH fitted on simulated portfolio returns with random weighted portfolios for 24 assets, and the DGP is reported in each column (BEKK, DCC, EDCC and SBEKK). The fitted MGARCH models are: the DCC-GARCH (first row, DCC) and the Scalar BEKK (second row, SBEKK). All values are averages across the 500 experiments.}
\end{figure}

\newpage
\section{Real data experiment: using a proxy}

\begin{sidewaystable}
	\centering
	\resizebox{\linewidth}{!}{
		
\begin{tabular}[t]{>{}l|cccc>{}c|cccc>{}c|ccccc}
\toprule
\multicolumn{1}{c}{\textbf{ }} & \multicolumn{5}{c}{\textbf{Day}} & \multicolumn{5}{c}{\textbf{Week}} & \multicolumn{5}{c}{\textbf{Month}} \\
\cmidrule(l{3pt}r{3pt}){2-6} \cmidrule(l{3pt}r{3pt}){7-11} \cmidrule(l{3pt}r{3pt}){12-16}
 & $base$ & $bu$ & $shr$ & $shr_{A}$ & $shr_{B}$ & $base$ & $bu$ & $shr$ & $shr_{A}$ & $shr_{B}$ & $base$ & $bu$ & $shr$ & $shr_{A}$ & $shr_{B}$\\
\midrule
\addlinespace[0.3em]
\multicolumn{16}{c}{\textbf{Base forecasts: DCC-GARCH}}\\
\hspace{1em}\textbf{MSE} & \textbf{2.056} & 2.535 & \em{2.293} & \em{2.293} & \em{2.293} & \em{2.376} & 2.376 & \textbf{2.350} & \textbf{2.350} & \textbf{2.350} & 2.858 & 2.835 & 2.805 & \textbf{2.805} & \em{2.805}\\
\hspace{1em}AvgRelMSE$_{base}$ & \textbf{1.000} & 1.233 & \em{1.115} & \em{1.115} & \em{1.115} & \em{1.000} & 1.000 & \textbf{0.989} & \textbf{0.989} & \textbf{0.989} & 1.000 & 0.992 & 0.981 & \textbf{0.981} & \em{0.981}\\
\hspace{1em}AvgRelMSE$_{bu}$ & \textbf{0.811} & 1.000 & \em{0.905} & \em{0.905} & \em{0.905} & \em{1.000} & 1.000 & \textbf{0.989} & \textbf{0.989} & \textbf{0.989} & 1.008 & 1.000 & 0.990 & \textbf{0.989} & \em{0.990}\\
\hspace{1em}$p$-value $dm_{base}$ &  & 0.890 & 0.819 & 0.819 & 0.819 &  & 0.501 & 0.412 & 0.412 & 0.412 &  & 0.371 & 0.162 & 0.162 & 0.162\\
\hspace{1em}$p$-value $dm_{bu}$ & 0.110 &  & 0.042 & 0.042 & 0.042 & 0.499 &  & 0.267 & 0.267 & 0.267 & 0.629 &  & 0.057 & 0.055 & 0.056\\
\hspace{1em}$p$-value MCS & 1.000 & 0.298 & 0.488 & 0.488 & 0.488 & 0.946 & 0.946 & 1.000 & 1.000 & 1.000 & 0.586 & 0.586 & 0.586 & 1.000 & 0.586\\
\midrule
\hspace{1em}\textbf{MAE} & \em{0.774} & 0.788 & \textbf{0.772} & \textbf{0.772} & \textbf{0.772} & 0.753 & \em{0.745} & \textbf{0.745} & \textbf{0.745} & \textbf{0.745} & 0.736 & 0.723 & 0.722 & \textbf{0.722} & \em{0.722}\\
\hspace{1em}AvgRelMAE$_{base}$ & \em{1.000} & 1.019 & \textbf{0.999} & \textbf{0.999} & \textbf{0.999} & 1.000 & \em{0.990} & \textbf{0.989} & \textbf{0.989} & \textbf{0.989} & 1.000 & 0.982 & 0.981 & \textbf{0.981} & \em{0.981}\\
\hspace{1em}AvgRelMAE$_{bu}$ & \em{0.981} & 1.000 & \textbf{0.980} & \textbf{0.980} & \textbf{0.980} & 1.010 & \em{1.000} & \textbf{0.999} & \textbf{0.999} & \textbf{0.999} & 1.018 & 1.000 & 0.999 & \textbf{0.999} & \em{0.999}\\
\hspace{1em}$p$-value $dm_{base}$ &  & 0.661 & 0.484 & 0.484 & 0.484 &  & 0.327 & 0.259 & 0.259 & 0.259 &  & 0.032 & 0.006 & 0.005 & 0.006\\
\hspace{1em}$p$-value $dm_{bu}$ & 0.339 &  & 0.068 & 0.068 & 0.068 & 0.673 &  & 0.453 & 0.453 & 0.453 & 0.968 &  & 0.382 & 0.358 & 0.377\\
\hspace{1em}$p$-value MCS & 0.979 & 0.701 & 1.000 & 1.000 & 1.000 & 0.740 & 0.935 & 1.000 & 1.000 & 1.000 & 0.166 & 0.886 & 0.886 & 1.000 & 0.886\\
\midrule
\hspace{1em}\textbf{QLIKE} & 1.659 & \em{1.611} & \textbf{1.604} & \textbf{1.604} & \textbf{1.604} & \textbf{1.523} & 1.584 & \em{1.559} & \em{1.559} & \em{1.559} & \textbf{1.530} & 1.581 & 1.556 & 1.556 & \em{1.556}\\
\hspace{1em}AvgRelQLIKE$_{base}$ & 1.000 & \em{0.971} & \textbf{0.967} & \textbf{0.967} & \textbf{0.967} & \textbf{1.000} & 1.040 & \em{1.024} & \em{1.024} & \em{1.024} & \textbf{1.000} & 1.033 & 1.017 & 1.017 & \em{1.017}\\
\hspace{1em}AvgRelQLIKE$_{bu}$ & 1.030 & \em{1.000} & \textbf{0.996} & \textbf{0.996} & \textbf{0.996} & \textbf{0.962} & 1.000 & \em{0.985} & \em{0.985} & \em{0.985} & \textbf{0.968} & 1.000 & 0.984 & 0.984 & \em{0.984}\\
\hspace{1em}$p$-value $dm_{base}$ &  & 0.204 & 0.113 & 0.113 & 0.113 &  & 0.994 & 0.969 & 0.969 & 0.969 &  & 1.000 & 0.990 & 0.990 & 0.990\\
\hspace{1em}$p$-value $dm_{bu}$ & 0.796 &  & 0.330 & 0.330 & 0.330 & 0.006 &  & {\footnotesize $<0.001$} & {\footnotesize $<0.001$} & {\footnotesize $<0.001$} & {\footnotesize $<0.001$} &  & {\footnotesize $<0.001$} & {\footnotesize $<0.001$} & {\footnotesize $<0.001$}\\
\hspace{1em}$p$-value MCS & 0.308 & 0.545 & 1.000 & 1.000 & 1.000 & 1.000 & 0.026 & 0.115 & 0.115 & 0.115 & 1.000 & {\footnotesize $<0.001$} & 0.024 & 0.024 & 0.024\\
\bottomrule
\end{tabular}

	}
	\caption{Forecast evaluation of DCC–GARCH base forecasts and reconciliation schemes across forecast horizons (Day, Week, Month). The table reports the Mean Squared Error (MSE), Mean Absolute Error (MAE), and QLIKE loss functions, together with average relative performance measures (AvgRelMSE, AvgRelMAE, AvgRelQLIKE) computed with respect to the base and bottom-up (\textit{bu}) benchmarks. The Diebold–Mariano ($p$-value $dm$) and Model Confidence Set ($p$-value MCS$)$ statistics are reported to examine the statistical significance of forecast differentials. The best result within each row is shown in bold, and the second-best in italics.}
	\label{tab:appDCC}
\end{sidewaystable}

\begin{sidewaystable}
	\centering
	\resizebox{\linewidth}{!}{
		
\begin{tabular}[t]{>{}l|cccc>{}c|cccc>{}c|ccccc}
\toprule
\multicolumn{1}{c}{\textbf{ }} & \multicolumn{5}{c}{\textbf{Day}} & \multicolumn{5}{c}{\textbf{Week}} & \multicolumn{5}{c}{\textbf{Month}} \\
\cmidrule(l{3pt}r{3pt}){2-6} \cmidrule(l{3pt}r{3pt}){7-11} \cmidrule(l{3pt}r{3pt}){12-16}
 & $base$ & $bu$ & $shr$ & $shr_{A}$ & $shr_{B}$ & $base$ & $bu$ & $shr$ & $shr_{A}$ & $shr_{B}$ & $base$ & $bu$ & $shr$ & $shr_{A}$ & $shr_{B}$\\
\midrule
\addlinespace[0.3em]
\multicolumn{16}{c}{\textbf{Base forecasts: SBEKK-GARCH}}\\
\hspace{1em}\textbf{MSE} & \textbf{2.056} & 2.556 & \em{2.113} & \em{2.113} & \em{2.113} & \textbf{2.376} & 2.679 & \em{2.438} & \em{2.438} & \em{2.438} & \textbf{2.858} & 3.102 & \em{2.870} & \em{2.870} & \em{2.870}\\
\hspace{1em}AvgRelMSE$_{base}$ & \textbf{1.000} & 1.243 & \em{1.028} & \em{1.028} & \em{1.028} & \textbf{1.000} & 1.128 & \em{1.026} & \em{1.026} & \em{1.026} & \textbf{1.000} & 1.086 & \em{1.004} & \em{1.004} & \em{1.004}\\
\hspace{1em}AvgRelMSE$_{bu}$ & \textbf{0.804} & 1.000 & \em{0.827} & \em{0.827} & \em{0.827} & \textbf{0.887} & 1.000 & \em{0.910} & \em{0.910} & \em{0.910} & \textbf{0.921} & 1.000 & \em{0.925} & \em{0.925} & \em{0.925}\\
\hspace{1em}$p$-value $dm_{base}$ &  & 0.853 & 0.597 & 0.597 & 0.597 &  & 0.931 & 0.706 & 0.706 & 0.706 &  & 0.983 & 0.573 & 0.573 & 0.573\\
\hspace{1em}$p$-value $dm_{bu}$ & 0.147 &  & 0.053 & 0.053 & 0.053 & 0.069 &  & 0.008 & 0.008 & 0.008 & 0.017 &  & {\footnotesize $<0.001$} & {\footnotesize $<0.001$} & {\footnotesize $<0.001$}\\
\hspace{1em}$p$-value MCS & 1.000 & 0.233 & 0.692 & 0.692 & 0.692 & 1.000 & 0.214 & 0.560 & 0.560 & 0.560 & 1.000 & 0.049 & 0.836 & 0.836 & 0.836\\
\midrule
\hspace{1em}\textbf{MAE} & \textbf{0.774} & 0.880 & \em{0.795} & \em{0.795} & \em{0.795} & \textbf{0.753} & 0.878 & \em{0.795} & \em{0.795} & \em{0.795} & \textbf{0.736} & 0.848 & \em{0.768} & \em{0.768} & \em{0.768}\\
\hspace{1em}AvgRelMAE$_{base}$ & \textbf{1.000} & 1.137 & \em{1.028} & \em{1.028} & \em{1.028} & \textbf{1.000} & 1.166 & \em{1.056} & \em{1.056} & \em{1.056} & \textbf{1.000} & 1.152 & \em{1.044} & \em{1.044} & \em{1.044}\\
\hspace{1em}AvgRelMAE$_{bu}$ & \textbf{0.879} & 1.000 & \em{0.903} & \em{0.903} & \em{0.903} & \textbf{0.857} & 1.000 & \em{0.906} & \em{0.906} & \em{0.906} & \textbf{0.868} & 1.000 & \em{0.906} & \em{0.906} & \em{0.906}\\
\hspace{1em}$p$-value $dm_{base}$ &  & 0.982 & 0.793 & 0.793 & 0.793 &  & 1.000 & 1.000 & 1.000 & 1.000 &  & 1.000 & 1.000 & 1.000 & 1.000\\
\hspace{1em}$p$-value $dm_{bu}$ & 0.018 &  & {\footnotesize $<0.001$} & {\footnotesize $<0.001$} & {\footnotesize $<0.001$} & {\footnotesize $<0.001$} &  & {\footnotesize $<0.001$} & {\footnotesize $<0.001$} & {\footnotesize $<0.001$} & {\footnotesize $<0.001$} &  & {\footnotesize $<0.001$} & {\footnotesize $<0.001$} & {\footnotesize $<0.001$}\\
\hspace{1em}$p$-value MCS & 1.000 & 0.103 & 0.537 & 0.537 & 0.537 & 1.000 & {\footnotesize $<0.001$} & {\footnotesize $<0.001$} & {\footnotesize $<0.001$} & {\footnotesize $<0.001$} & 1.000 & {\footnotesize $<0.001$} & {\footnotesize $<0.001$} & {\footnotesize $<0.001$} & {\footnotesize $<0.001$}\\
\midrule
\hspace{1em}\textbf{QLIKE} & \em{1.659} & 1.694 & \textbf{1.609} & \textbf{1.609} & \textbf{1.609} & \textbf{1.523} & 1.729 & \em{1.593} & \em{1.593} & \em{1.593} & \textbf{1.530} & 1.711 & \em{1.583} & \em{1.583} & \em{1.583}\\
\hspace{1em}AvgRelQLIKE$_{base}$ & \em{1.000} & 1.021 & \textbf{0.970} & \textbf{0.970} & \textbf{0.970} & \textbf{1.000} & 1.136 & \em{1.046} & \em{1.046} & \em{1.046} & \textbf{1.000} & 1.118 & \em{1.035} & \em{1.035} & \em{1.035}\\
\hspace{1em}AvgRelQLIKE$_{bu}$ & \em{0.979} & 1.000 & \textbf{0.950} & \textbf{0.950} & \textbf{0.950} & \textbf{0.881} & 1.000 & \em{0.921} & \em{0.921} & \em{0.921} & \textbf{0.895} & 1.000 & \em{0.925} & \em{0.925} & \em{0.925}\\
\hspace{1em}$p$-value $dm_{base}$ &  & 0.635 & 0.197 & 0.197 & 0.197 &  & 1.000 & 0.998 & 0.998 & 0.998 &  & 1.000 & 1.000 & 1.000 & 1.000\\
\hspace{1em}$p$-value $dm_{bu}$ & 0.365 &  & 0.059 & 0.059 & 0.059 & {\footnotesize $<0.001$} &  & {\footnotesize $<0.001$} & {\footnotesize $<0.001$} & {\footnotesize $<0.001$} & {\footnotesize $<0.001$} &  & {\footnotesize $<0.001$} & {\footnotesize $<0.001$} & {\footnotesize $<0.001$}\\
\hspace{1em}$p$-value MCS & 0.366 & 0.328 & 1.000 & 1.000 & 1.000 & 1.000 & 0.001 & 0.013 & 0.013 & 0.013 & 1.000 & {\footnotesize $<0.001$} & 0.001 & 0.001 & 0.001\\
\bottomrule
\end{tabular}

	}
	\caption{Forecast evaluation of SBEKK–GARCH base forecasts and reconciliation schemes across forecast horizons (Day, Week, Month). The table reports the Mean Squared Error (MSE), Mean Absolute Error (MAE), and QLIKE loss functions, together with average relative performance measures (AvgRelMSE, AvgRelMAE, AvgRelQLIKE) computed with respect to the base and bottom-up (\textit{bu}) benchmarks. The Diebold–Mariano ($p$-value $dm$) and Model Confidence Set ($p$-value MCS$)$ statistics are reported to examine the statistical significance of forecast differentials. The best result within each row is shown in bold, and the second-best in italics.}
	\label{tab:appSBE}
\end{sidewaystable}

\begin{sidewaystable}
	\centering
	\resizebox{\linewidth}{!}{
		
\begin{tabular}[t]{>{}l|cccc>{}c|cccc>{}c|ccccc}
\toprule
\multicolumn{1}{c}{\textbf{ }} & \multicolumn{5}{c}{\textbf{Day}} & \multicolumn{5}{c}{\textbf{Week}} & \multicolumn{5}{c}{\textbf{Month}} \\
\cmidrule(l{3pt}r{3pt}){2-6} \cmidrule(l{3pt}r{3pt}){7-11} \cmidrule(l{3pt}r{3pt}){12-16}
 & $base$ & $bu$ & $shr$ & $shr_{A}$ & $shr_{B}$ & $base$ & $bu$ & $shr$ & $shr_{A}$ & $shr_{B}$ & $base$ & $bu$ & $shr$ & $shr_{A}$ & $shr_{B}$\\
\midrule
\addlinespace[0.3em]
\multicolumn{16}{c}{\textbf{Base forecasts: EDCC-GARCH}}\\
\hspace{1em}\textbf{MSE} & \textbf{2.056} & 2.434 & \em{2.351} & 2.362 & 2.354 & 2.376 & 2.318 & \textbf{2.314} & 2.316 & \em{2.314} & 2.858 & 2.749 & \textbf{2.747} & 2.748 & \em{2.747}\\
\hspace{1em}AvgRelMSE$_{base}$ & \textbf{1.000} & 1.184 & \em{1.144} & 1.149 & 1.145 & 1.000 & 0.976 & \textbf{0.974} & 0.975 & \em{0.974} & 1.000 & 0.962 & \textbf{0.961} & 0.961 & \em{0.961}\\
\hspace{1em}AvgRelMSE$_{bu}$ & \textbf{0.844} & 1.000 & \em{0.966} & 0.970 & 0.967 & 1.025 & 1.000 & \textbf{0.998} & 0.999 & \em{0.998} & 1.040 & 1.000 & \textbf{0.999} & 1.000 & \em{0.999}\\
\hspace{1em}$p$-value $dm_{base}$ &  & 0.836 & 0.807 & 0.815 & 0.809 &  & 0.345 & 0.320 & 0.326 & 0.322 &  & 0.074 & 0.056 & 0.057 & 0.056\\
\hspace{1em}$p$-value $dm_{bu}$ & 0.164 &  & 0.051 & 0.078 & 0.056 & 0.655 &  & 0.364 & 0.424 & 0.378 & 0.926 &  & 0.403 & 0.433 & 0.410\\
\hspace{1em}$p$-value MCS & 1.000 & 0.454 & 0.554 & 0.526 & 0.551 & 0.697 & 0.697 & 1.000 & 0.697 & 0.697 & 0.140 & 0.837 & 1.000 & 0.837 & 0.837\\
\midrule
\hspace{1em}\textbf{MAE} & 0.774 & 0.768 & \textbf{0.765} & 0.768 & \em{0.766} & 0.753 & \textbf{0.727} & \em{0.730} & 0.730 & 0.730 & 0.736 & \textbf{0.711} & \em{0.713} & 0.713 & 0.713\\
\hspace{1em}AvgRelMAE$_{base}$ & 1.000 & 0.993 & \textbf{0.989} & 0.993 & \em{0.990} & 1.000 & \textbf{0.966} & \em{0.969} & 0.970 & 0.969 & 1.000 & \textbf{0.966} & \em{0.968} & 0.968 & 0.968\\
\hspace{1em}AvgRelMAE$_{bu}$ & 1.007 & 1.000 & \textbf{0.996} & 1.000 & \em{0.997} & 1.035 & \textbf{1.000} & \em{1.003} & 1.004 & 1.003 & 1.035 & \textbf{1.000} & \em{1.002} & 1.002 & 1.002\\
\hspace{1em}$p$-value $dm_{base}$ &  & 0.442 & 0.397 & 0.435 & 0.407 &  & 0.047 & 0.048 & 0.054 & 0.050 &  & {\footnotesize $<0.001$} & {\footnotesize $<0.001$} & {\footnotesize $<0.001$} & {\footnotesize $<0.001$}\\
\hspace{1em}$p$-value $dm_{bu}$ & 0.558 &  & 0.178 & 0.481 & 0.239 & 0.953 &  & 0.936 & 0.970 & 0.951 & 1.000 &  & 0.969 & 0.981 & 0.974\\
\hspace{1em}$p$-value MCS & 0.937 & 0.937 & 1.000 & 0.937 & 0.937 & 0.168 & 1.000 & 0.241 & 0.168 & 0.182 & 0.018 & 1.000 & 0.277 & 0.186 & 0.245\\
\midrule
\hspace{1em}\textbf{QLIKE} & 1.659 & 1.608 & \textbf{1.607} & 1.612 & \em{1.608} & 1.523 & 1.523 & \textbf{1.521} & 1.522 & \em{1.521} & \textbf{1.530} & 1.537 & \em{1.533} & 1.534 & 1.534\\
\hspace{1em}AvgRelQLIKE$_{base}$ & 1.000 & 0.969 & \textbf{0.969} & 0.972 & \em{0.969} & 1.000 & 1.000 & \textbf{0.999} & 0.999 & \em{0.999} & \textbf{1.000} & 1.004 & \em{1.002} & 1.002 & 1.002\\
\hspace{1em}AvgRelQLIKE$_{bu}$ & 1.032 & 1.000 & \textbf{0.999} & 1.003 & \em{1.000} & 1.000 & 1.000 & \textbf{0.998} & 0.999 & \em{0.998} & \textbf{0.996} & 1.000 & \em{0.998} & 0.998 & 0.998\\
\hspace{1em}$p$-value $dm_{base}$ &  & 0.141 & 0.119 & 0.150 & 0.123 &  & 0.506 & 0.446 & 0.473 & 0.449 &  & 0.711 & 0.616 & 0.626 & 0.618\\
\hspace{1em}$p$-value $dm_{bu}$ & 0.859 &  & 0.391 & 0.751 & 0.459 & 0.494 &  & 0.067 & 0.235 & 0.078 & 0.289 &  & {\footnotesize $<0.001$} & 0.002 & {\footnotesize $<0.001$}\\
\hspace{1em}$p$-value MCS & 0.330 & 0.909 & 1.000 & 0.464 & 0.909 & 0.925 & 0.843 & 1.000 & 0.925 & 0.925 & 1.000 & 0.334 & 0.790 & 0.750 & 0.787\\
\bottomrule
\end{tabular}

	}
	\caption{Forecast evaluation of EDCC–GARCH base forecasts and reconciliation schemes across forecast horizons (Day, Week, Month). The table reports the Mean Squared Error (MSE), Mean Absolute Error (MAE), and QLIKE loss functions, together with average relative performance measures (AvgRelMSE, AvgRelMAE, AvgRelQLIKE) computed with respect to the base and bottom-up (\textit{bu}) benchmarks. The Diebold–Mariano ($p$-value $dm$) and Model Confidence Set ($p$-value MCS$)$ statistics are reported to examine the statistical significance of forecast differentials. The best result within each row is shown in bold, and the second-best in italics.}
	\label{tab:appEDC}
\end{sidewaystable}

\begin{table}
	\resizebox{\linewidth}{!}{
		
\begin{tabular}[t]{l>{}l|cc>{}c|cc>{}c|ccc}
\toprule
\multicolumn{2}{c}{\textbf{ }} & \multicolumn{3}{c}{\textbf{MSE}} & \multicolumn{3}{c}{\textbf{MAE}} & \multicolumn{3}{c}{\textbf{QLIKE}} \\
\cmidrule(l{3pt}r{3pt}){3-5} \cmidrule(l{3pt}r{3pt}){6-8} \cmidrule(l{3pt}r{3pt}){9-11}
\textbf{Base forecasts} & \textbf{Approach} & Day & Week & Month & Day & Week & Month & Day & Week & Month\\
\midrule
GARCH & $base$ & 1.000 & 0.817 & 0.376 & 0.937 & 0.259 & 0.093 & 0.523 & 0.925 & 1.000\\
\addlinespace
EDCC-GARCH & $bu$ & 0.569 & 0.817 & 0.837 & 0.937 & 1.000 & 1.000 & 0.993 & 0.843 & 0.334\\
EDCC-GARCH & $shr$ & 0.599 & 1.000 & 1.000 & 1.000 & 0.259 & 0.277 & 0.993 & 1.000 & 0.790\\
EDCC-GARCH & $shr_{B}$ & 0.569 & 0.817 & 0.837 & 0.937 & 0.259 & 0.245 & 0.993 & 0.925 & 0.787\\
EDCC-GARCH & $shr_{A}$ & 0.569 & 0.817 & 0.837 & 0.937 & 0.259 & 0.186 & 0.959 & 0.925 & 0.750\\
\addlinespace
DCC-GARCH & $bu$ & 0.232 & 0.761 & 0.376 & 0.349 & 0.259 & 0.095 & 0.959 & 0.010 & {\footnotesize $<0.001$}\\
DCC-GARCH & $shr$ & 0.569 & 0.817 & 0.376 & 0.808 & 0.046 & 0.095 & 1.000 & 0.060 & 0.004\\
DCC-GARCH & $shr_{B}$ & 0.569 & 0.817 & 0.376 & 0.808 & 0.049 & 0.095 & 1.000 & 0.067 & 0.004\\
DCC-GARCH & $shr_{A}$ & 0.569 & 0.817 & 0.376 & 0.808 & 0.051 & 0.095 & 1.000 & 0.078 & 0.003\\
\addlinespace
SBEKK-GARCH & $bu$ & 0.140 & 0.097 & 0.087 & 0.011 & {\footnotesize $<0.001$} & {\footnotesize $<0.001$} & 0.198 & 0.001 & {\footnotesize $<0.001$}\\
SBEKK-GARCH & $shr$ & 0.692 & 0.113 & 0.376 & 0.270 & {\footnotesize $<0.001$} & {\footnotesize $<0.001$} & 0.988 & 0.007 & {\footnotesize $<0.001$}\\
SBEKK-GARCH & $shr_{B}$ & 0.692 & 0.114 & 0.376 & 0.272 & {\footnotesize $<0.001$} & {\footnotesize $<0.001$} & 0.990 & 0.007 & {\footnotesize $<0.001$}\\
SBEKK-GARCH & $shr_{A}$ & 0.692 & 0.121 & 0.376 & 0.278 & {\footnotesize $<0.001$} & {\footnotesize $<0.001$} & 0.993 & 0.007 & {\footnotesize $<0.001$}\\
\bottomrule
\end{tabular}

	}
	\caption{Model Confidence Set ($p$-value MCS$)$ for all the forecasting approach across forecast horizons (Day, Week, Month) and loss functions (MSE, MAE, QLIKE).}
\end{table}